\documentclass[a4paper,11pt]{article}
\usepackage[hmargin=0.6in,vmargin=0.8in]{geometry}
\usepackage{amsthm,amsmath,amsfonts,amssymb,mathrsfs}
\usepackage{graphicx,xcolor,color}
\usepackage{float,caption,subfig,grffile,wrapfig}
\usepackage{array,multirow,tabularx,tabulary,booktabs,diagbox,makecell}
\usepackage{tikzsymbols,adjustbox}
\usepackage{times}
\usepackage[T1]{fontenc}
\usepackage[numbers]{natbib}
\usepackage{url}
\usepackage{todonotes,changes}
\usepackage[normalem]{ulem}
\usepackage{comment}
\usepackage{setspace}
\usepackage[symbol]{footmisc}

\ifdefined\pdfminorversion
\fi
\makeatletter
\newtheoremstyle{thmstyleone}%
{6pt}%
{6pt}%
{\normalfont\itshape}%
{0pt}%
{\bfseries}%
{}%
{.5em}%
{\thmname{#1}\thmnumber{\@ifnotempty{#1}{ }\@upn{#2}}%
  \thmnote{ {\the\thm@notefont(#3)}}}%
\newtheoremstyle{thmstyletwo}%
{6pt}%
{6pt}%
{\itshape}%
{0pt}%
{\normalfont}%
{}%
{.5em}%
{\thmname{#1}\thmnumber{\@ifnotempty{#1}{ }{#2}}%
  \thmnote{ {\the\thm@notefont(#3)}}}%
\makeatother

\usepackage{aliascnt}
\usepackage{hyperref}
\hypersetup{colorlinks=true,linkcolor=blue,citecolor=blue,urlcolor=cyan}
\usepackage[nameinlink,capitalize]{cleveref}
\numberwithin{equation}{section}
\theoremstyle{thmstyleone}%
\newtheorem{theorem}{Theorem}%
\newaliascnt{proposition}{theorem}
\newtheorem{proposition}[proposition]{Proposition}
\aliascntresetthe{proposition}
\newaliascnt{lemma}{theorem}
\newtheorem{lemma}[lemma]{Lemma}
\aliascntresetthe{lemma}
\newaliascnt{definition}{theorem}
\newtheorem{definition}[definition]{Definition}
\aliascntresetthe{definition}
\newaliascnt{corollary}{theorem}
\newtheorem{corollary}[corollary]{Corollary}
\aliascntresetthe{corollary}
\newaliascnt{conjecture}{theorem}

\aliascntresetthe{conjecture}
\theoremstyle{thmstyletwo}%
\newaliascnt{example}{theorem}

\aliascntresetthe{example}
\newtheorem*{remark}{Remark}%

\newcommand{\HCCT}{T_{\mathrm{HCCT}}}
\newcommand{\HM}{T_{\mathrm{HM}}}
\newcommand{\FHCw}{F_{\mathrm{HC},\boldsymbol w}}
\newcommand{\FPw}{F_{\mathrm{Pareto},\boldsymbol w}}

\newcommand*{\dif}{\mathop{}\!\mathrm{d}}

\newcommand*{\iid}{\ifmmode\mathrel{\overset{\scalebox{.6}{ i.i.d.}}{\scalebox{1.2}[1]{$\sim$}}}\else i.i.d. \fi}

\newcommand*{\Parallel}{\mathrel{\rlap{$/$}{\kern2pt/}}}
\providecommand*{\Perp}{\mathrel{\rlap{$\perp$}{\kern2pt\perp}}}

\renewcommand{\Im}{\operatorname{Im}}
\renewcommand{\Re}{\operatorname{Re}}

\setaddedmarkup{\textcolor{red}{#1}}
\setdeletedmarkup{\textcolor{red}{\sout{#1}}}
\newcommand{\spacingset}[1]{}
\newcommand{\resetspacing}{}
\newcommand{\compactsection}[2]{\section{#1}\label{#2}}

\allowdisplaybreaks[4]
\renewcommand{\arraystretch}{1.2}
\makeatletter
\renewenvironment{proof}[1][\proofname]{%
  \par
  \pushQED{\qed}%
  \normalfont
  \topsep 6\p@ \@plus 6\p@
  \trivlist
  \item[\hskip\labelsep\itshape
        #1\@addpunct{.}]%
  \ignorespaces
}{%
  \popQED
  \endtrivlist
  \@endpefalse
}

\makeatother
\title{\bfseries A Heavily Right Strategy for Statistical Inference
with Dependent Studies in Arbitrary Dimensions}
\author{Tianle Liu, Xiao-Li Meng, and Natesh S. Pillai\\Harvard University}
\date{}
\begin{document}
\maketitle
\begin{abstract}

\hspace{-5pt}
We leverage recent advances in heavy-tail approximations for global hypothesis testing with dependent studies to construct approximate confidence regions without modeling or estimating their dependence structures. A non-rejection region is a confidence region but it may not be convex. Convexity is appealing because it ensures any one-dimensional linear projection of the region is a confidence interval, easy to compute and interpret. We show why convexity fails for nearly all heavy-tail combination tests proposed in recent years, including the influential Cauchy combination test. These insights motivate a \textit{heavily right} strategy: truncating the left half of the Cauchy distribution to obtain the Half-Cauchy combination test. The harmonic mean test also corresponds to a heavily right distribution with a Cauchy-like tail, namely a Pareto distribution with unit power. We prove that both approaches guarantee convexity when individual studies are summarized by Hotelling $T^2$ or $\chi^{2}$ statistics (regardless of the validity of this summary) and provide efficient, \textit{exact} algorithms for implementation. Applying these methods, we develop a divide-and-combine strategy for mean estimation in any dimension and construct simultaneous confidence intervals in a network meta-analysis for treatment effect comparisons across multiple clinical trials. We also present many open problems and conclude with epistemic reflections.
\end{abstract}
{\small\noindent\textbf{Key words:} Confidence region, Global testing,
Half-Cauchy combination, Harmonic mean, Network meta-analysis.}
\setcounter{footnote}{1}
\section{Dependence-Resilient Inference}\label{sec:intro}
\vspace*{-10pt}
\subsection{Addressing Dependence: Three Classes of Approaches}
\vspace*{-10pt}

In any theoretical or empirical investigation involving multiple entities—whether individual subjects, their characteristics, or related studies—assessing and accounting for their mutual influence is a key marker of scientific rigor. Conversely, a purely atomistic approach to analyzing multiple entities without valid justification often raises concerns about the credibility of results. In statistical studies, stochastic dependence encapsulates these interrelationships, making it essential for statistical validity. Realistically assessing dependence, however, is challenging, especially in high-dimensional settings, as it requires substantial data and/or information to ensure reliability. Numerous methods have been proposed to address stochastic dependence, and most fall into two broad categories.

\spacingset{1.75}

\begin{itemize}
    \item \textbf{\textit{Simplistic Assertive Approaches}} rely on strong assumptions to simplify dependence structures, such as assuming independencies or equal correlations. 
    \begin{itemize}
  \item \textbf{Pros}: Greatly simplified modeling and computation, making them more generally accessible.
	\item	\textbf{Cons}: Great risk of inaccuracies and challenges in scientific justification.
    \end{itemize}
    \item \textbf{\textit{Model-Intensive Approaches}} employ data-driven methods to estimate pre-specified dependence structures, relying on more flexible and realistic assumptions compared to the assertive approaches. 
    \begin{itemize}
        \item \textbf{Pros}: More principled approach with stronger validity and efficiency.
        \item \textbf{Cons}: Greater modeling and computational demand, and higher risk of overfitting. 
    \end{itemize}
\end{itemize}

Recently, a third class of methods has gained considerable attention, which we categorize as \emph{dependence-resilient} approaches because their validity is robust to dependence beyond what is specified by the model.
\begin{itemize}
    \item \textbf{\textit{Dependence-Resilient Approaches}} construct tests or estimates that are insensitive to dependence. 
    \begin{itemize}
        \item \textbf{Pros}: Principled and easy to apply, compute, and interpret.
        \item \textbf{Cons}: Can be overly conservative,  without careful constructions.
    \end{itemize}
\end{itemize}

Traditionally, approaches in the third category,  such as Bonferroni correction, are undesirable due to their overly conservative nature.  The development of dependence-resilient approaches with acceptable power began over a decade ago, largely motivated by a surprising observation by \citet{DrtonXiao14}. 
\vspace*{-10pt}
\subsection{A Cauchy Surprise and Its Inspiration}%
\vspace*{-10pt}
Let \(\boldsymbol{X} = (X_1, \ldots, X_m)^\top\) and \(\boldsymbol{Y} = (Y_1, \ldots, Y_m)^\top\) be two independent samples from \(\mathcal{N}(\boldsymbol{0}, \boldsymbol{\Sigma})\), where \(\boldsymbol{\Sigma}>0\) is $m\times m$. Based on simulations, \citet{DrtonXiao14} conjectured that for any $ \boldsymbol{w}=\{w_1, \ldots, w_m\}$ with \(\sum\limits_{j=1}^m w_j = 1\),
\begin{equation}\label{eq:cauchy}
\spacingset{1}
    T_{ \boldsymbol{w}}=\sum_{j=1}^m w_j \frac{X_j}{Y_j} \sim \mathrm{Cauchy}(0, 1) \quad [\text{Cauchy distribution with center 0 and scale 1}],
\resetspacing
\end{equation}
as long as  \(w_j \geq 0\). They provided a proof for  \(m = 2\), and left it as a conjecture for general $m>2$.

When \(\boldsymbol{\Sigma}\) is not diagonal, the ratios \(X_j / Y_j\) (for \(j = 1, \ldots, m\))—although each individually Cauchy distributed—are not independent, providing little reason to expect that $T_{ \boldsymbol{w}}$ follows Cauchy(0,1). However, \citet{pillai2016unexpected} proved that \eqref{eq:cauchy} indeed holds for arbitrary \(m\), based on a largely forgotten result that apparently generated the ``afterstat''---not aftermath---of this Cauchy surprise. Specifically, for any $\{u_1, \ldots, u_m\}\in \mathbb{R}^{m}$, and $\Theta_1 \sim \mathrm{Uniform}(-\pi, \pi]$ independent of $\{w_1, \ldots, w_m\}$ where $w_j \geq 0$ and $\sum_{j} w_j = 1$,  \citet{williams1969cauchy} reports that
\begin{equation}\label{eq:williams}
\spacingset{1}
\sum_{j=1}^{m} w_j \tan(\Theta_1 + u_j) \sim \mathrm{Cauchy}(0, 1).
\resetspacing
\end{equation}
\spacingset{1.9}
Writing $\{X_j=R_j\cos(\Theta_j), Y_j=R_j\sin(\Theta_j)\}$ and proving $\{u_i=(\Theta_j-\Theta_1)\operatorname{mod}(2\pi), j=2, \ldots, m\}$ is independent of $\Theta_1$ under the normal model, \citet{pillai2016unexpected} establishes \eqref{eq:cauchy} because $T_{\boldsymbol{w}}=\sum_{j=1}^{m} w_j \tan(\Theta_1 + u_j)$.

The result in \cref{eq:cauchy} has found applications in a variety of fields, from  financial portfolio management \citep{lindquist2021taylor} to genomewide epigenetic studies (\citealp{liu2022large}, \citealp{liu2024ensemble}), and to understanding post-processing noise in differentially private wireless federated learning \citep{wei2023differentially}. It also prompted theoretical work on heavy tail distributions \citep{cohen2020heavy,xu2022cauchy}, as well as suggested the existence of useful statistics that are ancillary to the dependence structure, giving rise to the potential power of Cauchy combination rules or tests (CCT). In particular, \citet{liu2020cauchy} proposed combining \(m\) possibly correlated $p$-values \(\{p_1, \ldots, p_m\}\) for testing the same null hypothesis \(H_0\) via
\begin{equation}\label{eq:liu}
\spacingset{1}
T_{\mathrm{CCT}} = \sum_{j=1}^m w_{j} \tan\bigl\{\pi(1/2 - p_{j}) \bigr\} = \sum_{j=1}^m w_{j} \cot(\pi p_{j}).
\resetspacing
\end{equation}
The power of \cref{eq:liu} is also demonstrated in \citet{liu2019acat} for using CCT in rare-variant analysis.

The same tangent function combining rule adopted by \cref{eq:liu,eq:williams} hints at the potential dependence resilience nature of $T_{\mathrm{CCT}}$. Indeed, 
\citet{liu2020cauchy} established the following results via representing \(p_{j} = 2\{1 - \Phi(|Z_{j}|)\}\), where \(\Phi(z)\) is the CDF of $\mathcal{N}(0,1)$. If for any \(i \neq j\), \((Z_i, Z_j)\) are bivariate normal with mean zero and mild constraints on \(\boldsymbol{\Sigma}\), the covariance matrix of \((Z_1, \ldots, Z_p)\), then
\begin{equation}
\spacingset{1}
\lim_{t \rightarrow \infty} \frac{\mathbb{P}(T_{\mathrm{CCT}}\ge t)}{\mathbb{P}(C \ge t)} = 1, \quad \text{where}\quad  C \sim \mathrm{Cauchy}(0, 1).
\resetspacing
\end{equation} 
Subsequently, \citet{vovk2020combining}, \citet{vovk2022admissible}, and \citet{fang2023heavy} showed that such robustness against dependence in \(\boldsymbol{\Sigma}\) can be extended to other combinations such as the harmonic mean $p$-value (HMP) $T_{\mathrm{HM}}=\sum_{j=1}^{m} w_{j} / p_{j}$
\citep{good1958significance,wilson2019harmonic}. A commonality of these methods is the use of quantile functions from heavy-tailed distributions—\(\cot(\pi p)\) for Cauchy and \(1 / p\) for \(\mathrm{Pareto}(1,1)\)—to transform individual $p$-values before combining them. The stability of  Cauchy facilitates tracking of the null distribution for independent studies, and inspires extensions  via other stable distributions \citep{wilson2021evy,ling2022stable}.
\vspace*{-10pt}
\subsection{A Heavily Right Strategy for Inference}\label{sec:heavy}
\vspace*{-20pt}
\setlength{\parskip}{13pt}
\spacingset{1.75}
 
Because $\tan(x)$ approaches $-\infty$ when $x \downarrow -\pi/2$, the CCT statistic in \cref{eq:liu} will approach $-\infty$ even if only one $p_j$ approaches 1 (and none of the $p_i$'s is extremely significant to compensate). This extreme sensitivity to large $p$-values is undesirable theoretically and practically \citep{fang2023heavy}.  For example, in genome-wide association studies, only a few SNPs (Single Nucleotide Polymorphisms) are likely related to the phenotype of interest, with most $p$-values close to one \citep{zeggini2009meta}. In such cases, CCT can cause numerical instability and substantial power loss. 

This instability indicates an issue that is rarely discussed---or even realized---when one focuses on $p$-values, but it is essential for constructing confidence region, at least from a practical perspective.
Whereas converting hypothesis tests to confidence regions is a classic approach, the conversion does not guarantee the resulting region is an interval for univariate cases or a convex region for multivariate parameters. Such is the case for CCT. That is, when we obtain a confidence set for a parameter \(\boldsymbol{\theta}\) by inverting a CCT based on multiple studies—each testing \(H_{0}: \boldsymbol{\theta} = \boldsymbol{\theta}_{0}\) against \(H_{1}: \boldsymbol{\theta} \neq \boldsymbol{\theta}_{0}\)—the non-rejection region for \(\boldsymbol{\theta}_{0}\) may be nonconvex or even disconnected, as illustrated in the following two examples.

\begin{figure}[tbp]
\vspace{-40pt}
  \centering
  \subfloat[Score functions with coverage thresholds]{%
    \includegraphics[width=0.337\textwidth]{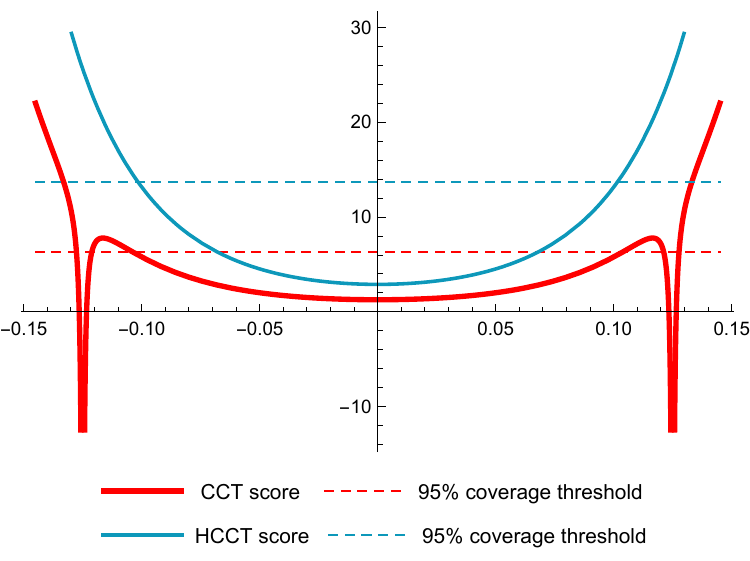}
    \label{fig:cctexample}
  }
  \hspace*{15pt}
  \subfloat[$95\%$ confidence regions in $2$-dimension]{%
    \includegraphics[width=0.4\textwidth]{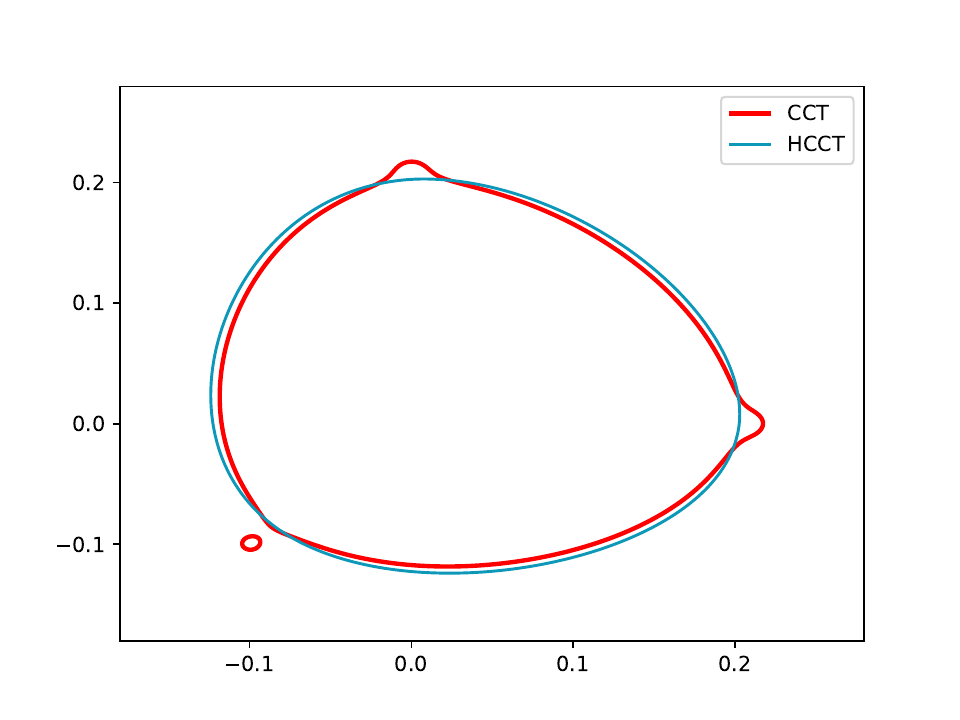}
    \label{fig:cctexample2}
  }
  \caption{Connectivity of confidence regions for CCT and HCCT.}\label{fig:cctbad}
\end{figure}

\vspace*{-10pt}
\begin{enumerate}
\setlength{\itemsep}{5pt}
  \item[\bf Ex 1]\label{thm:expdisconnect} Suppose we have two equally weighted studies with estimators from \(\mathcal{N}(\theta_{0}, 0.01)\) and obtain estimates of \(0.125\) and \(-0.125\). Inverting CCT at a $5\%$ significance level yields a disconnected $95\%$ confidence set: $
[-0.1277, -0.1212] \cup [-0.1038, 0.1038] \cup [0.1212, 0.1277]$, which includes the two individual estimates, as illustrated in \cref{fig:cctexample}.
  \item[\bf Ex 2]\label{thm:expdisconnect2} Suppose that we have three equally weighted studies with estimators from \(\mathcal{N}(\boldsymbol{\theta}_{0}, 0.01 \boldsymbol{I}_{2})\), and obtain estimates \((-0.10, -0.10)\), \((0.21, 0)\), and \((0, 0.21)\). Inverting CCT at a $5\%$ significance level yields disconnected $95\%$ confidence regions, which include all three individual estimates, as shown in \cref{fig:cctexample2}. 
\end{enumerate}
\vspace*{-10pt}
Later in \cref{sec:empty}, we will explain why any CCT region necessarily includes all individual estimates, irrespective of the confidence levels. This undesirable property, recognized in \citet{meng2024bffer}, along with other defects of inverting CCT for constructing confidence regions,  serves as a springboard for the present article.

Specifically, we truncate the entire left tail of the Cauchy distribution, resulting in the \emph{Half-Cauchy Combination Test (HCCT)}. This \textit{heavily right} strategy effectively resolves the two limitations of CCT revealed earlier, as demonstrated in \cref{fig:cctexample,fig:cctexample2}. However, it increases the computational demand, because the Half-Cauchy distribution, unlike the Cauchy distribution, is not a stable distribution.

Fortunately,  by leveraging Laplace transforms and numerical integration, we are able to compute exact tail probabilities for HCCT scores with independent studies. This resolves the computational issue because heavy-tail approximations rely on dependence-resilience to extend their applicability from independent studies to dependent ones, and hence avoiding modeling or computation all together for dealing with the dependence.  

\spacingset{2.0}
\setlength{\parskip}{3pt}

We remark that HCCT is a special case of a class of approaches to left-truncate or winsorize Cauchy methods for reducing sensitivity to large $p$-values \citep{gui2023aggregating,fang2023heavy}. These previous approaches did not provide sufficiently accurate distribution calculations for the test statistic, even with independent studies (see \cref{tab:wilson}), nor did they address the challenge of constructing confidence regions for parameter estimation. In fact, we show that Half-Cauchy is the only distribution in their proposed family of methods that guarantees convex confidence regions (see \cref{sec:conf}).

Another notable dependence-resilient approach for global testing is the HMP mentioned earlier, which  has been generalized to other averaging techniques \citep{vovk2020combining,fang2023heavy}, with a high-level theoretical analysis of this class provided by \citet{vovk2022admissible}. Because HMP corresponds to using Pareto$(1,1)$, which is heavily right with Cauchy-like tail, we are able to provide the same theoretical results and similar algorithms for computing the exact null distribution of $T_{\mathrm{HM}}$ with independent studies, but allowing for flexible weights. The resulting EHMP (Exact Harmonic Mean $p$-value) hence improves upon HMP,\footnote{Note: EHMP and HMP share the same score function $T_{\mathrm{HM}}$, where ``HM'' denotes ``harmonic mean''.} and behaves very similarly to HCCT throughout our investigation.  %

\subsection{The Presentation Flow of Our Article}

Because the primary goal of our article is to explore the use of heavily-right strategy for constructing confidence regions, not merely to improve CCT or HMP (which are happy byproducts), we start the rest of this article in \cref{sec:conf} with inverting HCCT and EHMP to obtain confidence regions, establish their convexity and compactness in common scenarios, and present algorithms for computing them. The study of HCCT and EHMP for testing purposes, as well as their comparisons to some other combination tests, will be deferred to \cref{sec:hctest}.

To demonstrate the potential of our approach, \cref{sec:divideac} then proposes a divide-and-combine strategy for  mean estimation in any dimension, providing a variety of set estimators that generalize Hotelling’s \(T^{2}\) approach. Notably, this strategy does not require estimating the full covariance matrix or even any matrix and can yield potentially more compact confidence regions (for some lower-dimensional estimands) with approximately valid coverage. As a concrete application, 
\cref{sec:networkmeta} examines the competitiveness of our approach to network meta-analysis in clinical trials, using both semi-synthetic and real-data examples. Since HCCT and EHMP yield very similar numerical results in these applications, we only report HCCT results to save space.

\resetspacing
\setlength{\parskip}{3pt}

The concluding \cref{sec:discussion} explicates practical limitations and theoretical open problems of our current proposals, which we hope will serve as a warm invitation to the statistical and broader data science community to fully explore and leverage the paradigm of heavy-tail approximation refined by the heavily-right strategy, just as we 
have for the large-sample approximations with a host of refinements throughout the history of statistical inference. 

Just to whet readers' appetites, especially those who care deeply about statistical foundations, here is a list of questions we invite readers to contemplate while diving deep into our article (and more broadly to the heavy-tail approximation literature): Why can dependence surrender to heavy marginal tails?  Is it the correct explanation or is there something more profound about stochastic behaviors that collectively we have failed to understand? Why heavily right is right? What would be an inferential principle that automatically prefers Half-Cauchy to Cauchy, because it prioritizes convexity as a desirable property?  When is convexity desirable epistemically? What are the consequences of having a $p$-value from a test statistic that does not lead to convex confidence regions?

To save space, some technical development and all proofs are in the supplemental material \citep{supsupplemental2024}, so is a section that briefly reviews the literature on other global testing procedures that are not necessarily dependence-resilient.

\vspace{0pt}

\section{Confidence Regions from Inverting Combination Tests}\label{sec:conf}
\vspace{-5pt}
\subsection{A General Strategy for Combining Dependent $p$-Values and Obtaining Confidence Intervals}\label{sec:unicase}

Let $p_{j}, j=1,2,\dots,m$ be individual $p$-values from testing a common null hypothesis, and we like to combine $p_{j}$'s into one test statistic. Given a random variable $\nu$ on $\mathbb{R}$ with continuous CDF $F_\nu(x)$,  define
\begin{equation}\label{eq:gct}
\spacingset{1}
  \textstyle T_{\nu, \boldsymbol{w}}=\sum_{j=1}^{m}w_{j}F_{\nu}^{-1}(1-p_{j}),\quad \text{where }\sum_{j=1}^{m}w_{j}=1,\quad w_{j}\geq 0\quad \forall j=1,\dots,m.
  \resetspacing
\end{equation}

We remark that the weights $w_{1},\ldots,w_{m}$ allow the analyst to encode prespecified differences in the relative contributions of studies. Equal weights are a neutral default; unequal weights can be useful when studies differ materially in data quality, design considerations, contributions to study inconsistency (see \cref{sec:empty}), etc. 

If $p_{j}$'s follow $\operatorname{Uniform}[0, 1]$, then $F^{-1}_\nu(1-p_{j})$'s are identically distributed as $\nu$.  Many choices are made in the literature, such as $\nu \sim \chi_{2}^{2}$ by Fisher's method and $\nu \sim \mathcal{N}(0,1)$ for Stouffer's Z-score method.  
For EHMP $\nu \sim \mathrm{Pareto}(1, 1)$ with density given by $f_{\nu}(x)={x^{-2}}\mathbb{I}_{x\geq 1}$, and for CCT $\nu\sim \mathrm{Cauchy}(0,1)$. Consequently, 
\begin{equation}\label{eq:cct}
\spacingset{1}
   \textstyle T_{\mathrm{HM}}=\sum_{j=1}^{m}\frac{w_{j}}{p_{j}},\quad T_{\mathrm{CCT}}=\sum_{j=1}^{m}w_{j}\cot (\pi p_{j}).
   \resetspacing
\end{equation}
Replacing Cauchy by Half-Cauchy amounts to replace $\pi$ by $\pi/2$ in the expression above, yielding 
\begin{equation}\label{eq:hcct}
\spacingset{1}
  \textstyle T_{\mathrm{HCCT}}= \sum_{j=1}^{m}w_{j}F^{-1}_{\mathrm{HC}} (1- p_j) = \sum_{j=1}^{m}w_{j}\cot \Bigl(\frac{\pi p_{j}}{2}\Bigr).
  \resetspacing
\end{equation}

\spacingset{1.7}

Suppose there are \(m\) possibly dependent studies, the $j$-th of which provides $\widehat{\theta}_j$ as its estimator of the common estimand $\theta \in \mathbb{R}$, together with a variance estimator $\widehat{\sigma}_j^2$.  For many common studies, it is acceptable to approximate the distribution of $(\widehat\theta_j - \theta)/\widehat{\sigma}_j$ by the normal distribution or \(t\)-distribution with \(k_j\) degrees of freedom. That is, we can compute the (two-sided) $p$-value as 
\begin{equation}\label{eq:pjunknownsigma}
\spacingset{1}
\textstyle
p_j = 2\bigl\{1 - F^{(j)}\bigl(\widehat{\sigma}_j^{-1}\lvert \widehat{\theta}_j - \theta \rvert\,\bigr)\bigr\},
\resetspacing
\end{equation}
where \(F^{(j)}\) is the CDF of the $t$ distribution with $k_j$ degrees of freedom, which includes $\mathcal{N}(0,1)$ when $k_j\to\infty$.

\spacingset{1.9}
\setlength{\parskip}{7pt}

For fixed $m$ and mutually independent $\{\widehat\theta_j, \widehat\sigma_j\}_{j=1}^{m}$, the convergence of each statistic to a continuous CDF \(F^{(j)}\) implies that \(p_j(\theta_0)\) converge jointly in distribution to independent uniform variables at the true value \(\theta_0\). The continuous-mapping theorem and test inversion \citep{casella2024statistical} therefore give asymptotic coverage \(1-p\)\footnote{We use $p$ instead of the common $\alpha$ to avoid a notation clash with the $\alpha$-stable law we shall discuss shortly.} for \cref{eq:invert}; the coverage is exact when the pivotal distributions are exact.
\begin{equation}\label{eq:invert}
\spacingset{1}
\textstyle\sum_{j=1}^{m} w_j F_{\nu}^{-1} \bigl\{2F^{(j)} \bigl( \widehat{\sigma}_j^{-1}\lvert \widehat{\theta}_j - \theta \rvert\,\bigr) - 1\bigr\} \leq F_{\nu, \boldsymbol{w}}^{-1}(1 - p).
\resetspacing
\end{equation}
Here \(F_{\nu}\) denotes the CDF of \(\nu\), and \(F_{\nu, \boldsymbol{w}}\) represents the CDF of \(T_{\nu, \boldsymbol{w}}\) for \textit{independent} studies, as defined in \eqref{eq:gct}. 

\begin{figure}[tbp]
  \centering
  \subfloat[Half-Cauchy \& Student's $t(10)$]{%
    \makebox[.25\textwidth][c]{%
      \includegraphics[width=.22\textwidth,height=.15\textwidth]{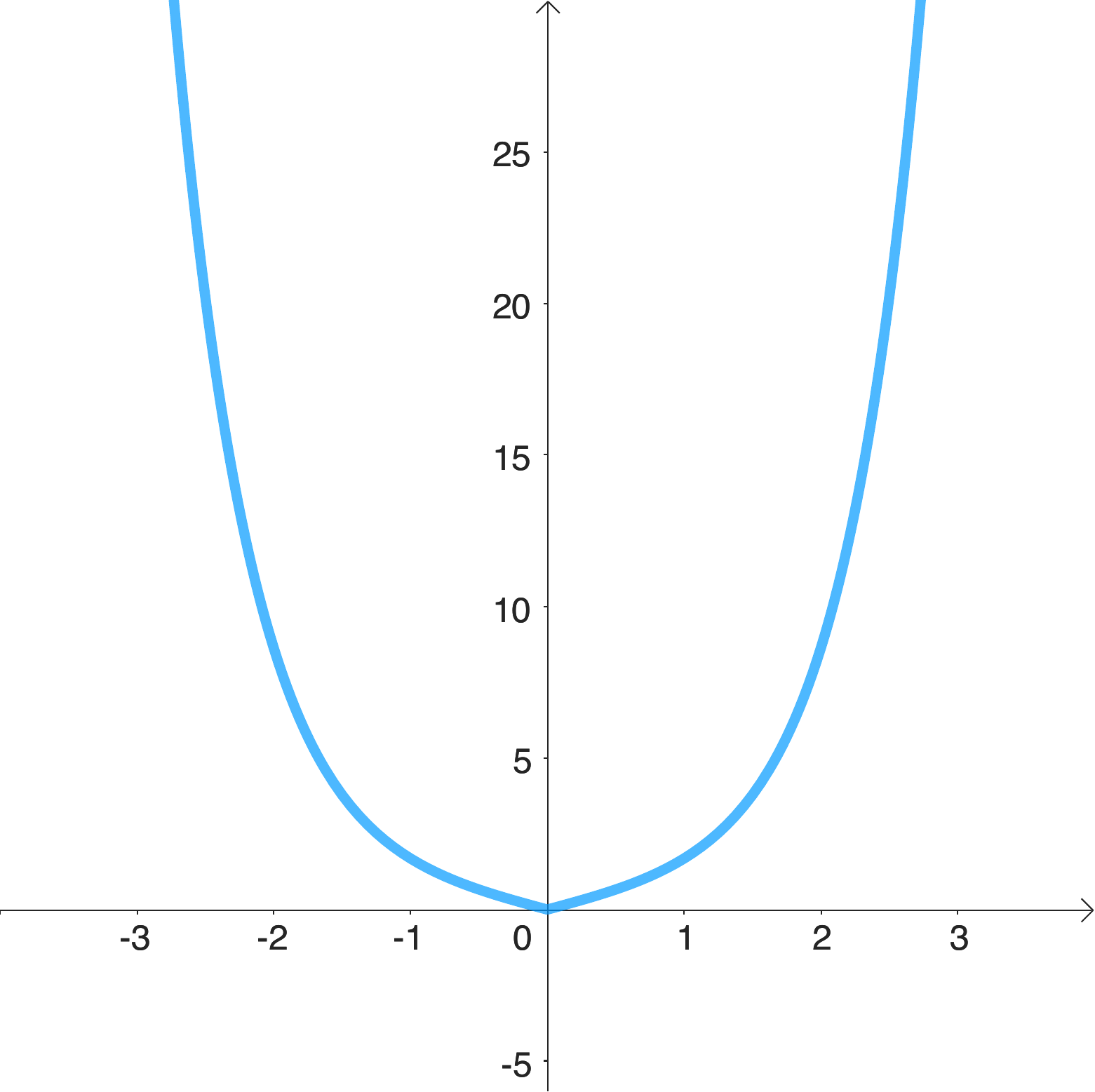}%
    }
    \label{fig:illhalfcauchy}
  }\hfill
  \subfloat[Cauchy \& Student's $t(10)$]{%
    \includegraphics[width=.22\textwidth,height=.15\textwidth]{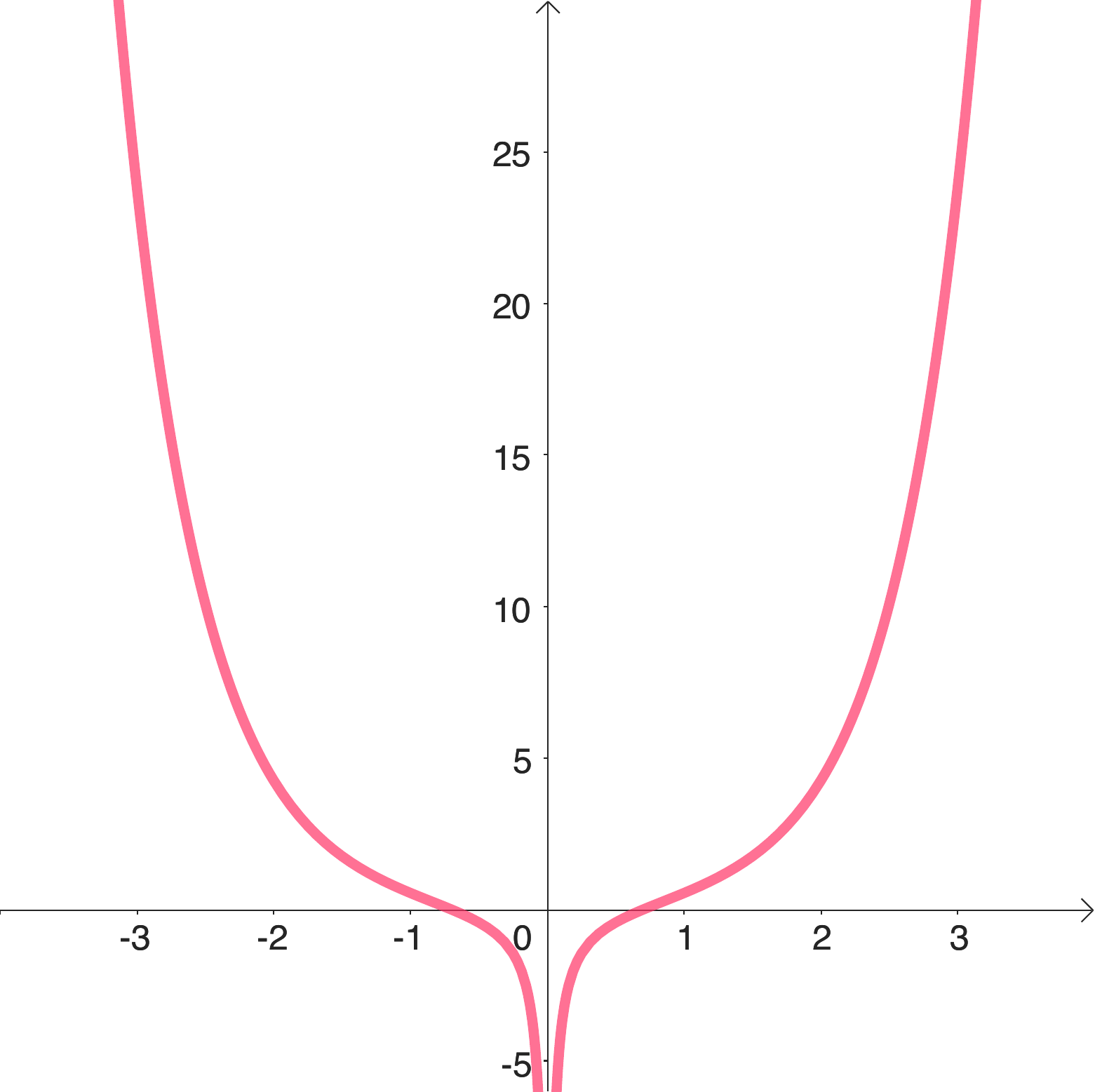}
    \label{fig:illcauchy}
  }\hfill
  \subfloat[$\chi^2_2$ \& normal]{%
    \includegraphics[width=.22\textwidth,height=.15\textwidth]{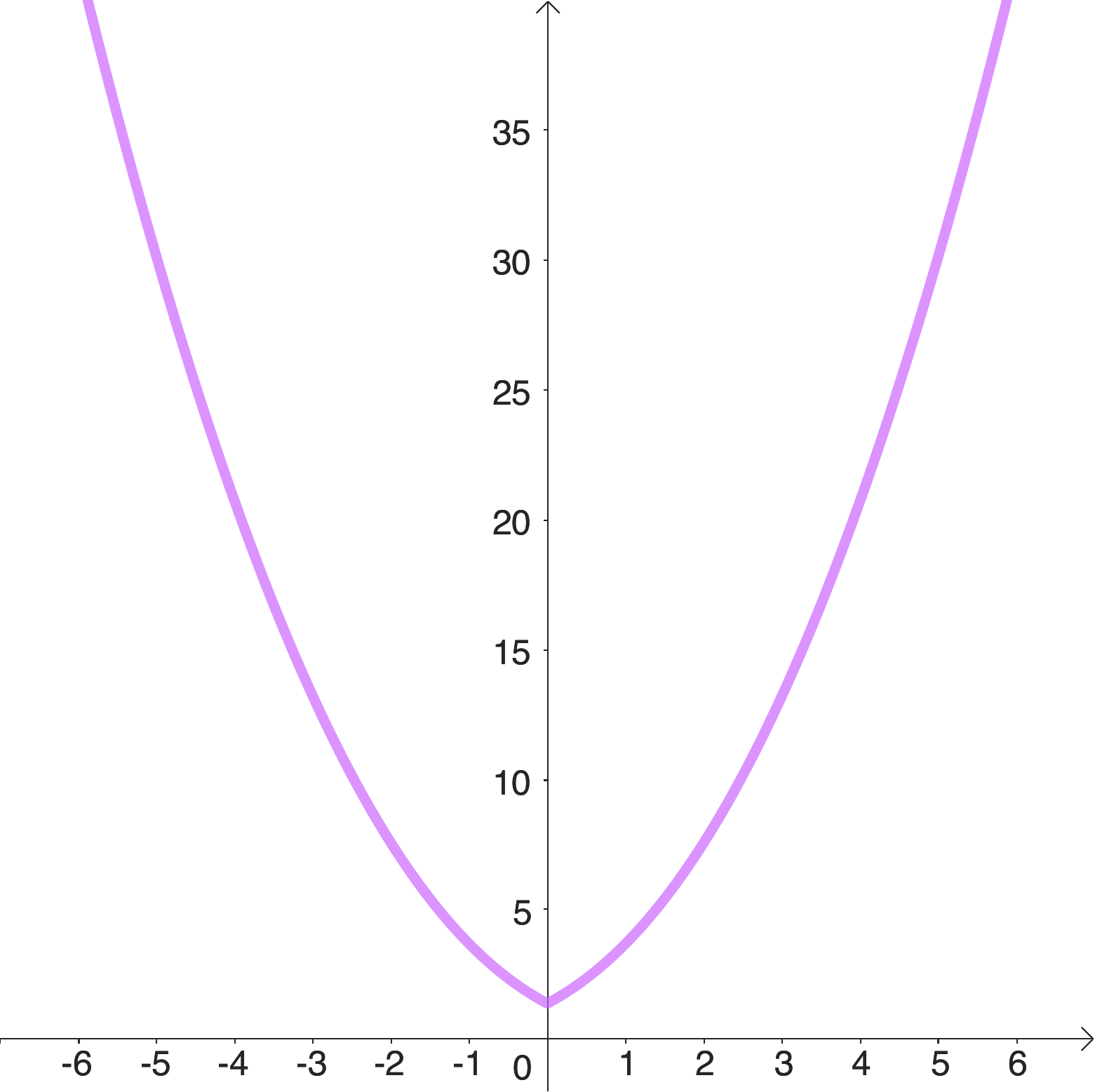}
    \label{fig:illfishernormal}
  }\hfill
  \subfloat[$\chi^2_2$ \& Student's $t(10)$]{%
    \includegraphics[width=.22\textwidth,height=.15\textwidth]{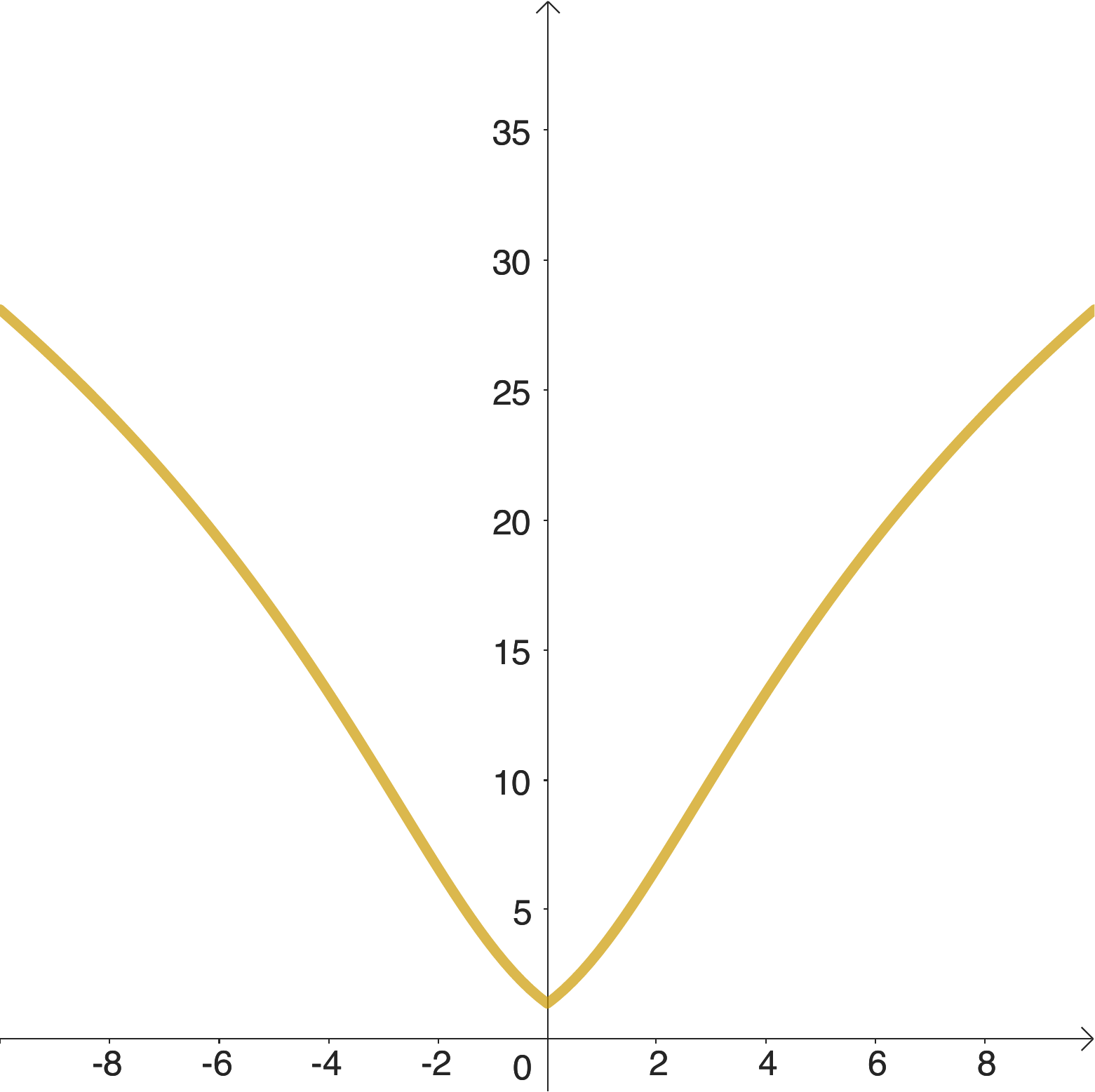}
    \label{fig:illfisherstudent}
  }
  \caption{\footnotesize Plots of $g_j(\theta) = F_{\nu}^{-1} \bigl\{2F^{(j)} (\lvert \theta \rvert ) - 1 \bigr\}$, where the first distribution in the caption refers to $F_v$, and the second to $F^{(j)}$. }
\end{figure}

When the studies are dependent, marginal validity of the individual \(p\)-values alone does not determine the null distribution of the combination statistic. Under suitable conditions on their pairwise lower-tail dependence, however, combination rules based on distributions with Cauchy-like right tails remain asymptotically calibrated by their independence reference as \(p\downarrow0\), and test inversion then yields asymptotically valid confidence regions. In \cref{sec:hctest}, we provide tractable sufficient conditions covering HCCT and EHMP; here we focus on the convexity of the resulting regions.
The following is an \textit{algebraic} result in the univariate case, meaning that it is guaranteed for any actual dataset, not depending on whether \cref{eq:pjunknownsigma} provides a valid $p$-value or not. %
Nevertheless, the validity of the $p$-value defined through \cref{eq:pjunknownsigma} is important in establishing the desired confidence coverage.
\begin{theorem}\label{thm:connectivity0}
  For HCCT or EHMP, the solution set of \eqref{eq:invert} is always a single (but possibly empty) finite interval.
\end{theorem}
We remark that this result does not hold for most other combination tests with general \(\nu\). We provide some intuition here, and defer the formal results to \cref{sec:insightconv} due to space limitations. Specifically, if we would like the solution set of \cref{eq:invert}
to be connected for arbitrary \(w_{j}\), \(\widehat{\theta}_j\) and \(\widehat{\sigma}_{j}\)'s, the function
\spacingset{1.65}
\[
\spacingset{1}
g_j(\theta)= F_{\nu}^{-1} \bigl\{2F^{(j)} \bigl(\widehat{\sigma}_j^{-1} \lvert \widehat{\theta}_j - \theta \rvert \bigr) - 1 \bigr\}
\resetspacing
\]
must be convex (see \cref{thm:lemmaconvexf}). To ensure the convexity of \(g_j\), two necessary conditions must be satisfied, the essence of which is again captured by the term \emph{``heavily right''}. 
First, the density \(f_{\nu}\) must be monotone decreasing on its support, as shown in \cref{thm:lemmanecessary}, since otherwise \(g_j(\theta)\) is nonconvex near \(\theta = 0\). Notably, this condition excludes all $\alpha$-stable distributions for \(\nu\). For example, as shown in \cref{fig:illhalfcauchy,fig:illcauchy}, the function \(g_j\) is convex when \(\nu\) follows a Half-Cauchy distribution, whereas it is nonconvex for \(\nu\) following a Cauchy distribution.

\spacingset{1.65}

Second, as established by  \cref{thm:lemmanecessary}, the convexity of \(g_j(\theta)\) (as \(\lvert \theta \rvert \to \infty\)) implies that the right tail of the density for \(F_{\nu}\) cannot be lighter than of the $F^{(j)}$. To ensure this property for any choice of $F^{(j)}$ in the $t_d$ family with integer degrees of freedom $d$, Half-Cauchy is near-optimal, because it is the same as $t_1$. As an illustration of this requirement, consider the Fisher’s combining rule, which sets \(\nu=\chi^2_2\). When \(\sigma_j\) is known, we can take $F^{(j)}$ as ${\mathcal N}(0, 1)$, hence $\nu=\chi^2_2$ is acceptable because its right tail is heavier than that of normal. \cref{fig:illfishernormal} shows the resulting $g_j(\theta)$ is convex, yielding a single confidence interval for $\theta$ for all confidence levels. In contrast, when \(\sigma_j\) is unknown and hence we must choose  $F^{(j)}$  from the $t_d$ family with $d<\infty$, say, $t_{10}$, then the density of \(F^{(j)}\) will have heavier tail than that of $\nu=\chi^2_2$. This will necessarily destroy the convexity of $g_j(\theta)$, as seen in \cref{fig:illfisherstudent}, potentially leading to disconnected confidence sets. (For this reason, we will assign a neutral rating to Fisher’s test regarding its performance on confidence regions; see \cref{tab:rates} of \cref{sec:hctest}.) 

\setlength{\parskip}{9pt}

\begin{figure}[tbp]
\vspace{-20pt}
  \centering
  \subfloat[Coverage (AR-$1$)]{%
    \includegraphics[height=0.23\textwidth,width=0.33\textwidth]{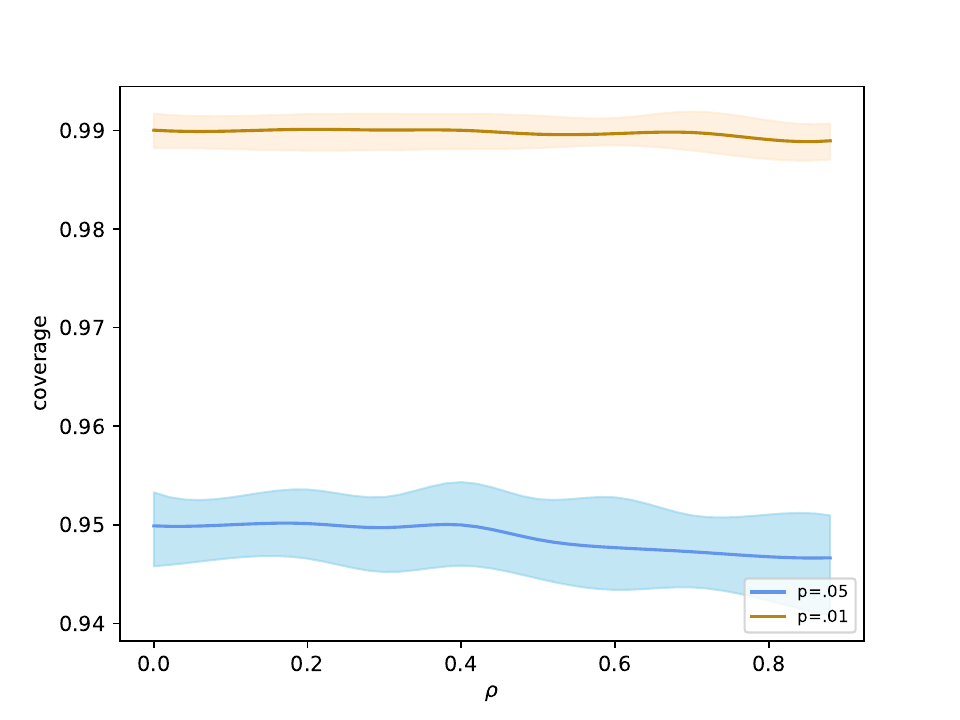}
  }~ 
  \subfloat[Coverage (equi-correlation)]{%
    \includegraphics[height=0.23\textwidth,width=0.33\textwidth]{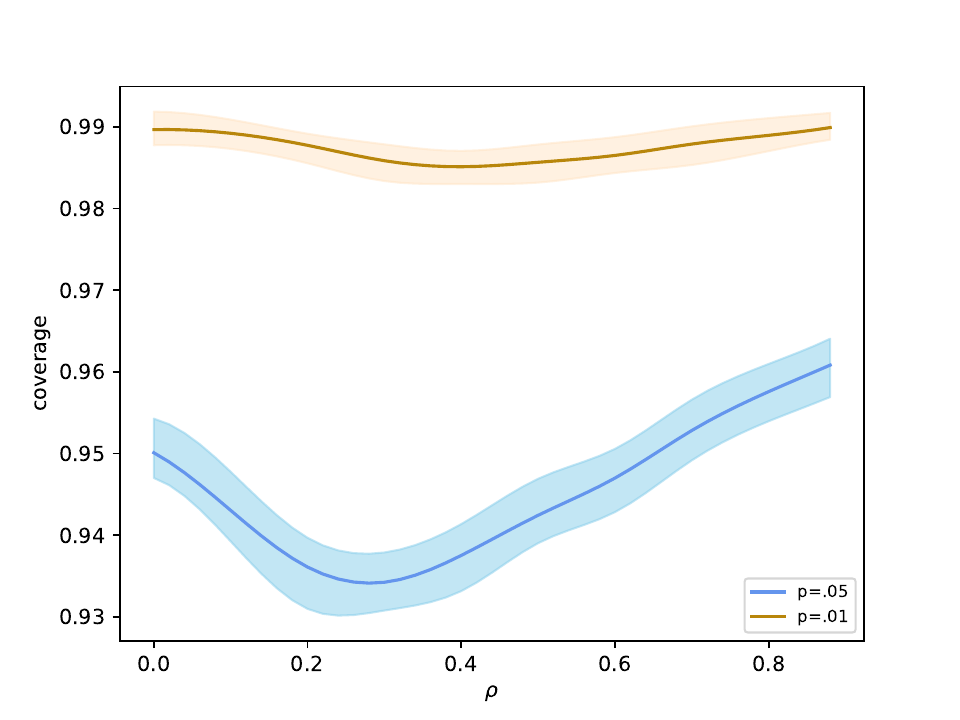}
    \label{fig:coverageequi}
  }\\
  \vspace*{-10pt}
  \subfloat[Width of $95\%$ CIs (AR-$1$)]{%
    \includegraphics[height=0.23\textwidth,width=0.33\textwidth]{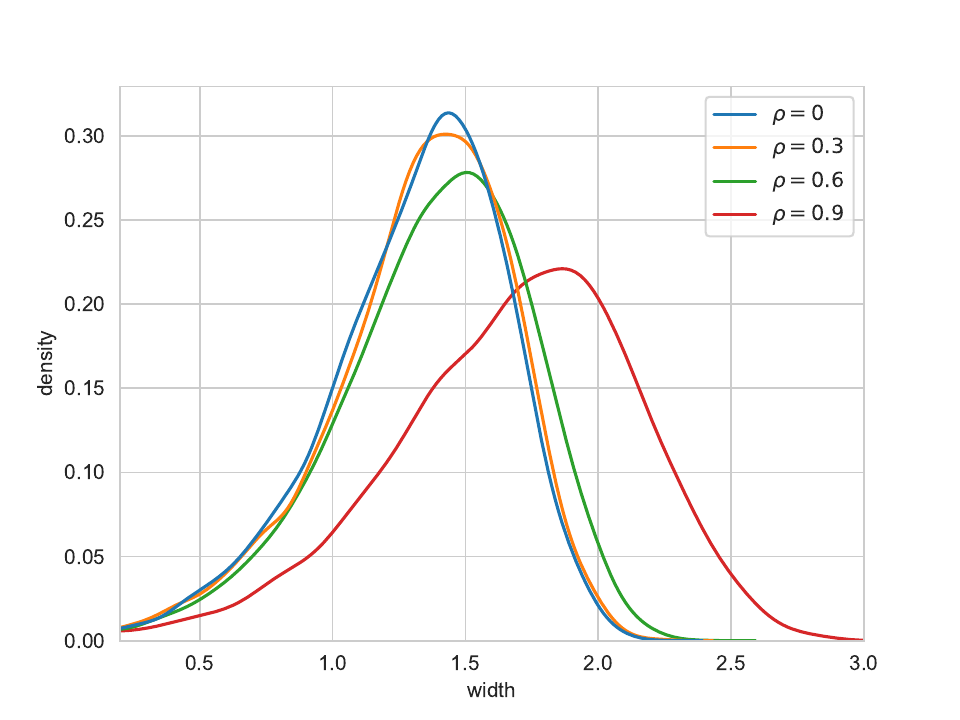}
    \label{fig:1dwidthar1}
  }~ 
  \subfloat[Width of $95\%$ CIs (equi-correlation)]{%
    \includegraphics[height=0.23\textwidth,width=0.33\textwidth]{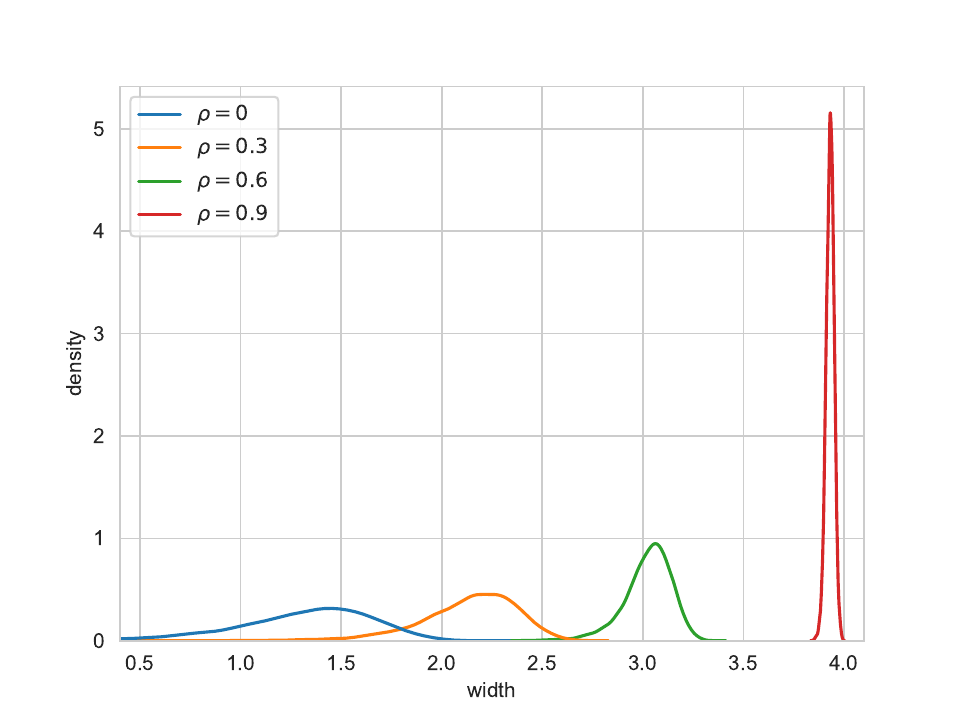}
    \label{fig:1dwidthequi}
  }
  \caption{Confidence intervals from $1$-dimensional HCCT.}
  \label{fig:1dimdensity}
\end{figure}

To compute the confidence intervals, we apply the method of \citet{brent1971algorithm}---the default optimization and root-finding algorithm for scalar functions in the Python package \textbf{SciPy}---to find both the minimizer of the score and the root of \cref{eq:invert}. We adopt the simulation settings from \citet{liu2020cauchy} to evaluate the performance of the confidence intervals obtained from HCCT (or EHMP): The vector of individual test statistics $\boldsymbol{X}$ is generated from $\mathcal{N}_{m}(\theta\,  \mathbf{1}_{m}, \boldsymbol{\Sigma})$ with $\theta=0$ under the null, where $m$ is the number of studies. We consider $m=20,100,500$ for each of the following correlation matrix $\boldsymbol{\Sigma}=(\sigma_{ij})$ to obtain confidence intervals for \(\theta\) using our approach:
\begin{itemize}
  \item AR-$1$ correlation: $\sigma_{ij}=\rho^{\lvert i-j \rvert}$ for $1\leq i,j\leq m$, where $\rho\in [0,1)$;
  \item Equi-correlation: $\sigma_{ij}=\rho$ for $1\leq i,j\leq m$, where $\rho\in [0,1)$.
\end{itemize}

\cref{fig:1dimdensity} presents the actual coverage and widths of the confidence intervals under two different correlation structures with $m=500$. We observe that, in general, the coverage for dependent studies is nearly as good as in the independent case. However, when the estimators are equally correlated with \(\rho\) around $0.25$, the coverage slightly falls below the desired level. (Note: Such a dip is more pronounced with the CCT as shown later in \cref{fig:fpreq}.) Additionally, HCCT demonstrates better robustness when conducted at a $99\%$ confidence level.

We also observe that the widths of the intervals increase as \(\rho\) grows. This effect is especially pronounced in the equi-correlation setup, demonstrating that our approach is robust to the underlying dependence structure by being adaptive to it. Intuitively, fixing the variance of each individual estimator, higher correlations between studies mean a smaller effective number of (independent) studies, and hence larger uncertainties and wider intervals. %

\vspace{-22pt}

\setlength{\parskip}{10pt}
\spacingset{1.8}
\subsection{Obtaining Approximate Confidence Regions in Arbitrary Dimensions}\label{sec:multidim}
\vspace{-12pt}

Next, we consider combining \(m\) studies to obtain a set estimate for \(\boldsymbol{\theta} \in \mathbb{R}^d\), where $d>1$. Suppose we have an estimator \(\widehat{\boldsymbol{\xi}}_j \in \mathbb{R}^{d_j}\) from the \(j\)-th study for $P_j\theta$, where \(\boldsymbol{P}_j \in \mathbb{R}^{d_j \times d}\) is a full-rank matrix with \(d_j \leq d\).  We also assume that the $j$-th study provides a positive definite covariance estimator \(\widehat{\boldsymbol{\Sigma}}_j\) for $\widehat{\boldsymbol{\xi}}_j$.  Because $P_j$ and  $d_j$ can vary with $j$, we can choose $d_j$'s and $\boldsymbol{P}_j$'s to form different lower dimensional projections, and to ensure \(\widehat{\boldsymbol{\Sigma}}_j>0\). For example, we can always choose $d_j=1$ for all $j$'s.

As a natural generalization from the $t$ approximation in the univariate case, here we adopt the Hotelling's $T^2$ distribution by assuming that it is acceptable to postulate that,  given the value of $\theta$,
\begin{equation}\label{eq:hotelling}
\spacingset{1}
  \textstyle (\widehat{\boldsymbol{\xi}}_j - \boldsymbol{P}_j \boldsymbol{\theta})^{\top} \widehat{\boldsymbol{\Sigma}}_j^{-1} (\widehat{\boldsymbol{\xi}}_j - \boldsymbol{P}_j \boldsymbol{\theta}) \sim\ T^2(d_j, k_j) = \frac{d_j k_j}{k_j + 1 - d_j} F(d_j, k_j + 1 - d_j),
  \resetspacing
\end{equation}
where \(T^2(d_j, k_j)\) is the Hotelling's \(T^2\)-distribution, related to the \(F\)-distribution as indicated, and the degrees of freedom with $ \widehat{\boldsymbol{\Sigma}}_j$, $k_j$ are supplied by the $j$-th study. Consequently, the $p$-value for testing $\theta$ from the $j$-th study is 
\begin{equation}\label{eq:rewrite2}
\spacingset{1}
  p_j = 1 - F^{(j)}\bigl\{(\widehat{\boldsymbol{\xi}}_j - \boldsymbol{P}_j \boldsymbol{\theta})^{\top} \widehat{\boldsymbol{\Sigma}}_j^{-1} (\widehat{\boldsymbol{\xi}}_j - \boldsymbol{P}_j \boldsymbol{\theta})\bigr\},
  \resetspacing
\end{equation}
where  $F^{(j)}$ is the CDF of $T^2(d_j, k_j)$ when $k_j<\infty$ or of $\chi^2$ with $d_j$ degrees of freedom when $k_j\to\infty$, which is applicable when $\widehat\Sigma_j$ is considered to be a fixed $\Sigma_j$ . 
The $(1-p)$-level confidence region for \(\boldsymbol{\theta}\) is then obtained via
\begin{equation}\label{eq:invert3}
\spacingset{1}
  \sum_{j=1}^{m} w_j F_{\nu}^{-1}\bigl[ F^{(j)} \bigl\{(\widehat{\boldsymbol{\xi}}_j - \boldsymbol{P}_j \boldsymbol{\theta})^{\top} \widehat{\boldsymbol{\Sigma}}_j^{-1} (\widehat{\boldsymbol{\xi}}_j - \boldsymbol{P}_j \boldsymbol{\theta})\bigr\}\bigr] \leq F_{\nu, \boldsymbol{w}}^{-1}(1 - p).
  \resetspacing
\end{equation}
The following result generalizes  \cref{thm:connectivity0}, irrespective to the validity of the distributional assumption \cref{eq:hotelling}.
\begin{theorem}\label{thm:connectivity2}
 For HCCT, the solution set of \cref{eq:invert3} is convex (which can be empty) if every $j\in J_+$ satisfies $k_j\ge1$ when $d_j=1$ and $k_j\ge d_j+1$ otherwise. For EHMP, $k_j\ge d_j$ for every $j\in J_+$ suffices. Furthermore, the confidence region is bounded if $\{\operatorname{Row} (\boldsymbol{P}_{j}),j \in J_+\}$ span $\mathbb{R}^{d}$, where $J_+=\{j:  w_j>0\}$.
\end{theorem}
\vspace*{-10pt}
Numerically, we evaluate the left-hand side of \cref{eq:invert3} to determine whether a point belongs to the confidence region. A point estimate can be obtained by minimizing this convex score and therefore belongs to the confidence region whenever the region is nonempty. Such a point estimate should be reported with an uncertainty assessment. Our confidence procedure is designed to reflect the potentially substantial uncertainty induced by dependence among studies; see the simulation and discussion below.

The current implementation uses Powell’s method \citep{powell1964efficient} by default for score minimization, with gradient-based methods such as L-BFGS-B \citep{fletcher1987practical} also available. It computes one-dimensional affine slices using scalar minimization and root-finding, and plots two-dimensional affine slices by evaluating the score on a grid and drawing its cutoff contours. When the parameter itself is two-dimensional, such a plane slice gives the full confidence region.

\begin{figure}[tbp]\label{fig:coverage1}
\vspace{-30pt}
  \centering
\includegraphics[width=.44\textwidth]{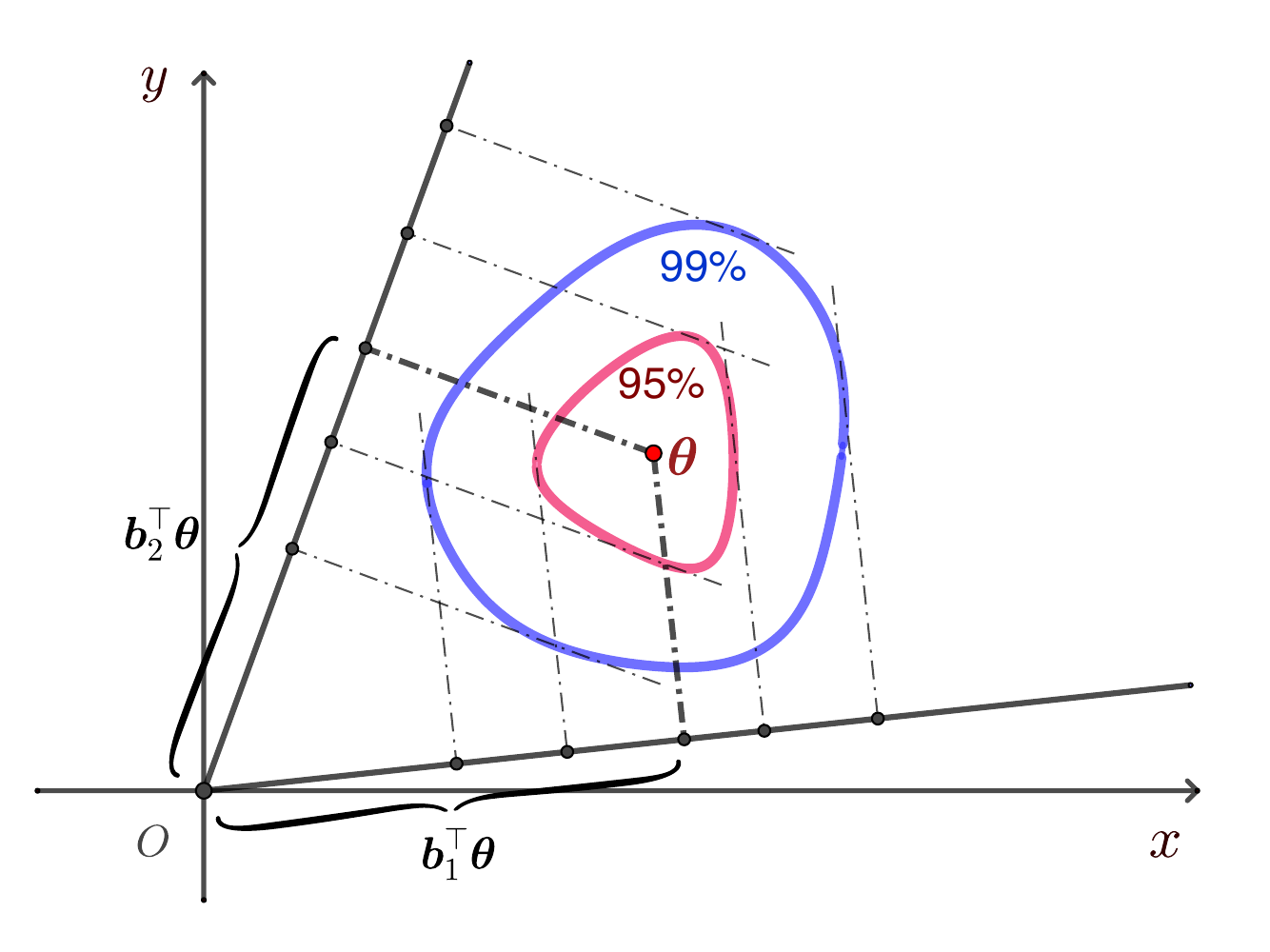}
\vspace*{-10pt}
  \caption{\footnotesize Illustration of obtaining simultaneous confidence intervals from confidence regions via projection. The plot shows $95\%$ and $99\%$ simultaneous confidence intervals for $\boldsymbol{b}_{i}^{\top}\boldsymbol{\theta}$ ($i=1,2$ with $\lVert \boldsymbol{b}_{i} \rVert_{2}=1$).}
  \label{fig:sim}
\end{figure}

\spacingset{1.65}

Another way to utilize multi-dimensional confidence regions is to obtain simultaneous confidence intervals for \(\boldsymbol{b}^{\top}\boldsymbol{\theta}\), given any \(\boldsymbol{b} \in \mathbb{R}^d\), by minimizing and maximizing \(\boldsymbol{b}^{\top}\boldsymbol{\theta}\) subject to \cref{eq:invert3}. A simultaneous confidence interval is one that provides joint coverage across multiple linear combinations of \(\boldsymbol{\theta}\). This means that the interval holds with a specified confidence level for all the directions \(\boldsymbol{b}\) considered. As illustrated in \cref{fig:sim}, confidence regions naturally induce simultaneous confidence intervals by projecting onto specific directions. 
Notably, when confidence regions are not accessible, practitioners often resort to using Bonferroni correction to obtain simultaneous confidence intervals from non-simultaneous ones, which tends to be overly conservative. In this sense, one can view our methods as providing a less conservative alternative to Bonferroni correction with a trade-off that the coverage is guaranteed only asymptotically as the confidence level approaches 100\% and it is not for all kinds of dependence structures. The empirical evidences so far suggest that is trade-off is wise for the common choices of confidence level of $99\%$ and even $95\%$ (see \cref{fig:comparetest}).

\setlength{\parskip}{5pt}

We compute these projection endpoints as support functions. When the positively weighted projected estimates share a common parameter center, central symmetry gives one endpoint by reflection; otherwise, the two directions are solved separately. Only connected coordinate components touched by the requested direction enter the optimization. Writing the standardized score and cutoff as \(G\) and \(c\), the equations \(\nabla G(\boldsymbol{z})=\eta\boldsymbol{b}\) and \(G(\boldsymbol{z})=c\) are the boundary KKT optimality conditions. For moderate-dimensional active components, Newton is the primary numerical method: it starts from a feasible radial boundary point and uses analytic score, gradient, and Hessian evaluations with a safeguarded merit-function line search. If Newton does not certify an endpoint, the first fallback uses numerically scaled SLSQP and, when needed, a Newton restart from the SLSQP point; the second uses quadratic-penalty continuation with optional Newton polishing. Every returned endpoint must pass the same checks on boundary error, positivity of \(\eta\), and the normalized KKT residual.

As an illustration, we simulate $m$ dependent studies for estimating \(\boldsymbol{\theta}\). Let \(\widehat{\boldsymbol{\theta}}^{(j)} =\bigl\{\widehat{\theta}_1^{(j)}, \dots,\widehat{\theta}_d^{(j)}\bigr\}^{\top}\) (\(j = 1, \dots, m\)) represent the estimator from the \(j\)-th study. For simplicity, we set \(\boldsymbol{\theta} = \boldsymbol{0}\) and generate:
\[
\spacingset{1}
\textstyle\bigl\{\widehat{\theta}_1^{(1)}, \dots, \widehat{\theta}_1^{(m)}, \dots \dots, \widehat{\theta}_d^{(1)}, \dots, \widehat{\theta}_d^{(m)}\bigr\}^{\top} \sim \mathcal{N}\bigl\{\boldsymbol{0}, \operatorname{diag}\bigl(\boldsymbol{M}_{\rho}, \dots, \boldsymbol{M}_{\rho}\bigr)\bigr\},
\resetspacing
\]
where \(\boldsymbol{M}_{\rho} = (1 - \rho) \boldsymbol{I}_m + \rho \mathbf{1} \mathbf{1}^{\top}\). Hence the between-study correlation is $\rho$, and the within-study correlation is 0.

\begin{figure}[tbp]\label{fig:contourplot1-4}
  \centering
  \subfloat[$\rho=0$]{%
    \includegraphics[width=.25\textwidth]{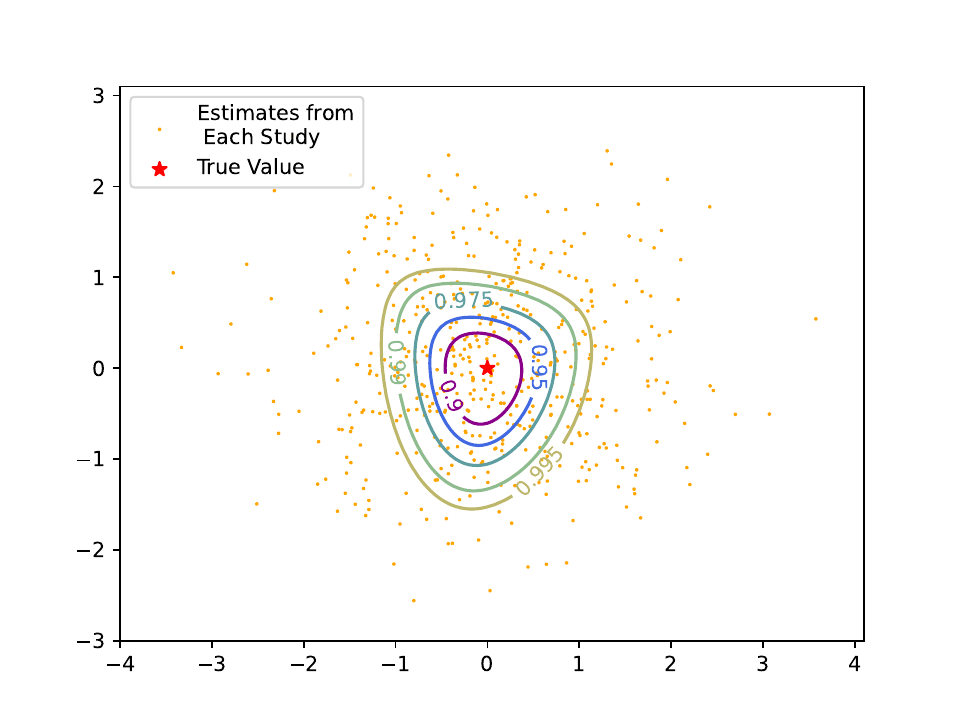}
  }\hspace*{-10pt}
  \subfloat[$\rho=0.3$]{%
    \includegraphics[width=.25\textwidth]{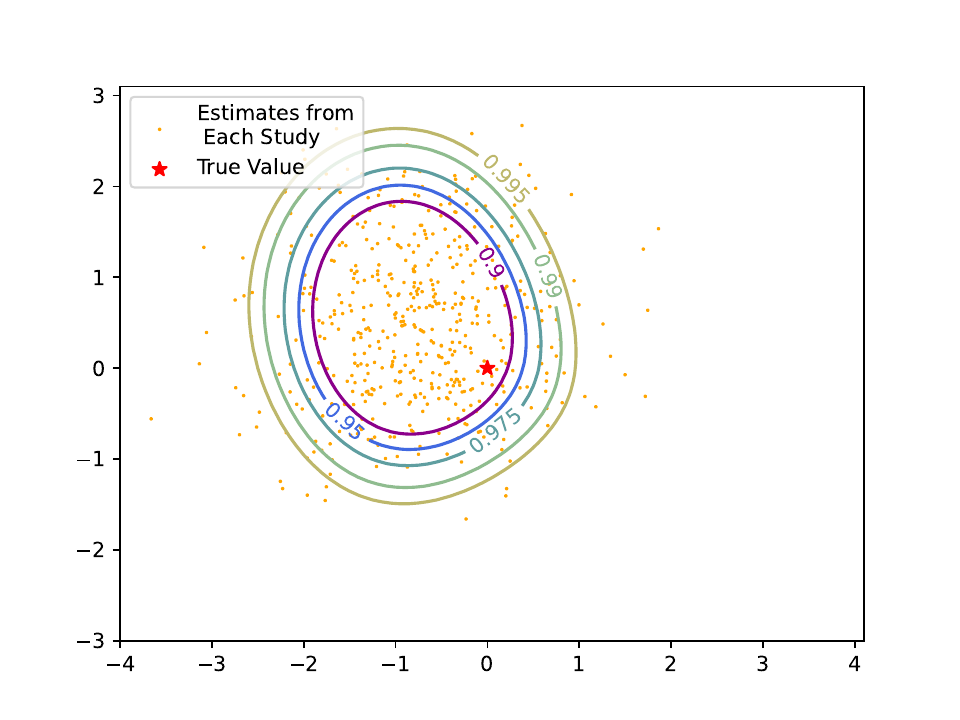}
  }\hspace*{-10pt}
  \subfloat[$\rho=0.6$]{%
    \includegraphics[width=.25\textwidth]{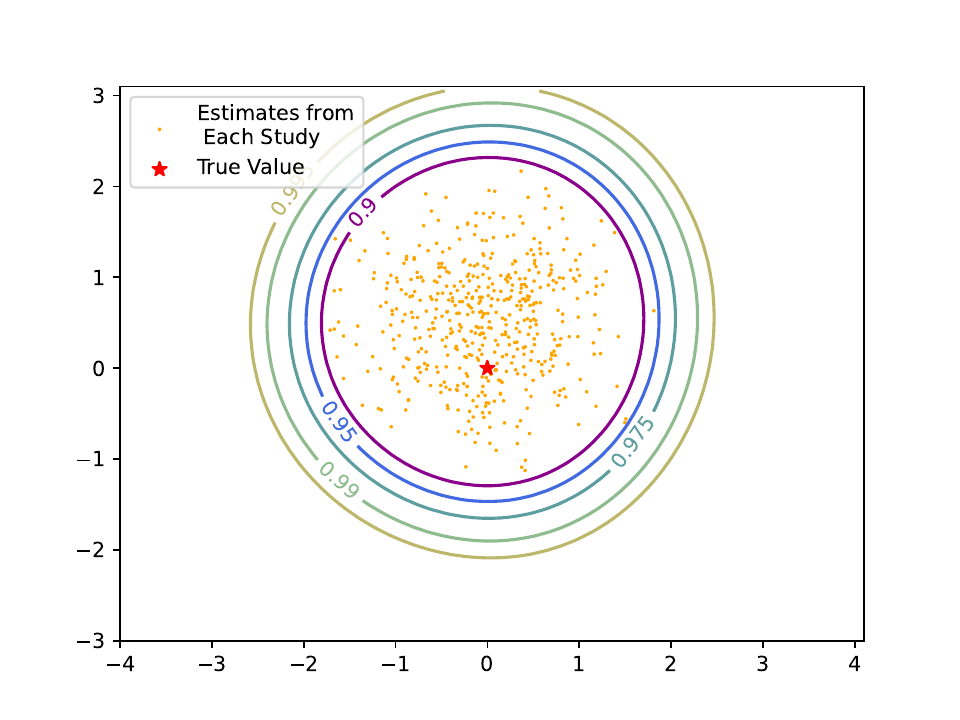}
  }\hspace*{-10pt}
  \subfloat[$\rho=0.9$]{%
    \includegraphics[width=.25\textwidth]{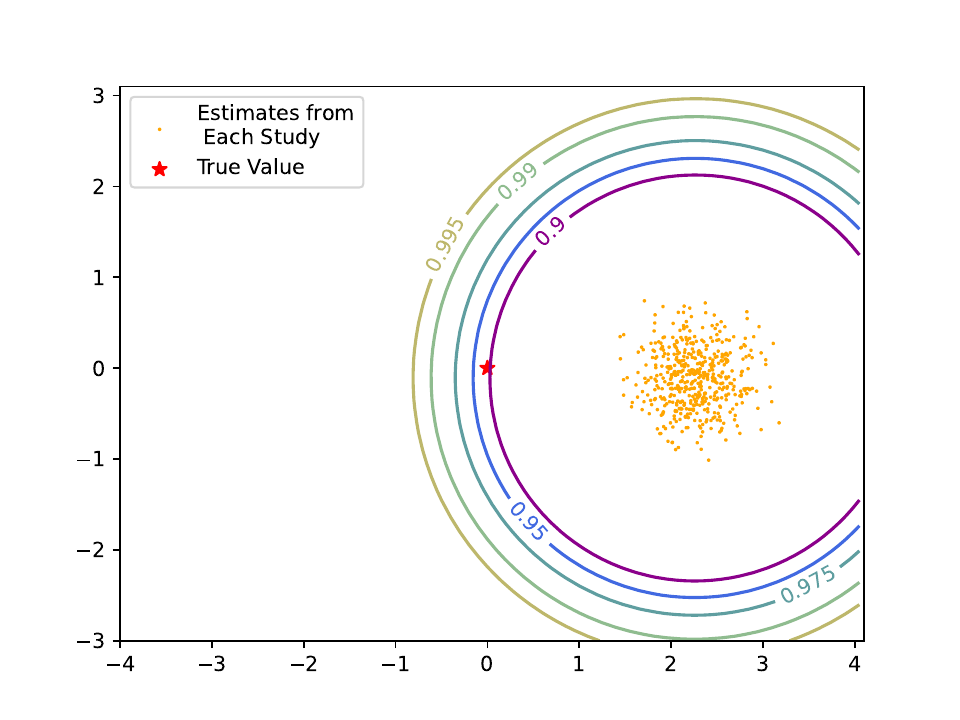}
    \label{fig:2dim9}
  }
  \caption{Contour plots of confidence regions from $2$-dimensional HCCT.}
  \label{fig:2dimcontour}
\end{figure}

We then apply HCCT approach with $P_j=I_d, j=1, \dots, m$. \cref{fig:2dimcontour} shows a single run with \(m = 500\), \(d = 2\), and \(\rho = 0, 0.3, 0.6, 0.9\), respectively. We observe that the confidence regions become larger as the correlation level increases, even though our approach does not involve incorporating correlations in the input or as part of the estimation process. This again suggests that the method is robust to correlation structures by adapting to them, a more efficient data-driven adjustment strategy than the conservative data-invariant level-adjustment approach of Bonferroni correction. For instance, when \(\rho = 0.9\), the individual estimates are often concentrated away from the true value. In \cref{fig:2dim9}, most estimates cluster around \((2.2, -0.1)\), while the true value of \(\boldsymbol{\theta}\) is \((0, 0)\). As a result, a larger confidence region is necessary to maintain 95\% coverage. This observation is consistent with the experimental results for \(d = 1\) shown in \cref{fig:1dwidthequi}. These observations also remind us the danger of using some center of a confidence region as a point estimator without taking into account its uncertainty.

\begin{figure}[tbp]
\vspace{-40pt}
  \centering
  \subfloat[$m=10$]{%
    \includegraphics[width=.33\textwidth]{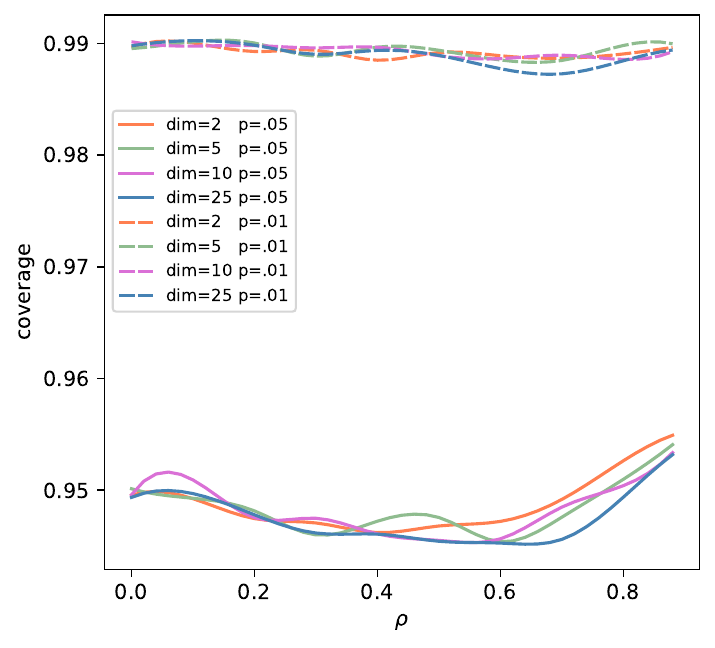}
  }\hspace*{15pt}
  \subfloat[$m=500$]{%
    \includegraphics[width=.33\textwidth]{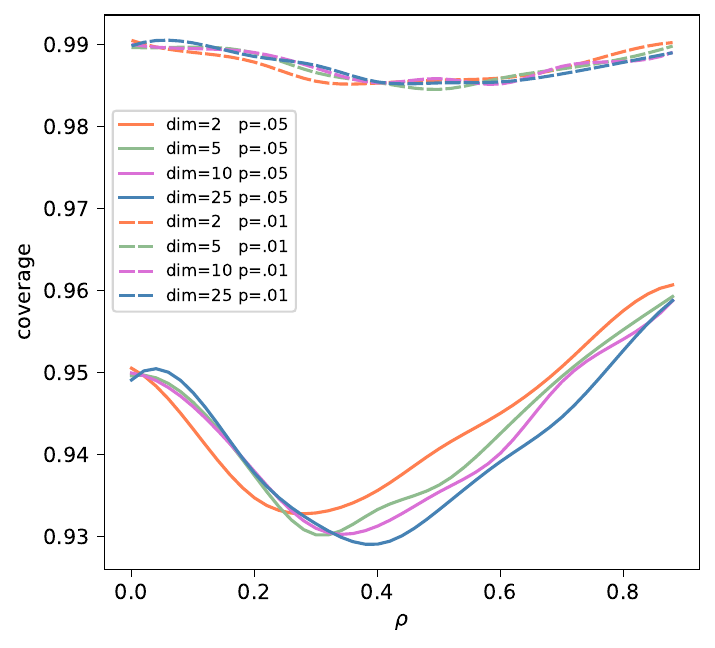}
  }
  \caption{Coverage of $d$-dimension confidence regions from HCCT.}
  \label{fig:coverco}
\end{figure}

We further examine the coverage of our constructed confidence regions in \cref{fig:coverco} with varying numbers of studies (\(m = 10, 500\)) and dimensions (\(d = 2, 5, 10, 25\)) across different levels of dependence \(\rho=0,0.1,\dots,0.9\). Specifically, the experimental results here are obtained from $1000$ different runs for each $\rho$, $m$ and $d$. In general, the behavior for \(d > 1\) is not significantly different from the univariate case (see \cref{fig:coverageequi}). All regions have essentially the nominal coverage at the $99\%$ level, though at the $95\%$ level, there is some small deterioration of  coverage when $m$ is large.  The fact that HCCT performs better at the 99\% level is consistent with our expectation from the nature of the tail approximation. The $U$-shape behavior in the amount of deterioration, as most visible at the $95\%$ level and with $m=500$, is also consistent with the fact that the Half-Cauchy approach is strictly valid when $\rho=0$ or $\rho=1$.
However, theoretically bounding the largest approximation error and locating the amount of dependence when it occurs are open problems.

\setlength{\parskip}{2pt}

\vspace{-5pt}
\subsection{Understanding and Dealing with Empty Confidence Sets}\label{sec:empty}

An important consideration is that the solution set of \eqref{eq:invert} or of  \eqref{eq:invert3}  can be empty when $\nu$ is Half-Cauchy or Pareto$(1,1)$ and $m>1$, a phenomenon that cannot occur when $\nu$ is Cauchy. To see this clearly, compare $T_{\mathrm{CCT}}$ of \eqref{eq:cct} with $T_{\mathrm{HCCT}}$ of \eqref{eq:hcct}, where $p_j$ is given by \eqref{eq:pjunknownsigma}, by explicating all three terms as functions of $\theta$, that is
\begin{equation}\label{eq:compare}
\spacingset{1}
\textstyle T_{\mathrm{CCT}}(\theta) = \sum_{j=1}^m w_{j} \cot(\pi p_j(\theta)), \quad  T_{\mathrm{HCCT}}(\theta) = \sum_{j=1}^m w_{j} \cot\bigl\{\frac{\pi}{2}p_j(\theta)\bigr\},  \quad p_j(\theta)= 2\bigl\{1 -  F^{(j)}\bigl(\frac{\lvert \widehat{\theta}_j- \theta \rvert}{\widehat{\sigma}_j}\bigr)\bigr\},
\resetspacing
\end{equation}
where $F^{(j)}$ is the CDF of a $t$ or normal distribution. Consequently, $p_j(\widehat\theta_j)=1$ for any $j$, which means $T_{\mathrm{CCT}}(\widehat\theta_j)=-\infty$ because $\lim_{x\uparrow \pi}\cot(x)=-\infty$. Hence any confidence region in the form of $C_K(\theta)=\{\theta: T_{\mathrm{CCT}}(\theta)\le K\}$ must contain all $\widehat{\theta}_j$'s, regardless of the value of cut-off $K<\infty$; we have seen two such examples in \cref{fig:cctbad}.

In contrast, because $\cot\left(\frac{\pi}{2}p_j(\theta)\right)\ge 0$ for all $\theta$, we see that $T_{\mathrm{HCCT}}(\theta)\ge 0$, and indeed it is possible for $\min_{\theta}T_{\mathrm{HCCT}}(\theta)=T_{\min}>K$, in which case, the set $C_K^{+}(\theta)=\{\theta: T_{\mathrm{HCCT}}(\theta)\le K\}$ will be empty. In particular, because $F^{(j)}(x) \ge F_{\mathrm{Cauchy}}(x)=\pi^{-1}\arctan(x)+0.5$ when $x\ge 0$, we have the following lower bound 
\begin{equation}\label{eq:Tbound}
\spacingset{1}
\textstyle T_{\mathrm{HCCT}}(\theta)\geq \sum_{j=1}^{m} w_j \cot \bigl\{\frac{\pi}{2} -\arctan\bigl( \frac{\lvert \widehat{\theta}_j - \theta \rvert}{\widehat{\sigma}_j}\bigr)\bigr\}= \sum_{j=1}^{m}\frac{w_{j}}{\widehat{\sigma}_{j}}\lvert \widehat{\theta}_{j}-\theta\rvert
\ge \sum_{j=1}^{m}\frac{w_{j}}{\widehat{\sigma}_{j}}\lvert \widehat{\theta}_{j}-\widehat\theta_{\mathrm{med}}\rvert,
\resetspacing
\end{equation}
where $\widehat\theta_{\mathrm{med}}$ is the median of the discrete distribution on $\{\widehat{\theta}_j, j=1, \ldots, m\}$ with $\mathbb{P}(\widehat{\theta}=\widehat{\theta}_j) \propto w_j/\widehat{\sigma}_j$. For weighted Euclidean-distance lower bounds, the multivariate analogue uses the weighted geometric median \citep{brimberg1995fermat}.

\spacingset{2}
\setlength{\parskip}{5pt}

The inequality \cref{eq:Tbound} is telling, since the lower bound is a measure of inconsistency among the $m$ studies, taking into account the weights.  Indeed, $T_{\min}$ is the smallest possible HCCT statistic against a common null from the $m$ studies, that is, by fitting the null to the minimizer of $T_{\mathrm{HCCT}}(\theta)$, $\theta^{\ast}$. If this fitted null still can be rejected at the level $p$, then what is being rejected at the level is not really the null value per se, but rather the existence of a common value across the $m$ studies at all. The increased probability for the occurrence of an empty set with the increased significance level $p$ can be understood intuitively from John Tukey's notion of ``outerval", the complement to the confidence interval. That is, constructing a confidence interval of $\theta$ for further consideration should be described as   ``constructing \textit{outerval} to eliminate implausible values as declared by our chosen criterion'', as discussed in \citet{meng_2022}. The larger the significance level $p$, the less stringent the criterion for implausibility, and hence the higher chance to declare that nothing is acceptable.

While an empty set is reasonable for ensuring declared  confidence coverage in repeated experiments, it is problematic in real-data analyses. To address this, we leverage the flexibility of HCCT (and EHMP) in assigning weights to different studies and propose a general adaptive procedure. Specifically, we can mitigate the problem by identifying studies that contribute most to the inconsistency and appropriately adjusting their weights in the combination test, potentially reducing some to zero.

For example, we can set $w_j=0$ if the largest change in the lower bound in \cref{eq:Tbound} occurs when we drop the $j$-th study, and continue such a process until a nonempty confidence set is found. Searching for a nonempty solution can only increase the (conditional) coverage. This intuition is formalized in the following result.
\begin{proposition}\label{thm:adaptiveprocedure}
Consider $\mathcal{W}=\{ \boldsymbol{w}=(w_{1},\cdots,w_{m})^{\top}: w_{j}\geq 0 \text{ for } 1\leq j\leq m,\  w_{1}+\cdots+w_{m}=1\}$ as the class of weight vectors. For any $\boldsymbol{w}\in \mathcal{W}$, let $z_{\boldsymbol{w}}$ be a weight-dependent threshold such that 
\(
  \mathbb{P}\bigl( T_{\boldsymbol{w}} \leq z_{\boldsymbol{w}}\bigr)\geq 1-p,
\)
where $T_{\boldsymbol{w}}$, defined by the left-hand-side of \cref{eq:invert} or \cref{eq:invert3} for HCCT or EHMP, also depends on the weight vector $\boldsymbol{w}$. Let $\tau$ be any stopping time for the random sequence: 
$
T_{\boldsymbol{w}^{(0)}},T_{\boldsymbol{w}^{(1)}},T_{\boldsymbol{w}^{(2)}},\ldots,
$ 
where $\boldsymbol{w}^{(k)}$ can be chosen adaptively based on the previous sequence and any data or statistic for individual studies for $k\geq 1$.
Then the following procedure produces a confidence region with at least $(1-p)$ coverage:
\begin{itemize}
  \item Start with a prespecified $\boldsymbol{w}^{(0)}\in \mathcal{W}$ and obtain the solution set $R^{(0)}$ of $T_{\boldsymbol{w}^{(0)}}\leq z _{ \boldsymbol{w}^{(0)} }$.
  \item For $1\leq k\leq \tau$, we choose $\boldsymbol{w}^{(k)}\in \mathcal{W}$ and get the solution set $R^{(k)}$ of $T_{\boldsymbol{w}^{(k)} }\leq z _{ \boldsymbol{w}^{(k)} }$.
  \item Report $R^{\ast}=\bigcup_{k=0}^{\tau}R^{(k)}$.
\end{itemize}
\end{proposition}

\resetspacing

\vspace*{10pt}
As an application of \cref{thm:adaptiveprocedure}, let $\tau$ be the first time a nonempty solution is found, assumed finite almost surely.  Then by construction, $R^{(k)}=\emptyset$ for all $k<\tau$, implying $R^{(\tau)}=R^{\ast}$.  Therefore, $R^{(\tau)}$, as an adaptive generating procedure, will have at least $1-p$ coverage. Intuitively, an empty solution set represents an extreme case where conditional coverage is zero, and the procedure addresses this by enhancing conditional coverage. Epistemically, we can view such an adaptive procedure either as removing studies that are too different to be considered as studying the same (preconceived) $\theta$ or as refining the meaning of $\theta$ as a synergistic summary of all the studies thereby retained.

We will empirically illustrate this procedure in the real-data application of \cref{sec:networkmeta}. In the simulation studies reported in \cref{sec:divideac}, empty sets occurred in approximately 0.5--3\% of the runs at the 95\% confidence level and 0.1--0.6\% at the 99\% level; there we do not use this adaptive procedure, but simply report empty sets as non-coverage confidence regions with zero volume.

\vspace*{5pt}

\vspace*{-10pt}
\section{A Divide-and-Combine Strategy for Mean Estimation in Arbitrary Dimensions}\label{sec:divideac}
\vspace*{0pt}
\subsection{Leveraging Hotelling's $T^2$ but Circumventing Its Curse of Dimension}
\vspace*{-5pt}
Many applications in practice involve hypothesis tests and point or set estimators for the mean vector \(\boldsymbol{\theta}\) from multivariate normal samples with an unknown covariance matrix \(\boldsymbol{\Sigma}\). A classical approach to this problem is Hotelling's \(T^2\)-test, which provides an ellipsoidal confidence region for \(\boldsymbol{\theta}\). However, Hotelling's test requires estimation of the full covariance (or precision) matrix, which poses significant numerical and statistical challenges when the dimension of $\theta$ can be arbitrarily large \citep{bai1996effect,pan2011central}. A considerable body of literature has focused on projection tests \citep{lopes2011more,srivastava2016raptt,Liu02012024} or covariance matrix estimation in high dimensions \citep{bickel2008covariance,cai2012adaptive,cai2016estimating,avella2018robust,lam2020high,liu2020minimax,goes2020robust}. Various approaches have been proposed to address these challenges, including the use of diagonal or block-diagonal matrices \citep{wu2006multivariate,srivastava2008test,tony2014two,dong2016shrinkage,feng2017composite}, U-statistics \citep{he2021asymptotically,li2023finite}, and regularization procedures \citep{chen2011regularized,li2020adaptable}.

\spacingset{2}

HCCT or EHMP provides a divide-and-combine strategy that circumvents the need for estimating the \textbf{\emph{full}} covariance matrix.  A key advantage of our method is that the resulting confidence regions are guaranteed to be convex and bounded, even when the sample size is smaller than the dimension \(d\), which contrasts with Hotelling's test, which is not applicable when the sample size goes below \(d\). Moreover, our approach can potentially yield smaller confidence regions compared to Hotelling's test, offering further practical benefits in some cases.

Our method leverages the same set of samples to construct \(m\) virtual sub-studies, where we estimate \(\boldsymbol{P}_j \boldsymbol{\theta}\) for \(j = 1, \dots, m\) using linear transformations of the original data. The matrices \(\boldsymbol{P}_j\) are \(d_j \times d\) matrices, where \(d_j\) can be much smaller than $d$. The estimator in each sub-study is then derived using the Student's \(t\)-test (for \(d_j = 1\)) or Hotelling’s \(T^2\)-test (for \(d_j \geq 2\)). These estimators are generally dependent, but HCCT or EHMP is exactly designed for such situations, permitting us to convert them into approximately valid confidence regions for \(\boldsymbol{\theta}\).

By \cref{thm:connectivity2}, the resulting region is convex and bounded if the positively weighted covariance estimates are positive definite, the corresponding row spaces of the $P_j$'s span \(\mathbb{R}^d\), and the sample size $n$ is at least \(c_0+\max_j d_j\), where $c_0=1$ for EHMP and $c_0=2$ for HCCT. With $d_j=1$, this permits $n=c_0+1\le3$ regardless of $d$; at least $m\ge d$ sub-studies are necessary, and their directions must span \(\mathbb{R}^d\). In contrast, the \(d\)-dimensional Hotelling’s test, corresponding to $m=1$ and $P_1=I_d$, requires $n\ge d+1$.

\spacingset{2}
\setlength{\parskip}{13pt}
In general, by choosing $d_j < < d$, we can turn an arbitrarily high dimensional problem into many low dimensional or even univariate problems, which are typically much easier to handle both statistically and computationally.  
Beyond simple coordinate projections, \(\boldsymbol{P}_j\)'s can also be derived from random projections or directions informed by a principal component analysis of the data. As permitted by \cref{thm:connectivity2}, redundancy due to non-orthogonal projections does not invalidate the convexity or the boundedness conclusions, allowing the number of (projected) sub-studies \(m\) to potentially exceed the dimension \(d\). Moreover, this divide-and-combine method remains effective even if the underlying distribution \(\mathcal{N}(\boldsymbol{\theta}, \boldsymbol{\Sigma})\) is degenerate with a low-rank \(\boldsymbol{\Sigma}\), provided that the sub-study covariance matrices \(\boldsymbol{\Sigma}_j\) are full rank. 

However, despite the versatility and robustness of our approach, it is desirable to choose $\boldsymbol{P}_{j}$'s that lead to smaller confidence regions, while maintaining the scalability and computational efficiency. Much research is needed to understand the impact of the choices of $m$ and \(\{\boldsymbol{P}_j, j=1, \ldots, m\}\) on the statistical and computational efficiencies of our method. We invite all interested to study and explore with us the full potential of this approach, and to seek optimal compromise between statistical efficiency, which tends to prefer smaller $m$ and larger $d_j$'s, and computational efficiency, which typically calls for larger $m$ and smaller $d_j$'s.

Generally speaking, divide-and-conquer methods fall into two broad classes. One divides a large dataset into many independent subsets, analyzes each for the whole problem, and combines the results via rules based on independence assumptions \citep{chen2021divide}. The other divides the problem itself into sub-problems, such as by breaking down high dimensions \citep{sabnis2016divide,gao2023divide}. Our divide-and-combine strategy belongs to the second class: it breaks the estimation problem into sub-problems via projections, using \textit{all} the data for each. The resulting modularized solutions likely have complex dependence, since they are derived from the same data. This is where HCCT, EHMP, or other dependence-resilient combination rules become handy and powerful. Using all data for each sub-problem also gives us a better chance to retain more statistical efficiency.

\spacingset{1.8}
\vspace{-15pt}
\subsection{Simulation Study with Normal Samples}\label{sec:highdimnormal}
\vspace{-15pt}
\begin{figure}[tbp]
\vspace{-40pt}
  \centering
  \subfloat[$d_{0}=1$; $\theta_{1}$ and $\theta_{51}$]{%
    \includegraphics[width=.245\textwidth]{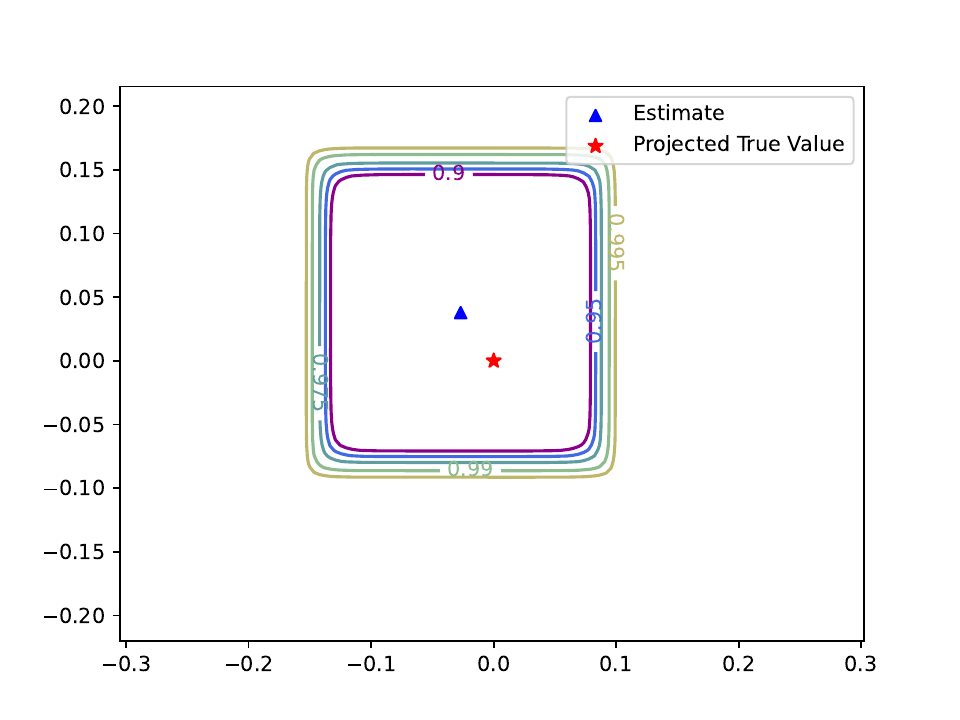}\label{fig:hdme-a}
  }\hspace*{-15pt}
  \subfloat[$d_{0}=5$; $\theta_{1}$ and $\theta_{51}$]{%
    \includegraphics[width=.245\textwidth]{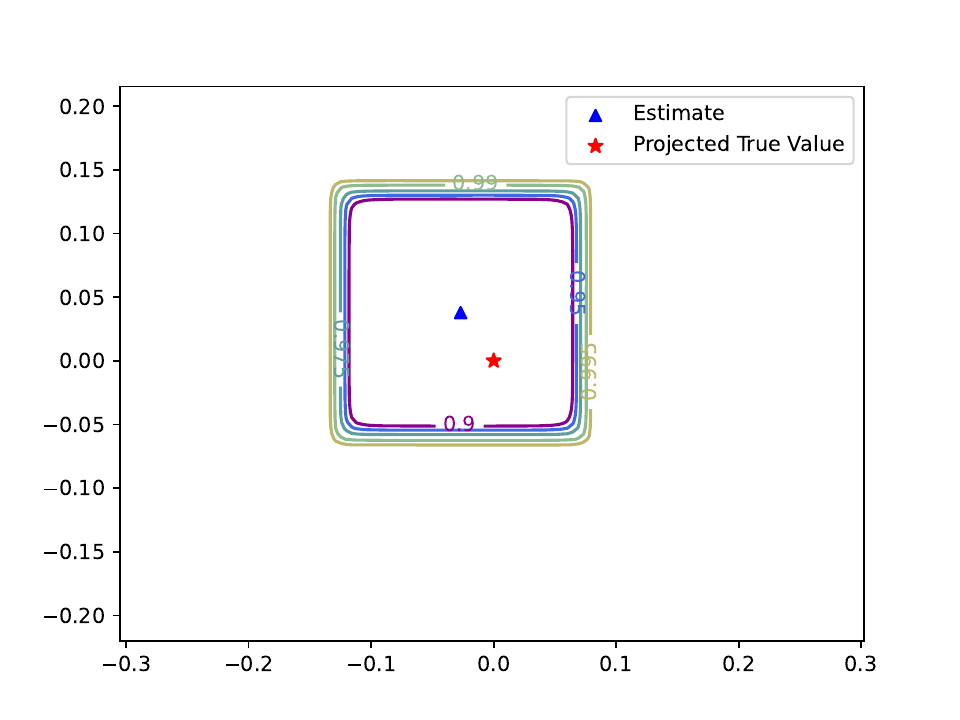}\label{fig:hdme-b}
  }\hspace*{-15pt}
  \subfloat[$d_{0}=25$; $\theta_{1}$ and $\theta_{51}$]{%
    \includegraphics[width=.245\textwidth]{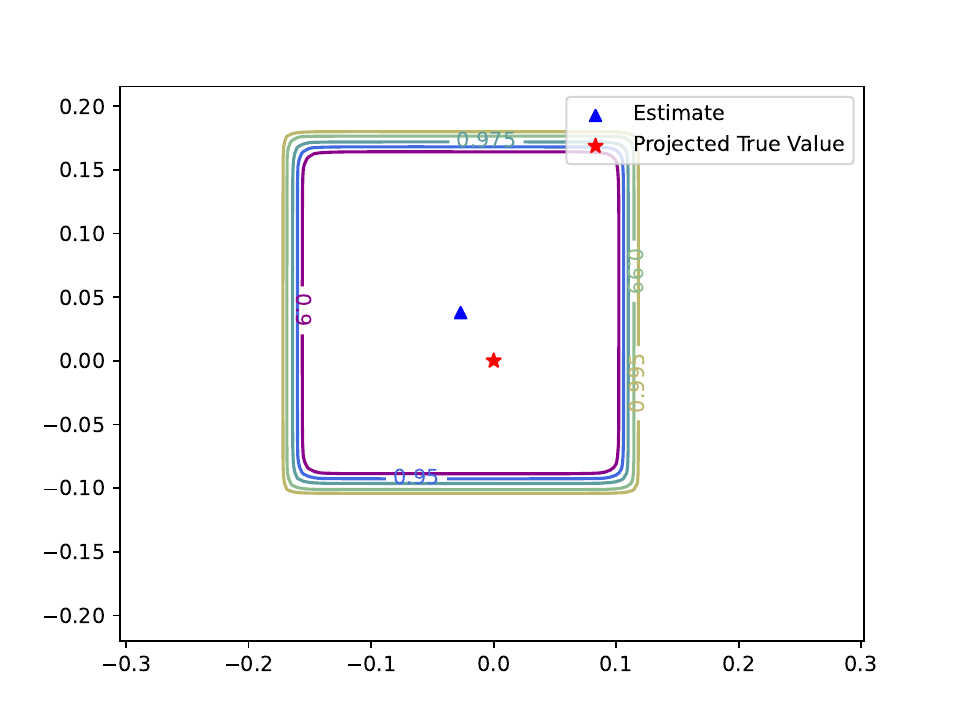}\label{fig:hdme-c}
  }\hspace*{-15pt}
  \subfloat[$d_{0}=100$; $\theta_{1}$ and $\theta_{51}$]{%
    \includegraphics[width=.245\textwidth]{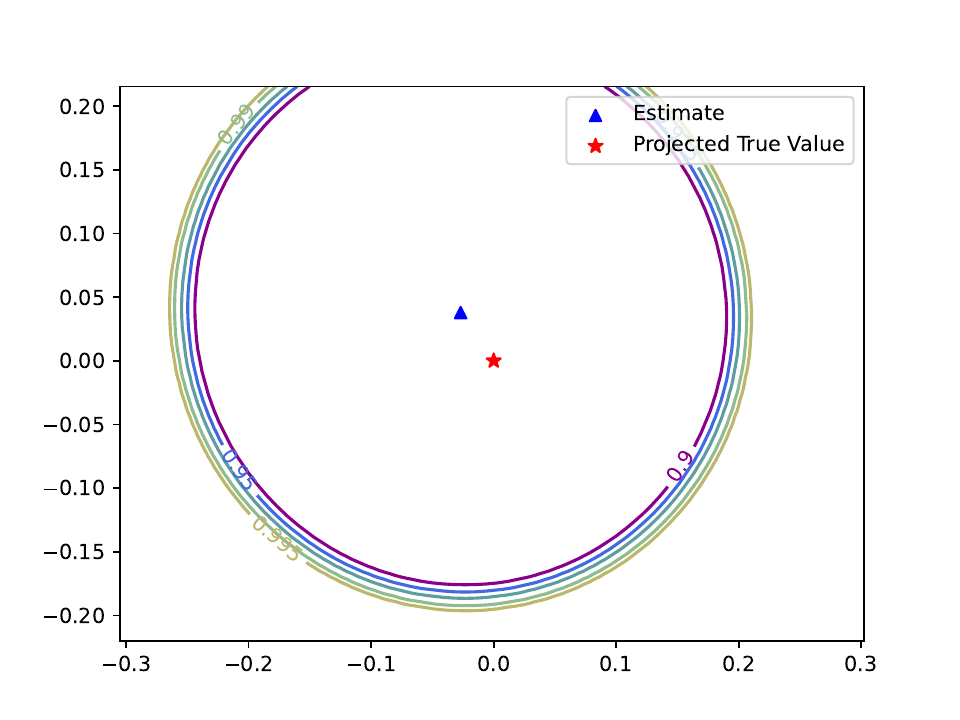}\label{fig:hdme-d}
  }\\
  \vspace*{-10pt}
  \subfloat[$d_{0}=1$; $\theta_{1}$ and $\theta_{2}$]{%
    \includegraphics[width=.245\textwidth]{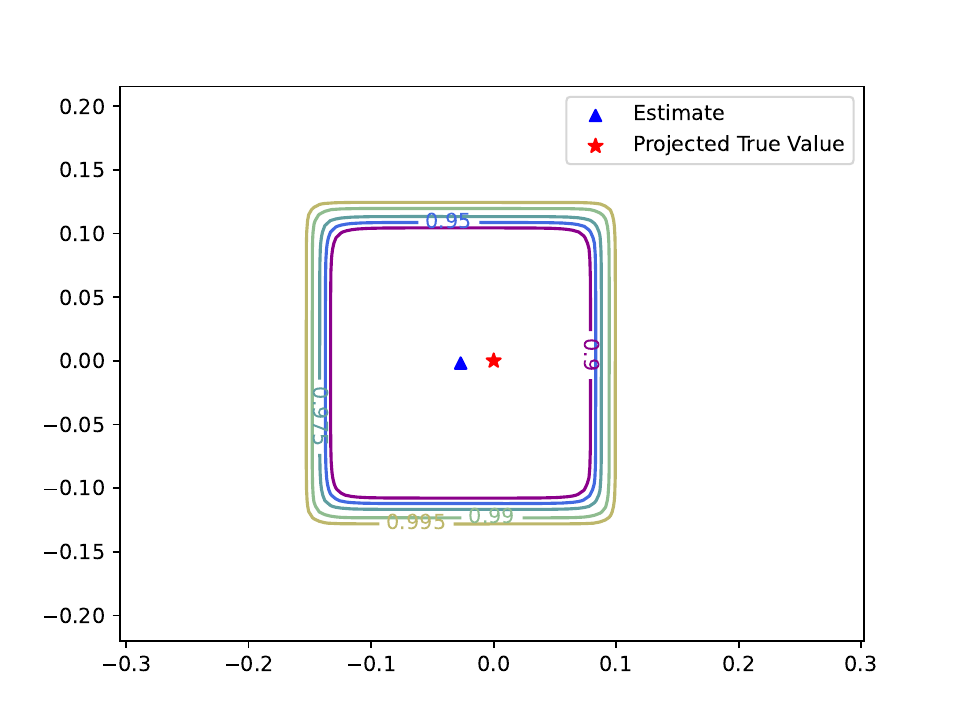}\label{fig:hdme-e}
  }\hspace*{-15pt}
  \subfloat[$d_{0}=5$; $\theta_{1}$ and $\theta_{2}$]{%
    \includegraphics[width=.245\textwidth]{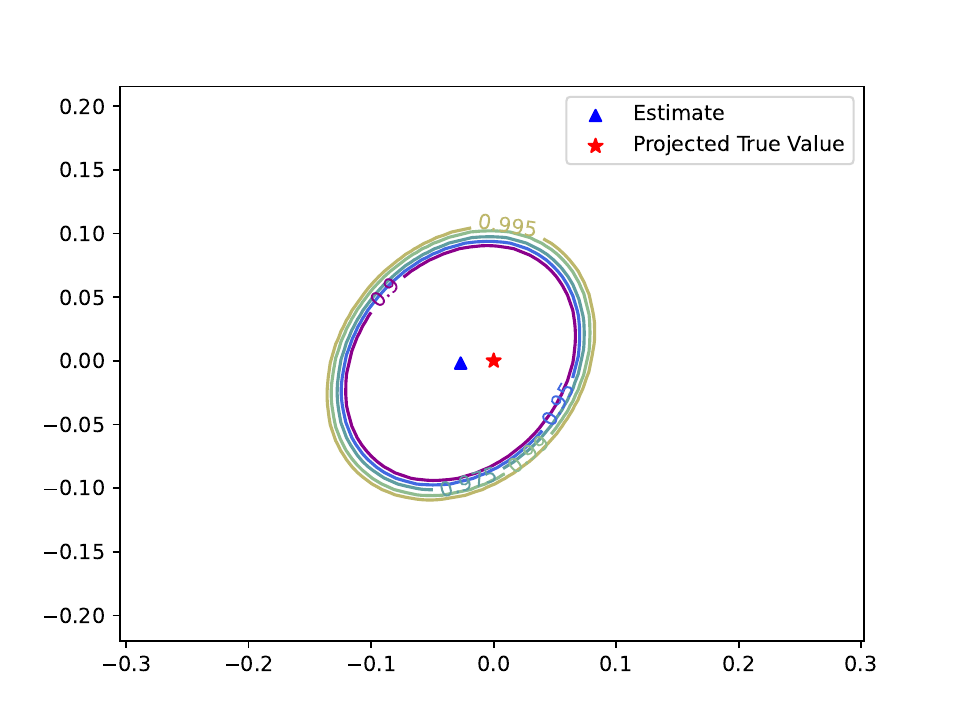}\label{fig:hdme-f}
  }\hspace*{-15pt}
  \subfloat[$d_{0}=25$; $\theta_{1}$ and $\theta_{2}$]{%
    \includegraphics[width=.245\textwidth]{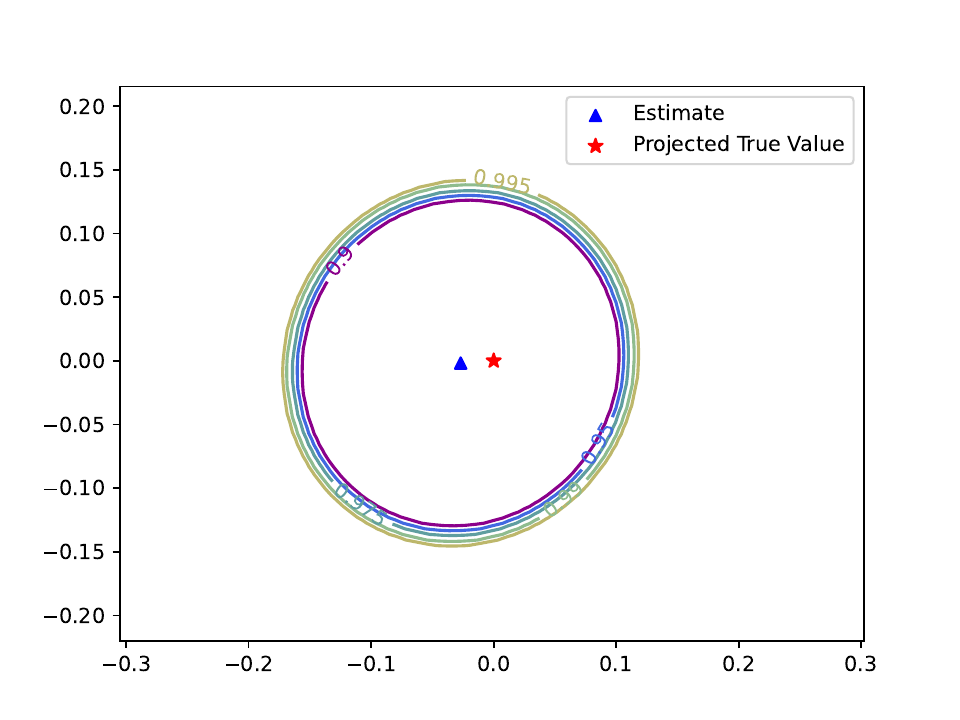}\label{fig:hdme-g}
  }\hspace*{-15pt}
  \subfloat[$d_{0}=100$; $\theta_{1}$ and $\theta_{2}$]{%
    \includegraphics[width=.245\textwidth]{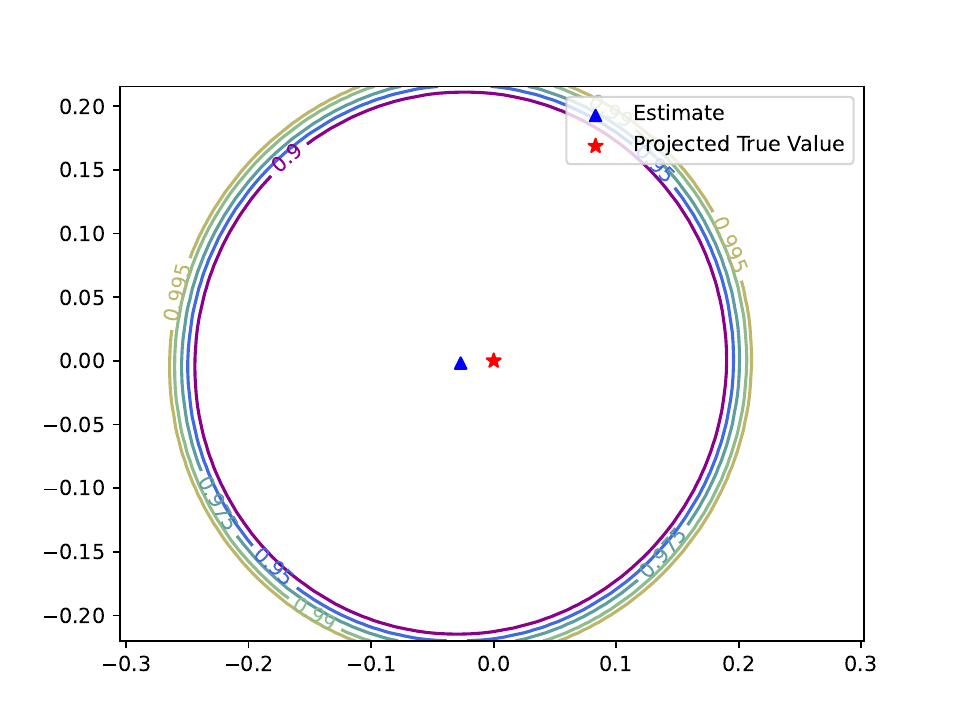}\label{fig:hdme-h}
  }
  \caption{\footnotesize 2d slices of confidence regions passing through the \textit{point estimate} with varying $d_{0}$ in the multivariate \textit{normal} study.  Project true values are the true mean value projected to the respective slices.}\label{fig:hdme}
\end{figure}

For our first simulation study, we generate \(n\) i.i.d. samples \(\boldsymbol{X}_1, \dots, \boldsymbol{X}_n \in \mathbb{R}^d\) from the ideal distribution \(\mathcal{N}(\boldsymbol{\theta}, \boldsymbol{M}_\rho)\), where \(\boldsymbol{\theta} = \boldsymbol{0}\) and \(\boldsymbol{M}_\rho = (1-\rho) \boldsymbol{I}_d + \rho \mathbf{1} \mathbf{1}^\top\). Our goal is to construct confidence regions for \(\boldsymbol{\theta}\) using the sample only; that is, without using any knowledge about \(\boldsymbol{M}_\rho\). We apply HCCT with \(\boldsymbol{P}_j\) being coordinate projections, i.e., we fix \(1 \leq d_0 \leq  d\), and split the \(d\)-dimensional study evenly into multiple non-overlapping sub-studies. Letting \(\boldsymbol{X}_i = (X_{i1}, \dots, X_{id})^\top\) and \(d = m d_0 - r\),  where \(0 \leq r \leq d_{0}-1\), we observe 
\(
\boldsymbol{P}_j \boldsymbol{X}_i = (X_{i,k_{j-1} + 1}, \dots, X_{i,k_j})^\top
\) with 
\( 
k_j = \min \{j d_{0}, d\}
\)
for \(i = 1, \dots, n\) and \(j=0,\dots,m\), which are i.i.d. from \(\mathcal{N}\bigl(\boldsymbol{P}_j \boldsymbol{\theta}, \boldsymbol{P}_j \boldsymbol{M}_\rho \boldsymbol{P}_j^\top\bigr)\) in the \(j\)-th sub-study for \(j = 1, \dots, m = \lceil d / d_0 \rceil\). We then conduct Hotelling's \(T^2\)-test for \(\boldsymbol{P}_j \boldsymbol{\theta}\) in each sub-study, and combine the results via HCCT.

For simplicity, we fix \(\rho = 0.6\), \(d = 100\), \(n = 1000\), and \(d_0 = 1, 5, 25, 100\). We repeat the experiments $2000$ times with a significance level of $0.05$ and find that the coverage of the confidence regions is $0.944$, $0.953$, $0.945$, and $0.956$, respectively, empirically validating our method regardless of the choice of $d_0$ in this ideal case.

\cref{fig:hdme} shows the intersection of an obtained confidence region with a plane passing through the same point estimate, using the same set of samples. In particular, \(d_0 = 100\) corresponds to Hotelling’s \(T^2\) test for the original \(d\)-dimensional problem. When the two axes in the plot are from different sub-studies (\cref{fig:hdme-a,fig:hdme-b,fig:hdme-c,fig:hdme-e}), the contour resembles squares but with rounded corners.
In contrast, when the two axes are from the same sub-study (\cref{fig:hdme-d,fig:hdme-f,fig:hdme-g,fig:hdme-h}), the contour captures the elliptical nature of the Hotelling $T^2$ distribution. 

\spacingset{1.6}
\setlength{\parskip}{2pt}

As the dimension of the sub-studies \(d_0\) increases, we have fewer sub-studies but need to estimate more entries from the unknown covariance matrix \(\boldsymbol{\Sigma}\) to compute Hotelling’s \(T^2\) statistics for each sub-study. For \(d_0 = 1\), only the variances are estimated, and we rely entirely on the dependence-resilient property of HCCT to obtain valid confidence regions. For \(d_0 = d\), there is a single sub-study where the full covariance matrix is estimated and utilized by Hotelling’s \(T^2\) statistic. 

The patterns in \cref{fig:hdme} do not contradict Hotelling’s optimality: Hotelling’s \(T^2\) is uniformly most powerful among invariant tests and admissible, whereas our blockwise procedure is not invariant to transformations mixing blocks and may therefore gain power for block-aligned alternatives while losing power somewhere else. Nor do smaller coordinate projections establish smaller \(d\)-dimensional volume or uniformly greater power. For example, a cube and its inscribed sphere have the same coordinatewise widths, yet their volume ratio grows rapidly with dimension because of the cube’s corners. Nevertheless, an intermediate \(d_0\) may balance stable low-dimensional covariance estimation against the costs of combining many sub-tests, while avoiding high-dimensional covariance estimation and omnibus testing. Determining whether such a choice improves volume or power remains open. Finally, because dependence resilience relies on a tail approximation, validity at a fixed positive significance level is only approximately guaranteed, so efficiency must be interpreted together with empirical coverage.

\vspace*{-5pt}
\subsection{Simulation Study with Log-Normal Samples}\label{sec:highdimlognormal}

\begin{figure}[tbp]
\vspace{-40pt}
  \centering
  \subfloat[$d_{0}=1$; $\theta_{1}$ and $\theta_{51}$]{%
    \includegraphics[width=.245\textwidth]{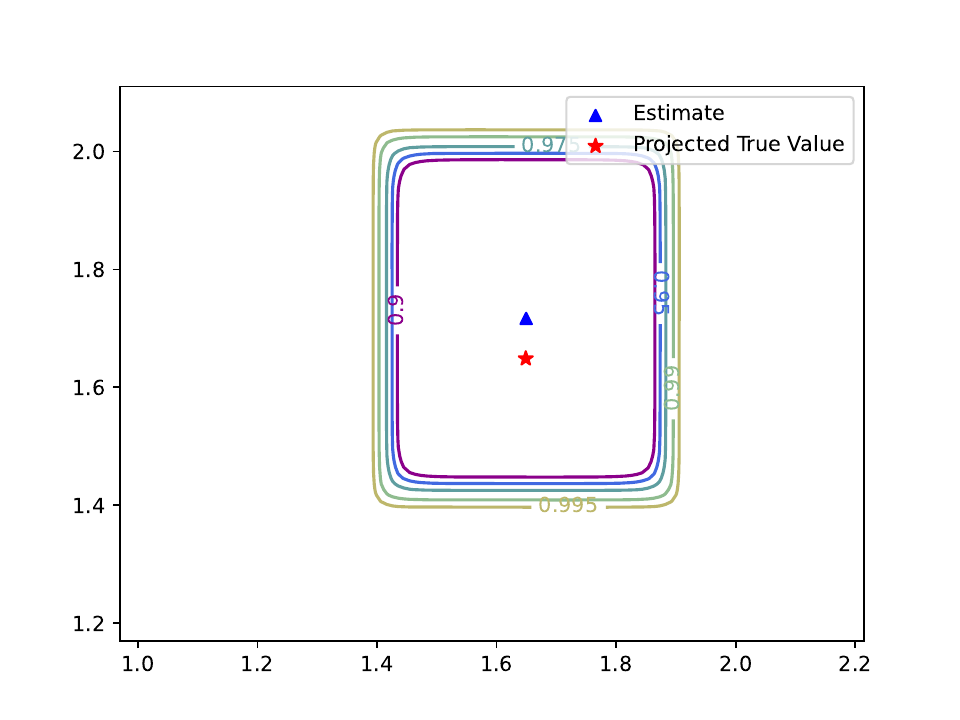}\label{fig:hdmelog-a}
  }\hspace*{-15pt}
  \subfloat[$d_{0}=5$; $\theta_{1}$ and $\theta_{51}$]{%
    \includegraphics[width=.245\textwidth]{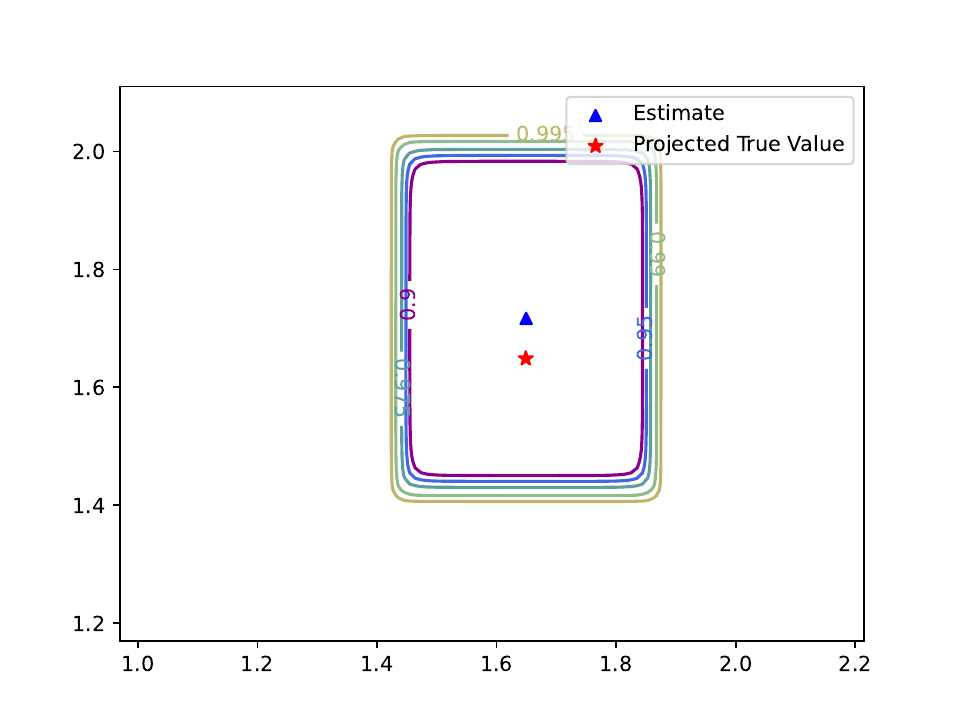}\label{fig:hdmelog-b}
  }\hspace*{-15pt}
  \subfloat[$d_{0}=25$; $\theta_{1}$ and $\theta_{51}$]{%
    \includegraphics[width=.245\textwidth]{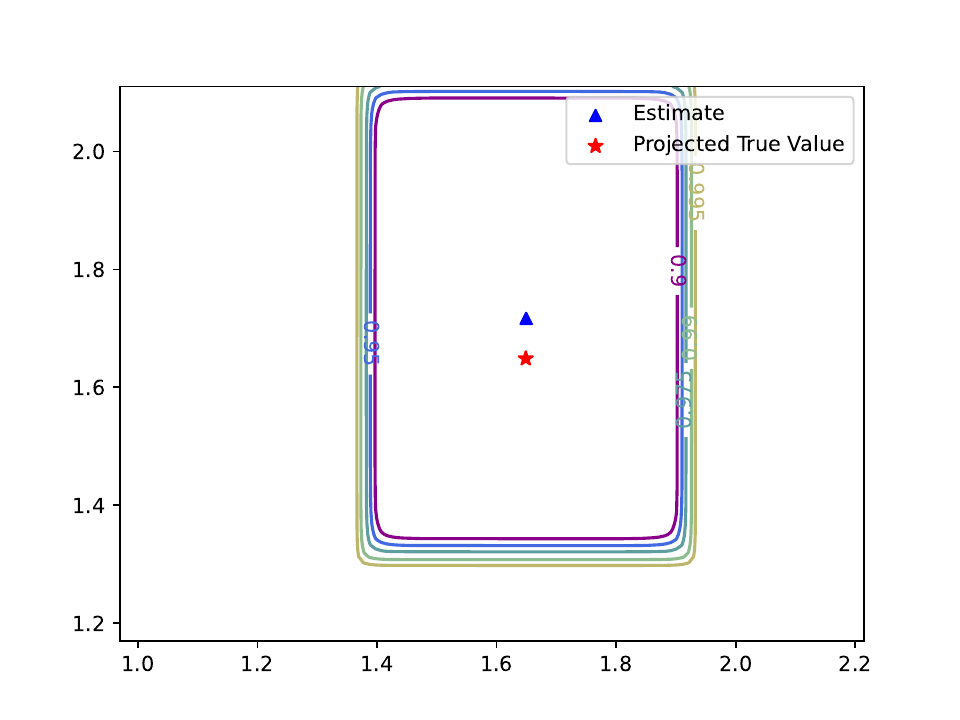}\label{fig:hdmelog-c}
  }\hspace*{-15pt}
  \subfloat[$d_{0}=100$; $\theta_{1}$ and $\theta_{51}$]{%
    \includegraphics[width=.245\textwidth]{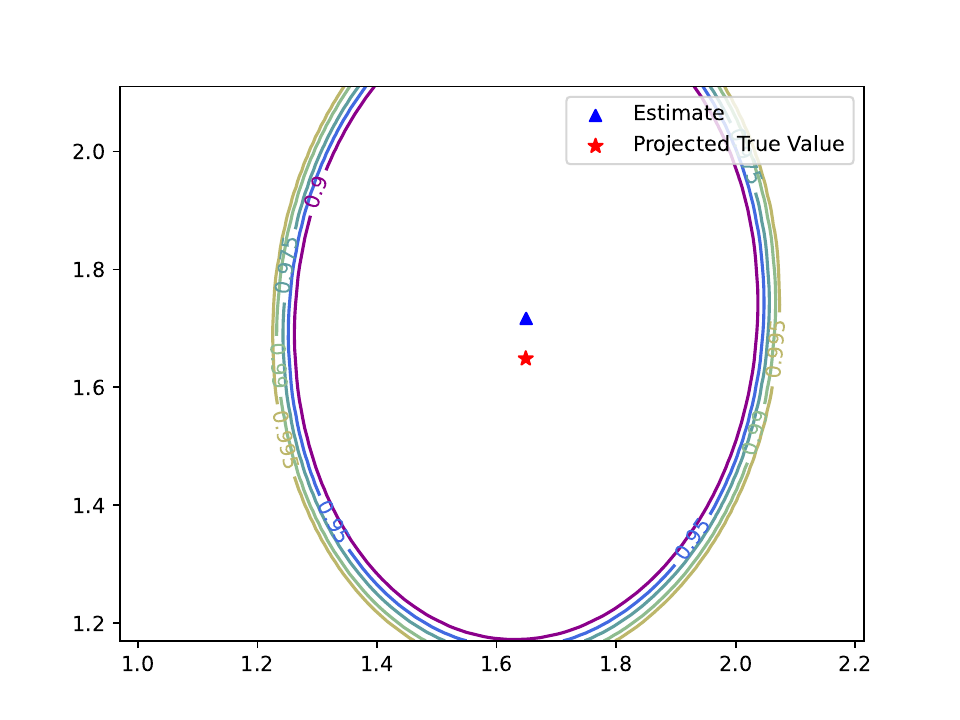}\label{fig:hdmelog-d}
  }\\
  \vspace*{0pt}
  \subfloat[$d_{0}=1$; $\theta_{1}$ and $\theta_{2}$]{%
    \includegraphics[width=.245\textwidth]{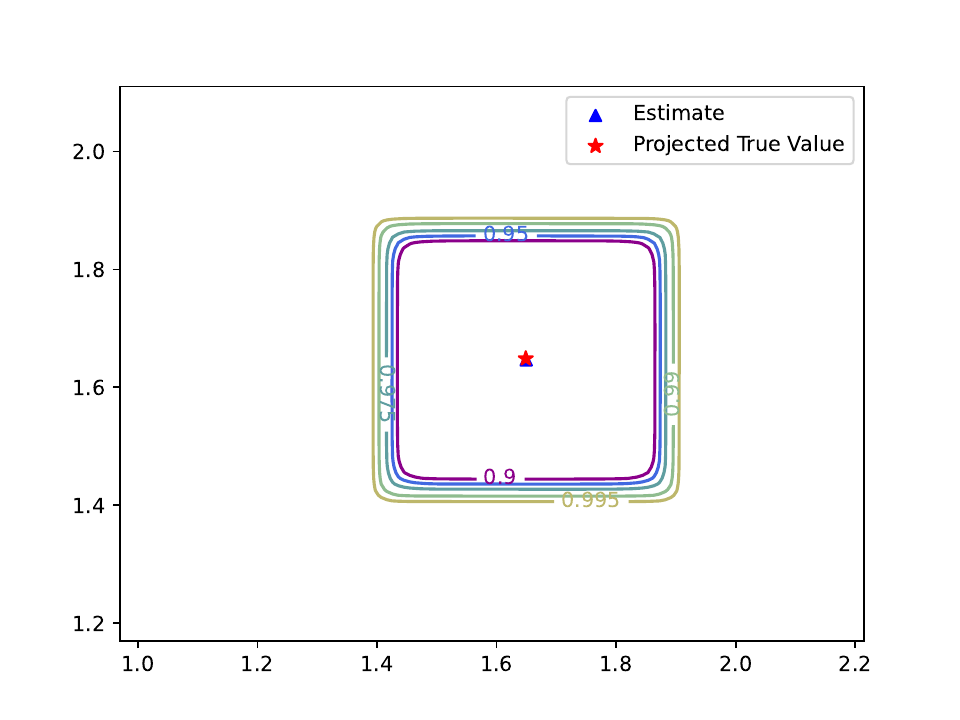}\label{fig:hdmelog-e}
  }\hspace*{-15pt}
  \subfloat[$d_{0}=5$; $\theta_{1}$ and $\theta_{2}$]{%
    \includegraphics[width=.245\textwidth]{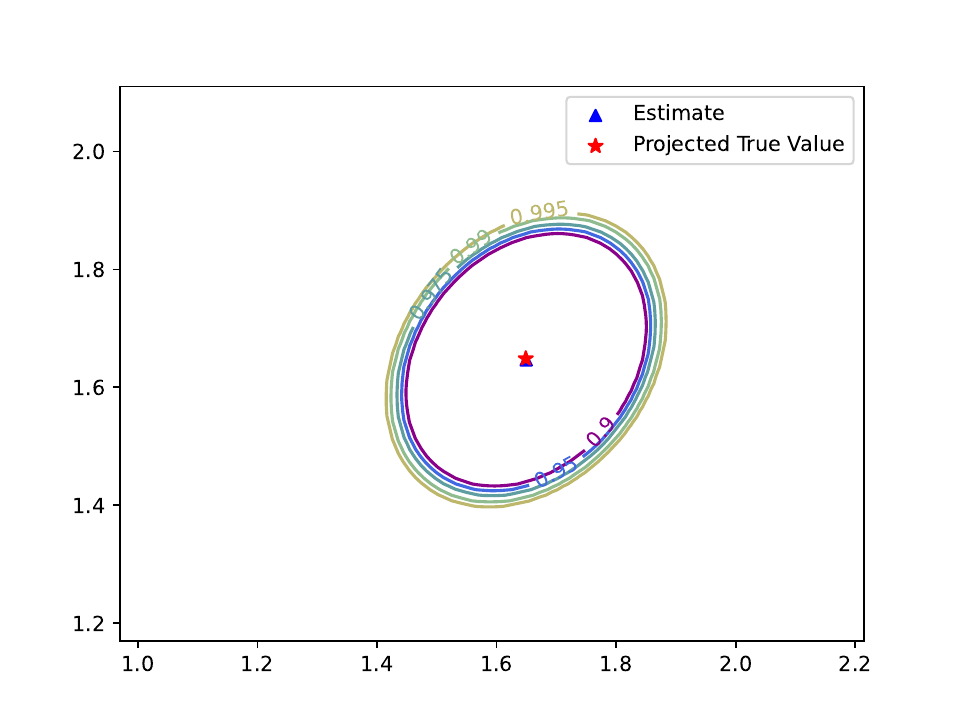}\label{fig:hdmelog-f}
  }\hspace*{-15pt}
  \subfloat[$d_{0}=25$; $\theta_{1}$ and $\theta_{2}$]{%
    \includegraphics[width=.245\textwidth]{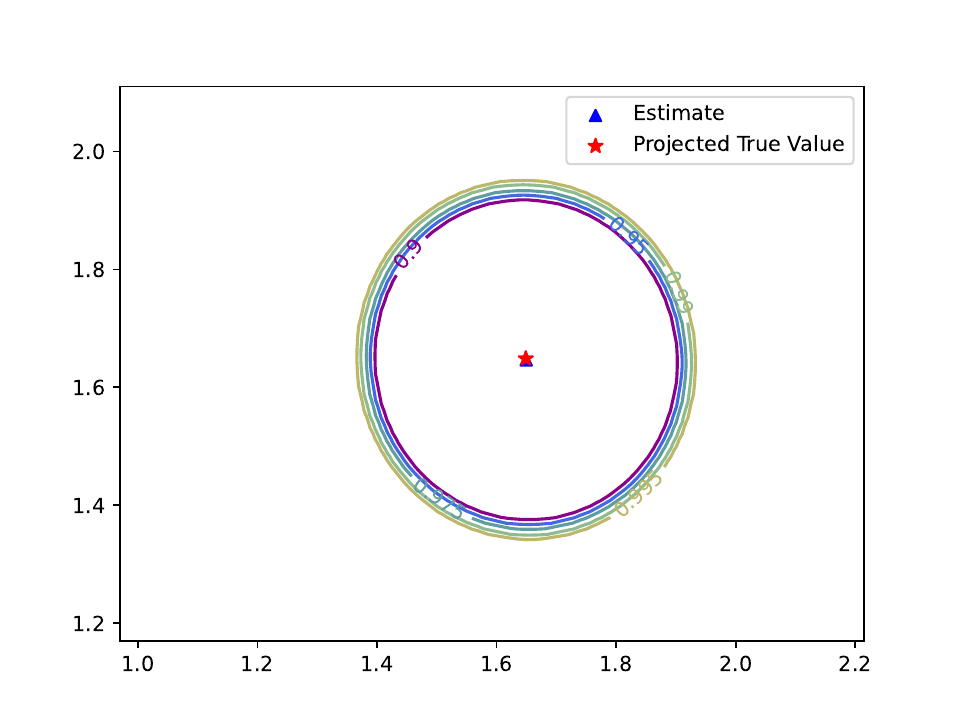}\label{fig:hdmelog-g}
  }\hspace*{-15pt}
  \subfloat[$d_{0}=100$; $\theta_{1}$ and $\theta_{2}$]{%
    \includegraphics[width=.245\textwidth]{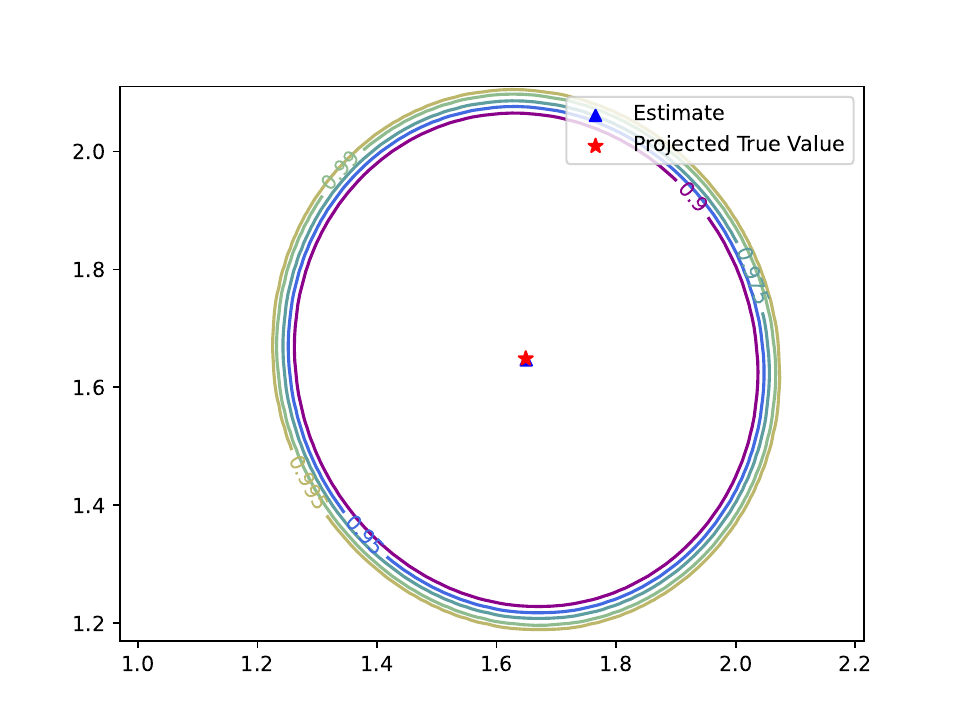}\label{fig:hdmelog-h}
  }
  \caption{\footnotesize 2d slices of confidence regions passing through the \textit{point estimate} with varying $d_{0}$ in the multivariate \textit{log-normal} study. Project true values are the true mean value projected to the respective slices.}\label{fig:hdmelog}
\end{figure}

Our key assumption \cref{eq:hotelling} does not require the underlying data to be normal, since it appeals to the usual large-sample approximations for parameter estimators.  Nevertheless, the fact that the assumption \cref{eq:hotelling} holds exactly for multivariate normal naturally raises the question if the good performance from the simulation studies in \cref{sec:highdimnormal} would be seen when the underlying data are not from normal. Our second simulation study is therefore designed to stress-test our method, by using a highly skewed distribution, log-normal, which is known to break common methods for constructing confidence intervals for the mean parameter, as in bootstrapping \citep{wood1999bootstrap}.  
Specifically, let $\boldsymbol{X}_1, \dots, \boldsymbol{X}_n \in \mathbb{R}^d$ be i.i.d. samples from the distribution $\mathcal{N}(\boldsymbol{\theta}, \boldsymbol{M}_\rho)$, as described in \cref{sec:highdimnormal}. Define $\boldsymbol{Y}_{i} = (e^{X_{i1}}, \ldots, e^{X_{id}})^\top$, such that $Y_{ij}$ is marginally log-normally distributed. Our goal is to estimate the mean of $\boldsymbol{Y}_{i}$, with the true value being $e^{1/2} \mathbf{1}_{d}$ (when $\boldsymbol{\theta}=\boldsymbol{0}$). 

\cref{fig:hdmelog} displays trends similar to those in \cref{fig:hdme}: the size of the confidence regions decreases initially and then increases as $d_{0}$ grows. However, unlike the multivariate normal case, $95\%$ coverage is not guaranteed by using the nominal significance level of $0.05$.  In over 2000 simulations,  the empirical coverage probabilities for $\boldsymbol{\theta}$ are $0.883$, $0.855$, $0.758$, and $0.322$ respectively with $d_{0} = 1, 5, 25, 100$. Therefore, our stress test does reveal the deterioration of our methods when the underlying data are log-normal, even with $d_0=1$. However, relative to the dramatic loss of coverage by the standard Hotelling's procedure $(d_0=d=100)$, the deterioration is significantly less.  
Because our HCCT approach relies on the tail approximation, we anticipated that the deterioration may be less at the $0.01$ level. Indeed, the respective empirical coverages are $0.959$, $0.948$, $0.887$, and $0.502$. While labeling $96\%$ confidence regions (when $d_0 = 1$) as $99\%$ may be excusable as an approximation, advertising $50\%$ confidence regions (when $d_0 = 100$) as $99\%$ surely is deceiving.

\begin{figure}[tbp]
\vspace{-40pt}
  \centering
  \subfloat[$d_{0}=1$; $\theta_{1}$ and $\theta_{51}$]{%
    \includegraphics[width=.245\textwidth]{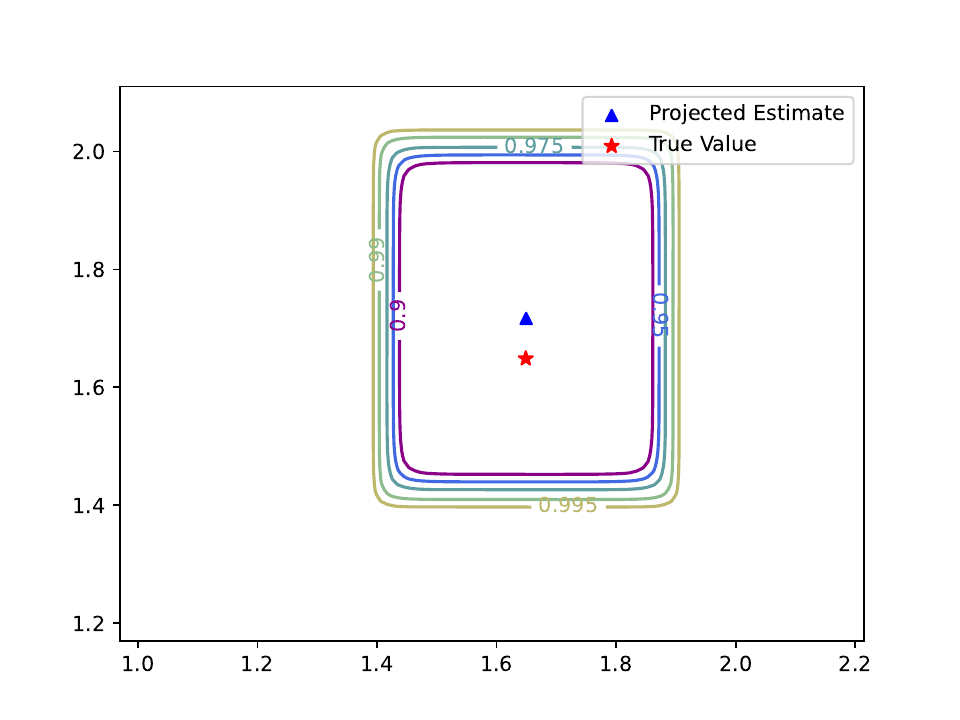}\label{fig:hdmelogtrue-a}
  }\hspace*{-15pt}
  \subfloat[$d_{0}=5$; $\theta_{1}$ and $\theta_{51}$]{%
    \includegraphics[width=.245\textwidth]{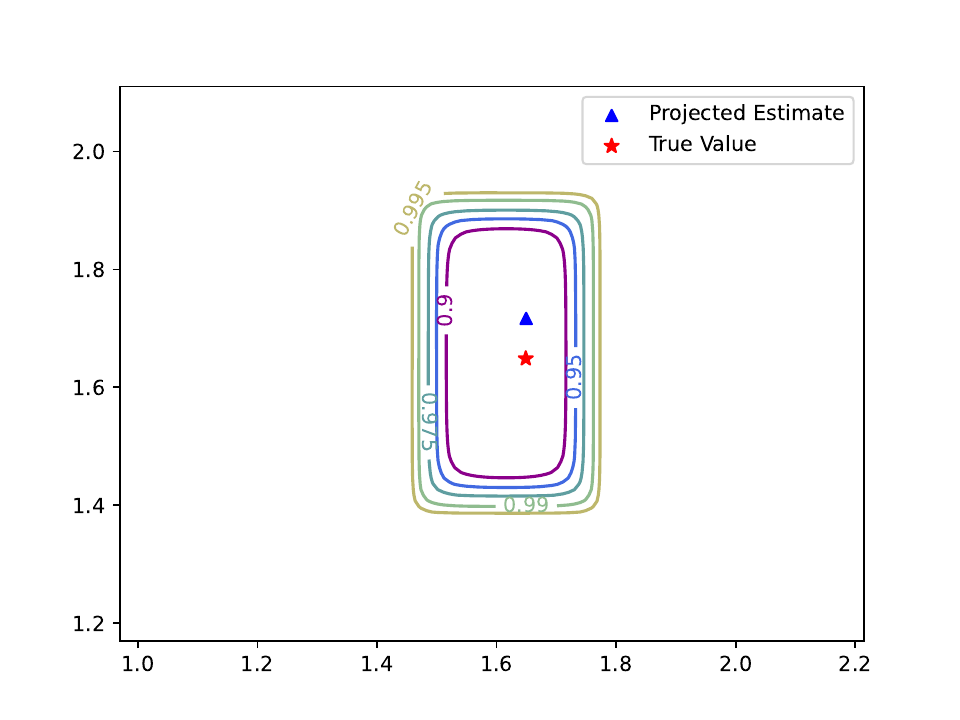}\label{fig:hdmelogtrue-b}
  }\hspace*{-15pt}
  \subfloat[$d_{0}=25$; $\theta_{1}$ and $\theta_{51}$]{%
    \includegraphics[width=.245\textwidth]{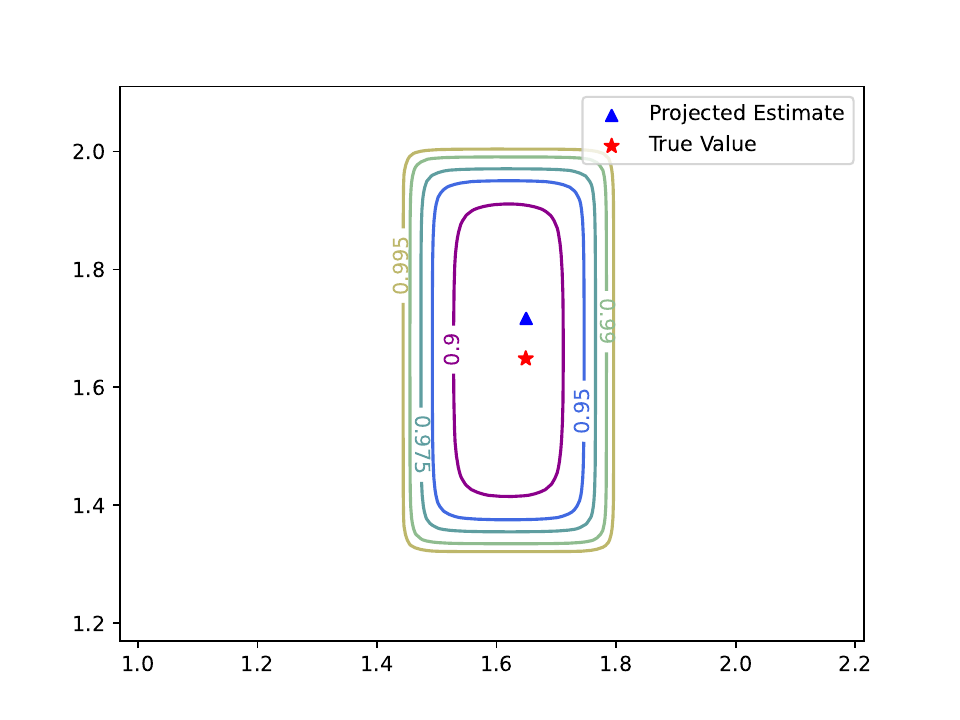}\label{fig:hdmelogtrue-c}
  }\hspace*{-15pt}
  \subfloat[$d_{0}=100$; $\theta_{1}$ and $\theta_{51}$]{%
    \includegraphics[width=.245\textwidth]{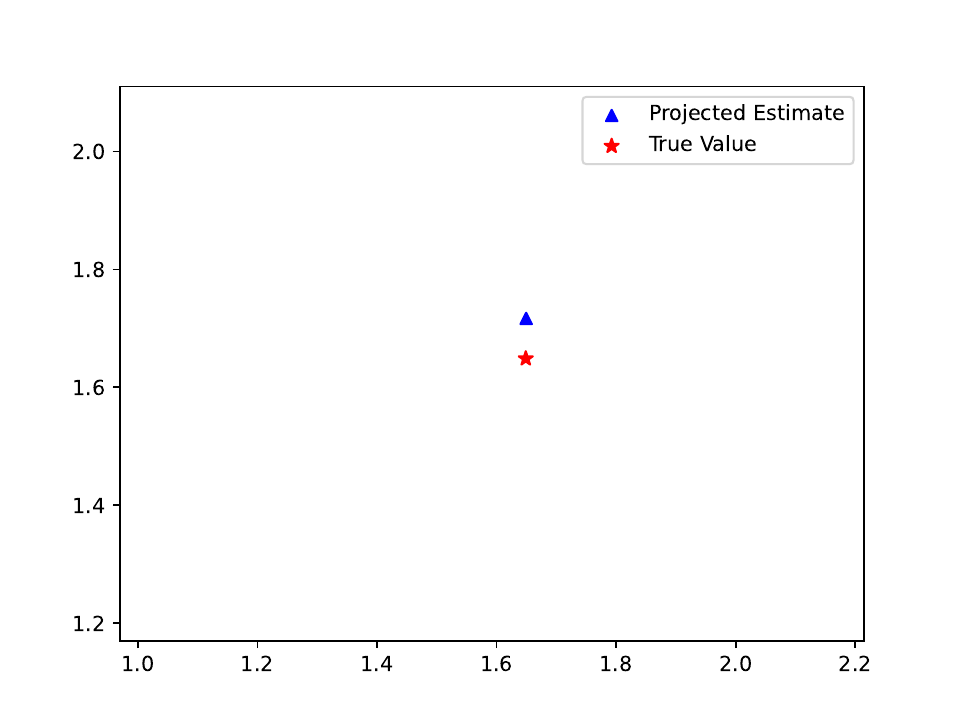}\label{fig:hdmelogtrue-d}
  }\\
  \vspace*{0pt}
  \subfloat[$d_{0}=1$; $\theta_{1}$ and $\theta_{2}$]{%
    \includegraphics[width=.245\textwidth]{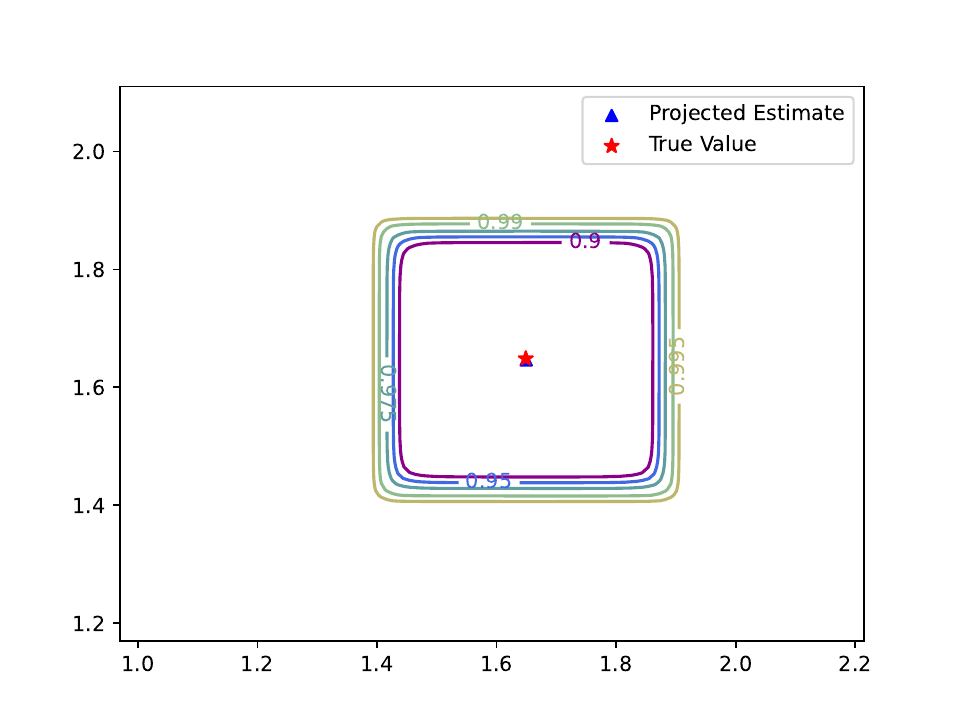}\label{fig:hdmelogtrue-e}
  }\hspace*{-15pt}
  \subfloat[$d_{0}=5$; $\theta_{1}$ and $\theta_{2}$]{%
    \includegraphics[width=.245\textwidth]{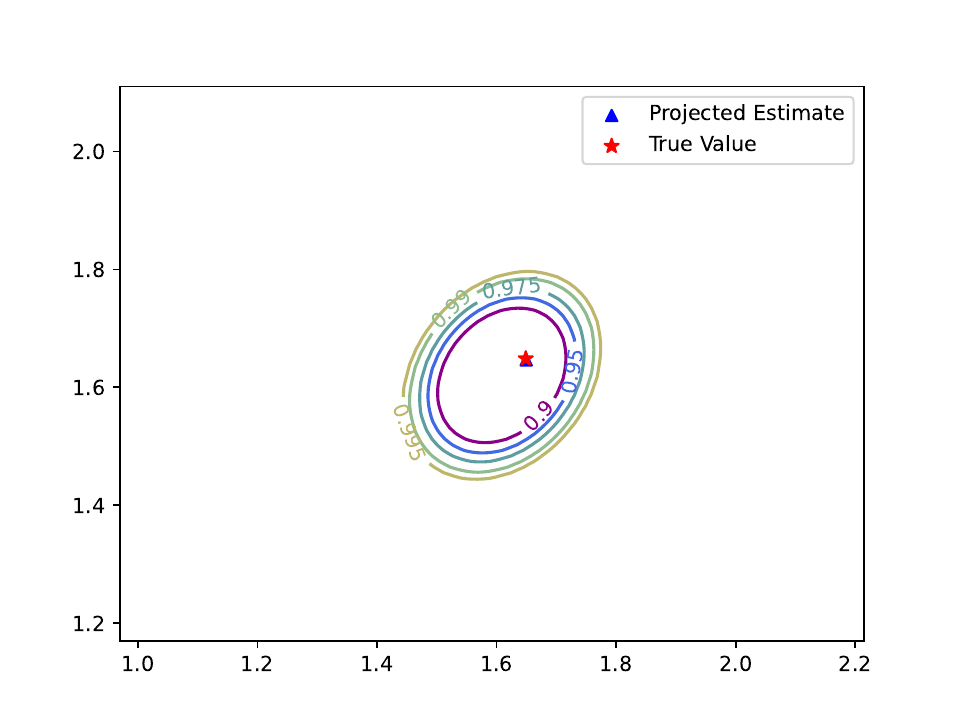}\label{fig:hdmelogtrue-f}
  }\hspace*{-15pt}
  \subfloat[$d_{0}=25$; $\theta_{1}$ and $\theta_{2}$]{%
    \includegraphics[width=.245\textwidth]{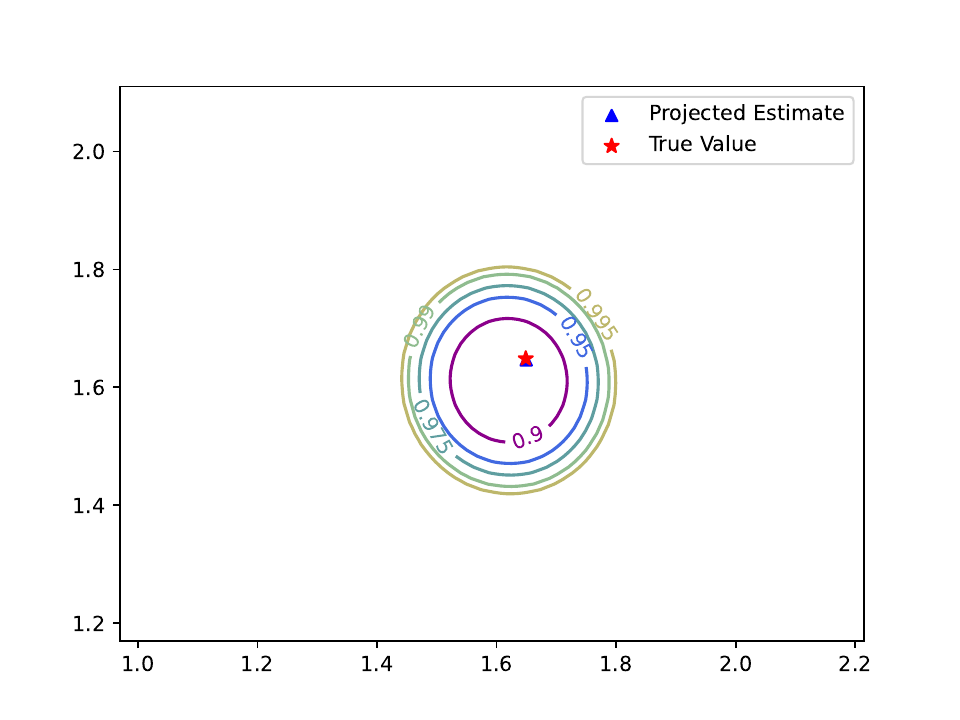}\label{fig:hdmelogtrue-g}
  }\hspace*{-15pt}
  \subfloat[$d_{0}=100$; $\theta_{1}$ and $\theta_{2}$]{%
    \includegraphics[width=.245\textwidth]{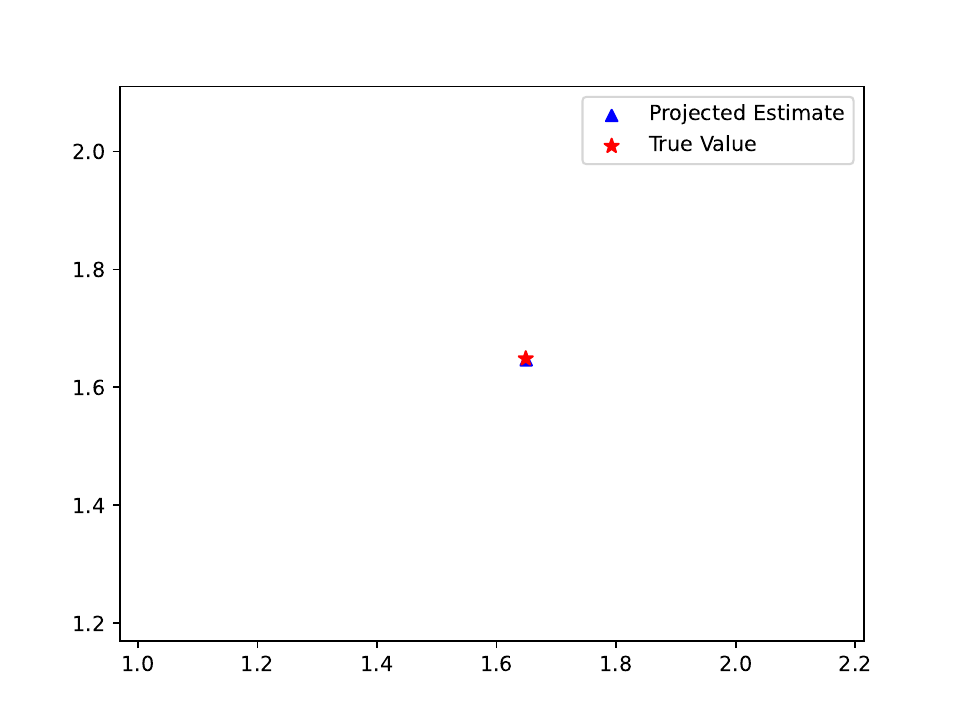}\label{fig:hdmelogtrue-h}
  }
  \caption{\footnotesize 2d slices of confidence regions passing through the \textit{true mean} with varying $d_{0}$ in the multivariate \textit{log-normal} study.  Project estimates are the estimated mean projected to the respective slices. Notably the true means are outside the confidence regions produced by Hotelling's $T^{2}$ approach in this run. }\label{fig:hdmelogtrue}
\end{figure}

We remark that the observed decay in validity as $d_{0}$ increases is likely due to the fact that, for a fixed sample size, the accuracy of Hotelling's $T^2$ approximation in \cref{eq:hotelling} diminishes as the dimension of the covariance matrix grows. 
This pattern is also evident in \cref{fig:hdmelogtrue}, which illustrates two-dimensional slices passing through the true mean rather than the empirical estimate in a single run. In particular, for $d_{0} = 100$, the confidence regions implied by Hotelling's $T^2$-test fail to contain the true mean altogether. General theoretical analysis for this phenomenon is another topic for further research.

\vspace{5pt}

\section{Application to Network Meta-Analysis}\label{sec:networkmeta}

\vspace*{-5pt}
\subsection{Simultaneous Inference and Comparisons of Multiple Treatment Effects}\label{sec:simultaneous}
\spacingset{2}
\setlength{\parskip}{8pt}

\vspace*{-10pt}
In network meta-analysis, we aim to combine evidence from clinical trials involving $d+1$ intervention arms, consisting of $d$ active treatments and a control arm. These treatments are represented as nodes in a network graph, with direct comparisons between treatments forming the edges. Trials may compare two or more arms. For multi-arm trials, we generate all possible pairwise comparisons between treatments and represent the trial as a set of two-arm studies. This decomposition allows each treatment comparison to be consistently evaluated across the network, enabling the synthesis of results from trials with varying designs and treatment combinations.

Our objective is to estimate the effects of \(d\) active treatments across all studies and provide simultaneous confidence intervals for any pairwise treatment comparison. By \emph{simultaneous}, we mean that the confidence intervals account for the uncertainty across all comparisons of interest, ensuring that the true effect sizes for all these pairs are captured with a specified overall confidence level. Let \(\boldsymbol{\theta}\) denote the \(d \times 1\) vector of treatment effects. We have data from \(m \geq d\) two-arm studies, represented by \(\widehat{\boldsymbol{\zeta}} = (\widehat{\zeta}_{1}, \dots, \widehat{\zeta}_{m})^{\top}\), where \(\widehat{\zeta}_j\) is the observed treatment effect in the \(j\)-th study (against the control treatment), and the associated standard errors are \(\sigma_1, \dots, \sigma_m\). 

\spacingset{2}
\setlength{\parskip}{8pt}

The fixed-effects model is given by
\(
\widehat{\boldsymbol{\zeta}} = \boldsymbol{\Omega} \boldsymbol{\theta} + \boldsymbol{\epsilon}
\) with \( 
\boldsymbol{\epsilon} \sim \mathcal{N}(\boldsymbol{0}, \boldsymbol{\Sigma})
\),
where \(\boldsymbol{\Sigma}\) is an unknown covariance matrix, with diagonal entries \(\sigma_1^2, \dots, \sigma_m^2\). The design matrix 
\(
\boldsymbol{\Omega} = (\boldsymbol{\omega}_{1}, \dots, \boldsymbol{\omega}_{m})^{\top} \in \mathbb{R}^{m \times d}
\)
 encodes the structure of the trials, where row \(\boldsymbol{\omega}_j^{\top}\) represents the design of the \(j\)-th study.
For a study comparing treatment \(\theta_k\) against the placebo, \(\boldsymbol{\omega}_j\) has \(\omega_{jk} = 1\) and \(\omega_{j\ell} = 0\) for all \(\ell \neq k\). For studies comparing two active treatments against one another, say \(\theta_{k_1}\) and \(\theta_{k_2}\), we set \(\omega_{jk_1} = 1\), \(\omega_{jk_2} = -1\), and \(\omega_{j\ell} = 0\) for all \(\ell \neq k_1, k_2\). We assume that the network graph is connected, ensuring that \(\boldsymbol{\Omega}\) is of full rank \(d\).

The traditional approach for estimating treatment effects in meta-analysis is to use the weighted least squares (WLS) estimator, assuming independence between different studies \citep{schwarzer2015meta}. The point estimator is given by
\(
\spacingset{1}
\widehat{\boldsymbol{\theta}} = (\boldsymbol{\Omega}^{\top} \widehat{\boldsymbol{W}} \boldsymbol{\Omega})^{-1} \boldsymbol{\Omega}^{\top} \widehat{\boldsymbol{W}} \widehat{\boldsymbol{\zeta}},
\resetspacing
\)
where \(\widehat{\boldsymbol{W}} = \operatorname{diag}\left( 1/\widehat{\sigma}_1^2, \dots, 1/\widehat{\sigma}_m^2 \right)\) is a diagonal matrix of inverse variance weights. Let \(\boldsymbol{L} = (\boldsymbol{\Omega}^{\top} \widehat{\boldsymbol{W}} \boldsymbol{\Omega})^{-1} = \{L_{ij}\}\). The variance for the \(j\)-th treatment effect is estimated by \(L_{jj}\), and the variance for the comparison between the \(i\)-th and \(j\)-th treatments is given by \(L_{ii} + L_{jj} - 2L_{ij}\). 
Using these variance estimates, one can construct asymptotic confidence intervals for each comparison. To obtain simultaneous confidence intervals across all comparisons, traditionally Bonferroni correction is applied to control the family-wise error rate. 

\setlength{\parskip}{5pt}
\spacingset{2.0}

For multi-arm trials, where multiple two-arm studies are derived from a single experiment, one can modify the approach by using a block-diagonal structure for \(\widehat{\boldsymbol{W}}\), with each block corresponding to the inverse of the estimated covariance matrix for the related two-arm studies. Such adjustments may require access to the original experimental data from the multi-arm trials.

In contrast to these traditional methods, we allow \(\boldsymbol{\Sigma}\) to have off-diagonal entries, accommodating many dependence structures between studies in practice (the theoretical conditions in \cref{thm:revalidity} of \cref{sec:hctest} are rather mild). Our approach only requires the estimated average treatment effects and their standard deviations from each study. The reasoning is straightforward: for each two-arm study, we have an estimate \(\widehat{\zeta}_j \sim \mathcal{N}(\boldsymbol{\omega}_j^{\top} \boldsymbol{\theta}, \sigma_j^2)\), where \(\boldsymbol{\omega}_j^{\top}\) is the \(j\)-th row of \(\boldsymbol{\Omega}\). This leads to the same setting introduced in \cref{sec:multidim}, where \(\boldsymbol{P}_j = \boldsymbol{\omega}_j^{\top}\) for \(j = 1, \dots, m\). Thus, we can obtain point estimates, confidence regions, and simultaneous confidence intervals via HCCT. 

Addressing dependence is crucial here, as dependence naturally arises when multi-arm studies are decomposed into two-arm comparisons or when there is overlap in datasets across studies. In particular, as demonstrated in \citet{abbas2023impact}, dependence between studies is common in genetic studies.
\vspace*{-10pt}
\spacingset{1.65}
\setlength{\parskip}{5pt}
\subsection{Empirical Demonstrations}\label{sec:clinicaldemo}

 \begin{table}[tbp]
 \vspace{-25pt}
  \centering
  \resizebox{\textwidth}{!}{\begin{minipage}
    {1.4\textwidth}
    \centering
  \begin{tabular}{ll|ccccccccc}
    \toprule
                             $\boldsymbol{\rho}$&& $\widehat{\theta}_{1}$   & $\widehat{\theta}_{2}$ & $\widehat{\theta}_{3}$   & $\widehat{\theta}_{4}$  & $\widehat{\theta}_{5}$  & $\widehat{\theta}_{6}$       & $\widehat{\theta}_{7}$ & $\widehat{\theta}_{8}$ & $\widehat{\theta}_{9}$ \\
    \midrule
    \multirow{2}{*}{$\textbf{0}$}&WLS & .0277& -.561& -.994& .0962& -.510& -1.02 & .137 & -.496 & -.949 \\
    &HCCT & .0349 & -.567 & -.985 & .114 & -.491 & -1.02 & .137 & -.496 & -.949 \\
    \multirow{2}{*}{$\textbf{0.3}$}&WLS&-.0334 & -.558& -1.09 & -.0927 & -.662 & -1.12 & -.0570 & -.430 & -.898\\
    &HCCT&-.0209 & -.579 & -1.09 & -.0923 & -.671 & -1.12 & -.0570 & -.422 & -.898\\
    \multirow{2}{*}{$\textbf{0.6}$}&WLS & .0850 & -.403 & -.926 & .113 & -.450 & -.949 & -.0325 & -.494 & -1.01\\
    &HCCT &.0857 & -.411 & -.928 & .104 & -.449 & -.951 & -.0325 & -.497 & -1.01\\
    \multirow{2}{*}{$\textbf{0.9}$}&WLS& -.0741 & -.679 & -1.13 & -.149 & -.762 & -1.18 & -.205 & -.501 & -1.14\\
    &HCCT& -.0666 & -.682 & -1.13 & -.139 & -.765 & -1.19 & -.205 & -.501 & -1.14 \\
\midrule
    \multicolumn{2}{l|}{\textbf{True Value} $\boldsymbol{\theta}$} & 0& -.5& -1&0&-.5&-1&0&-.5&-1\\
    \bottomrule
  \end{tabular}
\end{minipage}}
\caption{Average treatment effects against the placebo (simulation).}\label{tab:thetaressim}
\vspace{-5pt}
\end{table}

We illustrate the validity and utility of our approach by applying it to both semi-synthetic and real-world examples \citep{senn2013issues}, which compared different treatments for controlling blood glucose levels in patients with diabetes, using a meta-analysis of 26 previous medical studies, including $25$ two-arm clinical trials and $1$ three-arm trial. The analysis involved $10$ treatments, consisting of $9$ different drugs 
({\texttt{acar}, \texttt{benf}, \texttt{metf}, \texttt{migl}, \texttt{piog}, \texttt{rosi}, \texttt{sita}, \texttt{sulf}, \texttt{vild}}) 
and a placebo.\footnote{This dataset is available in the R package \textbf{netmeta} \citep{schwarzer2015meta} and also included in our supplemental material.} In total, $28$ two-way comparisons are available with reported means and standard deviations of the differences in glucose outcome levels.

To validate our approach and compare it with the traditional WLS method in the context of dependent studies, we consider a semi-synthetic experiment. The design matrix remains identical to that of the real-world example mentioned above, but the underlying average treatment effects and covariance structure are generated as follows:
\[
\spacingset{1}
\begin{aligned}
\boldsymbol{\theta} &= (0,-0.5,-1,0,-0.5,-1,0,-0.5,-1)^{\top}, \quad \boldsymbol{\Sigma} = (s_{ij}), \\ 
s_{ii} &=\operatorname{Var}(\widehat\zeta_j)= 0.01\  \text{ for }\ 1 \leq i \leq 28, \quad  s_{ij} =\operatorname{Cov}(\widehat\zeta_i,\widehat\zeta_j)= 0.01\rho\ \text{ for }\ i \neq j,
\end{aligned}
\resetspacing
\vspace*{5pt}
\]
where \(\rho=0,0.1,\dots,0.9\) is a hyperparameter controlling the dependence level between the studies.

\begin{figure}[tbp]
\vspace{-30pt}
\centering
  \subfloat[Coverage]{%
    \includegraphics[width=.27\textwidth]{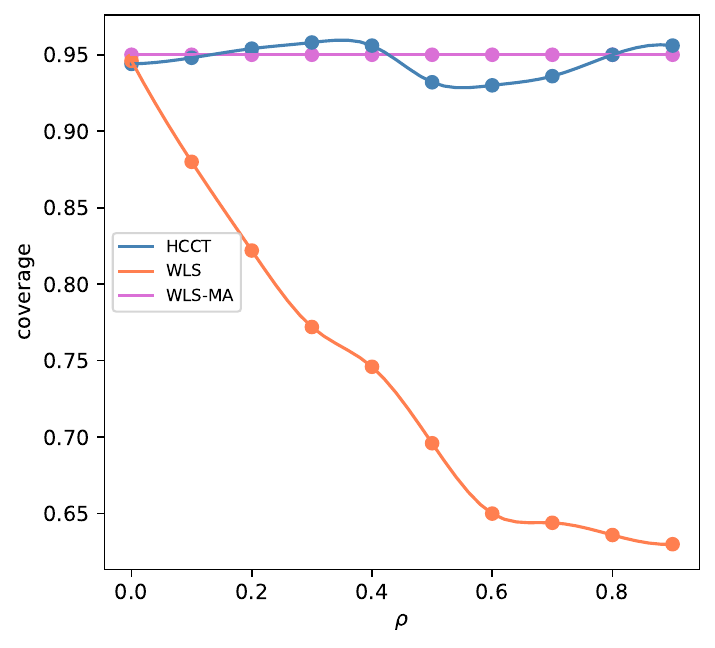}
  }~%
  \subfloat[Width for $\widehat{\theta}_{1}$]{%
    \includegraphics[width=.27\textwidth]{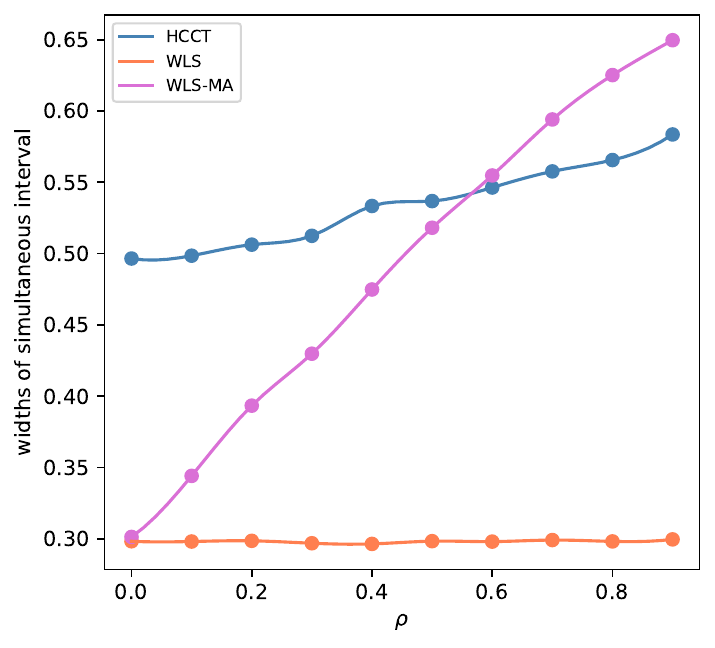}
  }~%
  \subfloat[Width for $\widehat{\theta}_{2}$]{%
    \includegraphics[width=.27\textwidth]{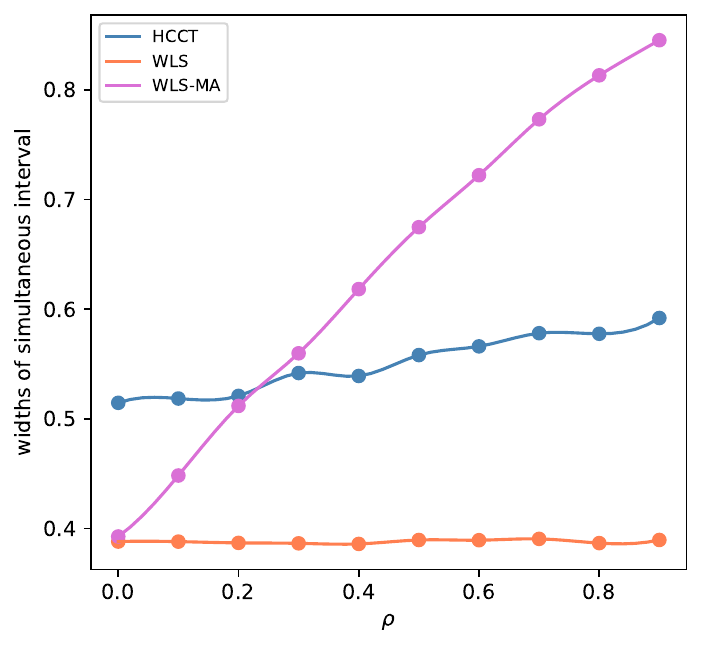}
  }
  \caption{\mbox{Coverage and width of simultaneous CIs (simulation).}}\label{fig:semisyn}
\end{figure}

\cref{tab:thetaressim} presents the point estimates from WLS and HCCT in a single run with correlation levels \(\rho = 0, 0.3, 0.6, 0.9\) respectively. \cref{fig:semisyn} shows the coverage of simultaneous confidence intervals and their average width for \(\theta_1\) and \(\theta_2\) at varying dependence levels, based on $500$ replications. These simultaneous intervals ensure joint coverage across all comparisons between each active treatment and the placebo at the $95\%$ confidence level. Additionally, we adjust the significance level for WLS by manually increasing the quantile multiplier in calculating confidence intervals until approximately $95\%$ coverage is achieved under dependence, and plot the widths of the resulting intervals (labeled “WLS-MA”). Such a manual adjustment is {\emph{not}} feasible in real applications, but it is included in our simulation both to ensure fair comparison of the power and to stress-test HCCT by pinning it against an impractical benchmark.

\vspace*{8pt}
As seen in \cref{tab:thetaressim}, both WLS and HCCT produce point estimates that are reasonably close to the ground truth. However, \cref{fig:semisyn} demonstrates that the simultaneous confidence intervals obtained from WLS, even with Bonferroni correction,
deteriorate rapidly as the dependence between studies increases. 
In this context, Bonferroni correction is not applied to the aggregation of dependent studies, but rather to construct simultaneous confidence intervals for the post-aggregation treatment effects or contrasts.
This shows that the validity of WLS depends critically on the  assumption of independence among studies.

In contrast, HCCT automatically accounts for the potential dependence between studies, and it does so using wider intervals, with width increases as the dependence level \(\rho\) increases. The fact that the WLS intervals remain narrower and are not affected by \(\rho\) is responsible for its deterioration in terms of validity. This point is also reflected by the fact that once we manually adjust the WLS to achieve the correct coverage, the width of the WLS intervals becomes much larger and exceeds those produced by HCCT when $\rho$ increases above a threshold. This threshold apparently depends on the components of $\boldsymbol{\theta}$, about $\rho =0.5$ for $\theta_1$ and $\rho=0.2$ for $\theta_2$, suggesting that the search for an adaptive optimal choice will be a complex matter. Using HCCT by itself is simpler and has built-in resilience to the (unknown) value of $\rho$.

Next, we consider the original real-world example, where we encounter the issue of empty confidence regions because of severe inconsistency in the studies.  We adopt the sequential elimination approach justified in \cref{sec:empty}, starting by including all studies. Once an empty solution is encountered, we can rank the studies according to an ``outlier score", such as the generalized heterogeneity statistic \citep{schwarzer2015meta}, $
Q_j = \bigl( \widehat{\zeta}_j - \boldsymbol{\omega}_j^{\top} \widehat{\boldsymbol{\theta}} \bigr)^2/\widehat{\sigma}_j^2, j=1, \ldots, m$ (or using the lower bound in \cref{eq:Tbound}).  We then give zero (or sufficiently small) weight to the study with the highest score and repeat our HCCT procedure (boundedness requires the positive-weight $P_j$'s to span $\mathbb R^d$). If an empty-set solution still occurs, we repeat the procedure, until a nonempty solution is found -- recall with $m=1$, the confidence region is always nonempty.

 \begin{table}[tbp]
  \centering
  \resizebox{\textwidth}{!}{\begin{minipage}
    {1.4\textwidth}
    \centering
  \begin{tabular}{l|ccccccccc}
    \toprule
                             & \texttt{acar}   & \texttt{benf} & \texttt{metf}   & \texttt{migl}  & \texttt{piog}  & \texttt{rosi}       & \texttt{sita} & \texttt{sulf} & \texttt{vild} \\
    \midrule
    WLS & -0.827&  -0.905&  -1.11&  -0.944&  -1.07& -1.20& -0.57&  -0.439&  -0.7\\
    HCCT & -0.806& -0.828& -1.01& -1.02& -1.02& -1.31& -0.57& -0.406& -0.7 \\
    \bottomrule
  \end{tabular}
\end{minipage}}
\caption{Average treatment effects against the placebo (real data).}\label{tab:thetares}
\end{table}

\begin{figure}[tbp]
\vspace{-10pt}
\centering
\subfloat[HCCT]{%
  \includegraphics[width=.32\textwidth]{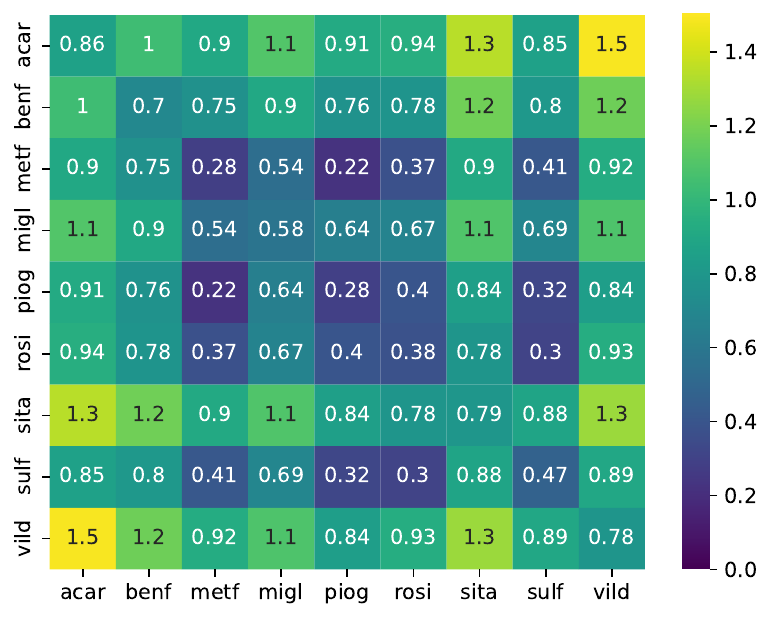}\label{fig:heatmap-1}
}~
\subfloat[WLS]{%
  \includegraphics[width=.32\textwidth]{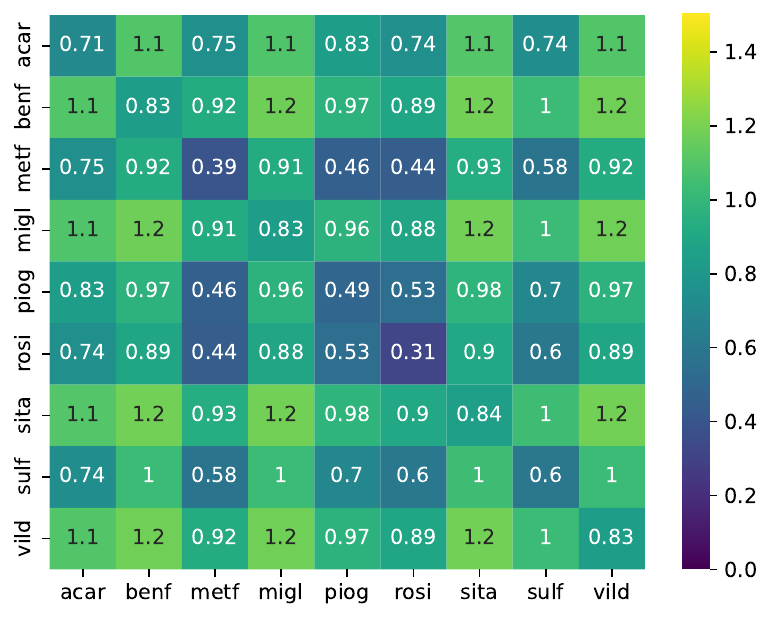}\label{fig:heatmap-3}
}
  \caption{Widths of simultaneous confidence intervals for all comparisons.}\label{fig:heatmaps}
\end{figure}

In the blood glucose control example, two studies were removed based on our approach.
The final point estimate from HCCT is quite close to that provided by WLS, as shown in \cref{tab:thetares}. However, the behavior of the simultaneous confidence intervals differs between the two methods, as seen in the width heatmaps of \cref{fig:heatmaps}. For WLS, Bonferroni correction is applied to all pairwise comparisons, including those involving placebo.

\begin{figure}[tbp]
\centering
\subfloat[{ \texttt{benf}} vs placebo]{%
  \includegraphics[width=.27\textwidth]{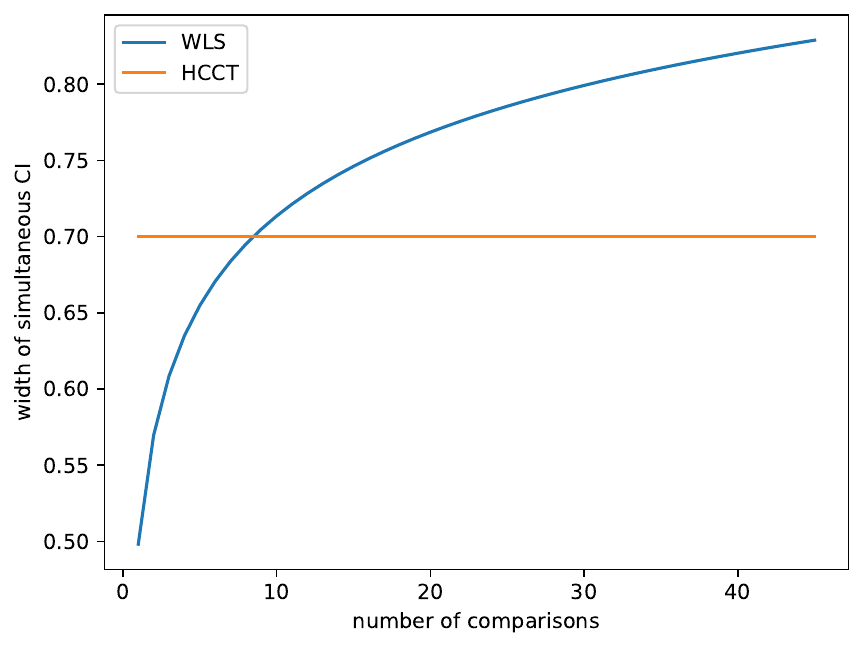}\label{fig:real_tr_1}
}~
\subfloat[{ \texttt{metf}} vs placebo]{%
  \includegraphics[width=.27\textwidth]{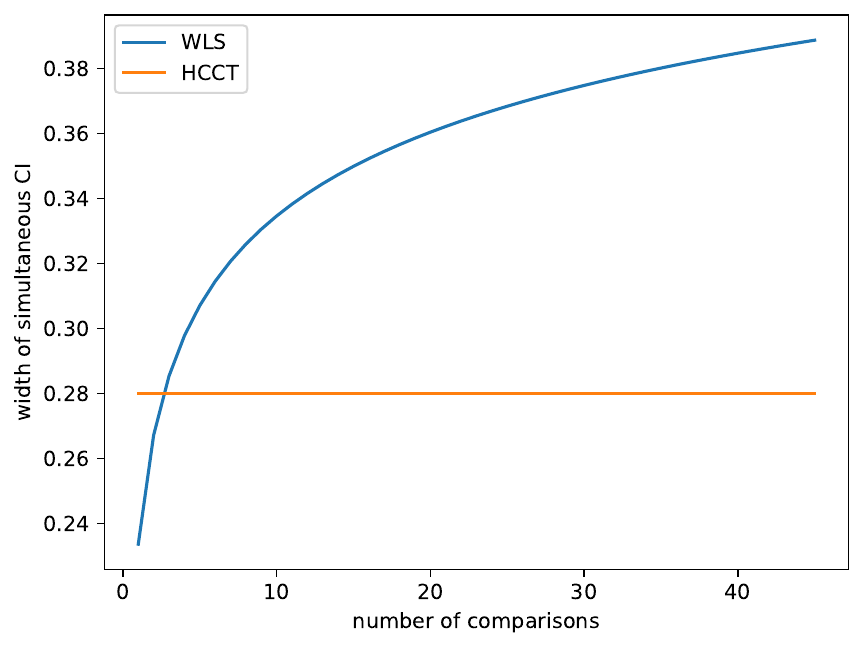}\label{fig:real_tr_2}
}~
\subfloat[{ \texttt{rosi}} vs placebo]{%
  \includegraphics[width=.27\textwidth]{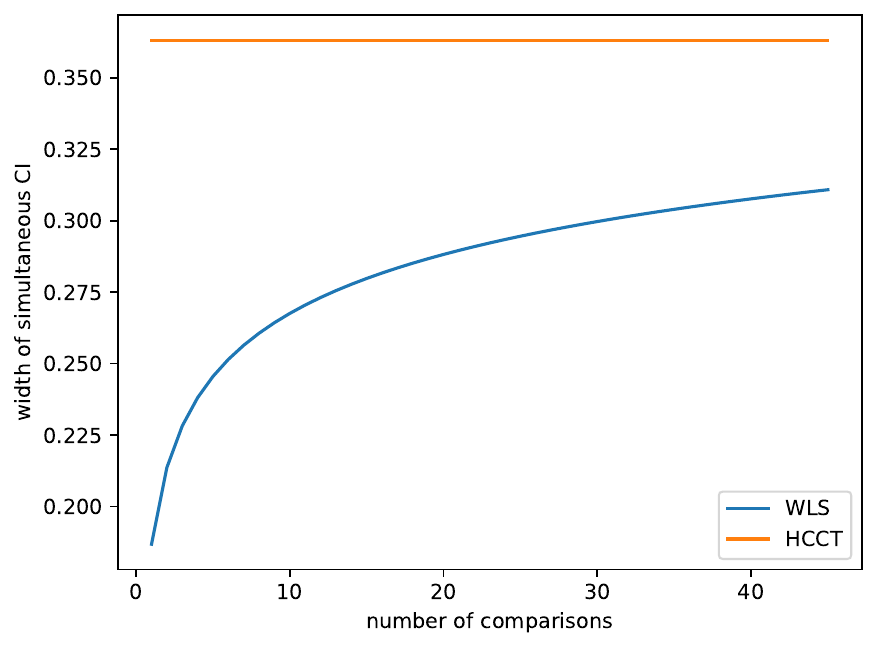}\label{fig:real_tr_4}
}
  \caption{\mbox{Widths of simultaneous CIs with increasing numbers of comparisons.}}\label{fig:real_tr}
\end{figure}

From \cref{fig:heatmaps}, we observe that the widths of simultaneous confidence intervals from HCCT are roughly comparable to those from WLS, though the former exhibit higher variability and can induce larger widths. As we alluded to earlier, the comparisons between the divide-and-combine approach and use Bonferroni correction will depend on how we choose the projections and the statistical information retained by them.  Nevertheless, \cref{fig:real_tr} highlights a key artifact of Bonferroni correction: the individual interval widths from WLS necessarily increase with the number of comparisons. This issue does not arise with our methods, as individual comparisons are derived from projections of $d$-dimensional confidence regions. In this sense, WLS intervals with the largest Bonferroni correction provide a more equitable comparison to the corresponding intervals obtained using HCCT. However, even these widest  WLS intervals may still fall (significantly) short in ensuring the nominal coverage, when there is dependence across studies, a problem that is overcome by HCCT (or EHMP)  without unduly widening the intervals.  Theoretically comparing HCCT or EHMP with Bonferroni correction is another open problem.

\vspace*{-20pt}
\section{Theoretical Understanding of Half-Cauchy and Harmonic Mean Combining Rules}\label{sec:hctest}
\vspace*{-15pt}
\subsection{Half-Cauchy and Pareto(1,1) are Attracted to the Landau Family}\label{sec:Landau}
\vspace*{-10pt}

\spacingset{1.9}

We start our theoretical study by first examining the asymptotic behaviors of the Half-Cauchy and Harmonic Mean combinations when the number of studies $m\to\infty$. Such approximations can provide efficient computations when $m$ is very large.  To present our findings,  we need a few basic concepts from extreme value theory. A distribution is called \emph{stable} if any positive linear combination of two independent random variables from this distribution results in a variable that has the same distribution, up to location and scale transformations. All continuous stable distributions $S(\alpha,\beta,c,\mu)$ can be obtained from the following parametrization of the characteristic function:
\begin{equation*}
  \spacingset{1}
  \textstyle\phi (t;\alpha,\beta,c,\mu)=\exp \bigl[it\mu-\lvert ct \rvert^{\alpha}\bigl\{1-i\beta\operatorname{sgn}(t)\chi (\alpha,t)\bigr\}\bigr], \quad\text{with}\quad \chi(\alpha,t)=\begin{cases}
    \tan \bigl(\frac{\pi\alpha}{2}\bigr)& \text{if }\alpha \neq 1\\
    -\frac{2}{\pi}\log \lvert t \rvert& \text{if }\alpha=1
  \end{cases},
  \resetspacing
\end{equation*}
where $\operatorname{sgn}(t)$ is the sign of $t$.
Here $\alpha \in (0,2]$ is the \emph{stability} parameter that controls the tail of the distribution, $\beta\in [-1,1]$ is called the \emph{skewness} parameter, $c\in(0,\infty)$ is the \emph{scale}, and $\mu\in (-\infty,\infty)$ is the \emph{location} parameter. Except for the normal distribution ($\alpha=2$), the stable family is always heavy-tailed. In particular, $\alpha=1$ and $\beta=0$ results in the Cauchy distribution, and $\alpha=\beta=1$ defines the Landau family \citep{zolotarev1986one} with the density function
\begin{equation*}
\spacingset{1}
  \textstyle f_{\mathrm{Landau}}(x;\mu,c)=\frac{1}{c\pi}\int_{0}^{\infty}\exp (-t)\cos \Bigl\{\frac{(x-\mu)t}{c}+\frac{2}{\pi}t\log\frac{t}{c}\Bigr\}\dif t.
  \resetspacing
\end{equation*}

Let $X_{1}, X_2, \dots,X_{n}$ be a sequence of random variables i.i.d. from $\nu$. If for suitably chosen real-number sequences $A_{n}$ and $B_{n}$, $
    \textstyle B_{n}^{-1}\sum_{i=1}^{n}X_{i}-A_{n}\mathrel{\xrightarrow{\mathrm{d}}}\mathcal{L},$
  we say that $\nu$ is \emph{attracted} to the limiting distribution $\mathcal{L}$. The totality of distributions attracted to $\mathcal{L}$ is called the \emph{domain of attraction} of $\mathcal{L}$. Nondegenerate limits of normalized i.i.d. sums are stable. For $0<\alpha<2$, a regularly varying total tail of index $-\alpha$ and limiting tail balance imply attraction to an $\alpha$-stable law, unique up to location and scale (See \citealp{gnedenko1954limit,zolotarev1986one,uchaikin2011chance} for details). Therefore, we can talk about $\alpha$ for any such distribution.

The following theorem shows that standard Half-Cauchy and Pareto$(1,1)$ both lie in the \emph{domain of attraction} of Landau distributions. The Half-Cauchy part of \cref{thm:gclthc} is new to the best of our knowledge, while the Pareto$(1,1)$ part is a generalization of \citet{wilson2019harmonic} by allowing for unequal weights (see \cref{sec:pfhctest}).
\begin{theorem}\label{thm:gclthc}
  Consider nonnegative weights $\{w_{j}^{(m)}, 1\leq j\leq m;  m\geq 1\}$ with $\sum_{j=1}^{m}w_{j}^{(m)}=1$ and $\max_{j}w_{j}^{(m)}\to 0$ as $m\to\infty$.
  Let $\{X_{j}, j=1, \ldots \}$ be a sequence of i.i.d. variables from standard Half-Cauchy, then we have (where we adopt the convention that $0\log0=0$)
    \begin{equation*}
    \spacingset{0}
      \textstyle\sum_{j=1}^{m}w_{j}^{(m)}X_{j}-\frac{2}{\pi}\bigl\{-\sum_{j=1}^{m}w_{j}^{(m)}\log w_{j}^{(m)}+1-C_\gamma\bigr\}\mathrel{\xrightarrow{\mathrm{d}}} S(1,1,1,0)=\mathrm{Landau}(0,1).
      \resetspacing
    \end{equation*}
    For $\mathrm{Pareto}(1,1)$ variables, we have
    \begin{equation*}
    \spacingset{0}
      \textstyle\sum_{j=1}^{m}w_{j}^{(m)}X_{j}-\bigl\{-\sum_{j=1}^{m}w_{j}^{(m)}\log w_{j}^{(m)}+1-C_\gamma\bigr\}\mathrel{\xrightarrow{\mathrm{d}}} S(1,1,\frac{\pi}{2},0)=\mathrm{Landau}(0,\frac{\pi}{2}),
      \resetspacing
    \end{equation*}
    where $C_\gamma=\lim_{m\to\infty}\bigl(\sum_{k=1}^{m}\frac{1}{k}-\log m\bigr)\approx 0.5772$ 
    is the Euler--Mascheroni constant \citep{campbell2003gamma}.
\end{theorem}

\vspace*{5pt}

\begin{figure}[tbp]\label{fig:pdfplot-4}
\vspace{-30pt}
  \centering
  \subfloat[$m=1$]{%
    \includegraphics[width=0.24\textwidth]{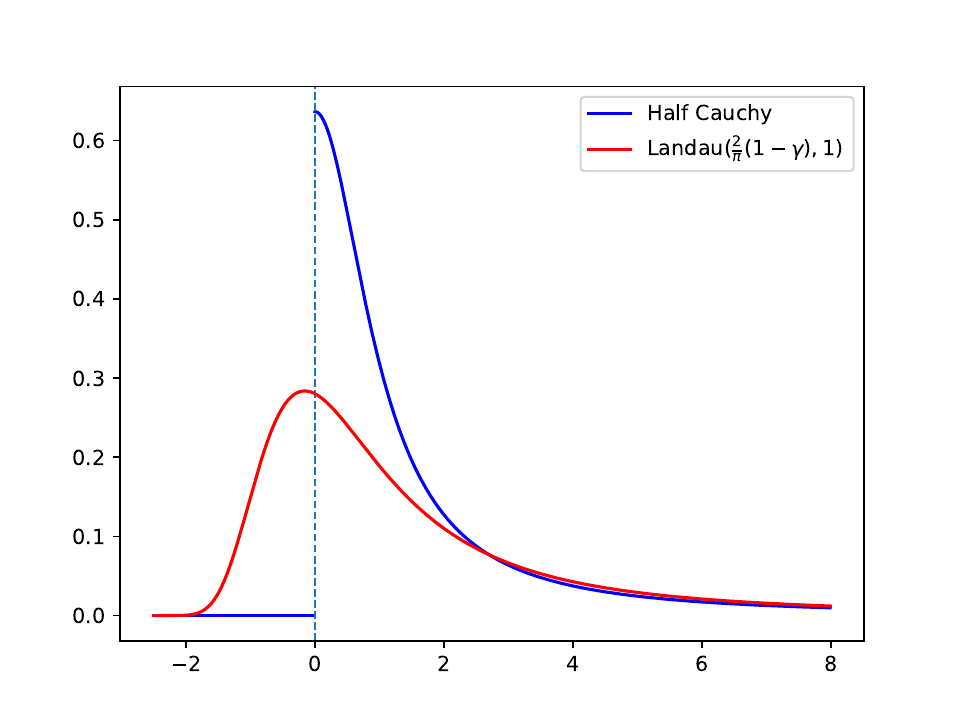}
  }\hspace*{-5pt}
  \subfloat[$m=10$]{%
    \includegraphics[width=0.24\textwidth]{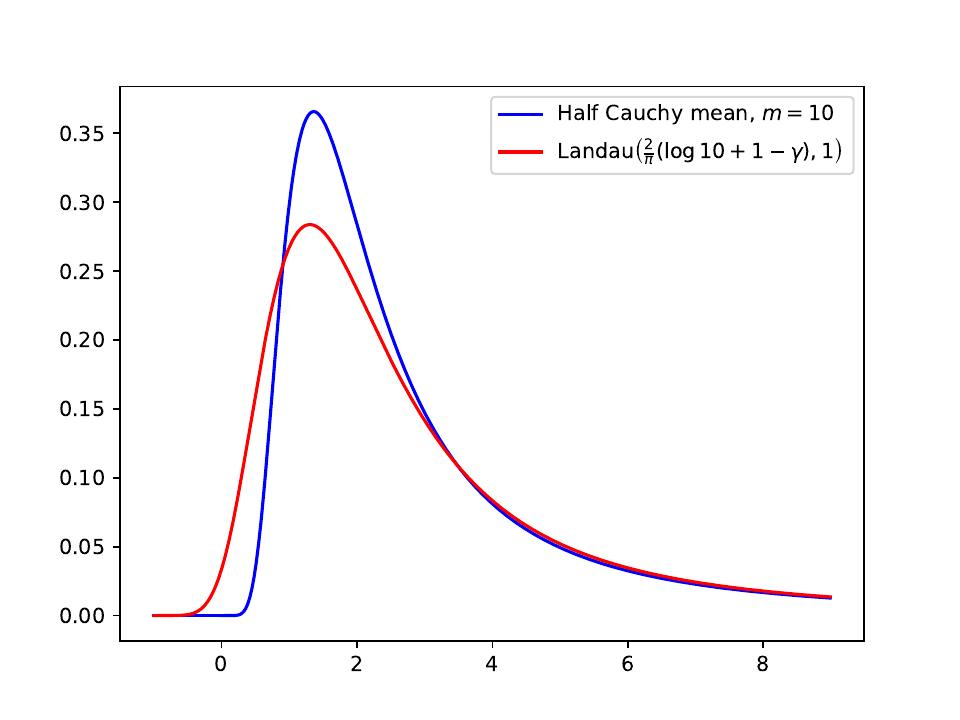}
  }\hspace*{-5pt}
  \subfloat[$m=100$]{%
    \includegraphics[width=0.24\textwidth]{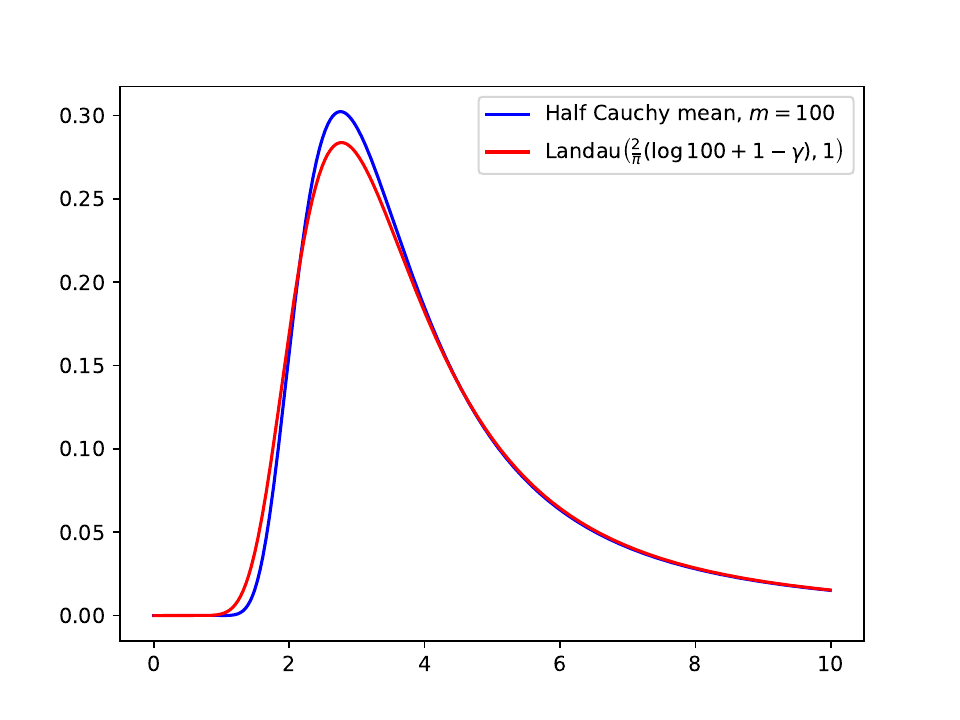}
  }\hspace*{-5pt}
  \subfloat[$m=1000$]{%
    \includegraphics[width=0.24\textwidth]{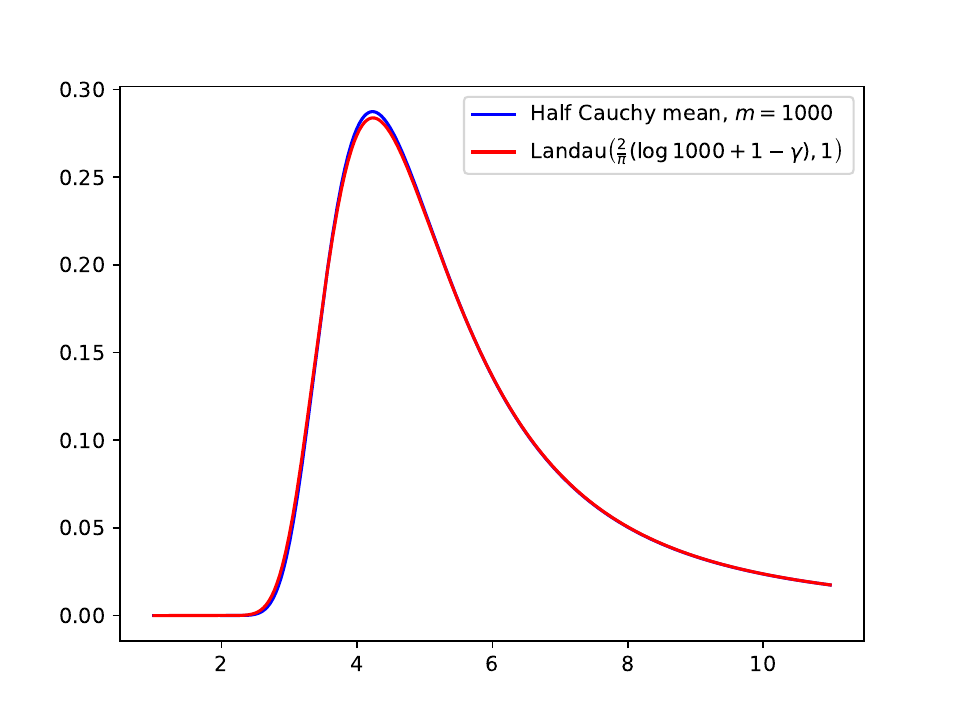}
  }
  \caption{Density functions for Landau distribution and Half-Cauchy means.}\label{fig:landau}
\end{figure}

To gain intuition from \cref{thm:gclthc}, \cref{fig:landau} compares the densities of  weighted Half-Cauchy sums and their Landau approximations. The Landau distribution is supported on $\mathbb{R}$, with a rapidly decaying left tail. The following proposition of \citet{zolotarev1986one} provides the stability property of Landau distributions:
\begin{proposition}\label{thm:landau}
  If $X\sim \mathrm{Landau}(\mu,c)$, then $aX+b\sim \mathrm{Landau}(a\mu+b-\frac{2c}{\pi}a\log a, ac)$ for any $a>0$. If $X\sim \mathrm{Landau}(\mu_{1},c_{1})\Perp Y\sim \mathrm{Landau}(\mu_{2},c_{2})$, then $X+Y\sim \mathrm{Landau}(\mu_{1}+\mu_{2},c_{1}+c_{2})$.
\end{proposition}
\noindent A caveat is that the Landau distribution is not \emph{strictly stable} in the sense that the location parameter does not change proportionally with rescaling. For example, if $X_{1},\dots,X_{m}$ is i.i.d. $\mathrm{Landau}(\mu,1)$, then we can check that
\begin{equation*}
\spacingset{1}
\textstyle  \sum_{j=1}^{m}w_{j}X_{j}\sim \mathrm{Landau}(-\frac{2}{\pi}\sum_{j=1}^{m}w_{j}\log w_{j}+\mu,1).
\resetspacing
\end{equation*}

\spacingset{1.65}
\setlength{\parskip}{5pt}
\vspace{-10pt}
\subsection{Numerical Calibration for Independent Studies}\label{sec:numeric}

\cref{thm:gclthc} hints that, unlike a weighted sum of independent Cauchy variables, which retains the Cauchy distribution, a weighted sum of independent Half-Cauchy or Pareto variables is not well-characterized.
Fortunately, we are able derive its density and CDF based on Laplace transform and contour integration, which enables us to provide an efficient and precise numerical method for computing its density, CDF, and quantile function.

\cref{thm:computeconvhc} gives an efficient numerical scheme for the density and CDF of weighted sums of i.i.d. Half-Cauchy or Pareto$(1,1)$ variables; quantiles follow by CDF inversion. The integrands in \cref{eq:numint0,eq:numint,eq:numint0hmp,eq:numinthmp} are continuous and decay exponentially as $z\to\infty$, so we can apply numerical integration methods to evaluate them as implemented in the Python package \textbf{SciPy}. The integrals can be computed with high precision (e.g., estimated absolute error below $10^{-8}$) using a moderate number of grid points, with empirically near-linear runtime in $m$. See \cref{sec:hmp} for implementation details.

\begin{theorem}\label{thm:computeconvhc} (I) For i.i.d. Half-Cauchy $\{X_1, \ldots, X_m\}$, 
  the density and CDF of $\sum_{j=1}^{m}w_{j}X_{j}$ are respectively
    \begin{align}
    \spacingset{1}
    &\begin{aligned}
    f_{\mathrm{HC}, \boldsymbol{w}}(x)
    =&\textstyle\frac{1}{2\pi i}\int_{0}^{\infty}\exp(-xz)\Bigl[\prod_{j=1}^{m}\{-f_{\mathrm{HC}}^{\ast}(w_{j}z)+2\cos (w_{j}z)+2i \sin (w_{j}z)\}\\*
    &\qquad\qquad\qquad\qquad\qquad\textstyle-\prod_{j=1}^{m}\{-f_{\mathrm{HC}}^{\ast}(w_{j}z)+2\cos (w_{j}z)-2i \sin (w_{j}z)\}\Bigr]\dif z,
    \end{aligned}\label{eq:numint0}\\
    &\begin{aligned}
    F_{\mathrm{HC}, \boldsymbol{w}}(x)
    =&\textstyle 1-\frac{1}{2\pi i}\int_{0}^{\infty}\frac{\exp(-xz)}{z}\Bigl[\prod_{j=1}^{m}\{-f_{\mathrm{HC}}^{\ast}(w_{j}z)+2\cos (w_{j}z)+2i \sin (w_{j}z)\}\\*
    &\qquad\qquad\qquad\qquad\qquad\textstyle -\prod_{j=1}^{m}\{-f_{\mathrm{HC}}^{\ast}(w_{j}z)+2\cos (w_{j}z)-2i \sin (w_{j}z)\}\Bigr]\dif z,
    \end{aligned}\label{eq:numint}
    \resetspacing
  \end{align} 
  where $f_{\mathrm{HC}}^{\ast}(z)$ denotes the Laplace transform of $f_{\mathrm{HC}}(x)$, which can be expressed as \begin{equation*}
  \spacingset{1}
    \textstyle f_{\mathrm{HC}}^{\ast}(z)=\frac{2}{\pi}\int_{0}^{+\infty}\frac{\exp(-xz)}{1+x^{2}}\dif x=-\frac{2}{\pi}\{\sin (z)\operatorname{ci}(z)+\cos (z)\operatorname{si}(z)\},\quad \operatorname{si}(z)=-\int_{z}^{\infty}\frac{\sin (\xi)}{\xi}\dif \xi,\  \operatorname{ci}(z)=\int_{z}^{\infty}\frac{\cos (\xi)}{\xi}\dif \xi.
    \resetspacing
  \end{equation*}
Here $\operatorname{si}(z)$ and $\operatorname{ci}(z)$ are known as the \emph{sine integral} and \emph{cosine integral} respectively \citep{abramowitz1968handbook}.

(II) For i.i.d. $\mathrm{Pareto}(1,1)$ $\{X_1, \ldots, X_m\}$, 
  the density and CDF of $\sum_{j=1}^{m}w_{j}X_{j}$ are respectively given by
    \begin{align}
    \spacingset{1}
    & f_{\mathrm{Pareto}, \boldsymbol{w}}(x)
    =\textstyle\frac{1}{2\pi i}\int_{0}^{\infty}\exp (-xz)\Bigl[\prod_{j=1}^{m}\{-\operatorname{Ei}_{2}(w_{j}z)+i\pi w_{j}z\}-\prod_{j=1}^{m}\{-\operatorname{Ei}_{2}(w_{j}z)-i\pi w_{j}z\}\Bigr]\dif z\label{eq:numint0hmp}\\
    &
    F_{\mathrm{Pareto}, \boldsymbol{w}}(x)
    =\textstyle 1-\frac{1}{2\pi i}\int_{0}^{\infty}\frac{\exp(-xz)}{z}\Bigl[\prod_{j=1}^{m}\{-\operatorname{Ei}_{2}(w_{j}z)+i\pi w_{j}z\}-\prod_{j=1}^{m}\{-\operatorname{Ei}_{2}(w_{j}z)-i\pi w_{j}z\}\Bigr]\dif z.
    \label{eq:numinthmp}
    \resetspacing
  \end{align} 
Here $\operatorname{Ei}_{2}$ is the second-order \emph{exponential integral};
the $\operatorname{Ei}$ integrals below are Cauchy principal values \citep{abramowitz1968handbook}.
\begin{equation*}
\spacingset{1}
  \textstyle\operatorname{Ei}_{2}(z):=-1+z(\log z+C_\gamma-1)+\sum_{j=2}^{\infty}\frac{z^{j}}{(j-1)j!}=z \operatorname{Ei}(z)-\exp (z),
\quad
  \operatorname{Ei}(z):=-\int_{-z}^{\infty} \frac{e^{-\xi}}{\xi} \dif \xi=\int_{-\infty}^z \frac{e^{\xi}}{\xi} \dif \xi.
  \resetspacing
\end{equation*}
\end{theorem}

\begin{table}[tbp]
\centering
    \resizebox{\textwidth}{!}{\begin{minipage}
    {1.4\textwidth}
    \centering
  \begin{tabular}{ll|cccccc}
    \toprule
     $m$ & weights                        &  \makecell{Previous Work\\{\footnotesize \citet{wilson2019harmonic}}} & \makecell{Exact\\{\footnotesize \citet{wilson2019harmonic}}} &  \makecell{Previous Work\\{\footnotesize \citet{fang2023heavy}}}  & \makecell{Exact\\{\footnotesize \citet{fang2023heavy}}} & \makecell{Previous Work\\{\footnotesize \citet{gui2023aggregating}}}  & \makecell{Exact\\{\footnotesize \citet{gui2023aggregating}}}\\
    \midrule
    2 &(.5, .5) & 23.57 & 21.73 & 6.33 & 6.32 & 12.71 & 13.69\\
    2 &(.8, .2) & 23.57 & 21.19 & 6.33 & 6.32 & 12.71 & 13.39\\
    5 &(.2, .2, .2, .2, .2) & 24.48 & 23.51 & 6.36 & 6.36 & 12.71 & 14.74\\
    5 & (.6, .1, .1, .1, .1) & 24.48 & 22.64 & 6.36 & 6.37& 12.71 & 14.24\\
    26 & (1/26, $\dots$, 1/26) & 26.13 & 25.85 & 6.86 & 6.62 & 12.71 & 16.19\\
    \bottomrule
  \end{tabular}
\end{minipage}}
\caption{\scriptsize This table shows the thresholds of \(T_{\nu, \boldsymbol{w}}\) for rejecting the global null at a significance level of $0.05$. ``Previous Work'' refers to the thresholds computed from the suggested approach in previous papers while ``Exact'' provides the calibrated thresholds based on the exact distribution of \(T_{\nu, \boldsymbol{w}}\) under independence. Following recommendations from \citet{fang2023heavy}, winsorization at the $1\%$-quantile of the Cauchy distribution is applied, and for \citet{gui2023aggregating}, left-truncation at zero is used to align with the Half-Cauchy.}\label{tab:wilson}
\end{table}

\spacingset{1.65}
\setlength{\parskip}{5pt}

The difficulty for computing the exact distribution of $T_{\nu,\boldsymbol{w}}$ has been one of the motivations for both Fisher's combination test \citep{fisher1925statistical} and the use of stable distributions in a similar context \citep{stouffer1949american,liu2020cauchy,wilson2021evy,ling2022stable}. For HMP, it has been a long-standing open problem, as discussed in \citet{wilson2019harmonic}, which adopted the Landau approximation. Similar concerns were raised by \citet{fang2023heavy} and \citet{gui2023aggregating}. The former proposed a hybrid approach that uses a Monte Carlo-based approach to compute the exact distribution when $m<25$ and switch to the asymptotic distribution when $m\geq 25$, while the latter suggested using the distribution of individual test score as a proxy. 

The resulting thresholds from these work can deviate from the exact ones, as shown in \cref{tab:wilson}. In particular, although the proxy in \citet{gui2023aggregating} is asymptotically sensible as the significance level goes to 0, it does not guarantee test validity at finite levels, even for independent studies. Indeed, \cref{tab:wilson} suggests that its thresholds are generally smaller than the exact ones, leading to inflated Type-I errors that worsen as the number of studies increases. In contrast, our calibration under independence ensures that our method is well-grounded before extending it to handle dependence. More generally, exact calibration in the i.i.d. setting provides an essential performance anchor and a logical prerequisite for meaningfully assessing robustness to dependence.

For equal weights, we recommend a hybrid approach: for $m\leq 1000$, compute the exact distribution using \cref{eq:numint0,eq:numint,eq:numint0hmp,eq:numinthmp}; for $m>1000$, use the Landau approximation from \cref{thm:gclthc}. See \cref{tab:precision,tab:precisionehmp} in \cref{sec:hmp} for numerical error, runtime, and the accuracy of the Landau approximation.

\vspace*{5pt}

\subsection{Tail Probability and Dependence-Resilient Property}\label{sec:validpower}

Following the approaches of \citet{long2023cauchy}, we establish the following justification for HCCT and EHMP. 

\vspace{-2pt}
\begin{theorem}[Fixed-$m$ HCCT \& EHMP validity]\label{thm:revalidity}
Let $m\geq1$ be fixed, let $p_1,\ldots,p_m$ have exact
$\operatorname{Uniform}(0,1)$ marginals, and let $w_i>0$ with
$\sum_iw_i=1$.  Suppose there exists $\delta_t\in(0,1)$ such that $\delta_t\to 0$, $t\delta_t\to\infty$, and that
\begin{equation}\label{eq:divmcondition2}
  \max_{1\leq i\neq j\leq m}
  \mathbb P\!\left(
    0<p_i<\frac{2w_i m}{\pi t},
    0<p_j<\frac{2w_j m}{\pi\delta_t t}
  \right)
  =o(t^{-1}).
\end{equation}
For $m=1$, condition \cref{eq:divmcondition2} is understood as
vacuous.  Then
\begin{equation}\label{eq:divmasympvalid2}
  \lim_{t\to\infty}
  \frac{\mathbb P(\HCCT>t)}{1-\FHCw(t)}
  =
  \lim_{t\to\infty}
  \frac{\mathbb P(\HCCT>t)}{1-(2/\pi)\arctan(t)}
  =1.
\end{equation}
and 
\begin{equation}\label{eq:divmasympvalidhmp}
\spacingset{1}
  \lim_{t\to\infty}\frac{\mathbb{P}(T_{\mathrm{HM}}>t)}{1-F_{\mathrm{Pareto}, \boldsymbol{w}}(t)}=\lim_{t\to \infty}\frac{\mathbb{P}(T_{\mathrm{HM}}>t)}{1/t}=1.
\resetspacing
\end{equation}
\end{theorem}

\spacingset{1.85}
\setlength{\parskip}{8pt}
\cref{thm:revalidity} suggests that, for a broad range of dependence structures, either \(F_{\mathrm{HC}, \boldsymbol{w}}(t)\) of \cref{eq:numint} or \(F_{\mathrm{HC}}(t) = \frac{2}{\pi} \arctan(t)\) can effectively approximate the CDF of \(T_{\mathrm{HCCT}}\). In practice, however, \(F_{\mathrm{HC}, \boldsymbol{w}}(t)\) tends to do better than \(F_{\mathrm{HC}}(t)\), unless the dependence is heavy. Ideally, we want the combination test to be exact or at least strictly valid for \textit{independent} studies: using the rejection threshold from the inverse of \(F_{\mathrm{HC}, \boldsymbol{w}}(t)\) ensures this requirement, whereas using \(F_{\mathrm{HC}}(t)\) compromises validity by a logarithmic term, as implied by \cref{sec:Landau,sec:numeric}.
 
Our assumption in \cref{thm:revalidity} follows from the first part of Assumption D1 in \citet{long2023cauchy}. For growing $m$, we provide similar results in \cref{sec:growingnumber}. 
Intuitively, it means negligible co-movement in the tails of the score distributions for any pair of studies, which is the case for many settings enumerated in \cref{sec:copula}. In particular, any random vector that is pairwise bivariate normal with bounded correlations satisfies the assumptions in both \cref{thm:revalidity} and its growing $m$ generalization, and we have the following proposition.

\begin{proposition}[Growing-$m$ scalar-normal case]\label{thm:nval}
Let $t\to\infty$ along a triangular array with $m=m(t)\to\infty$. Suppose
every pair of standardized variables is bivariate normal and $0<\max_{1\leq i\neq j\leq m}|\rho_{ij}| \leq\rho_{\max}<1$
uniformly along the array.  Let the $p_i$ be the one-sided or two-sided
normal $p$-values, and suppose
$\max_{1\leq i\leq m}w_i=O(m^{-1})$.
Put
$\gamma_0=
  \frac{1-\rho_{\max}}{1+\rho_{\max}}$.
If
$m=O(t^{\gamma_0/2})$,
then both \cref{eq:divmasympvalid2,eq:divmasympvalidhmp} hold.
\end{proposition}

However, there are common scenarios such as multivariate $t$-distributions for which the assumptions in \cref{thm:revalidity} are unmet, yet we still observe that HCCT (as well as EHMP) performs well in finite samples as shown in \cref{sec:copula}. This suggests that the assumptions in \cref{thm:revalidity} may be relaxed, which is another open problem. 

\vspace{-10pt}
\subsection{Bridging Independence and Perfect Dependence}\label{sec:extreme}
\vspace{-10pt}
An extreme dependence scenario occurs when \(p_1=\cdots=p_m\) always holds. Under our scaling, the combination statistic then equals the common individual score. \cref{thm:revalidity} suggests that HCCT and EHMP have the same tail scale under independence and perfect dependence. This is desirable: combining identical tests should make the evidence neither more nor less significant. The following proposition shows that, within a broad class of heavy-tailed rules, for any given set of non-negative and normalized weights  $\boldsymbol{w}=\{w_1, \ldots, w_m\}$ such that $\sum_{i=1}^m w_i=1$,  this property holds if and only if the tail index is \(\alpha=1\).

\begin{proposition}\label{thm:extremecase}
Suppose for some $c_{1}\geq 0$, $c_{2}>0$ and $0<\alpha<2$, the density function of $\nu$ satisfies that
  \begin{equation}\label{eq:densityalpha}
  \spacingset{1}
    \textstyle \lvert t\rvert^{\alpha+1}f_{\nu}(t)\longrightarrow \begin{cases}
      c_{1}& \text{as }t\to -\infty\\
      c_{2}&\text{as }t\to+\infty
    \end{cases},
    \resetspacing
  \end{equation}
   Let $F_{\nu, \boldsymbol{w}}$ be the CDF of $T_{\nu, \boldsymbol{w}}$ of \cref{eq:gct} when the $m$ studies are independent.  Then 
  \begin{equation}\label{eq:extre2}
  \spacingset{1}
\lim_{t\to\infty}\frac{\mathbb{P}_{\mathrm{identical}}(T_{\nu,\boldsymbol{w}}>t)}{\mathbb{P}_{\mathrm{independent}}(T_{\nu,\boldsymbol{w}}>t)}=\lim_{t\to\infty}\frac{1-F_{\nu}(t)}{1-F_{\nu, \boldsymbol{w}}(t)}=\frac{1}{\sum_{i=1}^{m}w_{i}^{\alpha}}.
    \resetspacing
  \end{equation}
  For $m\ge2$, the right-hand side of \cref{eq:extre2} is one if and only if $\alpha=1$.
\end{proposition}

\setlength{\parskip}{10pt}
Here the if and only if statement holds because for $m\ge 2$, $f(\alpha)\equiv\sum_{j=1}^m w_j^\alpha$ is a strictly monotone decreasing function of $\alpha$ and with $f(1)=1$. Condition \cref{eq:densityalpha} is sufficient for \(\nu\) to belong to the domain of attraction of \(S(\alpha,\beta,c,\mu)\). Related approaches rescale the statistic by
\(
\kappa=\bigl(\sum_{j=1}^m w_j^\alpha\bigr)^{1/\alpha}
\)
or constrain the weights so that \(\sum_jw_j^\alpha=1\). Such normalization matches the tail scales of an individual score and the independent combination, but when \(\alpha\ne1\), it produces different thresholds under independence and perfect dependence. In particular, defining \(\widetilde T_{\nu,\boldsymbol w}=T_{\nu,\boldsymbol w}/\kappa\), \citet{fang2023heavy} showed that
  \begin{equation}\label{eq:extre3}
  \spacingset{1}
    \lim_{t\to\infty}\frac{\mathbb{P}_{\mathrm{identical}}(\widetilde{T}_{\nu,\boldsymbol{w}}>t/\kappa)}{\mathbb{P}_{\mathrm{independent}}(\widetilde{T}_{\nu,\boldsymbol{w}}>t)}=\lim_{t\to\infty}\frac{1-F_{\nu}(t)}{1-F_{\nu, \boldsymbol{w}}(\kappa t)}=1.
    \resetspacing
  \end{equation}

\spacingset{1.9}
\setlength{\parskip}{7pt}

One may use either the independence-calibrated threshold or the larger threshold from the two extreme dependence scenarios. Under independence calibration, \(\alpha<1\) is conservative and \(\alpha>1\) is asymptotically invalid when the \(p\)-values are identical. Using the larger threshold protects both extremes, but is conservative under identical \(p\)-values when \(\alpha<1\) and under independence when \(\alpha>1\). In both cases, \(\alpha=1\) avoids this discrepancy.

Validity under these two extremes does not guarantee validity under every dependence structure. Bonferroni correction and calibrated generalized-mean \(p\)-values use more conservative thresholds to obtain universal validity, illustrating the trade-off with power. In some applications, accepting a small validity loss under pathological dependence may yield greater power under more typical dependence structures.

\spacingset{1.7}
\setlength{\parskip}{2pt}

\vspace*{-5pt}
\subsection{Comparisons with Other Combination Tests}\label{sec:sim1}
\vspace*{-5pt}
\cref{tab:rates} provides a summary of the comparisons of various combination tests, highlighting their pros and cons. The convexity property was discussed in \cref{sec:unicase}, and the issue on exact computation was addressed in \cref{sec:numeric}. Therefore, here we focus on the validity and power of different tests in the presence of dependence between studies, and conduct simulations following the conventional setups of \citet{liu2020cauchy} and \citet{wilson2021evy}. A further discussion on comparing sizes of confidence regions is deferred to \cref{sec:comparelrt,app:size-comparisons}.

\begin{table}[tbp]
\vspace{-20pt}
  \centering
  \resizebox{\textwidth}{!}{\begin{minipage}{1.45\textwidth}
    \centering
  \begin{tabular}{ll|ccccc}
    \toprule
     \diagbox{Procedure}{Property} &                        & \makecell{Validity {\footnotesize }\\{\footnotesize (dependent tests)}}   & \makecell{Power {\footnotesize }\\{\footnotesize (dependent tests)}}      & \makecell{Exactness {\footnotesize }\\{\footnotesize (independent tests)}}  & \makecell[c]{Insensitivity\\{\footnotesize to large $p$-values}}       & \makecell[c]{Convexity {\footnotesize of} \\ {\footnotesize confidence regions}}                                                           \\
    \midrule
    \multicolumn{2}{l|}{Fisher {\footnotesize \citep{fisher1925statistical}}} & \Sadey[][red] & \Smiley[][green] & \Smiley[][green]& \Smiley[][green]  & \Neutrey[][yellow]     \\
    \multicolumn{2}{l|}{Stouffer {\footnotesize \citep{stouffer1949american}}} & \Sadey[][red]  & \Smiley[][green]        & \Smiley[][green]& \Sadey[][red]  & \Sadey[][red] \\ 
    \multicolumn{2}{l|}{Bonferroni {\footnotesize \citep{dunn1961multiple}}}& \Smiley[][green] & \Sadey[][red]  & \Sadey[][red]&\Smiley[][green] & \Smiley[][green]\\
    \multicolumn{2}{l|}{Simes {\footnotesize \citep{simes1986improved}}}& \Smiley[][green] & \Sadey[][red]  & \Sadey[][red]&\Smiley[][green] & \Smiley[][green]\\
    \multicolumn{2}{l|}{HMP {\footnotesize \citep{good1958significance,wilson2019harmonic}}}& \Smiley[][green] & \Smiley[][green]  & \Sadey[][red] & \Smiley[][green]& \Smiley[][green]\\
    \multirow{2}{*}{\makecell[l]{GMP\\ {\footnotesize \citep{vovk2020combining}}}} & {\footnotesize $\alpha<1$} & \Smiley[][green] &\Sadey[][red]  & \Sadey[][red]& \Smiley[][green] & \Smiley[][green]\\
        & {\footnotesize $\alpha>1$} & \Sadey[][red] &\Smiley[][green]  & \Sadey[][red]& \Smiley[][green] & \Neutrey[][yellow]\\
    \multicolumn{2}{l|}{CCT {\footnotesize \citep{liu2020cauchy}}} & \Smiley[][green] & \Smiley[][green]& \Smiley[][green] & \Sadey[][red]  & \Sadey[][red]\\
    \multicolumn{2}{l|}{LCT {\footnotesize \citep{wilson2021evy}}}& \Smiley[][green] & \Sadey[][red] & \Smiley[][green] & \Smiley[][green] & \Sadey[][red]\\
    \multirow{2}{*}{\makecell[l]{SCT\\ {\footnotesize \citep{wilson2021evy,ling2022stable}}}} & {\footnotesize $\alpha<1$} & \Smiley[][green] &\Sadey[][red]  & \Smiley[][green]& \Smiley[][green] & \Sadey[][red]\\
        & {\footnotesize $\alpha>1$} & \Sadey[][red] &\Smiley[][green]  & \Smiley[][green]& \Smiley[][green] & \Sadey[][red]\\
    \multicolumn{2}{l|}{CAtr {\footnotesize \citep{fang2023heavy}}}& \Smiley[][green] & \Smiley[][green] & \Smiley[][green] & \Neutrey[][yellow] & \Sadey[][red]\\
    \multirow{3}{*}{\makecell[l]{Left-Truncated $t$\\ {\footnotesize \citep{gui2023aggregating}}}} & {\footnotesize $\alpha<1$} & \Smiley[][green] &\Sadey[][red]  & \Sadey[][red]& \Neutrey[][yellow] & \Neutrey[][yellow]\\
        & {\footnotesize $\alpha=1$} & \Smiley[][green] &\Smiley[][green]  & \Sadey[][red]& \Neutrey[][yellow] & \Neutrey[][yellow]\\
        & {\footnotesize $\alpha>1$} & \Sadey[][red] &\Smiley[][green]  & \Sadey[][red]& \Neutrey[][yellow] & \Neutrey[][yellow]\\
    \multicolumn{2}{l|}{\textbf{HCCT} [Proposed]} & \Smiley[][green] & \Smiley[][green] & \Smiley[][green] & \Smiley[][green] & \Smiley[][green]\\
    \multicolumn{2}{l|}{\textbf{EHMP} [Proposed]} & \Smiley[][green] & \Smiley[][green] & \Smiley[][green] & \Smiley[][green] & \Smiley[][green]\\
    \bottomrule
  \end{tabular}
\end{minipage}}
\caption{\scriptsize Comparison of different combination tests. The smiley (green) and sad (red) faces represent respectively positive and negative rating. The stoic (yellow) face means that the rating can change according to different situations. HMP is shorthand for Harmonic Mean P-value; GMP for Generalized Mean P-value; CCT for Cauchy Combination Test; LCT for Lévy Combination Test; SCT for Stable Combination Test; CAtr for CAuchy with truncation; HCCT for Half-Cauchy Combination Test; EHMP for Exact Harmonic Mean P-value.}\label{tab:rates}
\end{table}

\spacingset{1.7}
\setlength{\parskip}{5pt}
 
\begin{figure}[tbp]
\vspace{-30pt}
\centering
\subfloat[False positive rate (AR-$1$)]{%
  \includegraphics[width=.30\textwidth,height=.14\textheight]{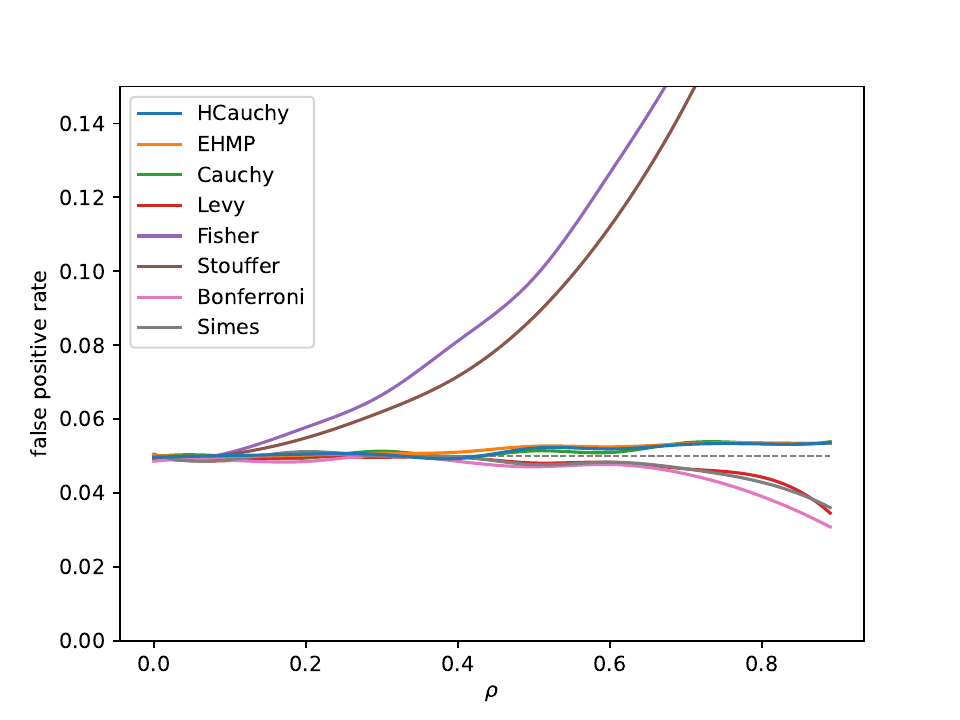}\label{fig:fprar1}
}~
\subfloat[False positive rate (equi-correlation)]{%
  \includegraphics[width=.30\textwidth,height=.14\textheight]{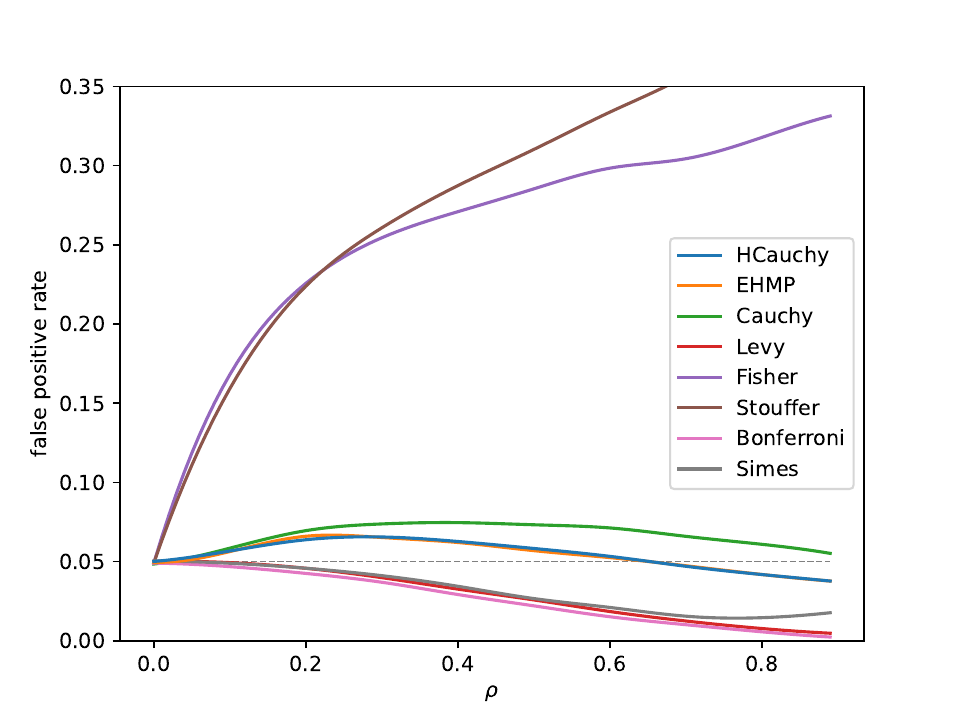}\label{fig:fpreq}
}\\
\vspace*{-15pt}
\subfloat[Power (weak signal)]{%
  \includegraphics[width=.30\textwidth,height=.14\textheight]{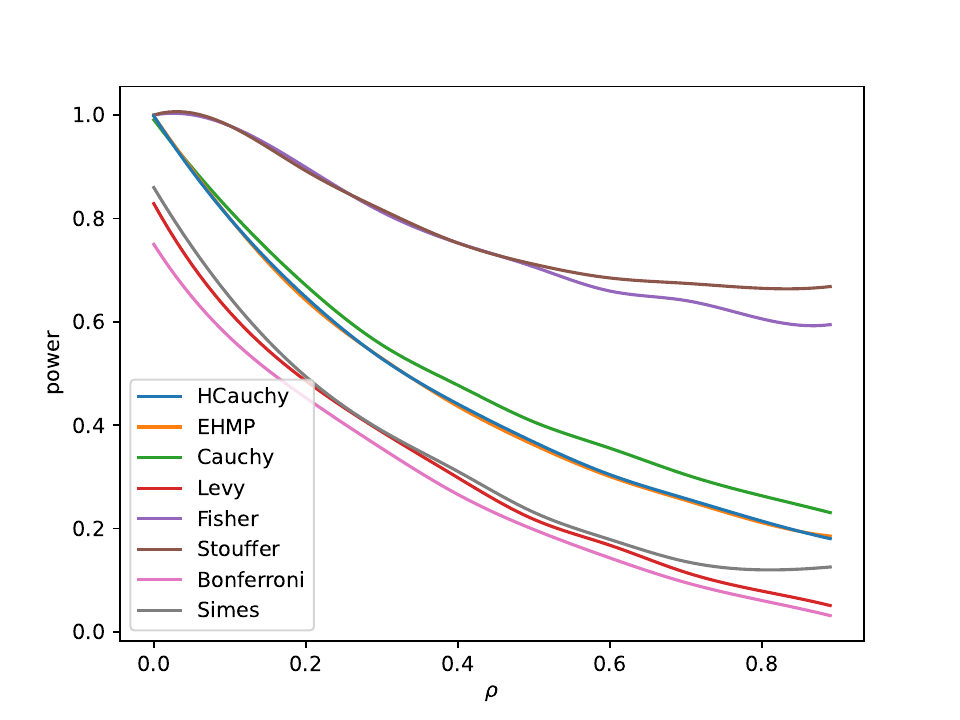}\label{fig:powerw}
}~
\subfloat[Power (sparse signal)]{%
  \includegraphics[width=.30\textwidth,height=.14\textheight]{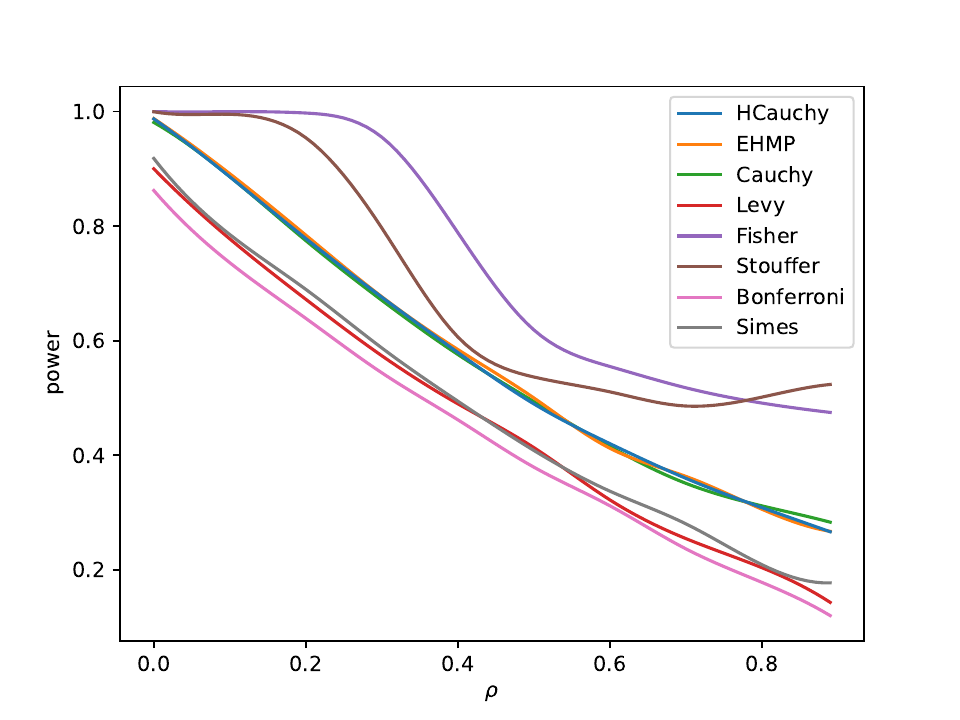}\label{fig:powers}
}
  \caption{Comparison of combination tests in false positive rate and power.}
  \label{fig:comparetest}
\end{figure}

We start by checking the validity with varying dependence structures and levels of dependence. For simplicity, we only present simulations under multivariate normal and leave simulations for other dependence structures such as multivariate $t$, and FGM and AMH copulas to \cref{sec:copula}. 
For multivariate normal, we adopt the setup of \cref{sec:unicase}, and show the false positive rates with the growth of $\rho$ for $m=500$ in \cref{fig:fprar1,fig:fpreq}.

Next, we investigate how signal strength and sparsity could influence the power of different tests along with levels of dependence. We consider the vector of individual test statistics $\boldsymbol{X}$ generated from the alternative $\mathcal{N}_{m}(\boldsymbol{\mu},\boldsymbol{\Sigma})$, where $\boldsymbol{\mu}=\{\mu_{i}\}$ and $\boldsymbol{\Sigma}=\{\sigma_{ij}\}$. Following the simulation setup of \citet{liu2020cauchy} and \citet{wilson2021evy}, we fix $\boldsymbol{\Sigma}$ to be the equi-correlation matrix as defined above and set
\begin{equation*}
\spacingset{1}
  \mu_{i}=
  \begin{cases}
    \sqrt{2r\log m} & 1\leq i\leq m_{0}=\lfloor m^{1-s}\rfloor\\
    0 & m_{0}+1\leq i\leq m
  \end{cases},
  \resetspacing
\end{equation*}
where $s\in [0,1)$ and $r>0$ are hyperparameters controlling the sparsity and strength. (The \(\sqrt{\log m}\) scale comes from Gaussian extreme-value theory and the classical sparse-signal detection framework.) \cref{fig:powerw} shows results for $s=0$ and $r=0.1$ (weak signal) and \cref{fig:powers} shows results for $s=r=0.3$ (sparse signal).

As shown in \cref{fig:comparetest}, Fisher's combination test and Stouffer's Z-score test corresponding to $\alpha=2$, tend to have inflated Type I error rates under dependence while Simes' test, Bonferroni correction ($\alpha\to 0$) and the Lévy Combination Test ($\alpha=1/2$) tend to have very low power. CCT, HCCT and EHMP corresponding to $\alpha=1$ strike a good balance between validity and power. In general, similar phenomena are observed in the GMP, SCT and Left-Truncated \(t\) approaches with different choices of \(\alpha\) (results not shown here). Specifically, \(\alpha < 1\) is conservative while \(\alpha > 1\) harms the validity. These observations align well with the theory in \cref{sec:extreme}. 

\spacingset{1.6}
\setlength{\parskip}{3pt}

Another important property of a combination test is its insensitivity to large $p$-values, which is crucial in applications where a large number of studies are combined. CCT, as discussed in \cref{sec:heavy}, is known to be sensitive to large $p$-values \citep{liu2020cauchy}, which is also the case for Stouffer's Z-score test. 
This sensitivity to large \(p_{j}\) values arises because the Cauchy distributions have equally heavy tails on both sides.  Placing more weight on the right tail than the left corresponds to using a larger skewness parameter \(\beta\) in the stable family \(S(\alpha, \beta, c, \mu)\). The choice of $\alpha=\beta=1$ leads exactly to the Landau family. See \cref{sec:insense} for more discussions. 

\spacingset{1.8}

For the five desirable properties considered here, HCCT and EHMP appear to be the most well-rounded methods as summarized in \cref{tab:rates}. However, we caution against over-interpreting this table as it does not capture all relevant aspects of these methods. For example, interpretability as Bayes factors, computation time, stability in distribution, and universal validity under arbitrary dependence are also important considerations.

\spacingset{1.75}
\setlength{\parskip}{4pt}

\compactsection{Reflections, Limitations, and Invitations}{sec:discussion}
A decade ago, the work on the Drton–Xiao conjecture—later published in \citet{pillai2016unexpected}—was motivated by theoretical curiosity \citep{meng2024bffer}. The rediscovery of the largely forgotten Cauchy combination result \cref{eq:cauchy} provided an elegant proof. At the time, its broader theoretical and practical implications were not fully recognized, beyond the intuition that heavy marginal tails might dominate joint stochastic behavior \citep[][Section 1]{pillai2016unexpected}. Subsequent work, including \citet{liu2020cauchy} and related developments summarized in \cref{sec:intro}, has substantially expanded these ideas into versatile heavy-tail approximation frameworks.

The appeal of heavy-tail approximations lies in their potential to play a role analogous to large-sample theory. Just as asymptotic approximations reduce sensitivity to infinite-dimensional distributional details, heavy-tail approximations can mitigate the burden of modeling complex dependence structures \citep{meng2024bffer}. As a proof of concept, we illustrate a divide-and-combine strategy in the elementary normal-mean setting. More broadly, the approach applies to any estimation problem permitting “lossless modularization,” where integrating modular components preserves the information integrity (e.g., identifiability) of the original problem.

\begin{figure}[tbp]
    \vspace{-15pt}
    \centering
    \subfloat[Bounded HCCT region.\label{fig:block-2d-hcct-main}]{%
      \includegraphics[width=0.25\linewidth]
        {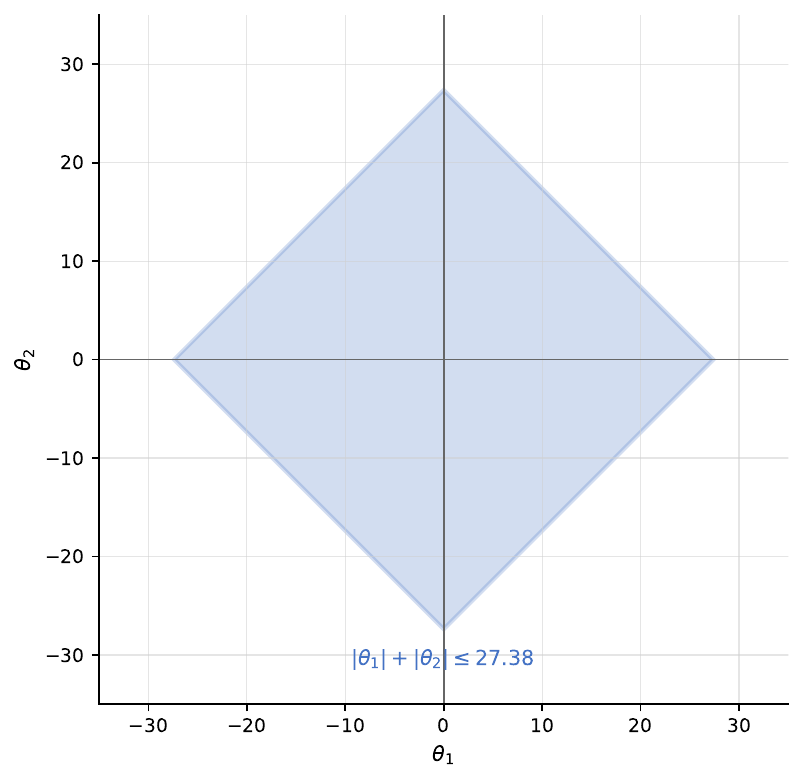}}
    ~~~
    \subfloat[CCT region, original scale.\label{fig:block-2d-cct-main}]{%
      \includegraphics[width=0.25\linewidth]
        {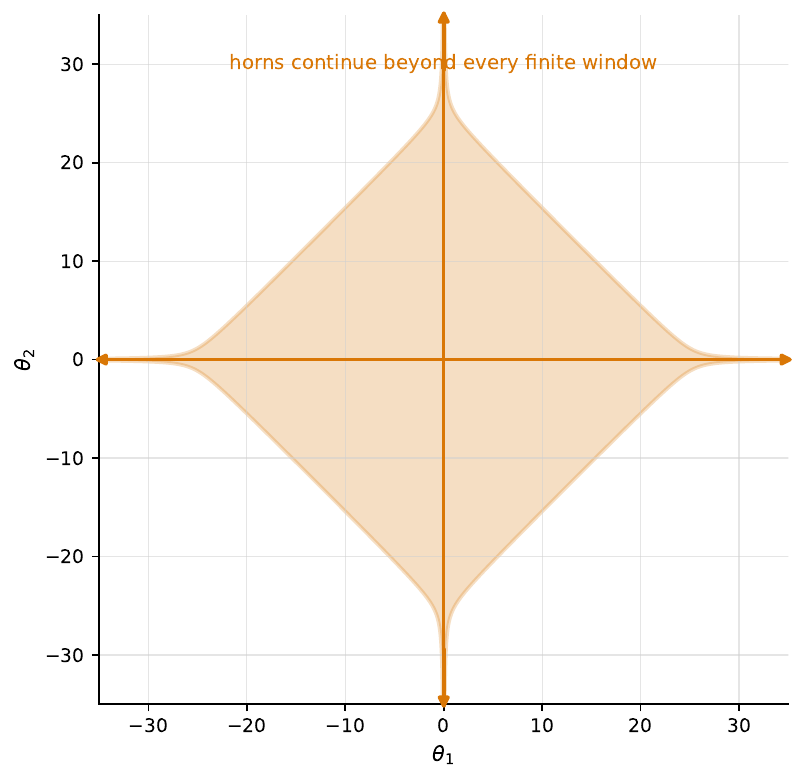}}
    ~~~
    \subfloat[Magnified upper CCT horn.\label{fig:block-2d-cct-horn-main}]{%
      \includegraphics[width=0.25\linewidth]
        {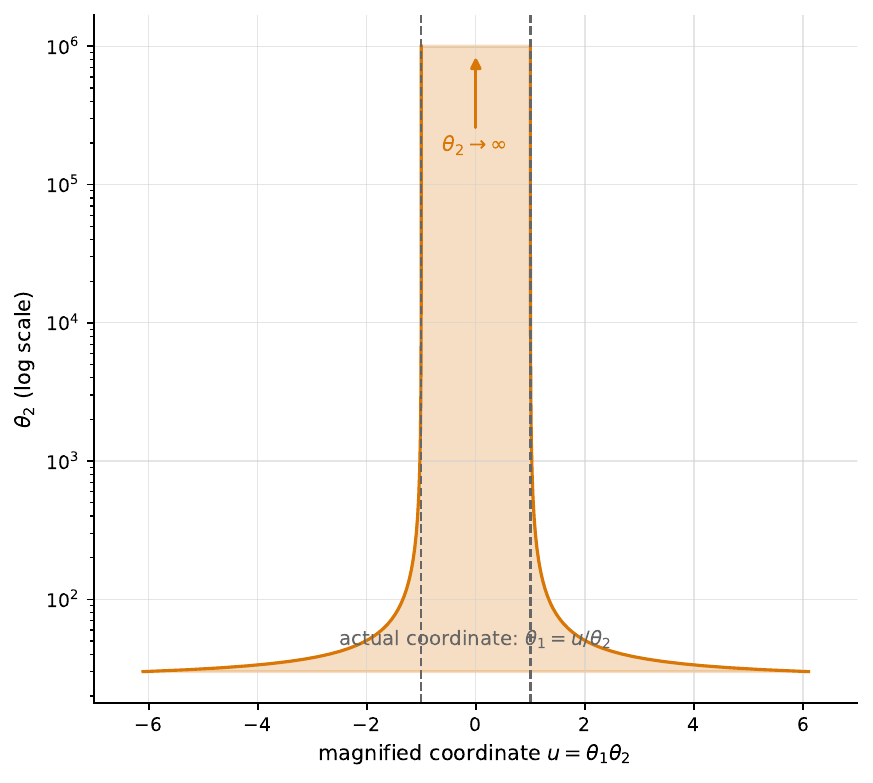}}
    \caption{\footnotesize Two-dimensional blockwise divide-and-combine geometry at
      $p=0.05$.  The first two panels use the same coordinate scale.
      The third uses $u=xy$ and a logarithmic vertical scale to make the
      narrowing but unbounded CCT horn visible.}
    \label{fig:block-2d-comparison-main}
\end{figure}

The divide-and-combine example also highlights a second, closely related theme: heavy-tail approximation and convex geometry can be interwoven within the same construction. For HCCT and EHMP, the transformations are chosen for their tail behaviors and that the resulting component scores are convex in the parameter, yielding convex confidence regions. Whereas the inferential principles for preferring convexity is yet to be established, its practical advantages are several. Computationally, as we demonstrated, convexity enables standard convex optimization methods for point estimates and simultaneous confidence intervals. Epistemically, it renders the usual confidence interval or well-shaped confidence regions for any linear transformations of the parameter vector. Furthermore, convexity, together with the spanning condition in \cref{thm:connectivity2}, ensures boundedness, whereas the corresponding nonconvex regions may be unbounded and even have infinite Lebesgue measure. In the CCT example shown in \cref{fig:block-2d-comparison-main}, unbounded horns extend along the coordinate directions, resulting in infinite volume; see \cref{app:size-comparisons} for details. In general, the interplay among heavy-tail approximation, modularization, and convex geometry opens a potentially fruitful direction for the design of confidence-set procedures.

How best to coordinate these ingredients remains a broad research question whose answer will depend on the inferential setting. Challenges arise, for example, with temporally or spatially dependent data. Even for the simpler problems considered here, we do not claim theoretical or practical optimality; rather, we demonstrate the feasibility and competitive gains of the heavily right strategy. Open problems include choosing sub-study dimensions and criteria for balancing statistical and computational efficiency; studying random dimension-reducing projections; bounding the difference between the actual and nominal coverages from the Half-Cauchy or Harmonic mean combinations; and incorporating partial information about the dependence structure.

With these and many more questions on our minds, we reiterate the invitations in previous sections to all interested parties to join us to explore this new paradigm of heavy-tail approximations for integrated dependent studies and especially for estimation in any dimension via the divide-and-combine strategy. Indeed, we will be most excited if the strategies, methods, and results presented in this article can all be improved significantly.

\clearpage
\numberwithin{theorem}{section}
\numberwithin{proposition}{section}
\numberwithin{lemma}{section}
\numberwithin{definition}{section}
\numberwithin{corollary}{section}
\numberwithin{conjecture}{section}
\numberwithin{example}{section}
\section*{Supplemental Material}

We provide the supplemental material for the paper ``A Heavily Right Strategy for Statistical Inference with Dependent Studies in Arbitrary Dimensions.'' \cref{sec:discconf} gives more insights on the convexity results in \cref{sec:conf} as well as comparisons of confidence interval widths between our approach and LRTs that assume known dependence structures, and comparisons of confidence region sizes between HCCT and CCT for several stylized setups. \cref{sec:furtherdis} expands the discussion in \cref{sec:hctest} by providing theoretical details on the dependence resilience properties of HCCT and EHMP, numerical details for computing the exact distribution under independence, simulations using other dependence structures beyond normal, and further discussion on sensitivity to large $p$-values. \cref{sec:pfconf,sec:pfhctest} contain all omitted proofs for the results in \cref{sec:conf,sec:hctest}, and \cref{sec:discconf,sec:furtherdis}. \cref{sec:global}  briefly reviews the literature on other global testing procedures that are not necessarily dependence-resilient.
\appendix
\crefalias{section}{appendix}
\crefalias{subsection}{appendix}
\crefalias{subsubsection}{appendix}
\section{Further Discussion for \cref{sec:conf}}\label{sec:discconf}

\subsection{More Insights on the Convexity Results}\label{sec:insightconv}

The following results explain when convexity of the component score is needed
for a guarantee across all study configurations and confidence levels. These
are deterministic statements; they do not establish the validity of the
individual or combined $p$-values.

\paragraph{Score convexity versus region convexity.}
Convexity of the defining score guarantees that every sublevel region is
convex. The converse does not hold for an individual region.
For example, a single identity-map study with a continuous, strictly
increasing, divergent radial score $\psi$ gives
\[
 \{\theta:\psi(\|\theta-\widehat\theta\|)\le t\},
\]
a ball (possibly empty or a singleton) at every finite cutoff, even if
$\psi$ is not convex.
The necessity statement below instead concerns a guarantee over all study
configurations and confidence levels: a nonconvex score permits suitable
shifted studies and a cutoff for which the combined region is nonconvex.
It does not imply failure for every observed configuration.

\begin{lemma}[Nonconvex scores and disconnected regions]\label{thm:lemmaconvexf}
Let $\psi:[0,\infty)\to[c,\infty)$ be continuous and strictly increasing,
where $c$ is finite, and set $g(\theta)=\psi(|\theta|)$.
If $g$ is not convex, there are $a>0$ and a finite $\delta>c$ such that
\[
 C_\delta=\left\{\theta:
 \tfrac12 g(\theta-a)+\tfrac12 g(\theta+a)\le\delta\right\}
\]
is disconnected. If $\psi=F_\nu^{-1}\circ H$, where $H$ is a continuous
strictly increasing CDF on $[0,\infty)$ and $F_\nu$ has a positive continuous
density on $(c,\infty)$ and support $[c,\infty)$, then $\delta$ can be
written as $F_{\nu,(1/2,1/2)}^{-1}(1-p)$ for some $0<p<1$.
\end{lemma}

\begin{proof}
A continuous function is convex if and only if it satisfies every midpoint
convexity inequality. Thus nonconvexity and symmetry give $a,h>0$ with
\[
 g(a)>\tfrac12\{g(a-h)+g(a+h)\}.
\]
The midpoint cannot be zero because $g(0)$ is its minimum; reflection
therefore allows $a>0$. Put
$M(\theta)=\{g(\theta-a)+g(\theta+a)\}/2$. Then
$M(0)=g(a)>M(h)=M(-h)$. Choose $\delta$ strictly between these values.
The sublevel set contains $-h$ and $h$ but not zero, so it is disconnected.
Moreover, $M(h)\ge c$, hence $\delta>c$.
Under the additional distributional assumptions, the mean of two independent
copies of $\nu$ has a continuous strictly increasing CDF on $(c,\infty)$.
Indeed, every nonempty open interval inside this support has positive
probability, and the convolution has no atoms. Consequently
$p=1-F_{\nu,(1/2,1/2)}(\delta)$ lies in $(0,1)$ and gives the stated quantile.
\end{proof}

The finite-continuous-score assumption matters. For a singular score, such as
the CCT score at radius zero, a local version suffices: if $\psi$ is twice
continuously differentiable near $a>0$ and $\psi''(a)<0$, then the same
symmetric construction has $M''(0)=\psi''(a)<0$. Choosing $0<h<a$
sufficiently small again gives $M(h)=M(-h)<M(0)$, with every evaluated score
finite. An interior cutoff in the support of the reference sum produces
the same failure of convexity.

\begin{lemma}[Necessary scalar score conditions]\label{thm:lemmanecessary}
Let $H$ be an absolutely continuous CDF on $[0,\infty)$ with positive,
nonincreasing, continuously differentiable density $h$ on $(0,\infty)$.
Let $F_\nu$ be absolutely continuous with support $[c,\infty)$, finite $c$,
and positive continuously differentiable density $f_\nu$ on $(c,\infty)$.
If $\psi=F_\nu^{-1}\circ H$, extended by $\psi(0)=c$, is convex, then
$f_\nu$ is nonincreasing and
\begin{equation}\label{eq:necessary-radial-tail}
 \lim_{u\uparrow1}\frac{F_\nu^{-1}(u)}{H^{-1}(u)}
 =\lim_{r\to\infty}\frac{\psi(r)}r
 \in(0,\infty].
\end{equation}
\end{lemma}

\begin{proof}
On $(0,\infty)$,
$\psi'=h/(f_\nu\circ\psi)>0$. Convexity makes $\psi'$ nondecreasing,
so $f_\nu\circ\psi=h/\psi'$ is nonincreasing. Since $\psi$ maps
$(0,\infty)$ onto $(c,\infty)$, $f_\nu$ is nonincreasing.
The secant slopes $(\psi(r)-c)/r$ are nondecreasing and strictly positive.
They have a limit in $(0,\infty]$, which is also the limit of $\psi(r)/r$.
Substituting $r=H^{-1}(u)$ proves \cref{eq:necessary-radial-tail}.
\end{proof}

For the scalar normal and Student reference laws in \cref{sec:unicase},
the relevant radial CDF is $H(r)=2F^{(j)}(r)-1$. Thus
$H^{-1}(1-\alpha)=\{F^{(j)}\}^{-1}(1-\alpha/2)$: the two-sided tail
probability must be matched when comparing quantiles.
\cref{eq:necessary-radial-tail} requires at least linear growth of the
score, not strictly faster growth. The Half-Cauchy score for a half-$t_1$
input is exactly linear. These are necessary conditions, not sufficient
conditions for convexity.

For the general construction, let $A_j\in\mathbb R^{d_j\times d}$ and
$b_j\in\mathbb R^{d_j}$ be fixed with respect to the candidate parameter
$\theta$. The matrices may be rectangular and need not be positive
semidefinite. Choose radial reference CDFs $\mathfrak F_j$ and define
\[
 p_j=1-\mathfrak F_j(\|A_j\theta+b_j\|).
\]
With $w_j\ge0$, $\sum_jw_j=1$, and $J_+=\{j:w_j>0\}$, the inverted region is
\begin{equation}\label{eq:invert2}
 \sum_{j\in J_+}w_jF_\nu^{-1}
 \{\mathfrak F_j(\|A_j\theta+b_j\|)\}
 \le F_{\nu,\boldsymbol w}^{-1}(1-p),\qquad 0<p<1.
\end{equation}
In the multivariate setup,
$A_j=\widehat\Sigma_j^{-1/2}P_j$ and
$b_j=-\widehat\Sigma_j^{-1/2}\widehat\xi_j$; in the scalar setup,
$A_j=1/\widehat\sigma_j$ and $b_j=-\widehat\theta_j/\widehat\sigma_j$.
This algebraic construction does not assume that the chosen reference CDF
is the actual sampling distribution of the norm.

\begin{lemma}[Convex transforms and normalized quantile curvature]
\label{thm:connectivity}
Let $F$ and $G$ be absolutely continuous CDFs with interval supports and
positive continuously differentiable densities $f$ and $g$ on their
support interiors.
For $Q_F=F^{-1}$, define the normalized quantile curvature
\begin{equation}\label{eq:quantile-curvature}
 \mathcal T_F(u)
 =-\frac{f'(Q_F(u))}{f(Q_F(u))^2}
 =\frac{Q_F''(u)}{Q_F'(u)}
 =(\log Q_F'(u))',\qquad 0<u<1.
\end{equation}
Then $F^{-1}\circ G$ is convex on the interior of the support of $G$
if and only if $\mathcal T_F(u)\ge\mathcal T_G(u)$ for every $0<u<1$.
A finite continuous extension of the transform to a finite support endpoint
preserves convexity.

More generally, let $\mathfrak H_j:\mathbb R^d\to I_j$ be finite convex
functions, where $I_j$ is an interval on which
$\psi_j=F_\nu^{-1}\circ\mathfrak F_j$ is finite, increasing, and convex,
including any required continuous endpoint extensions.
For nonnegative weights and any finite real $\delta$, the set
\begin{equation}\label{eq:invert2app}
 \left\{\theta:\sum_{j\in J_+}
 w_jF_\nu^{-1}\circ\mathfrak F_j\circ\mathfrak H_j(\theta)
 \le\delta\right\}
\end{equation}
is convex.
\end{lemma}

\begin{proof}
For $Q=F^{-1}\circ G$, the inverse-function and chain rules give
\begin{equation}\label{eq:convex-transform-identity}
 Q'(x)=\frac{g(x)}{f(Q(x))}>0,\qquad
 Q''(x)=Q'(x)g(x)
 \{\mathcal T_F(G(x))-\mathcal T_G(G(x))\}.
\end{equation}
All denominators are positive on the support interiors, so the displayed
identity proves the equivalence. Taking limits in the convexity inequality
extends it to any endpoint at which the transform has a finite continuous
extension; no density positivity or differentiability at that endpoint is
needed. Composition of an increasing convex function with a convex function
is convex. A nonnegative linear combination preserves convexity, and its
sublevel sets are convex, proving the final assertion.
\end{proof}

We use \emph{curvature} as shorthand for the normalized quantile curvature in
\cref{eq:quantile-curvature}. It measures the relative change in the quantile
slope; it is not the Euclidean curvature of a graph or of a region boundary.
It is invariant under positive rescaling, since the numerator and denominator
in $-f'/f^2$ acquire the same factor. The associated comparison is the
convex transform order.

For scalar half-normal and half-Student inputs,
$\mathcal T_H\ge0$, so convexity requires $\mathcal T_{F_\nu}\ge0$,
equivalently a nonincreasing target density. This observation is specific
to these scalar inputs: general radial Hotelling densities need not decrease.
Full-line targets, including Cauchy and Landau, also give a nonfinite score
at radius zero; their local behavior must not be inferred from a
finite-endpoint lemma. The proofs in \cref{sec:convexity-proofs} instead
establish the needed HCCT inequalities by six complete versions of three
related approaches and establish EHMP convexity directly. Their distinct
degree requirements and failure cases are discussed in
\cref{sec:convexity-boundaries}.

\subsubsection{Half-Cauchy and Pareto: Degree Boundaries and Sharpness}
\label{sec:convexity-boundaries}

The Half-Cauchy and Pareto comparisons are distinct requirements on the
radial input law. The analytic statements in this subsubsection allow
real $d\ge1$ and $k>d-1$; the applications to study dimensions and parameter
spaces use integer $d$. The thresholds concern convex score transforms,
and therefore uniform guarantees over configurations and finite cutoffs,
not a necessary condition for every particular observed region.

Write $H_{d,k}$ for the CDF of the positive square root of $T^2(d,k)$,
and $G_C$ and $G_P$ for the standard Half-Cauchy and Pareto CDFs.
The corresponding radial scores are $G_C^{-1}\circ H_{d,k}$ and
$G_P^{-1}\circ H_{d,k}$, with the continuous endpoint extensions in
\cref{eq:hc-pareto-score-transforms}.

\begin{lemma}[Pareto and Half-Cauchy curvature comparison]
\label{lem:hc-pareto-curvature}
For $0<u<1$,
\begin{equation}\label{eq:hc-pareto-curvature-comparison}
 \mathcal T_{G_P}(u)=\frac2{1-u}
 >\pi\tan(\pi u/2)=\mathcal T_{G_C}(u).
\end{equation}
\end{lemma}

\begin{proof}
Set $x=\pi(1-u)/2\in(0,\pi/2)$. The inequality becomes
$\cot x<1/x$, which follows from $\tan x>x$.
\end{proof}

Consequently, the Half-Cauchy curvature requirement implies the Pareto
requirement for the same input CDF. The converse need not hold.
This comparison explains the relationship between the methods but is
not used to prove the sharper EHMP theorem.
It implies neither nesting nor a volume ordering for the actual regions,
whose independence-reference cutoffs differ.

\begin{proposition}[Integer boundaries and failure cases]
\label{prop:convexity-integer-boundaries}
In dimension one, both the Half-Cauchy and Pareto scores are convex
exactly when $k\ge1$. For real $d>1$, neither score is convex when
$d-1<k<d$, and the Half-Cauchy score is not convex at $k=d$.
For integer $d\ge2$ and integer valid $k$, the Half-Cauchy score is therefore
convex exactly when $k\ge d+1$, whereas the Pareto score is convex exactly
when $k\ge d$.
\end{proposition}

\begin{proposition}[Real-degree improvements for Half-Cauchy]
\label{prop:convexity-real-degrees}
For $d=2$, the Half-Cauchy radial score is convex for every real
$k\ge5/2$. More generally, for each fixed real $d>1$ there exists
$\varepsilon_d\in(0,1)$ such that it is convex for every real
$k\ge d+1-\varepsilon_d$. Neither claim identifies the optimal real-degree
threshold.
\end{proposition}

The proofs of \cref{prop:convexity-integer-boundaries,prop:convexity-real-degrees}
are given in \cref{sec:convexity-boundary-proofs}.

\paragraph{Realizing the failure cases as nonconvex regions.}
The distinction between score convexity and an individual region's geometry
was explained in \cref{sec:insightconv}. We now show that the failure cases
above obstruct a uniform multi-study guarantee.
In an integer dimension choose $a_0>0$ with $\psi''(a_0)<0$,
two identity study maps with equal weights, and estimates $\pm a_0e_1$.
Along $\theta=te_1$, for $|t|<a_0$, the score is
\[
 J(t)=\tfrac12\psi(a_0-t)+\tfrac12\psi(a_0+t),\qquad
 J''(0)=\psi''(a_0)<0.
\]
For small $h>0$, choose a finite cutoff strictly between
$J(h)=J(-h)$ and $J(0)$. Both endpoint parameters are accepted and their
midpoint is not, so the full region is not convex, despite full active rank.
For HCCT and EHMP the cutoff is in the interior of the corresponding
reference-sum support and is an attainable quantile with $0<p<1$,
as in the cutoff argument of \cref{thm:lemmaconvexf}.
This establishes sharpness over configurations and cutoffs; it does not
assert failure for every data set or at every fixed confidence level.

\begin{table}[t]
\centering
\small
\renewcommand{\arraystretch}{1.15}
\begin{tabularx}{\linewidth}{@{}>{\raggedright\arraybackslash}p{0.26\linewidth}>{\raggedright\arraybackslash}X>{\raggedright\arraybackslash}X@{}}
\toprule
Setting & Half-Cauchy radial score & Pareto radial score\\
\midrule
$d=1$, real degrees
& Convex exactly for $k\ge1$; the score at \mbox{$k=1$} is linear.
& Convex exactly for $k\ge1$.\\
\addlinespace[4pt]
Integer $d\ge2$ and $k$
& Sharp uniform threshold $k\ge d+1$.
& Sharp threshold $k\ge d$.\\
\addlinespace[4pt]
$d>1$, real degrees
& $k\ge d+1$ is sufficient but not sharp; $k=d$ fails.
& Convex exactly for $k\ge d$.\\
\addlinespace[4pt]
$d=2$, real improvement
& $k\ge5/2$ is sufficient, not claimed optimal.
& Convex exactly for $k\ge2$.\\
\addlinespace[4pt]
Chi/normal limit
& Convex.
& Convex.\\
\bottomrule
\end{tabularx}
\caption{Different degree requirements for Half-Cauchy and Pareto.
Sharpness refers to score convexity and a uniform guarantee over study
configurations and finite cutoffs, not necessity for each observed region.
Finite-degree comparisons assume $k>d-1$. Real dimensions concern the
analytic laws; study dimensions are integers.}
\label{tab:hc-pareto-convexity}
\end{table}

\subsection{Comparison to LRT with Known Dependence Structures}\label{sec:comparelrt}

It would be interesting to compare the size of the confidence intervals to a gold standard approach that accounted for the dependence structure assuming it were known. Here we consider the univariate setting as in \cref{sec:unicase}, this gold standard is the likelihood ratio test (LRT) based on the joint distribution of \(\boldsymbol{X} = (X_{1}, \dots, X_{m})^{\top} \sim \mathcal{N}(\theta\, \mathbf{1}_{m}, \mathit{\Sigma}),\) where \(\mathit{\Sigma}\) is the known covariance matrix and $m$ is the number of studies. The LRT for testing \(H_{0}: \theta=0\) versus \(H_{1}: \theta\neq 0\) rejects \(H_{0}\) when
\[-2\log \Lambda=(\boldsymbol{X}-\theta \mathbf{1}_{m})^{\top}\mathit{\Sigma}^{-1}(\boldsymbol{X}-\theta \mathbf{1}_{m})-(\boldsymbol{X}-\widehat{\theta} \mathbf{1}_{m})^{\top}\mathit{\Sigma}^{-1}(\boldsymbol{X}-\widehat{\theta} \mathbf{1}_{m})=S (\widehat{\theta}-\theta)^{2}>c_{\alpha},\]
where 
\begin{equation*}
  \widehat{\theta} = \frac{\mathbf{1}^{\top} \mathit{\Sigma}^{-1} X}{\mathbf{1}^{\top} \mathit{\Sigma}^{-1} \mathbf{1}},\quad S = \mathbf{1}^{\top} \mathit{\Sigma}^{-1} \mathbf{1},
\end{equation*}
and \(c_{\alpha}\) is the \(1 - \alpha\) quantile of the \(\chi^{2}_{1}\) distribution. 

In the equi-correlation case, where $\mathit{\Sigma}$ is given by $(1-\rho)I_{m}+\rho \mathbf{1}\mathbf{1}^{\top}$ ($\rho$ is known), the LRT gives us that
\begin{equation*}
    \sqrt{m_{\mathrm{eff}}}(\bar{X}-\theta)\sim \mathcal{N}(0,1),\quad m_{\mathrm{eff}}=\frac{m}{1+(m-1)\rho}.
\end{equation*}
If $\rho=0$, the confidence interval shrinks at the rate of $1/\sqrt{m}$. If $\rho>0$, the confidence interval shrinks at the rate of $1/\sqrt{m_{\mathrm{eff}}}$, which converges to a positive constant as $m\to \infty$. In particular, we calculate that for $m=500$ and $\rho=0,0.3,0.6,0.9$, the corresponding widths of the LRT confidence intervals are approximately $0.18, 2.15,3.04,3.72$. Comparing this to \cref{fig:1dwidthequi}, we see that HCCT gives much larger confidence intervals when $\rho=0$ but roughly comparable intervals for $\rho>0$ without requiring the knowledge of $\rho$.

In the AR-1 correlation case, where $\mathit{\Sigma}$ is given by $(\rho^{|i-j|})_{1\leq i,j\leq m}$ ($\rho$ is known), the LRT gives us that
\begin{equation*}
    \sqrt{\frac{m-(m-2)\rho}{1+\rho}}(\widehat{\theta}-\theta)\sim \mathcal{N}(0,1),\quad \widehat{\theta}=\frac{X_{1}+X_{m}+(1-\rho)\sum_{i=2}^{m-1}X_{i}}{m-(m-2)\rho}.
\end{equation*}
In other words, the LRT confidence interval always shrinks at the rate of $1/\sqrt{m}$ for $0\leq \rho<1$.  In particular, we calculate that for $m=500$ and $\rho=0,0.3,0.6,0.9$, the corresponding widths of the LRT confidence intervals are approximately $0.18, 0.24,0.35,0.75$. Comparing this to \cref{fig:1dwidthar1}, we see that HCCT always gives larger confidence intervals for $\rho<1$ due to the relatively large $m$ and square root shrinkage in LRT intervals.

\subsection{Stylized Size Comparisons Between HCCT and CCT Regions}
\label{app:size-comparisons}

This section compares the sizes of confidence regions obtained by inverting
HCCT and CCT in two stylized settings.  The first is a two-focus model, which
makes the effect of disagreement between two studies transparent.  The second
is a blockwise divide-and-combine model, which mirrors the multivariate mean
construction in \cref{sec:divideac}.  The comparisons below are deterministic geometric
statements: once the individual estimates and critical values are fixed, they
do not require the working distribution used to produce the individual
$p$-values to be correctly specified.

We use $\lambda_d(A)$ for $d$-dimensional Lebesgue measure,
$\operatorname{diam}(A)$ for Euclidean diameter, and
\[
    \kappa_d=\frac{\pi^{d/2}}{\Gamma(d/2+1)}
\]
for the volume of the unit ball in $\mathbb R^d$.

\subsubsection{A Radial Model and a Useful Identity}

Let $r_j(\theta)\geq 0$ measure the discrepancy between $\theta$ and the
estimate from study $j$, and consider the two-sided $p$-value
\begin{equation}
    p_j(\theta)
    =1-\frac{2}{\pi}\arctan\{r_j(\theta)\}.
    \label{eq:radial-half-cauchy-pvalue}
\end{equation}
Equivalently, the null distribution of $r_j$ is standard Half-Cauchy.  This
choice is useful because
\begin{equation}
    \cot\left\{\frac{\pi p_j(\theta)}{2}\right\}=r_j(\theta),
    \qquad
    \cot\{\pi p_j(\theta)\}
    =\frac{1}{2}\left\{r_j(\theta)-\frac{1}{r_j(\theta)}\right\}.
    \label{eq:hcct-cct-radial-identities}
\end{equation}
The second identity also follows from
\[
    \cot(2x)=\frac{1}{2}\{\cot(x)-\cot(x)^{-1}\}.
\]
Thus HCCT produces a weighted sum of distances, whereas CCT subtracts a
reciprocal-distance term.  The latter creates a singular well at every
individual estimate.

Let
\(
    q_C(\alpha)=\cot(\pi\alpha)
\)
be the CCT critical value.  For weights $\boldsymbol w=(w_1,\ldots,w_m)$, let
$q_H(\alpha,\boldsymbol w)$ denote the $(1-\alpha)$-quantile of
$\sum_{j=1}^m w_j H_j$, where the $H_j$ are independent standard Half-Cauchy
variables.  This is the exact HCCT critical value under independent studies.

\subsubsection{A Two-Focus Setting}

Suppose $m=2$, $w_1=w_2=1/2$, and the two estimates are
\[
    a_1=-ce_1,\qquad a_2=ce_1,
\]
where $c>0$ and $e_1$ is the first coordinate vector.  Put
\[
    r_j(\theta)=\|\theta-a_j\|,
    \qquad
    q_H=q_H\{\alpha,(1/2,1/2)\},
    \qquad q_C=q_C(\alpha).
\]
By \cref{eq:hcct-cct-radial-identities}, the two confidence regions are
\begin{align}
    \mathcal R_H
    &=\left\{\theta:r_1(\theta)+r_2(\theta)\leq 2q_H\right\},
    \label{eq:equal-weight-hcct-region}\\
    \mathcal R_C
    &=\left\{\theta:
      \frac{1}{4}\left\{
      r_1(\theta)+r_2(\theta)
      -\frac{1}{r_1(\theta)}-\frac{1}{r_2(\theta)}
      \right\}\leq q_C\right\}.
    \label{eq:equal-weight-cct-region}
\end{align}
In \cref{eq:equal-weight-cct-region} and below, the value at $r_j=0$ is
understood by continuity as $-\infty$.

\begin{proposition}[Exact geometry of the HCCT region]
\label{prop:hcct-prolate-hyperspheroid}
The region $\mathcal R_H$ is empty if $c>q_H$.  If $c<q_H$, it is a prolate
hyperspheroid with focal semiaxis $q_H$ and $d-1$ orthogonal semiaxes
$\sqrt{q_H^2-c^2}$.  Consequently,
\begin{align}
    \lambda_d(\mathcal R_H)
    &=\kappa_d q_H(q_H^2-c^2)^{(d-1)/2},
    \label{eq:hcct-hyperspheroid-volume}\\
    \operatorname{diam}(\mathcal R_H)&=2q_H.
    \label{eq:hcct-hyperspheroid-diameter}
\end{align}
At $c=q_H$, the region degenerates to the line segment joining the two foci.
\end{proposition}

\begin{proof}
\cref{eq:equal-weight-hcct-region} is the focal definition of a
prolate hyperspheroid.  Its semiaxes and volume then follow from a linear
transformation of the unit ball.
\end{proof}

\begin{proposition}[When HCCT has smaller diameter]
\label{prop:equal-weight-projected-comparison}
Let $x_C>c$ be the unique solution of
\begin{equation}
    \frac{x_C}{2}-\frac{x_C}{2(x_C^2-c^2)}=q_C,
    \label{eq:cct-axial-endpoint}
\end{equation}
or, equivalently,
\begin{equation}
    (x_C-2q_C)(x_C^2-c^2)-x_C=0.
    \label{eq:cct-axial-cubic}
\end{equation}
Then $\pm x_Ce_1\in\mathcal R_C$, and hence
\begin{equation}
    \operatorname{diam}(\mathcal R_C)\geq 2x_C.
    \label{eq:cct-projection-and-diameter}
\end{equation}
Therefore, whenever $c<q_H$ and
\begin{equation}
    \frac{q_H}{2}-\frac{q_H}{2(q_H^2-c^2)}<q_C,
    \label{eq:equal-weight-diameter-condition}
\end{equation}
both regions are nonempty and
\begin{equation}
    \operatorname{diam}(\mathcal R_H)
    <\operatorname{diam}(\mathcal R_C).
    \label{eq:equal-weight-strict-size-comparison}
\end{equation}
If $q_H>2q_C$, the condition in \cref{eq:equal-weight-diameter-condition} is
equivalent to
\begin{equation}
    c^2>q_H^2-\frac{q_H}{q_H-2q_C}.
    \label{eq:equal-weight-diameter-threshold}
\end{equation}
\end{proposition}

\begin{proof}
For $x>c$, the CCT score at $xe_1$ is
\[
    \frac14\left\{(x-c)+(x+c)-\frac{1}{x-c}-\frac{1}{x+c}\right\}
    =\frac{x}{2}-\frac{x}{2(x^2-c^2)}.
\]
It is strictly increasing from $-\infty$ to $+\infty$, proving existence and
uniqueness of $x_C$.  By symmetry, both $\pm x_Ce_1$ lie on the boundary of
$\mathcal R_C$, which proves \cref{eq:cct-projection-and-diameter}.
Comparing $x_C$ with $q_H$ gives
\cref{eq:equal-weight-diameter-condition}; elementary algebra gives
\cref{eq:equal-weight-diameter-threshold}.
\end{proof}

For example, at $\alpha=0.05$, numerical evaluation gives
\[
    q_C\approx 6.31375,
    \qquad q_H\approx13.68751.
\]
The right side of \cref{eq:equal-weight-diameter-threshold} yields the
threshold $c_*\approx13.20740$, or $c_*/q_H\approx0.96492$.  Thus the strict
comparison in \cref{eq:equal-weight-strict-size-comparison} holds throughout
$c_*<c<q_H$, while the HCCT region remains nonempty.

This diameter comparison does not require the line segment between the two
extreme points to belong to $\mathcal R_C$.  Indeed, put
\[
    c_0=q_C+\sqrt{q_C^2+1}
       =\cot\left(\frac{\pi\alpha}{2}\right).
\]
When $c>c_0$, the CCT projection onto the focal axis is disconnected and is
given by
\[
    [-x_C,-x_0]\cup[x_0,x_C],
    \qquad
    x_0=
    \left(c^2-\frac{c}{c-2q_C}\right)^{1/2}.
\]
Thus $2x_C$ is relevant to diameter, but the Lebesgue measure of this
projection is $2(x_C-x_0)$.

We next give a lower bound for the CCT volume.  Define $f(r)=r-r^{-1}$, and let
$\rho_C(c)>0$ be the unique solution of
\begin{equation}
    f\{\rho_C(c)\}+f\{2c+\rho_C(c)\}=4q_C.
    \label{eq:equal-weight-inscribed-radius}
\end{equation}

\begin{proposition}[CCT volume lower bound]
\label{prop:equal-weight-cct-volume-lower-bound}
Up to their center points, at which the CCT score is defined by a limiting
value, the CCT region contains both balls
\[
    B_d\{a_1,\rho_C(c)\}\setminus\{a_1\}
    \quad\text{and}\quad
    B_d\{a_2,\rho_C(c)\}\setminus\{a_2\}.
\]
If $\rho_C(c)\leq c$, these balls have disjoint interiors and their
overlap has zero Lebesgue measure, so
\begin{equation}
    \lambda_d(\mathcal R_C)
    \geq 2\kappa_d\rho_C(c)^d.
    \label{eq:equal-weight-cct-volume-lower-bound}
\end{equation}
Consequently, for $0<c<q_H$ with $\rho_C(c)\leq c$, a sufficient
condition for HCCT to have smaller volume is
\begin{equation}
    2\rho_C(c)^d
    >q_H(q_H^2-c^2)^{(d-1)/2}.
    \label{eq:equal-weight-volume-comparison-condition}
\end{equation}
For every fixed $d\geq2$, there exists $c_d\in(0,q_H)$ such that
HCCT has smaller volume than CCT for all $c_d<c<q_H$.
\end{proposition}

\begin{proof}
If $\|\theta-a_1\|\leq\rho_C(c)$, then
$r_1(\theta)\leq\rho_C(c)$ and
$r_2(\theta)\leq2c+\rho_C(c)$. Since $f$ is increasing,
the CCT score is at most $q_C$ by
\cref{eq:equal-weight-inscribed-radius}. The argument for the other ball is
identical. Under $\rho_C(c)\leq c$, their overlap has zero Lebesgue measure,
which proves \cref{eq:equal-weight-cct-volume-lower-bound}.
Comparison with \cref{eq:hcct-hyperspheroid-volume} then gives
\cref{eq:equal-weight-volume-comparison-condition} under the same condition.
Regardless of overlap, either contained ball gives
\[
    \lambda_d(\mathcal R_C)\geq\kappa_d\rho_C(c)^d.
\]
As $c\uparrow q_H$, the HCCT volume converges to zero for $d\geq2$,
whereas continuity and \cref{eq:equal-weight-inscribed-radius} give
$\rho_C(c)\to\rho_C(q_H)>0$. The one-ball lower bound therefore proves
the final existence claim.
\end{proof}

For $0<c<q_H$ with $\rho_C(c)\leq c$, put $t=c/q_H$. The ratio of
the two-ball CCT lower bound to the exact HCCT volume is
\begin{equation}
    2\sqrt{1-t^2}
    \left\{
      \frac{\rho_C(c)}{q_H\sqrt{1-t^2}}
    \right\}^{d}.
    \label{eq:equal-weight-dimensional-amplification}
\end{equation}
Under these conditions, dimension amplifies the comparison once the radius
of the inscribed CCT balls exceeds the minor semiaxis of the HCCT hyperspheroid.  The volume conclusion
is deliberately conditional: it identifies fixed configurations, with both
regions nonempty, for which HCCT is smaller.  It does not by itself compare
expected volumes under repeated sampling.

\subsubsection{A Blockwise Divide-and-Combine Model}

We finally consider a stylized version of the divide-and-combine construction
from \cref{sec:divideac}.  Let $d=mk$ with $m\geq2$, and partition
$\theta\in\mathbb R^d$ and its estimate $\widehat\theta$ into $m$ blocks of
dimension $k$:
\[
    \theta=(\theta_1^\top,\ldots,\theta_m^\top)^\top,
    \qquad
    \widehat\theta=(\widehat\theta_1^\top,\ldots,
    \widehat\theta_m^\top)^\top.
\]
For a common scale $s>0$, define
\[
    r_j(\theta)=\frac{\|\theta_j-\widehat\theta_j\|}{s}
\]
and use the radial $p$-values in
\cref{eq:radial-half-cauchy-pvalue}, with equal weights $1/m$.  Let
\[
    q_H=q_H\{\alpha,(1/m,\ldots,1/m)\}.
\]

\begin{proposition}[Finite HCCT volume and infinite CCT volume]
\label{prop:block-divide-combine-volume}
The HCCT region is the block-$\ell_{2,1}$ ball
\begin{equation}
    \mathcal R_H
    =\left\{\theta:
      \sum_{j=1}^m\|\theta_j-\widehat\theta_j\|
      \leq smq_H
      \right\}.
    \label{eq:block-hcct-region}
\end{equation}
Writing $R=smq_H$, its diameter and volume are
\begin{align}
    \operatorname{diam}(\mathcal R_H)&=2R,
    \label{eq:block-hcct-diameter}\\
    \lambda_d(\mathcal R_H)
    &=
    \frac{
      \left\{2\pi^{k/2}\Gamma(k)/\Gamma(k/2)\right\}^{m}
    }{
      \Gamma(d+1)
    }
    R^d.
    \label{eq:block-hcct-volume}
\end{align}
In contrast, the CCT region
\begin{equation}
    \mathcal R_C
    =\left\{\theta:
      \frac{1}{2m}\sum_{j=1}^m
      \left\{r_j(\theta)-\frac{1}{r_j(\theta)}\right\}
      \leq q_C
      \right\}
    \label{eq:block-cct-region}
\end{equation}
has infinite diameter and infinite $d$-dimensional Lebesgue measure:
\begin{equation}
    \operatorname{diam}(\mathcal R_C)=\infty,
    \qquad
    \lambda_d(\mathcal R_C)=\infty.
    \label{eq:block-cct-infinite-size}
\end{equation}
\end{proposition}

The infinite-volume conclusion in \cref{prop:block-divide-combine-volume}
uses the radial Half-Cauchy model, whose CCT score grows only linearly away
from the estimates.  A broader conclusion does not require this particular
tail behavior.  For general divide-and-combine $p$-values, if a positively
weighted linear transformation $P_j$ has rank strictly smaller than $d$, then
\(
    \{\theta:P_j\theta=\widehat\xi_j\}
\)
is an unbounded affine subspace on which $p_j(\theta)=1$ and the CCT score is
$-\infty$.  Hence the inverted CCT region always has infinite diameter.  By
\cref{lem:convexity-active-rank}, the corresponding HCCT region is bounded
whenever the positively weighted row spaces of the $P_j$ span $\mathbb R^d$
and the active covariance estimates are positive definite. This boundedness
uses the nonnegative, diverging scores and does not require score convexity.
Thus the diameter
comparison persists for the Gaussian and Hotelling versions of the
divide-and-combine construction, even though infinite CCT Lebesgue measure is
not asserted without an additional tail condition.

\paragraph{A two-dimensional illustration.}
The mechanism is especially transparent when $m=2$, $k=1$, $s=1$, and
$\widehat\theta=(0,0)$.  Writing $\theta=(x,y)$ gives
\[
    r_1(\theta)=|x|,
    \qquad
    r_2(\theta)=|y|.
\]
The HCCT region is the bounded diamond
\[
    \mathcal R_H
    =\{(x,y):|x|+|y|\leq2q_H\},
\]
whereas the CCT region is
\[
    \mathcal R_C
    =\left\{(x,y):
      \frac14\left\{
        |x|-\frac1{|x|}+|y|-\frac1{|y|}
      \right\}\leq q_C
      \right\}.
\]
Here and below, the CCT score is interpreted as $-\infty$ when either
coordinate is zero.  At $\alpha=0.05$, the values used in the illustrations
are $q_H\approx13.68751$ and $q_C\approx6.31375$.  Thus
$\mathcal R_H=\{(x,y):|x|+|y|\leq27.37502\}$, with finite diameter
$54.75004$ and finite area $8q_H^2\approx1498.78$.  By contrast,
$\mathcal R_C$ contains both coordinate axes and therefore has infinite
diameter.

For $y\neq0$, the exact half-width of the vertical CCT horn is
\[
    x_{\max}(y)
    =f^{-1}\{4q_C-f(|y|)\},
    \qquad
    f^{-1}(a)=\frac{a+\sqrt{a^2+4}}2.
\]
As $|y|\to\infty$,
\[
    x_{\max}(y)\sim\frac1{|y|}.
\]
The horn therefore becomes arbitrarily thin but never terminates.
\cref{fig:block-2d-comparison-main} illustrates both regions and magnifies the
horizontal coordinate of the CCT horn by writing $u=xy$; in these coordinates
the upper horn approaches the strip $|u|\leq1$ as $y\to\infty$.

The axes alone have two-dimensional measure zero, so unboundedness does not
yet prove infinite area.  For dyadic $L=2^n$, consider the smaller tubes
\[
    \widetilde T_L
    =\left\{(x,y):
      \frac1{8L}\leq x\leq\frac1{4L},
      \quad L\leq y\leq2L
      \right\}.
\]
They are mutually disjoint, each is contained in $\mathcal R_C$ for all
sufficiently large $L$, and
\[
    \lambda_2(\widetilde T_L)
    =\left(\frac1{4L}-\frac1{8L}\right)L
    =\frac18.
\]
The infinitely many equal-area tubes therefore give
$\lambda_2(\mathcal R_C)=\infty$.

\begin{figure}[htbp]
    \centering
    \includegraphics[width=0.55\linewidth]
      {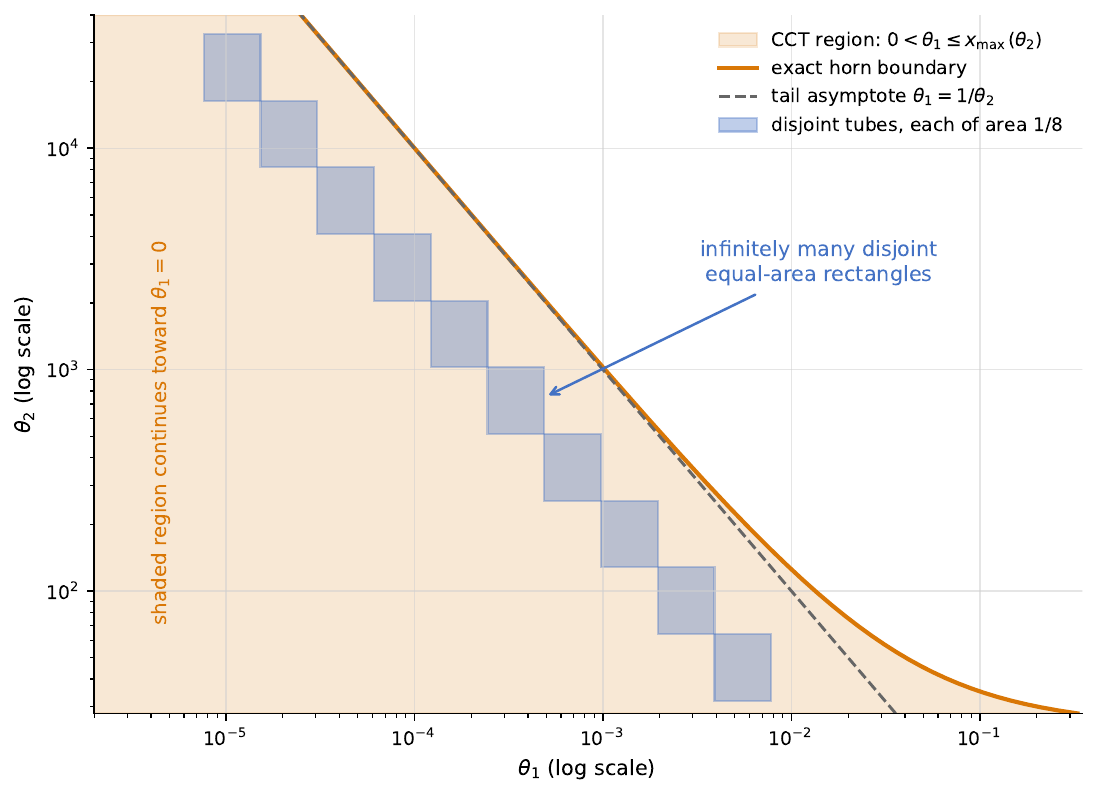}
    \caption{Disjoint tubes in the CCT horn, each of area $1/8$ in the
      original coordinates. Both axes use logarithmic scales.}
    \label{fig:cct-infinite-area-tubes}
\end{figure}

These examples also illustrate why no single notion of size is likely to
capture all relevant features of an irregular confidence region.  Diameter,
Lebesgue measure, and the lengths or measures of projections can rank the same
two procedures differently, particularly when one region is disconnected.
The calculations here are intended as tractable benchmarks rather than a
general efficiency ordering.  Developing expected-size comparisons under
sampling models, identifying other geometries that admit exact calculations,
and studying how the conclusions vary with dimension, dependence, and the
choice of projections are natural directions for future work.

\section{Further Discussion for \cref{sec:hctest}}\label{sec:furtherdis}

\subsection{Notation and Scope}

Let $p_1,\ldots,p_m$ be exact $\operatorname{Uniform}(0,1)$ marginal
$p$-values under the null, with arbitrary joint dependence unless stated
otherwise.  Let
\begin{equation}\label{eq:hcct-report}
  \HCCT
  =\sum_{i=1}^m w_i\cot\!\left(\frac{\pi p_i}{2}\right),
  \qquad
  w_i\geq 0,
  \qquad
  \sum_{i=1}^m w_i=1.
\end{equation}
The proof divides by $w_i$.  We therefore work on the active set
\[
  J_+=\{i:w_i>0\}.
\]
After deleting zero-weight coordinates and relabeling, we may and do assume
$w_i>0$ for every $i$.  Thus $m$ below is the number of active weights.

Write $\FHCw$ for the CDF of the statistic in
\cref{eq:hcct-report} when the $p_i$ are independent exact-uniform random
variables and the weights are kept fixed.  No independence is implicit in
$\mathbb P(\HCCT>t)$ without this explicit qualification.

For growing $m$, all limits are taken along an explicit sequence
$t\to\infty$ with
\[
  m=m(t),\qquad
  \boldsymbol w=\boldsymbol w(t),\qquad
  (p_1,\ldots,p_m)=(p_{1,t},\ldots,p_{m(t),t}).
\]
As in the manuscript, the row index $t$ is suppressed on $p_i$ and $w_i$
inside formulas.

\subsection{Half-Cauchy Marginal Identities}

Put
\[
  H_i=\cot\!\left(\frac{\pi p_i}{2}\right).
\]

\begin{lemma}[Exact Half-Cauchy tail]\label{lem:hc-tail}
For every $x>0$,
\begin{equation}\label{eq:exact-hc-tail}
  \mathbb P(H_i>x)
  =\frac{2}{\pi}\arctan\!\left(\frac{1}{x}\right).
\end{equation}
Consequently,
\begin{equation}\label{eq:hc-first-order}
  \mathbb P(H_i>x)
  =\frac{2}{\pi x}+O(x^{-3}).
\end{equation}
Moreover, for $x\geq0$,
\begin{equation}\label{eq:arctan-bounds}
  0\leq x-\arctan x\leq \frac{x^3}{3}.
\end{equation}
\end{lemma}

\begin{proof}
The event $H_i>x$ is equivalent to
\[
  0<p_i<\frac{2}{\pi}\arctan(1/x).
\]
Exact uniformity gives \cref{eq:exact-hc-tail}.  The remaining statements
follow from the alternating Taylor expansion of $\arctan x$.
\end{proof}

\subsection{Fixed Number of Studies}

We revisit \cref{thm:revalidity} and focus on HCCT for now.

\begin{theorem}[Fixed-$m$ HCCT validity]\label{thm:fixed-hcct}
Let $m\geq1$ be fixed, let $p_1,\ldots,p_m$ have exact
$\operatorname{Uniform}(0,1)$ marginals, and let $w_i>0$ with
$\sum_iw_i=1$.  Suppose there exists $\delta_t\in(0,1)$ such that
\begin{equation}\label{eq:fixed-delta}
  \delta_t\to0,
  \qquad
  t\delta_t\to\infty,
\end{equation}
and
\begin{equation}\label{eq:fixed-pair-condition}
  \max_{1\leq i\neq j\leq m}
  \mathbb P\!\left(
    0<p_i<\frac{2w_i m}{\pi t},
    0<p_j<\frac{2w_j m}{\pi\delta_t t}
  \right)
  =o(t^{-1}).
\end{equation}
For $m=1$, condition \cref{eq:fixed-pair-condition} is understood as
vacuous.  Then
\begin{equation}\label{eq:fixed-main-tail}
  \mathbb P(\HCCT>t)
  =\frac{2}{\pi t}+o(t^{-1}).
\end{equation}
Furthermore,
\begin{equation}\label{eq:fixed-ind-tail}
  1-\FHCw(t)
  =\frac{2}{\pi t}+o(t^{-1}),
\end{equation}
and hence
\begin{equation}\label{eq:fixed-ratios}
  \lim_{t\to\infty}
  \frac{\mathbb P(\HCCT>t)}{1-\FHCw(t)}
  =
  \lim_{t\to\infty}
  \frac{\mathbb P(\HCCT>t)}{1-(2/\pi)\arctan(t)}
  =1.
\end{equation}
\end{theorem}

\begin{remark}[Explicit fixed-$m$ remainder]\label{rem:master-fixed}
If
\[
  R_t=
  \max_{1\leq i\neq j\leq m}
  \mathbb P\!\left(
    0<p_i<\frac{2w_i m}{\pi t},
    0<p_j<\frac{2w_j m}{\pi\delta_t t}
  \right),
\]
the proof gives, for all sufficiently large $t$,
\begin{equation}\label{eq:master-error}
\begin{split}
  \left|
    \mathbb P(\HCCT>t)-\frac{2}{\pi t}
  \right|
  \leq{}&
  \frac{3m(m-1)}{2}R_t
  +\frac{4\delta_t}{\pi(1-\delta_t^2)t}\\
  &+\frac{2\delta_t}{\pi t}
  +\frac{2}{3\pi t^3}.
\end{split}
\end{equation}
This is the form needed for growing $m$.
\end{remark}

\subsection{Growing Number of Studies}\label{sec:growingnumber}

\begin{theorem}[Growing-$m$ HCCT validity]\label{thm:growing-hcct}
Let $t\to\infty$ along a triangular array with $m=m(t)\to\infty$.
For each $t$, let $p_1,\ldots,p_m$ have exact
$\operatorname{Uniform}(0,1)$ marginals and let $w_i>0$ with
$\sum_iw_i=1$.  Suppose there exists $\delta_t\in(0,1)$ such that
\begin{equation}\label{eq:growing-delta}
  \delta_t\to0,
  \qquad
  t\delta_t\to\infty,
\end{equation}
and
\begin{equation}\label{eq:growing-master-condition}
  m^2
  \max_{1\leq i\neq j\leq m}
  \mathbb P\!\left(
    0<p_i<\frac{2w_i m}{\pi t},
    0<p_j<\frac{2w_j m}{\pi\delta_t t}
  \right)
  =o(t^{-1}).
\end{equation}
Then
\begin{equation}\label{eq:growing-main-tail}
  \mathbb P(\HCCT>t)
  =\frac{2}{\pi t}+o(t^{-1}),
\end{equation}
and
\begin{equation}\label{eq:growing-half-cauchy-ratio}
  \frac{\mathbb P(\HCCT>t)}
       {1-(2/\pi)\arctan(t)}\longrightarrow1.
\end{equation}

If, in addition,
\begin{equation}\label{eq:growing-weight-and-m}
  \max_{1\leq i\leq m}w_i=O(m^{-1}),
  \qquad
  m=o(\sqrt t),
\end{equation}
then
\begin{equation}\label{eq:growing-ind-tail}
  1-\FHCw(t)
  =\frac{2}{\pi t}+o(t^{-1}),
\end{equation}
and therefore
\begin{equation}\label{eq:growing-both-ratios}
  \frac{\mathbb P(\HCCT>t)}{1-\FHCw(t)}
  \longrightarrow1,
  \qquad
  \frac{\mathbb P(\HCCT>t)}
       {1-(2/\pi)\arctan(t)}
  \longrightarrow1.
\end{equation}
\end{theorem}

\begin{corollary}[Polynomial rate formulation]\label{cor:gamma-rate}
Fix $0<\gamma\leq1$.  Suppose
\begin{equation}\label{eq:gamma-pair}
  \max_{1\leq i\neq j\leq m}
  \mathbb P\!\left(
    0<p_i<\frac{2w_i m}{\pi t},
    0<p_j<\frac{2w_j m}{\pi\delta_t t}
  \right)
  =o\!\left(t^{-(1+\gamma)}\right)
\end{equation}
uniformly along a triangular array, and suppose
\begin{equation}\label{eq:gamma-m}
  m=O(t^{\gamma/2}).
\end{equation}
Then \cref{eq:growing-main-tail} and
\cref{eq:growing-half-cauchy-ratio} hold.  If also
$\max_iw_i=O(m^{-1})$ and either $\gamma<1$ or $m=o(\sqrt t)$, then both
ratios in \cref{eq:growing-both-ratios} hold.
\end{corollary}

\begin{proof}
\cref{eq:gamma-pair,eq:gamma-m} imply
\[
  m^2\max_{i\neq j}\mathbb P(\text{pair event})
  =O(t^\gamma)o\!\left(t^{-(1+\gamma)}\right)
  =o(t^{-1}).
\]
Thus \cref{thm:growing-hcct} applies.  If $\gamma<1$, then
$m=O(t^{\gamma/2})=o(\sqrt t)$ automatically.
\end{proof}

\begin{remark}[The endpoint $\gamma=1$]
The one-big-jump proof above does not establish
$1-\FHCw(t)\sim2/(\pi t)$ uniformly when $m\asymp\sqrt t$.
Indeed, its independent pair remainder is of order
$m^2/(\eta_t t^2)$ and would require both $\eta_t\to0$ and
$m^2/(\eta_t t)\to0$, which are incompatible when $m\asymp\sqrt t$.
The endpoint may be retained only after proving a different uniform
large-deviation lemma for the independent weighted Half-Cauchy sum.
Without such a lemma, the correct elementary-proof statement is
$m=o(\sqrt t)$.
\end{remark}

\subsection{Pairwise Bivariate Normal $p$-Values}\label{sec:pairwisebinormal}

The following lemma replaces the unequal-threshold use of Berman's
equal-threshold asymptotic \citep{berman1962law}.

\begin{lemma}[Uniform bivariate-normal quadrant bound]\label{lem:gaussian-quadrant}
Let $(Z_1,Z_2)$ be standard bivariate normal with correlation $\rho$ and
$|\rho|\leq r<1$.  There is a constant $C_r<\infty$ such that, for all
sufficiently large $d$ and every sign pair
$(\epsilon_1,\epsilon_2)\in\{-1,1\}^2$,
\begin{equation}\label{eq:quadrant-bound}
  \mathbb P(\epsilon_1Z_1>d,\epsilon_2Z_2>d)
  \leq C_r d^{-2}\exp\!\left(-\frac{d^2}{1+r}\right).
\end{equation}
Let $K<\infty$, let $\delta_t\to0$ with $t\delta_t\to\infty$, and let
$u_{i,t}\leq K/t$ and $v_{j,t}\leq K/(\delta_t t)$.  If $p_i,p_j$ are
either the one-sided or the two-sided exact normal $p$-values, then
\begin{equation}\label{eq:normal-p-bound}
  \mathbb P(p_i<u_{i,t},p_j<v_{j,t})
  \leq
  C_{r,K}
  (\delta_t t)^{-2/(1+r)}
  \{\log(\delta_t t)\}^{-r/(1+r)}.
\end{equation}
The constant is uniform over all pairs satisfying $|\rho_{ij}|\leq r$.
\end{lemma}

\begin{corollary}[Fixed-$m$ scalar-normal case]\label{cor:normal-fixed}
Suppose $m$ is fixed and every pair
\[
  \left(
    \frac{X_i-\mu_i}{\sigma_i},
    \frac{X_j-\mu_j}{\sigma_j}
  \right)
\]
is standard bivariate normal.  Let $p_i$ be either
\[
  p_i=1-\Phi\!\left(\frac{X_i-\mu_i}{\sigma_i}\right)
\]
or
\[
  p_i=2\left\{
    1-\Phi\!\left(\frac{|X_i-\mu_i|}{\sigma_i}\right)
  \right\}.
\]
If
\[
  \rho_{\max}:=
  \max_{1\leq i\neq j\leq m}|\rho_{ij}|<1,
\]
then the conclusions of \cref{thm:fixed-hcct} hold.
\end{corollary}

\begin{proof}
Choose $0<\kappa<(1-\rho_{\max})/2$ and set
$\delta_t=t^{-\kappa}$.  \cref{lem:gaussian-quadrant} gives
\[
  \max_{i\neq j}\mathbb P(\text{pair event})
  =O\!\left(
    t^{-2(1-\kappa)/(1+\rho_{\max})}
    (\log t)^{-\rho_{\max}/(1+\rho_{\max})}
  \right)
  =o(t^{-1}).
\]
\cref{thm:fixed-hcct} applies.
\end{proof}

\begin{corollary}[Growing-$m$ scalar-normal case]\label{cor:normal-growing}
Consider a triangular array as in \cref{thm:growing-hcct}.  Suppose
every pair of standardized variables is bivariate normal and
\begin{equation}\label{eq:rho-bound}
  \max_{1\leq i\neq j\leq m}|\rho_{ij}|
  \leq\rho_{\max}<1
\end{equation}
uniformly along the array.  Let the $p_i$ be the one-sided or two-sided
normal $p$-values in \cref{cor:normal-fixed}, and suppose
\begin{equation}\label{eq:normal-weight-bound}
  \max_{1\leq i\leq m}w_i=O(m^{-1}).
\end{equation}
Put
\begin{equation}\label{eq:gamma0}
  \gamma_0=
  \frac{1-\rho_{\max}}{1+\rho_{\max}}.
\end{equation}
If
\begin{equation}\label{eq:normal-little-o-boundary}
  m=o(t^{\gamma_0/2}),
\end{equation}
then both ratios in \cref{eq:growing-both-ratios} tend to one.

For every fixed $0<\gamma<\gamma_0$, the simpler sufficient rate
\begin{equation}\label{eq:normal-strict-gamma}
  m=O(t^{\gamma/2})
\end{equation}
is therefore valid.  If $0<\rho_{\max}<1$, the boundary can be sharpened to
\begin{equation}\label{eq:normal-O-boundary}
  m=O(t^{\gamma_0/2}).
\end{equation}
Note that when $\rho_{\max}=0$ and only pairwise bivariate normality is assumed, this
argument proves $m=o(\sqrt t)$, not the endpoint $m=O(\sqrt t)$.
\end{corollary}

\begin{proof}
Write
\[
  a=\frac{2}{1+\rho_{\max}}=1+\gamma_0,
  \qquad
  b=\frac{\rho_{\max}}{1+\rho_{\max}}.
\]
Under \cref{eq:normal-weight-bound}, there is $K<\infty$ such that all
pair thresholds in \cref{thm:growing-hcct} are bounded by
$K/t$ and $K/(\delta_t t)$.

First assume \cref{eq:normal-little-o-boundary} and define
\[
  D_t=\frac{t^{\gamma_0}}{m^2}\longrightarrow\infty,
  \qquad
  \delta_t=D_t^{-1/(2a)}.
\]
Then $\delta_t\to0$.  Since $m\geq1$, $D_t\leq t^{\gamma_0}$, so
$t\delta_t\to\infty$.  By \cref{lem:gaussian-quadrant},
\begin{align*}
  &t m^2
  \max_{i\neq j}
  \mathbb P\!\left(
    0<p_i<\frac{2w_i m}{\pi t},
    0<p_j<\frac{2w_j m}{\pi\delta_t t}
  \right)\\
  &\quad\leq
  C\,m^2t^{1-a}\delta_t^{-a}
  \{\log(\delta_t t)\}^{-b}\\
  &\quad=
  C\,D_t^{-1}\delta_t^{-a}
  \{\log(\delta_t t)\}^{-b}
  =O(D_t^{-1/2})
  \longrightarrow0.
\end{align*}
Thus \cref{eq:growing-master-condition} holds.  Moreover,
$m=o(t^{\gamma_0/2})$ and $\gamma_0\leq1$ imply $m=o(\sqrt t)$.
\cref{thm:growing-hcct} proves both ratios.

The strict fixed-$\gamma$ statement follows immediately, since
$m=O(t^{\gamma/2})=o(t^{\gamma_0/2})$ whenever $\gamma<\gamma_0$.

Finally suppose $0<\rho_{\max}<1$ and
$m=O(t^{\gamma_0/2})$.  Choose
\[
  \delta_t=(\log t)^{-\kappa},
  \qquad 0<\kappa<\frac{\rho_{\max}}{2}.
\]
Then
\begin{align*}
  t m^2\max_{i\neq j}\mathbb P(\text{pair event})
  &=O\!\left(
    (\log t)^{\kappa a-b}
  \right)\\
  &=O\!\left(
    (\log t)^{(2\kappa-\rho_{\max})/(1+\rho_{\max})}
  \right)
  =o(1).
\end{align*}
Because $\gamma_0<1$ in this case,
$m=O(t^{\gamma_0/2})=o(\sqrt t)$, and both ratios again follow from
\cref{thm:growing-hcct}.
\end{proof}

\begin{remark}[Bivariate-normal formulas]
The bivariate standard-normal density has normalizing constant
\[
  \frac{1}{2\pi\sqrt{1-\rho_{ij}^2}}.
\]
For two-sided $p$-values with thresholds $a_i,a_j>0$,
\[
  \mathbb P(|Z_i|>a_i,|Z_j|>a_j)
  =2Q_{\rho_{ij}}(a_i,a_j)
   +2Q_{-\rho_{ij}}(a_i,a_j),
\]
where $Q_\rho(a,b)=\mathbb P_\rho(Z_1>a,Z_2>b)$.  Thus all four sign
quadrants and $|\rho_{ij}|$ must be accounted for.
\end{remark}

\subsection{Tail Independence Reformulations}\label{sec:tail-independence}

The next results use the current manuscript's variables $X_i$, marginal
CDFs $F_i$, and upper-tail conditional probabilities.  Continuity of
$F_i$ ensures that
\[
  p_i=1-F_i(X_i)
\]
is exact Uniform$(0,1)$.

\begin{proposition}[Fixed-$m$ tail-independence condition]
Let $m$ be fixed and suppose $F_1,\ldots,F_m$ are continuous.  Suppose
$r(v)\to0$, $r(v)/v\to\infty$ as $v\downarrow0$, and
\begin{equation}\label{eq:tail-ind-fixed}
  \max_{1\leq i\neq j\leq m}
  \mathbb P\!\left[
    X_j>F_j^{-1}\{1-r(v)\}
    \mid
    X_i>F_i^{-1}(1-v)
  \right]
  \longrightarrow0.
\end{equation}
Then the conclusions of \cref{thm:fixed-hcct} hold.
\end{proposition}

\begin{proof}
Let $w_{\max}=\max_iw_i$ and define
\[
  s_t=\frac{2mw_{\max}}{\pi t},
  \qquad
  \delta_t=\frac{s_t}{r(s_t)}.
\]
Then $\delta_t\to0$ and
\[
  t\delta_t=\frac{2mw_{\max}}{\pi r(s_t)}\to\infty.
\]
For every ordered pair,
\begin{align*}
  &\mathbb P\!\left(
    0<p_i<\frac{2w_i m}{\pi t},
    0<p_j<\frac{2w_jm}{\pi\delta_t t}
  \right)\\
  &\quad\leq
  \mathbb P\{0<p_i<s_t,\ 0<p_j<r(s_t)\}\\
  &\quad=s_t
  \mathbb P\!\left[
    X_j>F_j^{-1}\{1-r(s_t)\}
    \mid
    X_i>F_i^{-1}(1-s_t)
  \right].
\end{align*}
After multiplication by $t$, the right-hand side tends to zero by
\cref{eq:tail-ind-fixed}.  \cref{thm:fixed-hcct} applies.
\end{proof}

\begin{proposition}[Growing-$m$ matched tail-independence condition]
Let $m\to\infty$, let the marginal CDFs be continuous, and suppose the
positive weights are uniformly comparable, so that for fixed
$0<c<C<\infty$,
\begin{equation}\label{eq:comparable-weights}
  \frac{c}{m}\leq w_i\leq\frac{C}{m}
  \qquad(1\leq i\leq m).
\end{equation}
Suppose $v_m\to0$, $r(v_m)\to0$, $r(v_m)/v_m\to\infty$, and
\begin{equation}\label{eq:tail-ind-growing}
  m^2
  \max_{1\leq i\neq j\leq m}
  \mathbb P\!\left[
    X_j>F_j^{-1}\{1-r(v_m)\}
    \mid
    X_i>F_i^{-1}(1-v_m)
  \right]
  \longrightarrow0.
\end{equation}
Let $x_m\to\infty$ satisfy
\begin{equation}\label{eq:xm-vm-match}
  \frac{2mw_{\max}}{\pi x_m}\leq v_m,
  \qquad
  x_mv_m=O(1).
\end{equation}
Then
\begin{equation}\label{eq:tail-ind-growing-standard-ratio}
  \frac{\mathbb P(\HCCT>x_m)}
       {1-(2/\pi)\arctan(x_m)}
  \longrightarrow1.
\end{equation}
If also $m=o(\sqrt{x_m})$, then
\begin{equation}\label{eq:tail-ind-growing-exact-ratio}
  \frac{\mathbb P(\HCCT>x_m)}{1-\FHCw(x_m)}
  \longrightarrow1.
\end{equation}
\end{proposition}

\begin{proof}
Set
\[
  \delta_m=
  \frac{2mw_{\max}}{\pi x_m r(v_m)}.
\]
By \cref{eq:xm-vm-match},
\[
  0<\delta_m\leq\frac{v_m}{r(v_m)}\to0,
\]
whereas comparable positive weights imply $mw_{\max}\geq1$ and hence
\[
  x_m\delta_m
  =\frac{2mw_{\max}}{\pi r(v_m)}\to\infty.
\]
As in the fixed case, the ordered pair event is bounded by
\[
  v_m
  \max_{i\neq j}
  \mathbb P\!\left[
    X_j>F_j^{-1}\{1-r(v_m)\}
    \mid
    X_i>F_i^{-1}(1-v_m)
  \right].
\]
Multiplication by $x_mm^2$ gives a quantity tending to zero because
$x_mv_m=O(1)$ and \cref{eq:tail-ind-growing} holds.  Apply
\cref{thm:growing-hcct} with $t=x_m$.  The final exact-reference
ratio follows from the additional condition $m=o(\sqrt{x_m})$.
\end{proof}

\begin{remark}
A condition only of the form $\liminf x_mv_m>0$ is insufficient for the
preceding proof.  Either $x_mv_m=O(1)$ is needed, as above, or
\cref{eq:tail-ind-growing} must be strengthened by the factor $x_mv_m$.
Also, $r(v)\to0$ is needed to guarantee $t\delta_t\to\infty$.
\end{remark}

\subsection{Harmonic Mean Analogue}

The exact harmonic-mean procedure uses the score statistic
\begin{equation}\label{eq:hm-statistic}
  \HM=\sum_{i=1}^m\frac{w_i}{p_i}.
\end{equation}
Write $\FPw$ for its CDF under independent exact-uniform $p$-values with
the weights held fixed.  Since $Y_i=1/p_i$ is Pareto$(1,1)$,
\begin{equation}\label{eq:pareto-tail}
  \mathbb P(Y_i>x)=x^{-1},\qquad x\geq1.
\end{equation}

\subsubsection{Fixed Number of Studies}

\begin{theorem}[Fixed-$m$ EHMP validity]\label{thm:fixed-hm}
Let $m\geq1$ be fixed, let $p_1,\ldots,p_m$ have exact
$\operatorname{Uniform}(0,1)$ marginals, and let $w_i>0$ with
$\sum_iw_i=1$.  Suppose there exists $\delta_t\in(0,1)$ such that
$\delta_t\to0$, $t\delta_t\to\infty$, and
\begin{equation}\label{eq:fixed-hm-pair}
  R_t^{\mathrm{HM}}:=
  \max_{1\leq i\neq j\leq m}
  \mathbb P\!\left(
    0<p_i<\frac{w_i m}{t},
    0<p_j<\frac{w_j m}{\delta_t t}
  \right)
  =o(t^{-1}).
\end{equation}
For $m=1$, the pair condition is vacuous.  Then
\begin{equation}\label{eq:fixed-hm-tail}
  \mathbb P(\HM>t)=\frac1t+o(t^{-1}),
\end{equation}
\begin{equation}\label{eq:fixed-hm-ind-tail}
  1-\FPw(t)=\frac1t+o(t^{-1}),
\end{equation}
and consequently
\begin{equation}\label{eq:fixed-hm-ratios}
  \frac{\mathbb P(\HM>t)}{1-\FPw(t)}\longrightarrow1,
  \qquad
  \frac{\mathbb P(\HM>t)}{1/t}\longrightarrow1.
\end{equation}
\end{theorem}

\begin{corollary}[The main fixed-$m$ HCCT condition also proves EHMP]
\label{cor:hcct-condition-implies-hm}
Under the assumptions of \cref{thm:fixed-hcct}, the conclusions
\cref{eq:fixed-hm-tail,eq:fixed-hm-ind-tail,eq:fixed-hm-ratios} also hold.
\end{corollary}

\begin{proof}
Apply \cref{eq:fixed-pair-condition} at $s=2t/\pi$ and define
$\widetilde\delta_t=\delta_{2t/\pi}$.  Then
$\widetilde\delta_t\to0$, $t\widetilde\delta_t\to\infty$, and the two
HCCT thresholds at $s$ become exactly
\[
  \frac{w_i m}{t},
  \qquad
  \frac{w_j m}{\widetilde\delta_t t}.
\]
Since $o(s^{-1})=o(t^{-1})$, \cref{thm:fixed-hm} applies.
\end{proof}

\subsubsection{Growing Number of Studies}

\begin{theorem}[Growing-$m$ EHMP validity]\label{thm:growing-hm}
Let $t\to\infty$ along a triangular array with $m=m(t)\to\infty$,
exact-uniform marginal $p$-values, and positive normalized weights.  Suppose
$\delta_t\to0$, $t\delta_t\to\infty$, and
\begin{equation}\label{eq:growing-hm-master}
  m^2 R_t^{\mathrm{HM}}=o(t^{-1}),
\end{equation}
where $R_t^{\mathrm{HM}}$ is defined in \cref{eq:fixed-hm-pair} row by
row.  Then
\begin{equation}\label{eq:growing-hm-tail}
  \mathbb P(\HM>t)=\frac1t+o(t^{-1}),
  \qquad
  \frac{\mathbb P(\HM>t)}{1/t}\longrightarrow1.
\end{equation}
If, in addition,
\begin{equation}\label{eq:growing-hm-weight-m}
  \max_iw_i=O(m^{-1}),
  \qquad
  m=o(\sqrt t),
\end{equation}
then
\begin{equation}\label{eq:growing-hm-ind}
  1-\FPw(t)=\frac1t+o(t^{-1}),
  \qquad
  \frac{\mathbb P(\HM>t)}{1-\FPw(t)}\longrightarrow1.
\end{equation}
\end{theorem}

\begin{corollary}[Polynomial EHMP rate]\label{cor:hm-gamma-rate}
Fix $0<\gamma\leq1$.  Suppose $\delta_t\to0$ and
$t\delta_t\to\infty$.  If, uniformly along the triangular array,
\[
  R_t^{\mathrm{HM}}=o\{t^{-(1+\gamma)}\},
  \qquad
  m=O(t^{\gamma/2}),
\]
then \cref{eq:growing-hm-tail} holds.  If also
$\max_iw_i=O(m^{-1})$ and either $\gamma<1$ or $m=o(\sqrt t)$, then
\cref{eq:growing-hm-ind} holds.
\end{corollary}

\begin{proof}
The pair contribution is
$m^2R_t^{\mathrm{HM}}=O(t^\gamma)o\{t^{-(1+\gamma)}\}=o(t^{-1})$.
The independence-reference assertion follows from
\cref{thm:growing-hm}; when $\gamma<1$,
$m=O(t^{\gamma/2})=o(\sqrt t)$.
\end{proof}

\subsubsection{Tail-Independence Formulations for EHMP}

\begin{proposition}[Fixed-$m$ EHMP under tail independence]
Under the assumptions of the fixed-$m$ tail-independence proposition in
\cref{sec:tail-independence}, the conclusions of
\cref{thm:fixed-hm} hold.
\end{proposition}

\begin{proof}
Let $s_t=mw_{\max}/t$ and $\delta_t=s_t/r(s_t)$.  Then
$\delta_t\to0$, $t\delta_t=mw_{\max}/r(s_t)\to\infty$, and each natural
EHMP pair event is bounded by
\[
  s_t\max_{i\neq j}
  \mathbb P\!\left[
    X_j>F_j^{-1}\{1-r(s_t)\}
    \mid X_i>F_i^{-1}(1-s_t)
  \right]
  =o(t^{-1}).
\]
Apply \cref{thm:fixed-hm}.
\end{proof}

\begin{proposition}[Growing-$m$ matched tail independence for EHMP]
\label{prop:hm-tail-ind-growing}
Assume the comparable-weight and conditional tail-independence conditions
\cref{eq:comparable-weights,eq:tail-ind-growing}, and suppose
$v_m\to0$, $r(v_m)\to0$, and $r(v_m)/v_m\to\infty$.  Let $x_m\to\infty$
satisfy
\begin{equation}\label{eq:hm-xm-vm-match}
  \frac{mw_{\max}}{x_m}\leq v_m,
  \qquad
  x_mv_m=O(1).
\end{equation}
Then
\[
  \frac{\mathbb P(\HM>x_m)}{1/x_m}\longrightarrow1.
\]
If also $m=o(\sqrt{x_m})$, then
\[
  \frac{\mathbb P(\HM>x_m)}{1-\FPw(x_m)}\longrightarrow1.
\]
\end{proposition}

\begin{proof}
Set
\[
  \delta_m=\frac{mw_{\max}}{x_mr(v_m)}.
\]
Condition \cref{eq:hm-xm-vm-match} gives
$0<\delta_m\leq v_m/r(v_m)\to0$, while
$x_m\delta_m=mw_{\max}/r(v_m)\to\infty$.  Each ordered natural pair
event is bounded by $v_m$ times the maximum conditional probability in
\cref{eq:tail-ind-growing}.  Multiplication by $x_mm^2$ gives $o(1)$
because $x_mv_m=O(1)$. \cref{thm:growing-hm} completes the proof.
\end{proof}

\subsection{Details on Numerical Computation}\label{sec:hmp}

Evaluation of either the density or the CDF in \cref{thm:computeconvhc}
requires a one-dimensional numerical integral. Each integrand evaluation
uses $O(m)$ scalar operations and special-function evaluations. Total
runtime also depends on the number of adaptive quadrature evaluations,
which varies with $m$, $x$, the weights, and the requested tolerance.
The Python package \textbf{NumPy} supports the complex arithmetic.
We evaluate products through sums of logarithms to reduce overflow and
underflow.

In particular, for $f_{\mathrm{HC}, \boldsymbol{w}}(x)$ and $F_{\mathrm{HC}, \boldsymbol{w}}(x)$ we compute the integrand using the following formula
\begin{equation*}
\spacingset{1}
\begin{aligned}
  & \exp(-xz)\Bigl[\prod_{j=1}^{m}\{-f_{\mathrm{HC}}^{\ast}(w_{j}z)+2\cos (w_{j}z)+2i \sin (w_{j}z)\}\\ 
  &\quad-\prod_{j=1}^{m}\{-f_{\mathrm{HC}}^{\ast}(w_{j}z)+2\cos (w_{j}z)-2i \sin (w_{j}z)\}\Bigr]\\
    =& 2i\Im \Bigl[\exp(-xz)\prod_{j=1}^{m}\{-f_{\mathrm{HC}}^{\ast}(w_{j}z)+2\cos (w_{j}z)+2i \sin (w_{j}z)\}\Bigr]\\
    =& 2i\Im \exp \Bigl[-xz+\sum_{j=1}^{m}\log\{-f_{\mathrm{HC}}^{\ast}(w_{j}z)+2\cos (w_{j}z)+2i \sin (w_{j}z)\}\Bigr],
\end{aligned}
\resetspacing
\end{equation*}
where $\log$ denotes the pointwise principal logarithm, with argument
in $(-\pi,\pi]$. Exponentiating the sum recovers the product; each factor
is nonzero on the real integration path. Here
$f_{\mathrm{HC}}^{\ast}(z)$ is the Laplace transform of
$f_{\mathrm{HC}}(x)$, given for real $z>0$ by
\begin{gather*}
  \spacingset{1}
    f_{\mathrm{HC}}^{\ast}(z)=\frac{2}{\pi}\int_{0}^{+\infty}\frac{\exp(-xz)}{1+x^{2}}\dif x=-\frac{2}{\pi}\{\sin (z)\operatorname{ci}(z)+\cos (z)\operatorname{si}(z)\},\\
     \operatorname{si}(z)=-\int_{z}^{\infty}\frac{\sin (\xi)}{\xi}\dif \xi,\  \operatorname{ci}(z)=\int_{z}^{\infty}\frac{\cos (\xi)}{\xi}\dif \xi.
    \resetspacing
  \end{gather*}

For $z>0$, \texttt{scipy.special.sici} returns
$\operatorname{Si}$ and $\operatorname{Ci}$, related to our conventions by
\[
    \operatorname{si}(z)=\operatorname{Si}(z)-\frac{\pi}{2},
    \qquad
    \operatorname{ci}(z)=-\operatorname{Ci}(z).
\]
The combined transform extends continuously to
$f_{\mathrm{HC}}^{\ast}(0)=1$.

For the Pareto$(1,1)$ variables in the HMP method, a similar expression can be derived as follows.
\begin{align*}
\spacingset{1}
  &\exp(-xz)\Bigl[\prod_{j=1}^{m}\{-\operatorname{Ei}_{2}(w_{j}z)+i\pi w_{j}z\}-\prod_{j=1}^{m}\{-\operatorname{Ei}_{2}(w_{j}z)-i\pi w_{j}z\}\Bigr]\\
  =&2i\exp(-xz)\Im \prod_{j=1}^{m}\{-\operatorname{Ei}_{2}(w_{j}z)+i\pi w_{j}z\}
  =2i\Im \exp \Bigl[-xz+\sum_{j=1}^{m}\log\{-\operatorname{Ei}_{2}(w_{j}z)+i\pi w_{j}z\}\Bigr],
  \resetspacing
\end{align*}
where, for real $z>0$, the auxiliary function
$\operatorname{Ei}_{2}(z)$ is defined by
\citep{abramowitz1968handbook}
\begin{align*}
\spacingset{1}
  &\operatorname{Ei}_{2}(z):=-1+z(\log z+C_{\gamma}-1)+\sum_{j=2}^{\infty}\frac{z^{j}}{(j-1)j!}
    =z\operatorname{Ei}(z)-\exp(z),
\\
  &\operatorname{Ei}(z):=-\operatorname{PV}\int_{-z}^{\infty}
       \frac{e^{-\xi}}{\xi}\dif\xi
    =\operatorname{PV}\int_{-\infty}^z
       \frac{e^{\xi}}{\xi}\dif\xi,\qquad z>0.
  \resetspacing
\end{align*}
Here $\operatorname{PV}$ denotes the Cauchy principal value.

Direct evaluation of $\operatorname{Ei}(z)$ can overflow for large
positive $z$. For moderate $z$, we evaluate the scaled quantity
$e^{-z}\operatorname{Ei}(z)$ directly using \texttt{scipy.special.expi}; for
large $z$, we use the optimally truncated asymptotic expansion
\[
  e^{-z}\operatorname{Ei}(z)
  \sim \frac{1}{z}\sum_{k=0}^{K}\frac{k!}{z^k},
  \qquad z\to\infty.
\]
This avoids overflow of the unscaled exponential integral
\citep[Equation~5.1.51]{abramowitz1968handbook}.

Although $e^{-z}\operatorname{Ei}(z)$ diverges logarithmically as
$z\downarrow0$, the combined Pareto factor has a finite limit:
\[
    e^{-t}\{-\operatorname{Ei}_{2}(t)+i\pi t\}
    =1-te^{-t}\operatorname{Ei}(t)+i\pi te^{-t}
    \longrightarrow1\qquad(t\downarrow0).
\]
Since $te^{-t}\operatorname{Ei}(t)\to0$, zero arguments, including
zero weights, use this continuous value.

\cref{tab:precision,tab:precisionehmp} report timings for the current
implementation. Compiling the scaled exponential-integral routine could
reduce EHMP overhead, but any speedup requires benchmarking.

As noted in the main text, the computational challenges have arisen for the HMP \citep{wilson2019harmonic} and the left-truncated or winsorized Cauchy method \citep{gui2023aggregating,fang2023heavy}. \citet{wilson2019harmonic} used the limiting Landau distribution as an approximation, which works well for large $m$ as in their assumption but proves inaccurate for small \(m\). As a side note, they only obtained the asymptotic distribution of the test statistics with $m\to\infty$ and $w_{1}=\dots=w_{m}=1/m$ while we allow for unequal weights both in the generalized CLT and the numerical approach for calculating the exact distribution with finite $m$.

\citet{fang2023heavy}, on the other hand, introduced an iterative importance sampling scheme for small \(m\), and switched to the Landau approximation only when \(m\) exceeds a set threshold \(m_0\). However, this approach is computationally intensive and unstable without a very large sample size, requiring at least \(10^5 m\) samples per iteration. As a result, \(m_0\) cannot be set too high, and they recommend \(m_0 = 25\); yet, accuracy declines noticeably for \(m = 26\).

\begin{table}[tbp]
  \centering
  \resizebox{\textwidth}{!}{\begin{minipage}
      {1.4\textwidth}
      \centering
      \begin{tabular}{cccccccc}
        \toprule
        $m$                   & $x$ & PDF (Err)  & Time (s) & Landau Approx (Err)  & CDF (Err) & Time (s) & Landau Approx (Err)                                                        \\
        \midrule
        \multirow{4}{*}{2}    & .2  & .292879165 ($\pm$9E-9) & .043     & .282722127 ($-$2E-2)         & .030804228 ($\pm$1E-8) & .028 & .223733981 ($+$2E-1)\\
                              & 2   & .164879638 ($\pm$8E-9) & .011     & .139681018 ($-$3E-2)         & .639966151 ($\pm$4E-9) & .011 & .621681447 ($-$2E-2)\\
                              & 10  & .007305301 ($\pm$5E-9) & .012     & .008434884 ($+$2E-3)         & .930504308 ($\pm$2E-9) & .011 & .923528833 ($-$7E-3)\\
                              & 50  & .000267851 ($\pm$6E-9) & .006     & .000282679 ($+$2E-5)         & .986896089 ($\pm$3E-9) & .013 & .986491736 ($-$5E-4)\\
        \midrule
        \multirow{4}{*}{10}   & 1   & .298436871 ($\pm$1E-9) & .019     & .267219180 ($-$4E-2)         & .084662651 ($\pm$3E-9) & .018 & .161603641 ($+$8E-2)\\
                              & 4   & .081183591 ($\pm$9E-9) & .011     & .083422558 ($+$3E-3)         & .740788721 ($\pm$2E-9) & .013 & .727771746 ($-$2E-2)\\
                              & 10  & .009975760 ($\pm$1E-9) & .012     & .010582384 ($+$7E-4)         & .916911594 ($\pm$4E-9) & .016 & .913846326 ($-$4E-3)\\
                              & 50  & .000290372 ($\pm$2E-9) & .010     & .000295108 ($+$5E-6)         & .986315767 ($\pm$1E-9) & .014 & .986195804 ($-$2E-4)\\
        \midrule
        \multirow{4}{*}{100}  & 2   & .158076048 ($\pm$4E-9) & .045     & .169847092 ($+$2E-2)         & .040232564 ($\pm$6E-9) & .050 & .056630205 ($+$2E-2)\\
                              & 5   & .105381463 ($\pm$1E-9) & .021     & .106135365 ($+$1E-3)         & .687530806 ($\pm$1E-8) & .021 & .683873904 ($-$4E-3)\\
                              & 10  & .015109635 ($\pm$1E-9) & .012     & .015315611 ($+$3E-4)         & .895973685 ($\pm$7E-9) & .017 & .895170441 ($-$9E-4)\\
                              & 50  & .000313579 ($\pm$5E-9) & .016     & .000314359 ($+$2E-6)         & .985767643 ($\pm$1E-9) & .022 & .985749325 ($-$2E-5)\\
        \midrule
        \multirow{4}{*}{1000} & 4   & .277750260 ($\pm$4E-9) & .162     & .274061911 ($-$4E-3)         & .177916458 ($\pm$5E-9) & .185 & .180088077 ($+$3E-3)\\
                              & 7   & .080390569 ($\pm$9E-9) & .096     & .080617466 ($+$3E-4)         & .733973017 ($\pm$1E-8) & .079 & .733369559 ($-$6E-4)\\
                              & 10  & .023685955 ($\pm$2E-9) & .055     & .023750783 ($+$1E-4)         & .867373631 ($\pm$9E-9) & .073 & .867174483 ($-$2E-4)\\
                              & 50  & .000335429 ($\pm$1E-8) & .068     & .000335545 ($+$2E-7)         & .985275813 ($\pm$2E-9) & .100 & .985273239 ($-$3E-6)\\
        \bottomrule
      \end{tabular}
    \end{minipage}}
\caption{Precision and runtime cost of HCCT with equal weights. The
$\pm$ entries in the PDF and CDF columns are SciPy estimates of absolute
quadrature error. Signed Landau errors are the reported discrepancies
relative to the finite-$m$ numerical evaluations.}\label{tab:precision}
\vspace{10pt}
\end{table}

\begin{table}[tbp]
  \centering
  \resizebox{\textwidth}{!}{\begin{minipage}
      {1.4\textwidth}
      \centering
      \begin{tabular}{cccccccc}
        \toprule
        $m$                   & $x$ & PDF (Err)  & Time (s) & Landau Approx (Err)  & CDF (Err) & Time (s) & Landau Approx (Err)                                                       \\
        \midrule
        \multirow{3}{*}{2}
                              & 2   & .303993203 ($\pm$5E-9) & .150     & .150080964 (-2E-1)         & .362673464 ($\pm$2E-8) & .046  & .433900891 (+7E-2)\\
                              & 10  & .012418123 ($\pm$2E-8) & .039     & .014947778 (+3E-3)         & .885277805 ($\pm$4E-9) & .035  & .868002274 (-2E-2) \\
                              & 50  & .000432721 ($\pm$2E-9) & .049     & .000471188 (+4E-5)         & .979080976 ($\pm$7E-9) & .030  & .978043335 (-1E-3)\\
        \midrule
        \multirow{3}{*}{10}
                              & 4   & .155679561 ($\pm$1E-9) & .039     & .133578865 (-2E-2)         & .492596674 ($\pm$2E-8) & .028  & .489298321 (-3E-3)\\
                              & 10  & .019829249 ($\pm$3E-9) & .019     & .021397821 (+2E-3)         & .847965230 ($\pm$8E-9) & .040  & .839184630 (-9E-3)\\
                              & 50  & .000491781 ($\pm$6E-9) & .023     & .000505060 (+1E-5)         & .977583372 ($\pm$1E-9) & .034  & .977258199 (-3E-4)\\
        \midrule
        \multirow{4}{*}{100}  & 2   & .000000387 ($\pm$1E-8) & .445     & .000554016 (+6E-4)         & .000000015 ($\pm$4E-9) & .272  & .000068807 (+7E-5)\\
                              & 5   & .191884746 ($\pm$1E-8) & .097     & .179262887 (-1E-2)         & .274570971 ($\pm$2E-8) & .096  & .281827251 (+7E-3)\\
                              & 10  & .038837066 ($\pm$6E-9) & .097     & .039487463 (+7E-4)         & .774900747 ($\pm$8E-9) & .086  & .771927461 (-3E-3)\\
                              & 50  & .000557767 ($\pm$2E-8) & .045     & .000560181 (+2E-6)         & .976086590 ($\pm$2E-9) & .045  & .976033423 (-5E-5)\\
        \midrule
        \multirow{4}{*}{1000} & 4   & .000009348 ($\pm$2E-8) & 1.565    & .000043914 (+4E-5)         & .000000671 ($\pm$1E-9) & 1.405 & .000004086 (+4E-6)\\
                              & 7   & .182779813 ($\pm$1E-8) & .455     & .180180123 (-3E-3)         & .225626049 ($\pm$3E-9) & .501  & .227272659 (+2E-3)\\
                              & 10  & .083072268 ($\pm$2E-8) & .337     & .083192398 (+1E-4)         & .639103576 ($\pm$2E-8) & .377  & .638216812 (-9E-4)\\
                              & 50  & .000624345 ($\pm$2E-9) & .323     & .000624742 (+4E-7)         & .974679223 ($\pm$5E-9) & .317  & .974671236 (-8E-6)\\
        \bottomrule
      \end{tabular}
    \end{minipage}}
\caption{Precision and runtime cost of EHMP with equal weights. The
$\pm$ entries in the PDF and CDF columns are SciPy estimates of absolute
quadrature error. Signed Landau errors are the reported discrepancies
relative to the finite-$m$ numerical evaluations.}\label{tab:precisionehmp}
\end{table}

\citet{gui2023aggregating} directly applied the left-truncated Cauchy proxy to all cases regardless of $m$. While this does make sense asymptotically as the significance level goes to 0, it does not guarantee the validity of the test at finite levels even for independent studies. In fact, it introduces substantial bias and undermines validity for large \(m\). For a detailed comparison of the accuracy and limitations across different values of \(m\) for these three approaches, see \cref{tab:wilson} of the main text.  

The integral formulas allow unequal weights, while
\cref{tab:precision,tab:precisionehmp} report equal-weight calculations.
Following the empirical recommendation in \cref{sec:numeric}, we use
numerical evaluation of the exact finite-$m$ formulas for $m\leq1000$
and the Landau approximation in \cref{thm:gclthc} for $m>1000$ with equal
weights. The cutoff is a computational choice. At the listed HCCT point
$m=1000$, $x=50$, the absolute CDF discrepancy is below $0.0002$;
the table does not establish that bound uniformly over the tail.
For unequal weights, a large $m$ alone does not ensure accuracy:
\cref{thm:gclthc} requires the maximum weight to vanish and gives no
uniform error rate. These quadrature estimates do not certify the total numerical error,
which also depends on special-function evaluation, series truncation,
and floating-point arithmetic.
For computing Landau distributions, we adopted the Pad\'e approximants;
see the source code of the C++ numerical framework \textbf{ROOT} for
implementation \citep{kolbig1983program}.
For further references on the computation of Landau distributions see \citet{chambers1976method,weron1996chambers,nolan1997numerical,amindavar2008novel,ament2018accurate}.

\subsection{Simulations Using Other Copulas}\label{sec:copula}

The concept of \emph{copulas} is an important tool in studying tail independence \citep{embrechts2001modelling}. Consider a random vector $\boldsymbol{X}=(X_{1},\dots,X_{m})^{\top}$. Suppose its marginal CDFs $F_{j}(x)=\mathbb{P}(X_{j}\leq x)$ are continuous. By applying the probability integral transform to each component, the random vector
\begin{equation*}
\spacingset{1}
  (U_{1},\dots,U_{m})=\{F_{1}(X_{1}),\dots,F_{m}(X_{m})\}
  \resetspacing
\end{equation*}
has marginals that are uniformly distributed on the interval $[0,1]$. 
\begin{definition}[Copula]\label{thm:defcopula}
  The \emph{copula} of $\boldsymbol{X}$ is defined as the joint cumulative distribution of $(U_{1},\dots, U_{m})$ given by
\begin{equation*}
\spacingset{1}
  C(u_{1},\dots,u_{m})=\mathbb{P}(U_{1}\leq u_{1},\dots,U_{m}\leq u_{m}).
  \resetspacing
\end{equation*}
\end{definition}
Sklar's theorem \citep{Skla59,durante2013topological} shows that every multivariate CDF of a random vector $\boldsymbol{X}$ can be expressed in terms of its marginals $F_{j}(x_{j})$ ($j=1,\dots,m$) and a copula $C$, i.e.,
\begin{equation*}
\spacingset{1}
  H(x_{1},\dots,x_{m})=\mathbb{P}(X_{1}\leq x_{1},\dots,X_{m}\leq x_{m})=C\{F_{1}(x_{1}),\dots,F_{m}(x_{m})\}.
  \resetspacing
\end{equation*}
In other words, the copula contains all information on the dependence structure between the components of $(X_{1},\dots,X_{m})$ whereas the marginal CDFs contain all information on the marginal distributions of $X_{j}$.

As established in \citet{long2023cauchy} the assumptions of \cref{thm:fixed-hcct,thm:growing-hcct} are satisfied by a number of commonly-used bivariate copulas, including but not limited to the independence copula and the normal copula:
\begin{itemize}
  \setlength\itemsep{1em}
  \item Independence Copula: 
  \vspace*{-10pt}
  \begin{equation*}
  \spacingset{1}
  C(u,v)=uv;
  \resetspacing
  \end{equation*}
  \item Normal Copula:
  \begin{equation*}
  \spacingset{1}
    \textstyle C(u,v)=\frac{1}{2\pi\sqrt{1-\rho^{2}}}\int_{-\infty}^{\Phi^{-1}(u)}\int_{-\infty}^{\Phi^{-1}(v)}\exp \biggl\{
      -\frac{x^{2}-2\rho xy+y^{2}}{2(1-\rho^{2})}
    \biggr\}\dif x\dif y,
    \resetspacing
  \end{equation*}
  where $\rho$ is the correlation between the two normal variables;
  \item Gumbel--Barnett Copula: 
  \begin{equation*}
  \spacingset{1}
  C(u,v)=uv\exp (-\theta \log u\log v),\quad\theta\in [0,1];
  \resetspacing
  \end{equation*}
  \item Farlie--Gumbel--Morgenstern (FGM) Copula:
  \begin{equation*}
  \spacingset{1}
    C(u,v)=uv\{1+\theta(1-u)(1-v)\},\quad \theta\in [-1,1];
    \resetspacing
  \end{equation*}
  \item Cuadras--Augé Copula\footnote{Auditing \citet{long2023cauchy}: $\theta=1$ does not satisfy the assumption.}:
  \begin{equation*}
  \spacingset{1}
    C(u,v)=(\min \{ u,v \})^{\theta}(uv)^{1-\theta},\quad \theta\in [0,1);
    \resetspacing
  \end{equation*}
  \item Ali--Mikhail--Haq (AMH) Copula\footnote{Auditing \citet{long2023cauchy}: $\theta=1$ does not satisfy the assumption.}:
  \begin{equation*}
  \spacingset{1}
    \textstyle C(u,v)= \frac{uv}{1-\theta (1-u)(1-v)},\quad \theta \in [0,1).
    \resetspacing
  \end{equation*}
\end{itemize}

\begin{figure}[tbp]
  \centering
  \subfloat[HCCT with FGM copula]{%
    \includegraphics[width=0.3\textwidth]{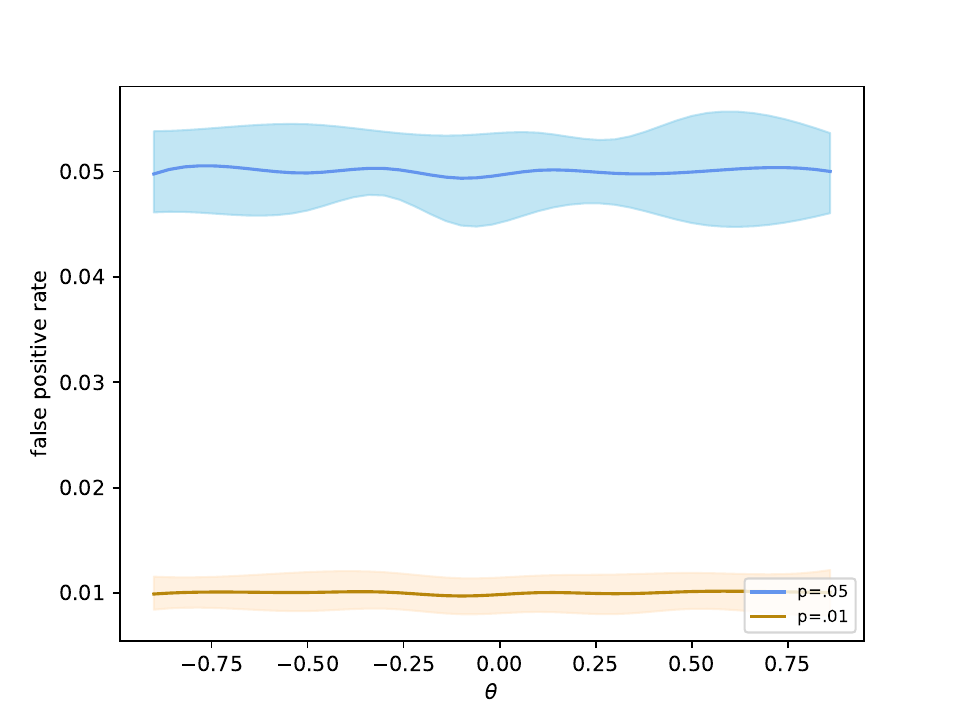}
  }\hspace*{15pt}
  \subfloat[Fisher with FGM copula]{%
    \includegraphics[width=0.3\textwidth]{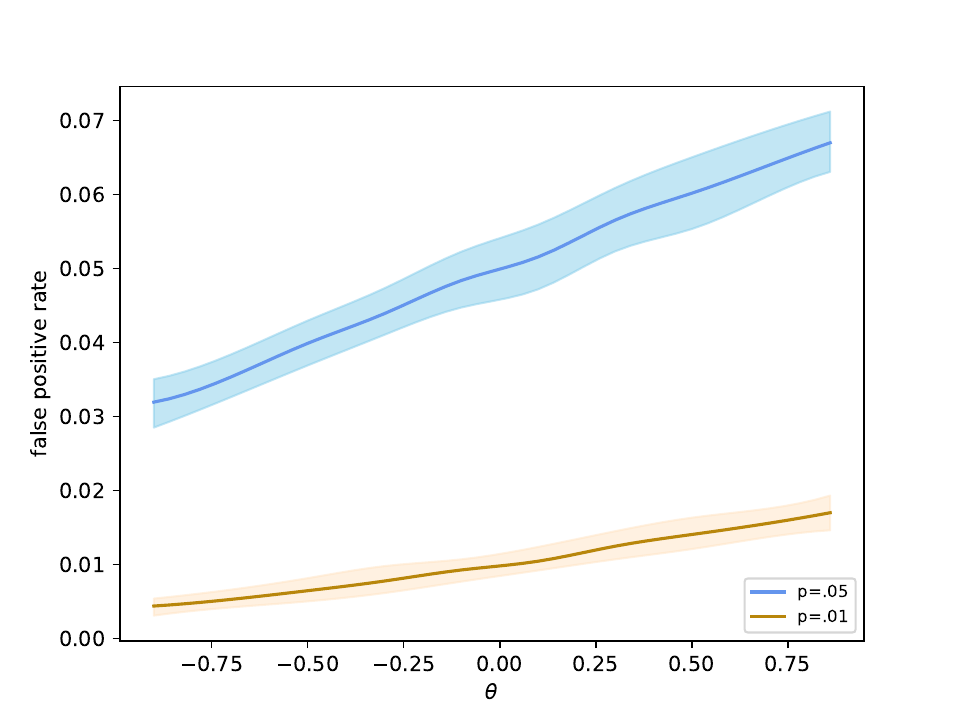}
  }\\
  \vspace*{-10pt}
  \subfloat[HCCT with AMH copula]{%
    \includegraphics[width=0.3\textwidth]{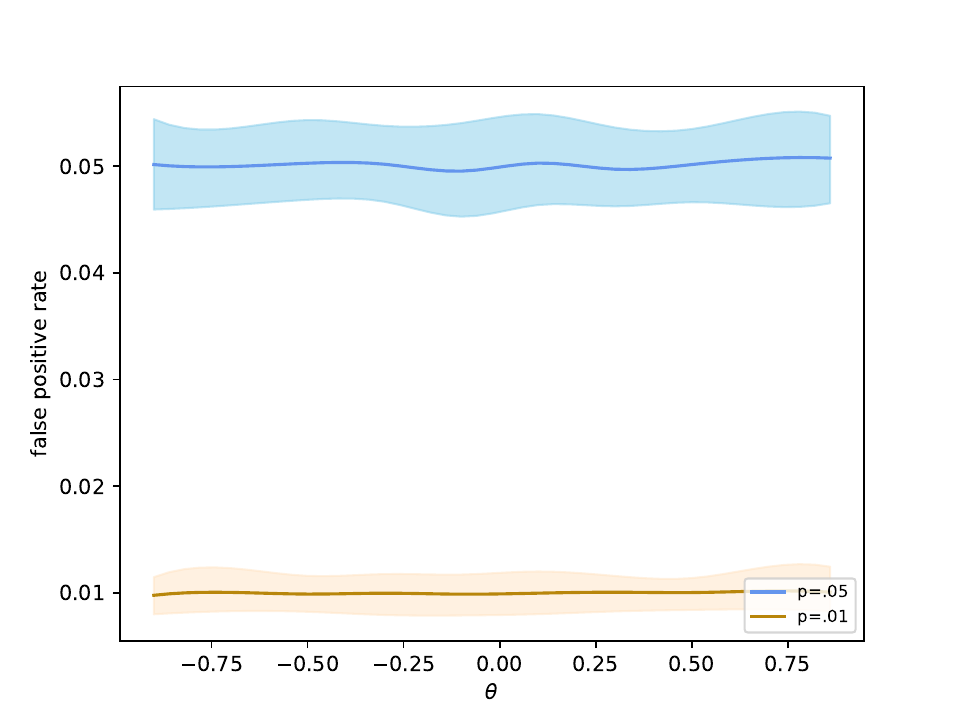}
  }\hspace*{15pt}
  \subfloat[Fisher with AMH copula]{%
    \includegraphics[width=0.3\textwidth]{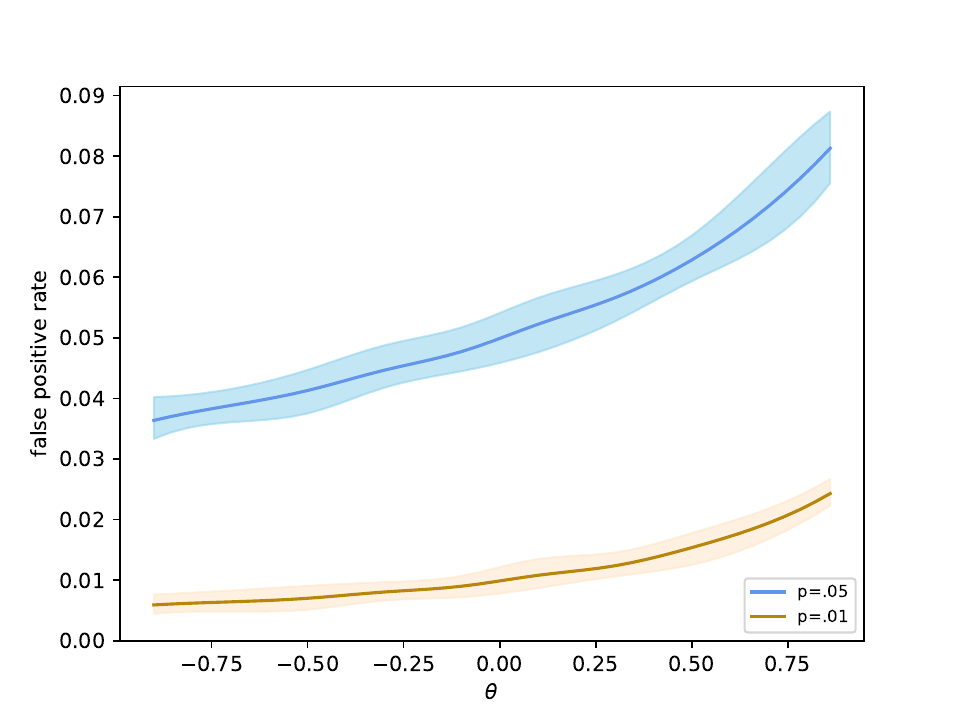}
  }
  \caption{Comparison of false positive rates with AMH and FGM copulas.}
  \vspace*{-10pt}
  \label{fig:copulamore}
\end{figure}

To illustrate, we show more simulation results on the validity of HCCT using dependency structures other than the multivariate normal of \cref{sec:sim1}. First, we check the FGM and AMH copulas as mentioned above using the following setup from \citep{long2023cauchy}:
\begin{itemize}
  \item FGM copula mixed with product copula model: 
  \[
  \spacingset{1}
  (p_{j},p_{j+1})^{\top}\sim C(u_{j}, v_{j+1})=
  \begin{cases}
  u_{j}v_{j+1}\bigl\{1+\theta (1-u_{j})(1-v_{j+1})\bigr\} & j=1,3,\dots, 2\lfloor m/2\rfloor-1\\
  u_{j}v_{j+1} & \text{else}
  \end{cases},
  \resetspacing
  \]
  \item AMH copula mixed with product copula model:
  \begin{equation*}
  \spacingset{1}
  (p_{j},p_{j+1})^{\top}\sim C(u_{j}, v_{j+1})=
  \begin{cases}
  \frac{u_{j}v_{j+1}}{1-\theta (1-u_{j})(1-v_{j+1})} & j=1,3,\dots, 2\lfloor m/2\rfloor-1\\
  u_{j}v_{j+1} & \text{else}
  \end{cases}.
  \resetspacing
  \end{equation*}
\end{itemize}
The p-values are generated from the null hypothesis based on the above two models with $m=500$. \cref{fig:copulamore} reports the false positive rate from $10000$ runs for HCCT and Fisher's combination test in these two settings. We can see that the combination test has roughly the correct size for HCCT while the actual size for Fisher's combination test changes monotonically with the hyperparameter $\theta$. As a result, Fisher's combination test is less valid with large positive $\theta$'s.

\begin{figure}[tbp]
  \centering
  \subfloat[AR-$1$ correlation]{%
    \includegraphics[width=0.3\textwidth]{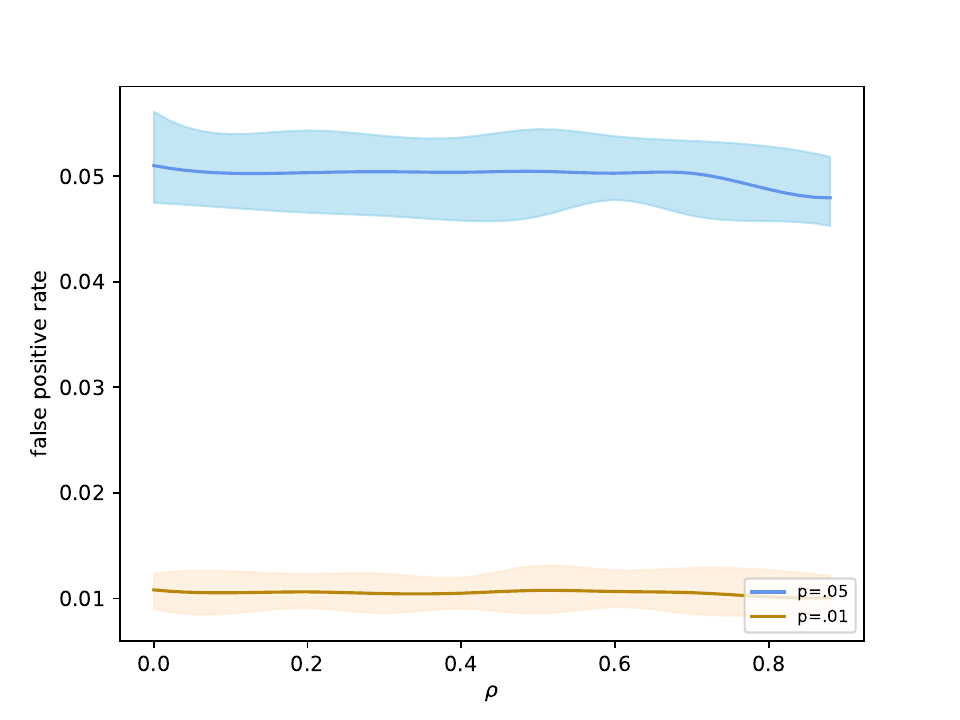}
  }\hspace*{15pt}
  \subfloat[Equi-correlation]{%
    \includegraphics[width=0.3\textwidth]{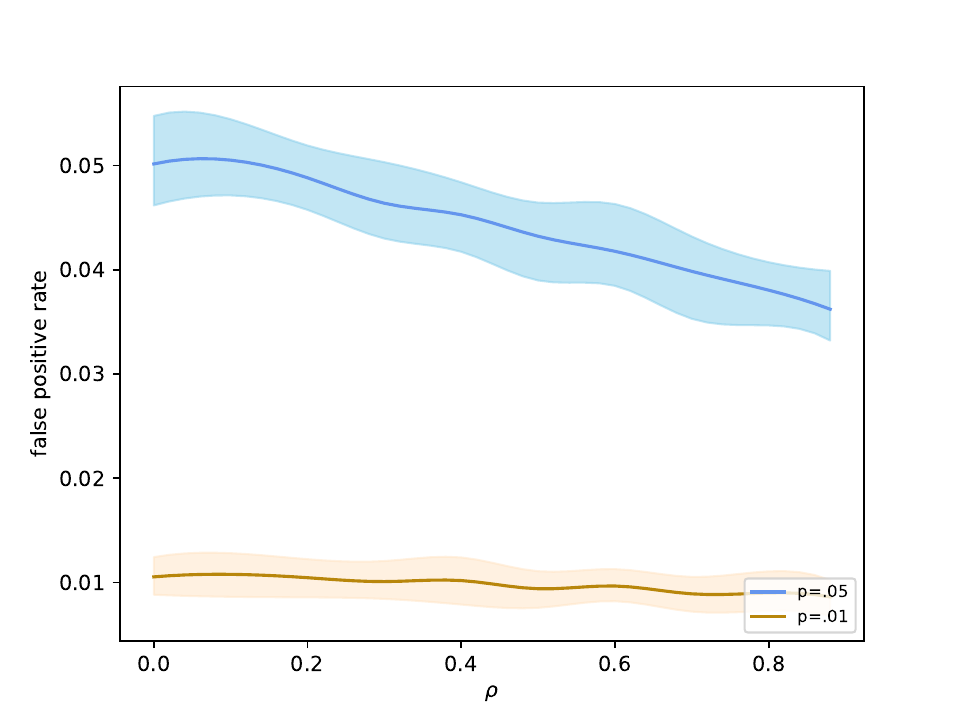}
  }
  \caption{False positive rates of HCCT with multivariate $t$ copulas.}
  \label{fig:copulat}
\vspace*{10pt}
\end{figure}

As shown in the proof of \cref{cor:normal-fixed,cor:normal-growing}, the bivariate normal distribution is tail independent. However, there are other distributions that are tail dependent including the bivariate $t$-distribution as shown in \citet{schmidt2002tail}. Its tail dependence coefficient has been extended to multivariate cases and extensively studied in \citet{frahm2006extremal,chan2007tail}.

Next we consider replacing the normal distribution in \cref{sec:sim1} by the multivariate $t$-distribution $t_{m,k}(\boldsymbol{\theta},\mathit{\Sigma})$ with degrees of freedom $k=10$ and dimension $m=500$, the density of which is given by
\begin{equation*}
\spacingset{1}
  \frac{\Gamma\{(k+m) / 2\}}{\Gamma(k / 2) k^{m / 2} \pi^{m / 2}|\mathit{\Sigma}|^{1 / 2}}\left\{1+\frac{1}{k}(\boldsymbol{x}-\boldsymbol{\theta})^{\mathrm{T}} \mathit{\Sigma}^{-1}(\boldsymbol{x}-\boldsymbol{\theta})\right\}^{-(k+m) / 2}.
  \resetspacing
\end{equation*}
The individual p-values here are calculated from the tail probabilities of those marginal Student's $t$-distributions with degrees of freedom $k=10$. We set $\boldsymbol{\theta}=\boldsymbol{0}$ under the null and compute the false positive rates from $10000$ runs with $\mathit{\Sigma}$ being either AR-$1$ correlation or equi-correlation matrices as defined in \cref{sec:sim1}. The results are shown in \cref{fig:copulat}. We can see that the HCCT is almost always of the correct size with AR-$1$ correlations and is slightly conservative with equi-correlations as $\rho$ grows.

\subsection{Sensitivity to Large \texorpdfstring{$p$}{p}-Values \& Heavily Right Strategy}\label{sec:insense}

In global testing we care mostly about the small $p$-values and would like the combined $p$-values to be insensitive to large individual ones. However, as mentioned in \cref{sec:intro}, the Cauchy combination test is quite sensitive to large $p_{j}$'s and does not address this concern well enough. In this section we aim to present the comparison of these combination tests in terms of sensitivity to large $p$-values.

\begin{table}[tbp]
  \centering
  \resizebox{\textwidth}{!}{\begin{minipage}
    {1.25\textwidth}
    \centering
  \begin{tabular}{l|ccccccc}
    \toprule
     $p$-values                        & Fisher   & Stouffer & Bonferroni   & CCT  & CAtr  & HCCT       & EHMP \\
    \midrule
    (.02, .03, .96) & .021 & .104 & .060 & .051 & .051 & .039 & .039\\
    (.02, .03, .98) & .021 & .139 & .060 & .088 & .088 & .039 & .039\\
    (.02, .03, .99) & .021 & .177 & .060 & .837 & .837 & .039 & .039\\
    (.015, .9, .96) & .192 & .691 & .045 & .091 & .091 & .050 & .049\\
    (.02, .02, .8, .98) & .040 & .272 & .080 & .086 & .086 & .045 & .045 \\
    (.01, .05, .3, .5, .99) & .040 & .166 & .050 & .197 & .197 & .046 & .046\\
    \bottomrule
  \end{tabular}
\end{minipage}}
\caption{Examples of $p$-value combinations.}\label{tab:expcombine}
\end{table}

\begin{figure}[tbp]
  \centering
  \subfloat[Cauchy Combination Test]{%
    \includegraphics[width=0.4\textwidth]{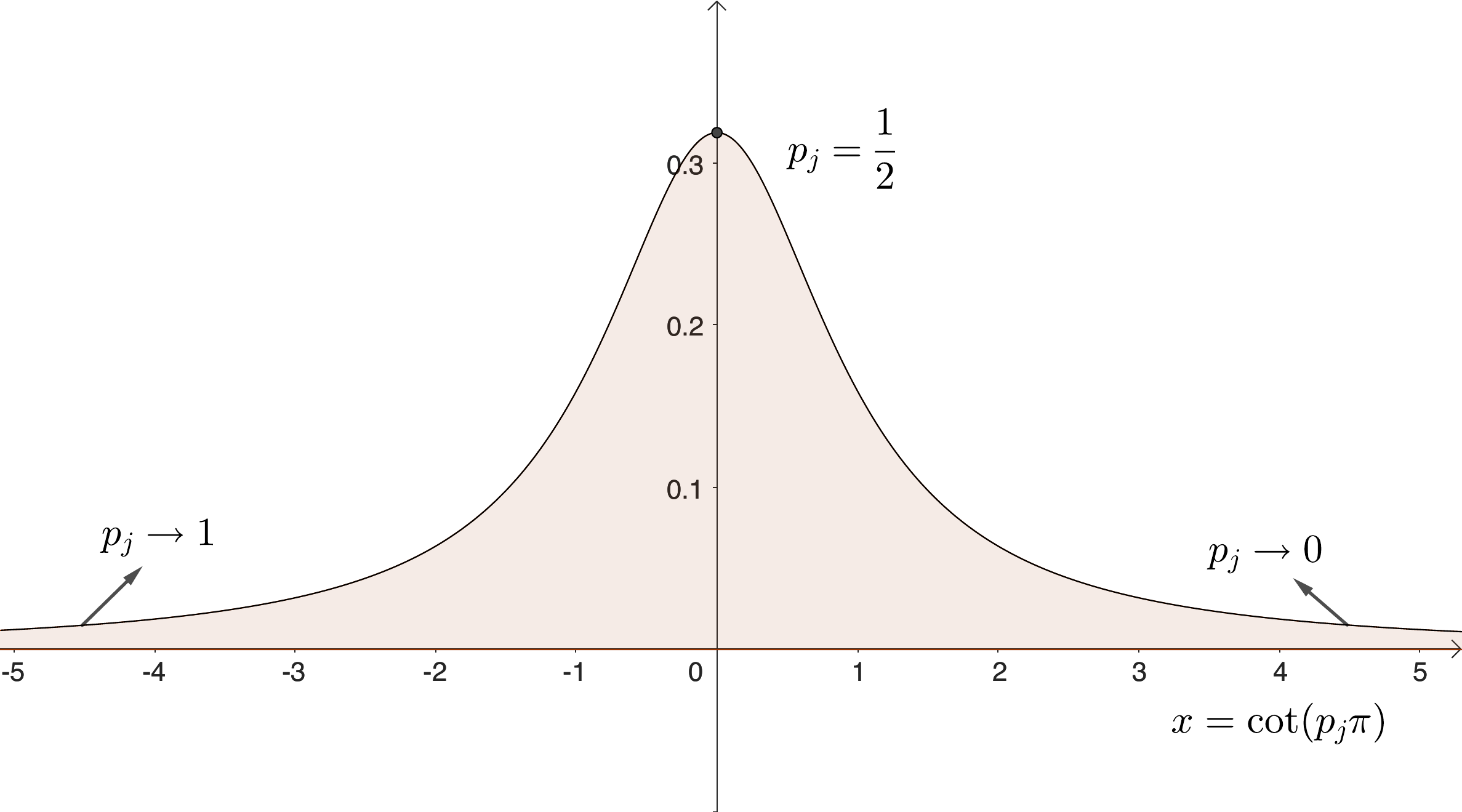}
  }\hspace*{15pt}
  \subfloat[Half-Cauchy Combination Test]{%
    \includegraphics[width=0.4\textwidth]{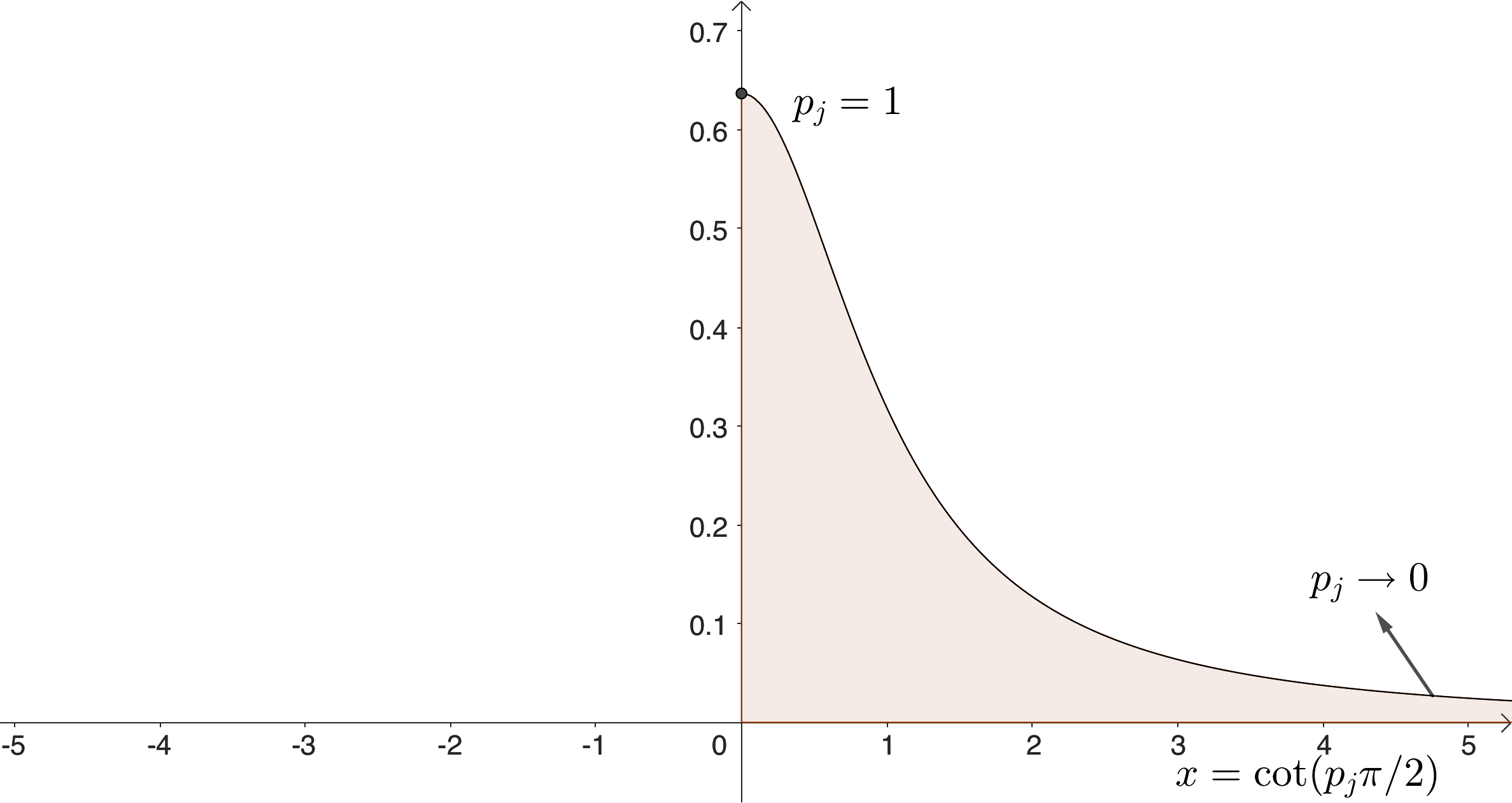}
  }
  \caption{Cauchy vs Half-Cauchy.}
  \label{fig:cvshc}
\end{figure}

\cref{tab:expcombine} gives some tuples of $p$-values where it is more reasonable to reject the global null at significance level $0.05$ yet several previous approaches including CCT fail to do so because of their sensitivity to large $p_{j}$'s. Our proposed Half-Cauchy combination test (HCCT) and exact harmonic mean $p$-value (EHMP) along with Fisher's test and Bonferroni correction perform well in these extreme cases while Stouffer's Z-score test, CCT, and CAtr do not work as expected. 

\begin{figure}[tbp]
\centering
\subfloat[Fisher's test]{%
  \includegraphics[width=.24\textwidth]{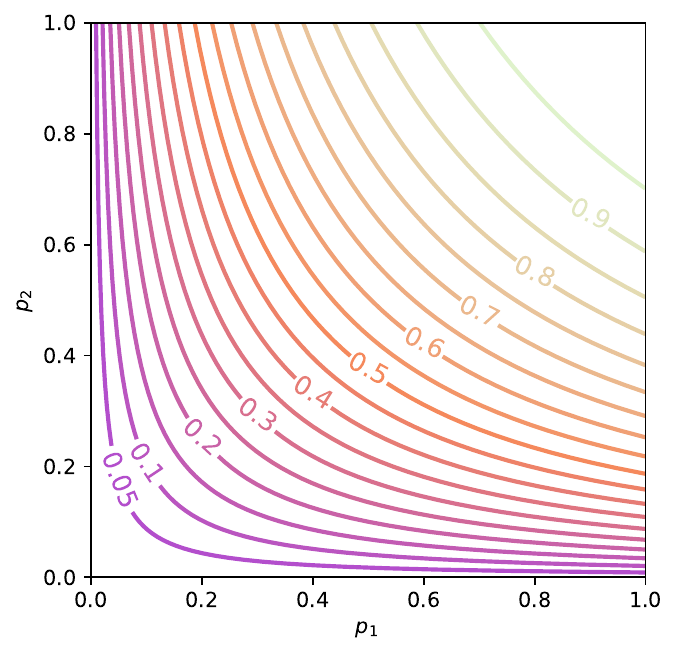}
}~
\subfloat[Stouffer's test]{%
  \includegraphics[width=.24\textwidth]{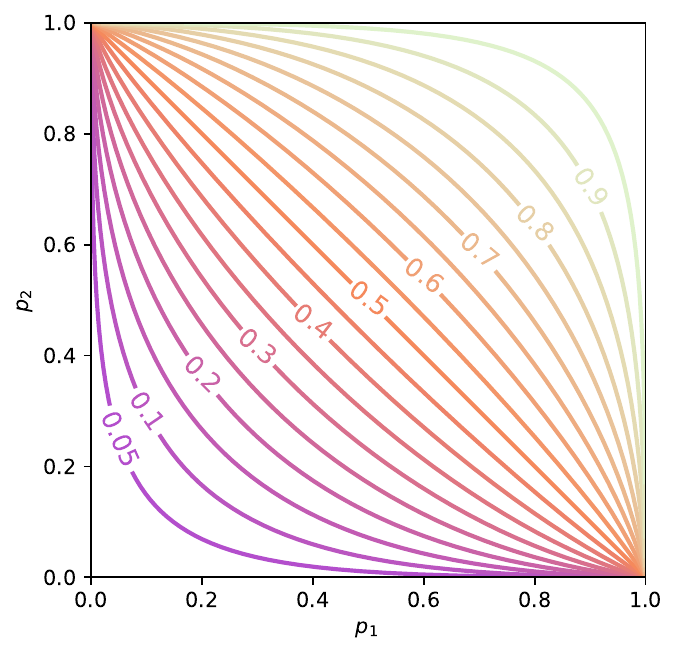}
}~
\subfloat[Bonferroni correction]{%
  \includegraphics[width=.24\textwidth]{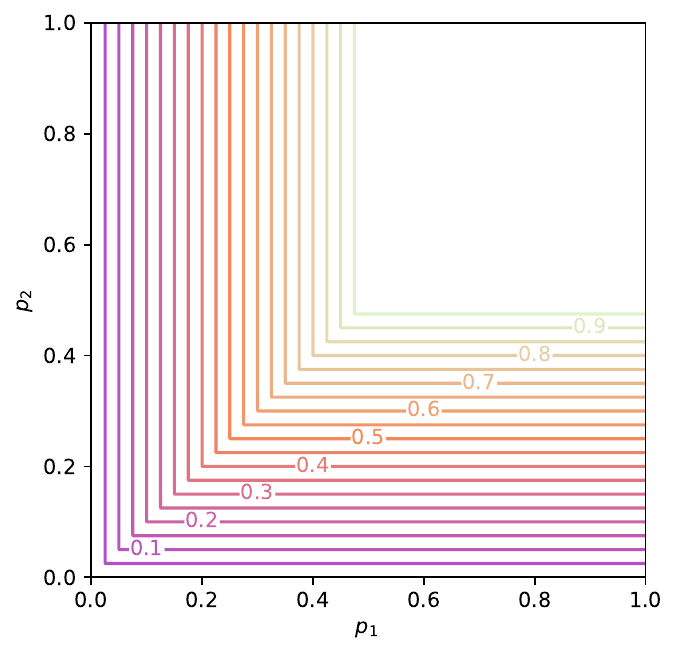}
}\\
\subfloat[CCT]{%
  \includegraphics[width=.24\textwidth]{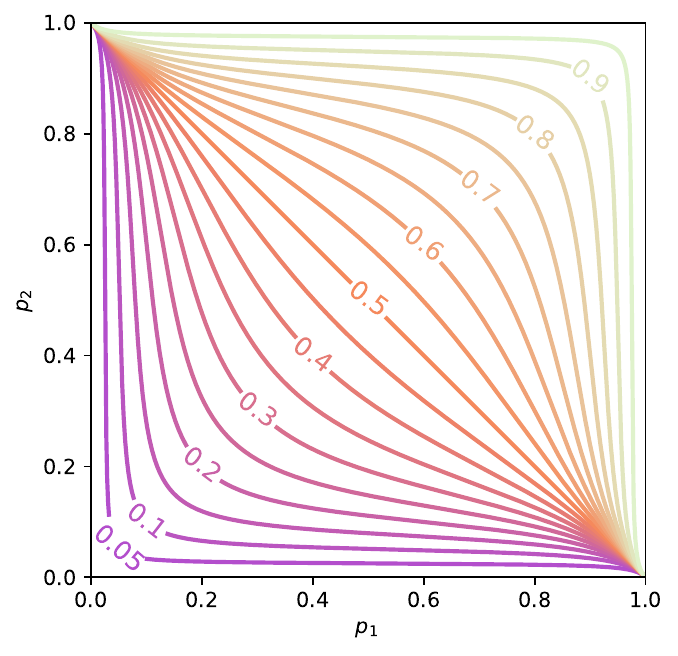}
}~
\subfloat[HCCT]{%
  \includegraphics[width=.24\textwidth]{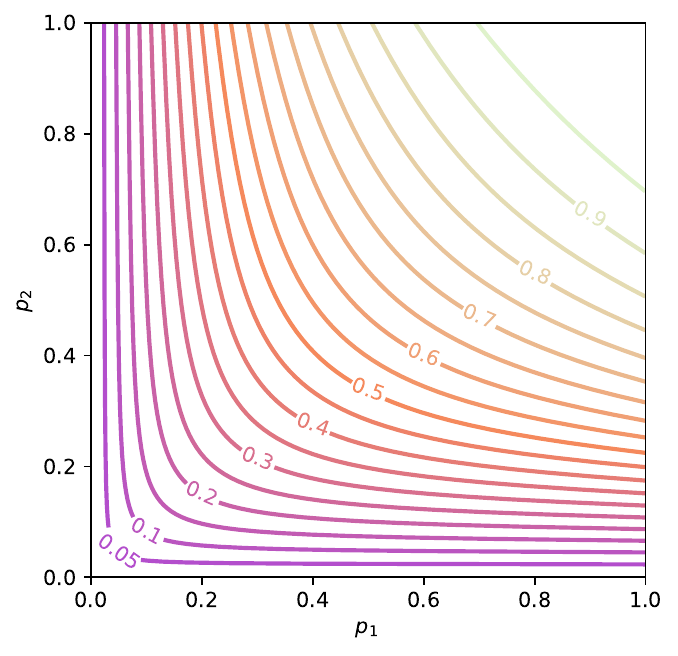}
}~
\subfloat[EHMP]{%
  \includegraphics[width=.24\textwidth]{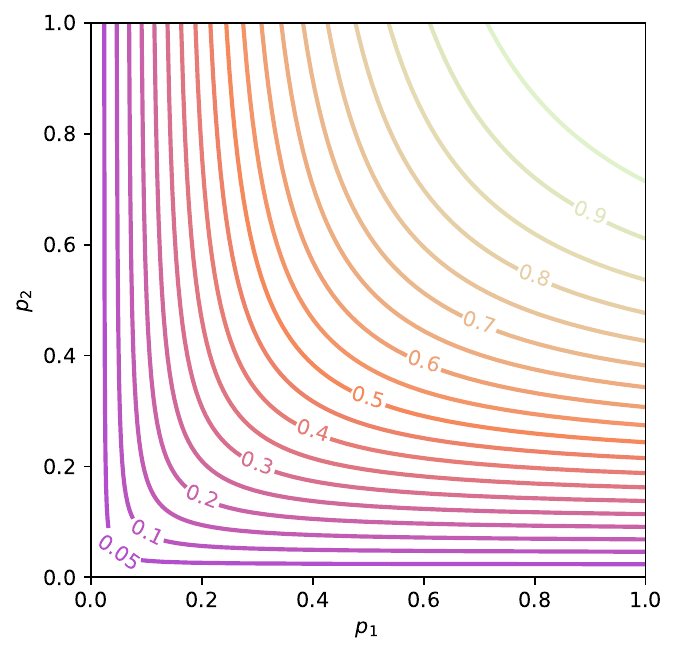}
}
  \caption{Combining two $p$-values with equal weights.}\label{fig:squareplot}
\end{figure}

\cref{fig:squareplot} shows the contour plot when we combine two $p$-values. We can see that for Stouffer's Z-score test and CCT, the contour lines get close together near the point $(1,0)$ in both $p_{1}$ and $p_{2}$ directions, which signifies that the combined $p$-value is sensitive to both $p_{1}$ and $p_{2}$. However, for the other approaches, the contour lines are close in the $p_{2}$ direction around $(1,0)$ but are at a distance away from one another in the $p_{1}$ direction, meaning that the combined $p$-value is sensitive to the smaller $p_{2}$ but insensitive to the larger $p_{1}$.

\cref{fig:cvshc} reveals a key observation that problematic large \(p_{j}\) values are mapped to the negative tail of the Cauchy (or normal) distribution when calculating scores for CCT (or Stouffer's Z-score test). Specifically, if a \(p_{j}\) is close to one, the corresponding component \(\cot(\pi p_{j})\) in \cref{eq:cct} will be far below zero, making it harder to reject the global null. Both the Cauchy and normal distributions have symmetric tails, so negative scores from large \(p_j\) values can offset positive scores from small ones. A potential remedy is to use a distribution \(\nu\) with a heavier right tail than the negative tail. In the stable family \(S(\alpha, \beta, c, \mu)\), this imbalance is controlled by the skewness parameter \(\beta \in [-1, 1]\), where a larger \(\beta\) gives a relatively heavier right tail. Ideally, we select \(\nu\) attracted to \(S(\alpha, \beta, c, \mu)\) with \(\beta = 1\).

The tail agreement between perfect dependence and independence in \cref{thm:extremecase} motivates \(\alpha=1\). Together with the preceding preference for \(\beta=1\), this leads us to distributions attracted to the Landau family (\(\alpha=\beta=1\)). Examples from this class include \(\mathrm{Pareto}(1,1)\), left-truncated or winsorized Cauchy, and the Landau family itself. Moreover, for a small number of studies, if the truncation threshold is far below $0$, the left-truncated or winsorized Cauchy methods of \citet{gui2023aggregating,fang2023heavy} are still sensitive to large $p$-values (see \cref{tab:expcombine}). Finally, we will show in the next section that the Half-Cauchy and $\mathrm{Pareto}(1,1)$ are the only two among all these choices that lead to connected confidence regions when we invert the combination test.

\section{Omitted Proofs for \cref{sec:conf} and \cref{sec:discconf}}\label{sec:pfconf}

This section contains proofs for \cref{thm:adaptiveprocedure,thm:connectivity0,thm:connectivity2,prop:block-divide-combine-volume}. 

\begin{proof}[Proof of \cref{thm:adaptiveprocedure}]
  Fix the true parameter $\boldsymbol{\theta}$ and assume all membership
  events below are measurable. Write $\mathbb P_{\boldsymbol{\theta}}$
  for probabilities under its sampling law.
  By definition of $z_{\boldsymbol{w}^{(0)}}$, we have
  \begin{equation*}
    \mathbb P_{\boldsymbol{\theta}}\{ \boldsymbol{\theta}\in R^{(0)} \}=\mathbb P_{\boldsymbol{\theta}}\bigl\{T_{\boldsymbol{w}^{(0)}}\leq z_{\boldsymbol{w}^{(0)}}\bigr\}\geq 1-p.
  \end{equation*}
  Since $R^{\ast}=\bigcup_{k=0}^{\tau}R^{(k)}\supseteq R^{(0)}$, we have
  \begin{equation*}
    \mathbb P_{\boldsymbol{\theta}}(\boldsymbol{\theta}\in R^{\ast})\geq \mathbb P_{\boldsymbol{\theta}}\{ \boldsymbol{\theta}\in R^{(0)} \}\geq 1-p,
  \end{equation*}
  meaning that the procedure yields a confidence region with at least $(1-p)$ coverage.
\end{proof}

\subsection{Convexity of HCCT and EHMP Regions}\label{sec:convexity-proofs}

We give six versions of the HCCT proofs, three scalar and then three
multivariate, followed immediately by a comparison of their scope.
EHMP is proved separately by direct arguments.
For every active study, the finite-degree assumptions in
\cref{thm:connectivity0,thm:connectivity2} are $k_j\ge1$ when $d_j=1$;
when $d_j\ge2$, they are $k_j\ge d_j+1$ for HCCT and $k_j\ge d_j$ for
EHMP. Chi-square references are also permitted.
All inputs and weights are fixed with respect to the candidate parameter.
The geometric assertions below do not require validity of the working
sampling distributions or independence of the studies.

\subsubsection{Common ingredients}\label{sec:convexity-common}

Write $G_k$ for the standard half-Student CDF with $k>0$ degrees of freedom,
and $G_\infty$ for the half-normal CDF. Write $H_{d,k}$ for the CDF of the
positive square root of $T^2(d,k)$, defined for $k>d-1$, and $H_d$ for the
$\chi_d$ CDF. Thus, in the notation of the main paper,
\[
 G_{k_j}(r)=2F^{(j)}(r)-1
 \quad\hbox{in \cref{sec:unicase}},\qquad
 H_{d_j,k_j}(r)=F^{(j)}(r^2)
 \quad\hbox{in \cref{sec:multidim}}.
\]
In particular, $G_k=H_{1,k}$ and $G_\infty=H_1$.
The dimension $d$ in the analytic radial comparisons may be any real
$d\ge1$; study dimensions and the ambient parameter dimension are integers
when these comparisons are applied to regions.

For $a,b>0$, define
\[
 B(a,b)=\int_0^1t^{a-1}(1-t)^{b-1}\,dt,\qquad
 I_x(a,b)=\frac1{B(a,b)}\int_0^xt^{a-1}(1-t)^{b-1}\,dt.
\]
The scale-free radial density and its CDF are
\begin{equation}\label{eq:radial-beta-family}
 f_{a,b}(x)=\frac{2}{B(a,b)}x^{2a-1}(1+x^2)^{-a-b},
 \qquad
 F_{a,b}(x)=I_{x^2/(1+x^2)}(a,b),\qquad x\ge0.
\end{equation}
For the Hotelling family,
\begin{equation}\label{eq:radial-hotelling-scale}
 a=\frac d2,\qquad b=\frac{k+1-d}{2},\qquad
 H_{d,k}(r)=F_{a,b}(r/\sqrt{k}).
\end{equation}
All curvature comparisons are unaffected by this positive scale change.
When $a$ is fixed we also write $F_b,f_b$.
Let $q_b(u)=I_u^{-1}(a,b)$, meaning $I_{q_b(u)}(a,b)=u$, so that
\[
 F_{a,b}^{-1}(u)=\left\{\frac{q_b(u)}{1-q_b(u)}\right\}^{1/2}.
\]
Every derivative of an inverse beta function below is taken at $0<u<1$.

The reference laws and score endpoints are
\[
 G_C(x)=\frac2\pi\arctan x\quad(x\ge0),\qquad
 G_P(x)=1-\frac1x\quad(x\ge1),
\]
\begin{equation}\label{eq:hc-pareto-score-transforms}
 G_C^{-1}(u)=\tan(\pi u/2),\qquad
 G_P^{-1}(u)=\frac1{1-u},\qquad 0\le u<1.
\end{equation}
The inverse values at zero mean their continuous extensions, zero and one,
respectively, rather than an unrestricted generalized-inverse convention.

\begin{lemma}[Active-rank geometry]\label{lem:convexity-active-rank}
Let $A_j\in\mathbb R^{d_j\times D}$ and $b_j\in\mathbb R^{d_j}$, where
$D\ge1$, and let $J_+$ be a finite nonempty set with weights $w_j>0$.
Suppose $\psi_j:[0,\infty)\to[0,\infty)$ is continuous, nondecreasing,
and tends to infinity. For a finite real $t$, put
\[
 \mathcal C_t=\left\{\theta\in\mathbb R^D:
 \sum_{j\in J_+}w_j\psi_j(\|A_j\theta+b_j\|)\le t\right\}.
\]
Then $\mathcal C_t$ is closed. If nonempty, it is bounded, hence compact,
if and only if the vertically stacked active matrix has full column rank.
If that rank condition fails, every nonempty $\mathcal C_t$ contains
an affine line. If each $\psi_j$ is also convex, then $\mathcal C_t$ is
convex, possibly empty.
\end{lemma}

\begin{proof}
Continuity gives closedness. On a nonempty sublevel set every active term
is at most $t$, because all terms are nonnegative. Since $w_j>0$ and
$\psi_j$ diverges, each $\|A_j\theta+b_j\|$ is bounded there, and hence
so is each $\|A_j\theta\|$.
If the stacked matrix $A$ has full column rank, then
\[
 \sigma_{\min}(A)\|\theta\|
 \le\|A\theta\|
 =\left\{\sum_{j\in J_+}\|A_j\theta\|^2\right\}^{1/2}
\]
bounds $\theta$. Closedness now gives compactness.
If instead $Av=0$ for some $v\ne0$, the score is unchanged by
$\theta\mapsto\theta+sv$ for any real $s$. Every accepted point therefore
generates an entire affine line.
Finally, affine norms are convex; composition with a nondecreasing convex
$\psi_j$ and a nonnegative weighted sum preserves convexity.

Alternatively, nonconstant convex $\psi_j$ have finite minima by
monotonicity and grow at least linearly by the secant-slope inequality.
Full column rank then makes the sum diverge as $\|\theta\|\to\infty$.
Our proof instead uses lower-boundedness and divergence separately;
nonnegativity supplies only the former, which fails for CCT.
\end{proof}

For the manuscript's matrices,
$A_j=\widehat\Sigma_j^{-1/2}P_j$,
positive-definite $\widehat\Sigma_j$ makes the active rank condition
equivalent to spanning by the active row spaces of $P_j$.
For a scalar parameter and positive study scales it holds automatically.
Consequently a convex scalar sublevel is empty, a singleton, or a closed
bounded interval. A nonempty compact multivariate region has finite diameter
and finite Lebesgue measure, and every scalar linear projection attains finite
minimum and maximum values. Rank deficiency alone does not imply infinite
ambient Lebesgue measure.
For $0<p<1$, the reference quantile in the main theorem is finite:
it is the quantile of a finite weighted sum of almost surely finite
reference variables. No dependence-validity conclusion is used here.

\begin{lemma}[Pointwise chi-limit completion]\label{lem:convexity-pointwise-limit}
For fixed real $d\ge1$, $H_{d,k}(x)\to H_d(x)$ for every $x\ge0$ as
$k\to\infty$. Let $Q$ be finite and continuous on $[0,1)$, including its
specified endpoint value at zero. If $Q\circ H_{d,k}$ is convex on
$[0,\infty)$ for all sufficiently large $k$, then $Q\circ H_d$ is convex.
\end{lemma}

\begin{proof}
Let $Z\sim\chi_d^2$ and, independently,
$V_k\sim\chi_{k+1-d}^2$. The radial Hotelling variable is
$\sqrt{kZ/V_k}$. Since
\[
 \mathbb E(V_k/k)=\frac{k+1-d}{k}\to1,\qquad
 \operatorname{Var}(V_k/k)=\frac{2(k+1-d)}{k^2}\to0,
\]
it converges in distribution to $\sqrt Z$.
The limiting CDF is continuous, so the asserted CDF convergence holds.
For finite $x$, $H_d(x)<1$; continuity of $Q$ gives pointwise convergence
of the transforms. Taking limits in their convexity inequalities proves
the result. The statement at zero uses the specified continuous endpoint,
not differentiation there.
\end{proof}

This argument uses the correctly scaled family
\cref{eq:radial-hotelling-scale}; weak convergence alone is not invoked
to conclude convergence of density derivatives.
For HCCT we take $Q=G_C^{-1}$. The EHMP chi endpoint will instead be
proved directly.

\begin{lemma}[Half-Cauchy boundary comparison]\label{lem:hc-boundary-curvature}
For every real $d\ge1$ and $0<u<1$,
\begin{equation}\label{eq:hc-boundary-curvature}
 \mathcal T_{H_{d,d+1}}(u)
 <\frac{4u}{1-u^2}
 <\pi\tan(\pi u/2)=\mathcal T_{G_C}(u).
\end{equation}
\end{lemma}

\begin{proof}
At $k=d+1$ the beta parameter $b$ equals one. Inverting
\cref{eq:radial-beta-family} and using scale invariance gives
\[
 T_d(u):=\mathcal T_{H_{d,d+1}}(u)
 =\frac{3-(d+2)(1-u^{2/d})}{du(1-u^{2/d})}.
\]
Set $v=-\log u>0$, $z=2v/d$, and
$\Phi(z)=3z/(1-e^{-z})-2z$, with $\Phi(0)=3$.
Then
\[
 uT_d(u)+1=\frac{\Phi(z)}{2v},\qquad 0<z\le2v.
\]
For $h(z)=z/(1-e^{-z})$,
\[
 h''(z)=
 \frac{e^{-z}\{(z-2)+(z+2)e^{-z}\}}{(1-e^{-z})^3}>0.
\]
The expression in braces vanishes at zero and has derivative
$1-(z+1)e^{-z}>0$ for $z>0$. Thus $\Phi$ is convex, and its maximum
on $[0,2v]$ is attained at an endpoint. Consequently
\[
 T_d(u)\le
 \max\left\{\frac{3u}{1-u^2},\frac{3/(2v)-1}{u}\right\}.
\]
The first term is strictly below $4u/(1-u^2)$.
For the second, with $z=2v$, the required inequality is equivalent to
\[
 z-3+3(z+1)e^{-z}>0.
\]
It is immediate when $z\ge3$. For $0<z<3$, put
$R(z)=\log\{3(1+z)/(3-z)\}-z$. We have $R(0)=0$ and
\[
 R'(z)=\frac{(z-1)^2}{(1+z)(3-z)}\ge0.
\]
The derivative is positive except at one isolated point, so $R(z)>0$.
Exponentiation proves the inequality and hence the first strict bound
in \cref{eq:hc-boundary-curvature}.

For the second, define
\[
 A(x)=(\pi^2-4x^2)\sin x-8x\cos x,\qquad 0\le x\le\pi/2.
\]
Both endpoint values are zero, while
$A'(x)=(\pi^2-8-4x^2)\cos x$ changes sign exactly once, from positive
to negative. Hence $A(x)>0$ in the interior. Substitute $x=\pi u/2$
to obtain $4u/(1-u^2)<\pi\tan(\pi u/2)$.
\end{proof}

The equivalent cleared-denominator inequality is
\begin{equation}\label{eq:hc-boundary-cleared}
 (d+2)(1-u^{2/d})
 +\pi du(1-u^{2/d})\tan(\pi u/2)>3,\qquad 0<u<1.
\end{equation}
Its left side has continuous endpoint limits $d+2$ and $4$.
The non-strict extension therefore holds, with endpoint equality only
at $d=1,u=0$. Curvatures themselves are compared only on $(0,1)$.
This analytic lemma replaces the former numerical verification.
It completes the multivariate degree-order arguments below; the scalar
proofs retain their sharper anchor $H_{1,1}=G_C$.

\begin{lemma}[An affine-crossing criterion]\label{lem:convexity-crossings}
Let \(F\) and \(G\) be continuous, strictly increasing CDFs on
\([0,\infty)\), with \(F(0)=G(0)=0\) and positive continuous densities
\(f\) and \(g\) on \((0,\infty)\). For \(s,h>0\), define
\[
\begin{aligned}
D_s(x)&=F(sx)-G(x),&
\rho_s(x)&=\frac{s f(sx)}{g(x)},\qquad x>0,\\
D_{s,h}(z)&=F(sz)-G(z+h),&
\rho_{s,h}(z)&=\frac{s f(sz)}{g(z+h)},\qquad z>0.
\end{aligned}
\]
Suppose the following hold for every \(s,h>0\).
\begin{enumerate}
\item \(D_s(x)<0\) for all sufficiently large \(x\), and
\(D_{s,h}(z)<0\) for all sufficiently large \(z\).
\item \(\rho_s\) is first nonincreasing and then nondecreasing, and
\(\rho_s(x)\to\infty\) as \(x\to\infty\).
\item \(\rho_{s,h}\) is successively nondecreasing, nonincreasing, and
nondecreasing, and \(\rho_{s,h}(z)\to\infty\) as \(z\to\infty\).
\end{enumerate}
Any of these monotonicity phases may be empty. Then
\(Q=F^{-1}\circ G\), with \(Q(0)=0\), is increasing and convex on
\([0,\infty)\).
\end{lemma}

\begin{proof}
The transform \(Q\) is continuous and strictly increasing. We show that
\(\{x>0:Q(x)<L(x)\}\) is an interval, possibly empty, for every affine
function \(L\).

First consider lines through the origin. Since
\[
D_s'(x)=g(x)\{\rho_s(x)-1\},
\]
the possible strict signs of \(D_s'\) are \(-,+\) or \(+,-,+\), with
phases possibly absent. An everywhere nonnegative derivative is
incompatible with \(D_s(0)=0\) and eventual negativity. Also
\(D_s(x)\to0\) at infinity. For the pattern \(-,+\), the function stays
nonpositive: its final increasing phase approaches zero from below.
For the pattern \(+,-,+\), only the middle decreasing phase can cross
zero, and that crossing is from positive to negative. These statements
remain valid if zeros occupy intervals. Thus a negative-to-positive
strict sign change of \(D_s\) is impossible.

Because
\[
\operatorname{sgn}D_s(x)=\operatorname{sgn}\{sx-Q(x)\},
\]
the ratio \(Q(x)/x\) is nondecreasing. Indeed, a decrease of this ratio
would, by choosing \(s\) strictly between two of its values, produce the
forbidden negative-to-positive sign change of \(D_s\). Consequently, for
\(s>0\) and \(t\ge0\),
\[
\frac{Q(x)-(sx+t)}{x}=\frac{Q(x)}x-s-\frac tx
\]
is nondecreasing. Its strict negative set is an interval, which handles
positive-slope lines with nonnegative intercepts.

For a positive-slope line with negative intercept, write
\(L(x)=s(x-h)\), where \(h>0\). There are no points with \(Q(x)<L(x)\)
when \(0<x\le h\). For \(x=z+h>h\), this inequality is equivalent to
\(D_{s,h}(z)>0\). Now
\[
D_{s,h}'(z)=g(z+h)\{\rho_{s,h}(z)-1\}.
\]
The possible strict derivative signs are \(-,+,-,+\), with phases
possibly absent. The initial decreasing phase stays negative because
\(D_{s,h}(0)=-G(h)<0\). The final increasing phase also stays negative,
because the function is eventually negative and converges to zero.
Only the middle increasing and decreasing phases can contain positive
values. Hence \(\{z>0:D_{s,h}(z)>0\}\) is an interval. This handles the
remaining positive-slope lines. If \(L\) has nonpositive slope, then
\(Q-L\) is increasing, so its strict negative set is again an interval.

To conclude, if \(Q\) were nonconvex on \((0,\infty)\), some chord would
lie strictly below \(Q\) at an interior point. Shifting that chord upward
by a sufficiently small positive constant would make \(Q-L\) negative
at the two chord endpoints and positive at the intervening point,
contradicting the interval property just proved. Thus \(Q\) is convex
on \((0,\infty)\). Its continuous extension at zero is convex by taking
limits in the convexity inequality.
\end{proof}

\subsubsection{Proof 1: Univariate inverse-beta argument}
\label{sec:hcct-proof1}

We first establish a beta-quantile comparison and then apply the original
curvature-ratio argument.  The comparison is stated for an arbitrary first
beta parameter because its multivariate counterpart uses the same notation.

\begin{lemma}[Beta-quantile ratio]
\label{lem:hcct-beta-quantile-ratio}
Fix $a>0$ and $0<b<c$, and write $q_b(u)=I_u^{-1}(a,b)$.
Then $0<q_c(u)<q_b(u)<1$ for $0<u<1$, and $q_c(u)/q_b(u)$ is strictly
increasing on $(0,1)$.  Its limits at zero and one are, respectively,
\[
  \left\{\frac{B(a,c)}{B(a,b)}\right\}^{1/a}
  \quad\text{and}\quad 1.
\]
\end{lemma}

\begin{proof}
Write $p_{a,b}(x)=x^{a-1}(1-x)^{b-1}/B(a,b)$ for the beta density.
These densities satisfy
\[
  \frac{p_{a,c}(x)}{p_{a,b}(x)}
  =\frac{B(a,b)}{B(a,c)}(1-x)^{c-b},\qquad 0<x<1.
\]
This ratio strictly decreases from a value greater than one to zero.
Consequently $p_{a,c}-p_{a,b}$ changes sign exactly once, from positive to
negative.  Its integral over $(0,1)$ is zero, so
$I_x(a,c)>I_x(a,b)$ for every $0<x<1$.  Inverting gives the asserted strict
ordering of the quantiles.

To compare their ratio, fix $0<\lambda<1$ and define
\[
  \Xi(x)=I_x(a,b)-I_{\lambda x}(a,c),\qquad 0\le x\le1.
\]
We have $\Xi(0)=0$ and $\Xi(1)=1-I_\lambda(a,c)>0$.
For $0<x<1$, the sign of $\Xi'(x)$ is the sign of
\[
  \Theta(x)=\log
  \frac{(1-x)^{b-1}/B(a,b)}
       {\lambda^a(1-\lambda x)^{c-1}/B(a,c)}.
\]
Direct differentiation yields
\begin{equation}
\label{eq:hcct-beta-ratio-theta-derivative}
  \Theta'(x)=
  \frac{(1-b)(1-\lambda)+\lambda(c-b)(1-x)}
       {(1-x)(1-\lambda x)}.
\end{equation}
If $b\le1$, the numerator is strictly positive.  Thus $\Theta$ is strictly
increasing, and $\Xi'$ has at most the sign pattern $-,+$, with an interval
possibly absent.  Together with the endpoint values of $\Xi$, this shows
that either $\Xi$ is positive throughout $(0,1)$ or it has exactly one
interior zero, crossing from negative to positive.

If $b>1$, the numerator in
\cref{eq:hcct-beta-ratio-theta-derivative} decreases strictly and is
negative at $x=1$.  Therefore $\Theta$ either decreases throughout or
increases and then decreases; moreover $\Theta(x)\to-\infty$ as
$x\uparrow1$.  The possible sign patterns for $\Xi'$ are consequently
$-$, $+,-$, and $-,+,-$, allowing absent intervals.  The first is
incompatible with $\Xi(0)=0<\Xi(1)$.  In the remaining cases, $\Xi$ stays
positive on its final decreasing interval, since $\Xi(1)>0$.
The initial increasing interval, if present, is also positive because
$\Xi(0)=0$.  Otherwise the initial decreasing interval is negative, and
only the intervening increasing interval can contain a zero.  Thus again
there is at most one interior zero, necessarily a crossing from negative
to positive.  The strict monotonicity on the relevant intervals also
excludes an additional zero without a sign change.

When $x=q_b(u)$, strict monotonicity of $I_x(a,c)$ gives
\[
  \operatorname{sgn}\Xi(q_b(u))
  =\operatorname{sgn}\{q_c(u)/q_b(u)-\lambda\}.
\]
Testing every $\lambda\in(0,1)$ proves that the quantile ratio is
nondecreasing.  Equality of its values at two distinct arguments would,
at the common level $\lambda\in(0,1)$, produce two zeros of $\Xi$.
The ratio is therefore strictly increasing.

For the endpoint calculations, expansion of the smooth factor in the
beta integral gives
\[
\begin{aligned}
  I_x(a,b)&=\frac{x^a}{aB(a,b)}\{1+O(x)\}
             &&(x\downarrow0),\\
  1-I_x(a,b)&=\frac{(1-x)^b}{bB(a,b)}\{1+O(1-x)\}
             &&(x\uparrow1).
\end{aligned}
\]
In particular $q_b(u)\sim\{aB(a,b)u\}^{1/a}$ at zero, proving the
first ratio limit.  At the other endpoint both quantiles tend to one.
The restriction $0<\lambda<1$ in the fixed-level argument is essential
for its strict endpoint inequality; no such inequality is asserted at
$\lambda=1$.
\end{proof}

\begin{lemma}[Half-Student curvature ordering]
\label{lem:hcct-half-student-curvature-order}
For every real $0<k<\ell$ and $0<u<1$,
\[
  \mathcal T_{G_k}(u)>\mathcal T_{G_\ell}(u).
\]
\end{lemma}

\begin{proof}
Set $a=1/2$, $b=k/2$, and $c=\ell/2$.  Define
\[
  U_b=\frac{q_b'}{q_b},\qquad
  V_b=\frac{q_b'}{1-q_b},\qquad
  T_b=\mathcal T_{G_{2b}}.
\]
Inverse-beta differentiation gives
\[
  q_b'=B(1/2,b)q_b^{1/2}(1-q_b)^{1-b}.
\]
The half-Student quantile is a positive constant times
$\{q_b/(1-q_b)\}^{1/2}$.  Scale invariance of the curvature and
logarithmic differentiation therefore give
\begin{equation}
\label{eq:hcct-half-student-curvature}
  T_b=(b+\tfrac12)V_b>0,
  \qquad
  (\log T_b)'=\tfrac12 U_b+bV_b.
\end{equation}
Since $q_c/q_b$ is increasing and differentiable,
\cref{lem:hcct-beta-quantile-ratio} implies $U_b\le U_c$.
Equivalently, in the original half-Student scale,
\[
  g_k(x)=\frac{2}{\sqrt{k}B(1/2,k/2)}
          (1+x^2/k)^{-(k+1)/2},\qquad x>0,
\]
and direct differentiation gives the identity
\[
  -\frac{g_k'(x)}{g_k(x)^2}
  =\frac{(k+1)x}{2\sqrt{k}}B(k/2,1/2)
     (1+x^2/k)^{(k-1)/2}.
\]

Let $R(u)=T_b(u)/T_c(u)$.  At the upper endpoint the beta-tail expansion
and the exact formula for $q_b'$ give
\[
  1-u\sim\frac{\{1-q_b(u)\}^b}{bB(1/2,b)},
  \qquad
  T_b(u)\sim\frac{b+1/2}{b(1-u)}.
\]
Consequently
\begin{equation}
\label{eq:hcct-half-student-ratio-upper}
  R(1-)=\frac{c(b+1/2)}{b(c+1/2)}>1.
\end{equation}
For completeness, the lower-endpoint expansion is
\[
  T_b(u)\sim\frac12(b+\tfrac12)B(b,1/2)^2u.
\]
To compare these coefficients, put $K(b)=(b+1/2)B(b,1/2)^2$.
Differentiating the beta integral, with
$Z\sim\operatorname{Beta}(b,1/2)$, gives
\begin{equation}
\label{eq:beta-endpoint-monotonicity}
\begin{aligned}
  \frac{K'(b)}{K(b)}
  &=\frac1{b+1/2}+2\mathbb E\log Z\\
  &<\frac1{b+1/2}+2(\mathbb E Z-1)=0,
    \qquad b>0.
\end{aligned}
\end{equation}
Here $\log Z<Z-1$ almost surely and
$\mathbb E Z=b/(b+1/2)$.  Differentiation under the beta integral is
justified on every compact subinterval of $b>0$: its derivative is
dominated by a constant times
$z^{b_0-1}(1-z)^{-1/2}|\log z|$ for some $b_0>0$, an integrable
function on $(0,1)$.  Thus $K'(b)<0$, and
$R(0+)=K(b)/K(c)>1$.  The upper endpoint alone will suffice for the
following contradiction; the lower-endpoint calculation also supplies
the beta bound used in the degree-boundary analysis.

Whenever $R<1$, positivity of the curvatures and $b<c$ imply
\[
  bV_b=\frac{b}{b+1/2}T_b
  <\frac{c}{c+1/2}T_c=cV_c.
\]
Combining this with $U_b\le U_c$ and
\cref{eq:hcct-half-student-curvature} yields
\[
  (\log R)'=\tfrac12(U_b-U_c)+bV_b-cV_c<0.
\]
Suppose a connected component $(\alpha,\beta)$ of $\{u:R(u)<1\}$
exists, where either endpoint may be an endpoint of $(0,1)$.
The function $R$ is strictly decreasing on this component.
If $\beta<1$, continuity gives $R(\beta)=1$, whereas monotonicity
gives $R(\beta)\le R(u_0)<1$ for any $u_0\in(\alpha,\beta)$,
a contradiction.  If $\beta=1$, the same argument contradicts
\cref{eq:hcct-half-student-ratio-upper}.  Hence $R\ge1$ throughout.
At an interior point with $R=1$, the strict inequality
$b/(b+1/2)<c/(c+1/2)$ still gives $(\log R)'<0$.
Such a point would instead be a minimum of the differentiable function
$R\ge1$, with derivative zero.  Thus $R>1$, proving the lemma.
\end{proof}

\begin{proof}[HCCT interval conclusion]
The identity $G_1=G_C$ and
\cref{lem:hcct-half-student-curvature-order,thm:connectivity} show that
$G_C^{-1}\circ G_k$ is increasing and convex for every real $k\ge1$.
At $k=1$ this transform is the identity.  Its value tends to zero at the
left endpoint, so its continuous extension to $[0,\infty)$ is convex.

The half-normal case follows without differentiating a weak limit.
If $Z$ is standard normal and $V_k\sim\chi_k^2$ is independent of $Z$,
then $|Z|/\sqrt{V_k/k}$ has CDF $G_k$.
Since $\mathbb E(V_k/k)=1$ and $\operatorname{Var}(V_k/k)=2/k$,
$V_k/k$ converges in probability to one, and
$G_k(x)\to H_1(x)$ for every $x\ge0$.
The inverse $G_C^{-1}$ is continuous at $H_1(x)<1$ for every finite
$x$, including its continuous value at zero.  Hence
\cref{lem:convexity-pointwise-limit} gives convexity of
$G_C^{-1}\circ H_1$ on $[0,\infty)$.

Write $G_\infty=H_1$.  For the active studies, the functions
\[
  r\longmapsto G_C^{-1}\{G_{k_j}(r)\},\qquad
  k_j\in[1,\infty],
\]
are continuous, nonnegative, increasing, convex, and tend to infinity
with $r$.  Apply \cref{lem:convexity-active-rank} with
$A_j=1/\widehat\sigma_j$ and
$b_j=-\widehat\theta_j/\widehat\sigma_j$.
At least one study is active, and its scale is positive and finite, so
the active stacked matrix has full column rank.
Thus the solution set of \cref{eq:invert}, at its finite cutoff for
$0<p<1$, is empty or a closed bounded interval, possibly a singleton.
Zero-weight studies are omitted throughout.
This proves the HCCT assertion in \cref{thm:connectivity0}.
\end{proof}

\subsubsection{Proof 2: Univariate convexity by density and CDF crossings}
\label{sec:hcct-proof2}

We prove the all-degrees comparison
\[
\mathcal T_{G_k}(u)\ge\mathcal T_{G_\ell}(u),
\qquad 0<k<\ell,\quad 0<u<1,
\]
directly from the half-Student densities, without inverse-beta
differentiation.

\begin{proof}
Write
\[
g_k(x)=C_k\left(1+\frac{x^2}{k}\right)^{-(k+1)/2},
\qquad
C_k=\frac{2}{\sqrt{k}B(1/2,k/2)},\qquad x\ge0.
\]
Fix \(0<k<\ell\), and set \(Q=G_k^{-1}\circ G_\ell\). We verify
\cref{lem:convexity-crossings} with \(F=G_k\) and \(G=G_\ell\).
Integration of the explicit density tail gives
\[
1-G_k(x)\sim A_kx^{-k},\qquad A_k>0.
\]
It follows that, for every fixed \(s,h>0\), both
\(G_k(sx)-G_\ell(x)\) and \(G_k(sz)-G_\ell(z+h)\) are eventually
negative. Their limits at infinity are zero. The corresponding density
ratios tend to infinity, since each is asymptotic to a positive constant
times \(x^{\ell-k}\), or \(z^{\ell-k}\), respectively.

For a line through the origin, put
\[
\Theta_s(x)=\log\frac{s g_k(sx)}{g_\ell(x)}.
\]
Direct differentiation gives
\[
\Theta_s'(x)=
\frac{x\{k(\ell+1)-\ell(k+1)s^2+(\ell-k)s^2x^2\}}
{(k+s^2x^2)(\ell+x^2)}.
\]
The expression in braces increases strictly with \(x^2\), so the density
ratio first decreases and then increases, with the decreasing phase
possibly absent.

For a negative-intercept line \(s(x-h)\), put \(z=x-h>0\) and
\[
\Theta(z)=\log\frac{s g_k(sz)}{g_\ell(z+h)}.
\]
After multiplication by the positive denominator
\((k+s^2z^2)\{\ell+(z+h)^2\}\), the derivative \(\Theta'\) has numerator
\[
\begin{aligned}
N(z)={}&k(\ell+1)h\\
&+\{k(\ell+1)-(k+1)s^2(\ell+h^2)\}z\\
&+s^2h(\ell-2k-1)z^2+s^2(\ell-k)z^3.
\end{aligned}
\]
The useful normalization is
\[
\left(\frac{N(z)}z\right)''
=\frac{2k(\ell+1)h}{z^3}+2s^2(\ell-k)>0.
\]
Moreover, \(N(z)/z\to+\infty\) at both zero and infinity. Its negative
set is therefore an interval, possibly empty, and its only possible
strict sign pattern is \(+,-,+\). Multiplication by positive factors
does not change signs, so \(\Theta'\) has the same pattern. Thus the
shifted density ratio successively increases, decreases, and increases,
with phases possibly absent.

All hypotheses of \cref{lem:convexity-crossings} now hold. Consequently
\(Q\) is convex on \([0,\infty)\), and \cref{thm:connectivity} gives the
claimed curvature comparison.

Since \(G_1=G_C\), taking \(k=1\) shows that
\(\psi_\ell=G_C^{-1}\circ G_\ell\) is increasing and convex for every
\(\ell\ge1\); at \(\ell=1\) it is the identity. The half-normal CDF is
the pointwise limit of \(G_\ell\), so
\cref{lem:convexity-pointwise-limit} gives the same conclusion for
\(\psi_\infty=G_C^{-1}\circ H_1\). These scores are nonnegative,
continuous at zero with value zero, and tend to infinity with their
argument.

For the active studies in \cref{thm:connectivity0}, apply
\cref{lem:convexity-active-rank} with
\[
A_j=\widehat\sigma_j^{-1},
\qquad b_j=-\widehat\theta_j/\widehat\sigma_j.
\]
Every active scale is positive and finite, and there is at least one
active study, so the stacked scalar matrix has full column rank.
Therefore the HCCT sublevel set in \cref{eq:invert}, at its finite
reference cutoff, is empty or a closed bounded interval, possibly a
singleton. This proves the HCCT assertion of \cref{thm:connectivity0}.
\end{proof}

\subsubsection{Proof 3: A direct univariate Half-Cauchy comparison}
\label{sec:hcct-proof3}

This proof establishes the required scalar score convexity directly,
without ordering all pairs of degrees of freedom.

\begin{proof}
Positive rescaling of the argument preserves convexity. Thus, for a
half-Student distribution with \(k\ge1\), it suffices to use the density
and CDF
\[
h(x)=C(1+x^2)^{-\kappa},\qquad
\kappa=\frac{k+1}{2}\ge1,\qquad x\ge0,
\]
where \(C\) is the normalizing constant and \(H\) is the corresponding
CDF. Define
\[
T(x)=-\frac{h'(x)}{h(x)^2}
=\frac{2\kappa}{C}x(1+x^2)^{\kappa-1},
\qquad
W(x)=\frac2\pi\arctan\left(\frac{T(x)}\pi\right).
\]
The Half-Cauchy score \(Q=G_C^{-1}\circ H\) satisfies
\[
Q''(x)=\frac\pi2\sec^2\{\pi H(x)/2\}
\left[h'(x)+\pi h(x)^2\tan\{\pi H(x)/2\}\right].
\]
Because \(T(x)\ge0\) and \(0\le H(x)<1\), the inequality \(Q''(x)\ge0\)
is equivalent to \(H(x)\ge W(x)\).

Put \(D=H-W\). Differentiating gives
\[
D'(x)=
\frac{h(x)\{\pi^2+T(x)^2\}-2T'(x)}{\pi^2+T(x)^2},
\qquad
T'(x)=\frac{2\kappa}{C}(1+x^2)^{\kappa-2}
       \{1+(2\kappa-1)x^2\}.
\]
After multiplication of the numerator by the positive factor
\(C(1+x^2)^\kappa\), its sign is the sign of
\[
V(x)=\pi^2C^2
-4\kappa(1+x^2)^{2\kappa-2}\{1+(\kappa-1)x^2\}.
\]
For \(\kappa>1\), this function decreases strictly to minus infinity.
Indeed, with \(z=x^2\),
\[
\frac{d}{dz}\left[(1+z)^{2\kappa-2}\{1+(\kappa-1)z\}\right]
=(\kappa-1)(1+z)^{2\kappa-3}
  \{3+(2\kappa-1)z\}>0.
\]
On the other hand,
\[
D(0)=0,\qquad \lim_{x\to\infty}D(x)=0,
\]
because \(H(x)\to1\) and \(T(x)\to\infty\). These two endpoint values
force \(V(0)>0\). Otherwise \(V(x)<0\), and hence \(D'(x)<0\), for every
\(x>0\), contradicting the equal endpoint values of \(D\).
It follows that \(D'\) changes sign exactly once, from positive to
negative. Thus \(D\) first increases strictly from zero and then
decreases strictly to zero. In particular, \(D(x)>0\) for every \(x>0\),
so \(Q''(x)>0\) whenever \(k>1\).

When \(\kappa=1\), normalization gives \(C=2/\pi\). Then
\(T(x)=\pi x\), \(H(x)=W(x)=2\arctan(x)/\pi\), and \(Q(x)=x\).
This includes the equality case \(k=1\). Rescaling back to the
half-Student argument proves the finite-degree assertion. Continuity
at zero extends convexity to \([0,\infty)\); no differentiability of an
absolute-value composition at its cusp is asserted.

The half-normal case admits the same direct argument. Now
\[
h(x)=\sqrt{2/\pi}\,e^{-x^2/2},\qquad
T(x)=\frac{x}{h(x)},\qquad
T'(x)=\frac{1+x^2}{h(x)}.
\]
With the same definitions of \(H,W,D\), the sign of \(D'\) is the sign
of
\[
\pi^2h(x)^2-x^2-2=2\pi e^{-x^2}-x^2-2.
\]
This function is positive at zero, strictly decreasing on
\((0,\infty)\), and tends to minus infinity. Also \(D(0)=D(\infty)=0\).
The preceding one-sign-change argument therefore gives \(D(x)>0\),
and hence strict Half-Cauchy score convexity, for every \(x>0\).

All the resulting scores are increasing, nonnegative, continuous at
zero with value zero, and divergent at infinity. For the active
studies, their compositions with
\(|\widehat\theta_j-\theta|/\widehat\sigma_j\) therefore satisfy
\cref{lem:convexity-active-rank}. Positive finite active scales and the
existence of at least one active study give full stacked rank in one
parameter. Thus the HCCT sublevel set in \cref{eq:invert} is empty or a
closed bounded interval, possibly a singleton, proving
\cref{thm:connectivity0} for HCCT.
\end{proof}

\subsubsection{Proof 4: Multivariate compensated inverse-beta argument}
\label{sec:hcct-proof4}

The inverse-beta approach in \cref{sec:hcct-proof1} extends to the radial
family after compensating for a term that vanishes in one dimension.
We prove the comparison between $F_{a,1}$ and $F_{a,c}$ for
every $a\ge1/2$ and $c>1$.  This is the boundary-reference comparison
needed for HCCT; no ordering of arbitrary pairs $1<b<c$ is used or
claimed in this argument.

For a fixed $a>0$, write
\[
  q_b(u)=I_u^{-1}(a,b),\qquad
  U_b=\frac{q_b'}{q_b},\qquad
  V_b=\frac{q_b'}{1-q_b},\qquad 0<u<1.
\]
Since $F_{a,b}^{-1}(u)=\{q_b(u)/(1-q_b(u))\}^{1/2}$,
\[
  \log (F_{a,b}^{-1})'
  =\log q_b'-\tfrac12\log q_b-\tfrac32\log(1-q_b)-\log2.
\]
Also $q_b''/q_b'=-(a-1)U_b+(b-1)V_b$.  Differentiating the preceding
display therefore gives
\begin{equation}
\label{eq:hcct-radial-beta-curvature}
  T_b:=\mathcal T_{F_{a,b}}
  =(b+\tfrac12)V_b-AU_b,
  \qquad A=a-\tfrac12.
\end{equation}
When $a=1/2$, the second term vanishes, recovering the positive-curvature
structure of \cref{sec:hcct-proof1}.  If $a>1/2$, the beta expansion at
zero gives $U_b\sim1/(au)$ and $V_b=o(1/u)$, so
\[
  T_b(u)\sim-\frac{a-1/2}{au}
  =-\frac{d-1}{du},\qquad d=2a,\quad u\downarrow0.
\]
The ratio $T_1/T_c$ is consequently not a globally usable positive
comparison function.  We replace it by a compensated positive ratio.
The additional beta information required for this step is as follows.

\begin{lemma}[Quantitative beta-quantile derivative comparison]
\label{lem:hcct-beta-derivative-gap}
For every $a>0$ and $c>1$, put
\[
  E=U_c-U_1,\qquad W=\frac{E}{V_c},
  \qquad q_1(u)=u^{1/a},\quad U_1(u)=\frac1{au}.
\]
Then
\[
  W'(u)<0,\qquad
  \lim_{u\downarrow0}W(u)=\frac{c-1}{a+1},\qquad
  \lim_{u\uparrow1}W(u)=0.
\]
In particular,
\begin{equation}
\label{eq:hcct-beta-gap-bound}
  0<U_c-U_1<\frac{c-1}{a+1}V_c,\qquad 0<u<1.
\end{equation}
Moreover, $q_c/q_1$ is strictly increasing, and its logarithm is strictly
concave as a function of $s=-\log(1-q_c(u))$.
\end{lemma}

\begin{proof}
Set $x=q_c(u)$ and $z=x/(1-x)$.  With
\[
  p_c(x)=\frac{x^{a-1}(1-x)^{c-1}}{B(a,c)},
\]
inverse differentiation and $U_1=1/(au)$ yield
\begin{equation}
\label{eq:hcct-beta-gap-explicit}
  W=\frac{1-x}{x}-\frac{(1-x)p_c(x)}{au}.
\end{equation}
Integration by parts on $(0,x)$ gives
\[
  a\int_0^x t^{a-1}(1-t)^{c-1}\,\mathrm{d}t
  =x^a(1-x)^{c-1}
   +(c-1)\int_0^x t^a(1-t)^{c-2}\,\mathrm{d}t.
\]
The boundary term at zero vanishes because $a>0$; all integrands are
integrable on $(0,x)$ for every $x<1$.
Substitute this identity into \cref{eq:hcct-beta-gap-explicit}, and then
make the changes of variables $y=t/(1-t)$ and $y=zv$.  This gives
\begin{equation}
\label{eq:hcct-beta-gap-moment}
\begin{aligned}
  W
  &=\frac{c-1}{az}
    \frac{\int_0^z y^a(1+y)^{-a-c}\,\mathrm{d}y}
         {\int_0^z y^{a-1}(1+y)^{-a-c}\,\mathrm{d}y}\\
  &=\frac{c-1}{a}\,\mathbb E Z_z,
\end{aligned}
\end{equation}
where $Z_z$ is a random variable on $(0,1)$ with density
\[
  \frac{v^{a-1}(1+zv)^{-a-c}}
       {\int_0^1 t^{a-1}(1+zt)^{-a-c}\,\mathrm{d}t}.
\]
Thus $W>0$.  Differentiating the normalized expectation gives
\begin{equation}
\label{eq:hcct-beta-gap-covariance}
  \frac{\mathrm{d}}{\mathrm{d}z}\mathbb E Z_z
  =-(a+c)\operatorname{Cov}
       \left(Z_z,\frac{Z_z}{1+zZ_z}\right)<0.
\end{equation}
To justify the differentiation, on every bounded interval of $z\ge0$
the unnormalized density and its derivative are dominated by constants
times $v^{a-1}$; the positive normalizing integral is bounded away from
zero on that interval.  For strictness of the covariance sign, both
$v$ and $v/(1+zv)$ are strictly increasing in $v$, and $Z_z$ has a
positive density throughout $(0,1)$.  If $\widetilde X$ is an independent
copy of $X$, then
\[
  2\operatorname{Cov}(X,\phi(X))
  =\mathbb E\bigl[(X-\widetilde X)
                   \{\phi(X)-\phi(\widetilde X)\}\bigr]>0
\]
for this nondegenerate $X$ and strictly increasing $\phi$.
Finally, $z$ has strictly positive derivative with respect to $u$, so
\cref{eq:hcct-beta-gap-covariance} implies $W'(u)<0$.

As $z\downarrow0$, dominated convergence gives the limiting density
$av^{a-1}$ and hence $\mathbb E Z_z\to a/(a+1)$.  The moment
representation proves the stated limit of $W$ at zero.  At the other
endpoint, \cref{eq:hcct-beta-gap-explicit} and $W>0$ give
\[
  0<W<\frac{1-x}{x}\longrightarrow0\qquad(x\uparrow1).
\]
Strict decrease and these endpoint limits prove
\cref{eq:hcct-beta-gap-bound}.

The first strict inequality in \cref{eq:hcct-beta-gap-bound} implies
$(\log(q_c/q_1))'=E>0$.  Moreover, $\mathrm{d}s/\mathrm{d}u=V_c>0$, so
\[
  \frac{\mathrm{d}}{\mathrm{d}s}\log\frac{q_c(u)}{q_1(u)}=W(u),
  \qquad
  \frac{\mathrm{d}^2}{\mathrm{d}s^2}
       \log\frac{q_c(u)}{q_1(u)}=\frac{W'(u)}{V_c(u)}<0.
\]
This proves the remaining assertions.
\end{proof}

\begin{corollary}[Complementary beta-quantile comparison]
\label{cor:hcct-beta-complementary-ratio}
For every $a>0$ and $c>1$, the function $x\mapsto I_x(a,c)^{1/a}$ is
strictly concave on $(0,1)$, and
\[
  R:=V_1-V_c>0.
\]
Equivalently, $(1-q_c(u))/(1-q_1(u))$ is strictly increasing on $(0,1)$.
\end{corollary}

\begin{proof}
Put $g(x)=I_x(a,c)^{1/a}$.  For $u=I_x(a,c)$,
logarithmic differentiation of $g'$ gives the exact identity
\[
  (1-x)\frac{g''(x)}{g'(x)}
  =(a-1)W(u)-(c-1).
\]
The right-hand side is strictly negative if $a\le1$ because $W>0$.
If $a>1$, the bound $W<(c-1)/a$ follows immediately from
\cref{eq:hcct-beta-gap-moment}, since $0<Z_z<1$, and again makes the
right-hand side strictly negative.  Thus $g''<0$ on $(0,1)$.
Also $g$ has continuous endpoint values $g(0)=0$ and $g(1)=1$.
Its strict tangent inequality at $x$, extended to the right endpoint,
gives
\[
  1-g(x)<(1-x)g'(x),\qquad 0<x<1.
\]
For example, this strict inequality also follows by integrating the
strictly decreasing derivative $g'$ from $x$ to one.
Since $g(x)=q_1(u)$ and $x=q_c(u)$, the chain rule yields
\[
  \frac{V_1}{V_c}
  =\frac{(1-x)g'(x)}{1-g(x)}>1.
\]
This proves $R>0$.  Finally,
$\{\log((1-q_c)/(1-q_1))\}'=V_1-V_c$, proving the ratio assertion.
\end{proof}

\begin{lemma}[Radial boundary-reference comparison]
\label{lem:hcct-beta-radial-boundary-order}
For every $a\ge1/2$ and $c\ge1$,
\[
  \mathcal T_{F_{a,1}}(u)\ge\mathcal T_{F_{a,c}}(u),
  \qquad 0<u<1.
\]
The inequality is strict if $c>1$, and equality holds identically if
$c=1$.
\end{lemma}

\begin{proof}
The case $c=1$ is immediate.  Suppose $c>1$, and retain
$A=a-1/2\ge0$, $E=U_c-U_1$, $W=E/V_c$, and $R=V_1-V_c$.
By \cref{lem:hcct-beta-derivative-gap,cor:hcct-beta-complementary-ratio},
we have $E>0$, $R>0$, and $W'<0$.
Set
\[
  P_1=\tfrac32V_1,\qquad P_c=(c+\tfrac12)V_c,
  \qquad h(u)=\frac{P_1+AE}{P_c}>0.
\]
Using \cref{eq:hcct-radial-beta-curvature}, we obtain
\begin{equation}
\label{eq:hcct-beta-compensated-ratio}
  h=1+\frac{T_1-T_c}{P_c}
   =\frac{P_1}{P_c}+\frac{A}{c+1/2}W.
\end{equation}
Thus $h\ge1$ is exactly the required curvature inequality, although
the curvatures themselves may be negative.

Inverse-beta differentiation gives
\[
  (\log V_b)'=-(a-1)U_b+bV_b,
\]
and therefore
\begin{equation}
\label{eq:hcct-beta-compensated-derivative}
  \left(\log\frac{P_1}{P_c}\right)'
  =(a-1)E+V_1-cV_c
  =(a-1)E+(P_1-P_c)-\tfrac12R.
\end{equation}
Whenever $h<1$, we have $P_1-P_c<-AE$.  Since
$(a-1)-A=-1/2$, it follows that
\[
  \left(\log\frac{P_1}{P_c}\right)'
  <-\tfrac12(E+R)<0.
\]
Consequently \cref{eq:hcct-beta-compensated-ratio}, $A\ge0$, and $W'<0$
give
\[
  h'=\left(\frac{P_1}{P_c}\right)'
       +\frac{A}{c+1/2}W'<0
  \qquad\text{wherever }h<1.
\]

For any fixed $b>0$, the beta-tail expansion and exact inverse
derivative give
\[
  1-u\sim\frac{\{1-q_b(u)\}^b}{bB(a,b)},
  \qquad V_b(u)\sim\frac1{b(1-u)}.
\]
Together with $W(u)\to0$, these imply
\[
  h(1-)=\frac{3c}{2c+1}>1.
\]
As in the univariate ratio argument, a connected component of
$\{u:h(u)<1\}$ is impossible.  On any such component $h$ is strictly
decreasing.  At a right endpoint below one, continuity would require
the value one, contradicting this decrease from a value below one;
a component extending to one contradicts the displayed upper-endpoint
limit.  Hence $h\ge1$ throughout.
If $h=1$ at an interior point, then $P_1-P_c=-AE$, and
\cref{eq:hcct-beta-compensated-derivative} equals
$-\tfrac12(E+R)<0$.  Thus $h'<0$ even at this point, whereas an interior
equality point of $h\ge1$ must have derivative zero.  This contradiction
proves $h>1$ and hence $T_1>T_c$.
\end{proof}

\begin{proof}[HCCT multivariate region conclusion]
Put $a=d/2$ and $c=(k+1-d)/2$.
Scale invariance, \cref{lem:hcct-beta-radial-boundary-order}, and the
independent boundary comparison in \cref{lem:hc-boundary-curvature}
give, for every $d\ge1$, $k\ge d+1$, and $0<u<1$,
\[
  \mathcal T_{G_C}(u)
  >\mathcal T_{H_{d,d+1}}(u)
  \ge\mathcal T_{H_{d,k}}(u).
\]
By \cref{thm:connectivity}, $G_C^{-1}\circ H_{d,k}$ is increasing and
convex on $(0,\infty)$, with a finite continuous extension of value zero
at the left endpoint.  It is therefore convex on $[0,\infty)$.

For the chi limit, use the correctly scaled radial Hotelling variable
$\sqrt{kZ/V_k}$, where $Z\sim\chi_d^2$ and
$V_k\sim\chi_{k+1-d}^2$ are independent.  Since
\[
  \mathbb E(V_k/k)=\frac{k+1-d}{k}\longrightarrow1,
  \qquad
  \operatorname{Var}(V_k/k)=\frac{2(k+1-d)}{k^2}\longrightarrow0,
\]
$H_{d,k}(x)\to H_d(x)$ for every $x\ge0$.
For every finite $x$, $H_d(x)<1$, and the continuous endpoint convention
at zero also applies.  Thus $G_C^{-1}\circ H_{d,k}$ converges pointwise
to $G_C^{-1}\circ H_d$.  Apply
\cref{lem:convexity-pointwise-limit} to obtain the chi convexity
conclusion.  This step uses neither an arbitrary-pair radial ordering
nor convergence of density derivatives.

Apply these comparisons separately to each active study with dimension
$d_j\ge2$ and $k_j\ge d_j+1$, or with a chi reference.  For scalar
studies, \cref{lem:hcct-half-student-curvature-order} gives the sharper
range $k_j\ge1$, including the identity score at $k_j=1$.
The present boundary-reference calculation alone uses $k\ge2$ when
$d=1$; it does not replace that sharper scalar comparison.

Each resulting radial score
$\psi_j=G_C^{-1}\circ H_{d_j,k_j}$, with $H_{d_j,\infty}=H_{d_j}$,
is continuous, nonnegative, increasing, convex, and diverges at infinity.
Set
\[
  A_j=\widehat\Sigma_j^{-1/2}P_j,\qquad
  b_j=-\widehat\Sigma_j^{-1/2}\widehat\xi_j.
\]
The positive-definite active covariance estimates make these maps
well defined.  The score in \cref{eq:invert3} is
$\sum_{j\in J_+}w_j\psi_j(\|A_j\theta+b_j\|)$, because its squared
pivot CDF satisfies $F^{(j)}(r^2)=H_{d_j,k_j}(r)$.
The cutoff is finite for $0<p<1$.  By
\cref{lem:convexity-active-rank}, the inverted region is closed and
convex, possibly empty.  If nonempty, it is bounded, hence compact,
if and only if the active row spaces of the $P_j$ span the ambient
parameter space; otherwise it contains an affine line.
This proves the HCCT assertion in \cref{thm:connectivity2}, independently
of any dependence-validity claim.
\end{proof}

\subsubsection{Proof 5: Multivariate convexity by degree-five sign analysis}
\label{sec:hcct-proof5}

We first prove an all-pairs comparison for the scale-free radial family.
Fix \(a=d/2\ge1/2\), and abbreviate \(f_b=f_{a,b}\), \(F_b=F_{a,b}\).
Then
\begin{equation}\label{eq:radial-curvature-order}
\mathcal T_{F_b}(u)\ge\mathcal T_{F_c}(u),
\qquad 0<b<c,\quad 0<u<1.
\end{equation}
Equivalently, \(F_b^{-1}\circ F_c\) is increasing and convex. This
comparison holds for every real \(d\ge1\).

\begin{proof}
Write \(r=d-1\ge0\), \(c=b+\delta\), and \(q=2b+1>1\). First suppose
\[
0<\delta\le q/4.
\]
Any interval \([b,c]\subset(0,\infty)\) has a finite partition into
increments at most \(1/4\). At every left endpoint of such a partition,
\(q=2b+1>1\), so each increment satisfies the displayed restriction.
It therefore suffices to establish the comparison for one increment
and then chain the resulting curvature inequalities.

We verify \cref{lem:convexity-crossings} for \(F=F_b\), \(G=F_c\).
The densities are
\[
f_b(x)=\frac{2}{B(a,b)}x^r(1+x^2)^{-a-b},\qquad x>0,
\]
and integration of their tails gives
\[
1-F_b(x)\sim\frac{x^{-2b}}{bB(a,b)}.
\]
Consequently, for every \(s,h>0\), the CDF differences
\(F_b(sx)-F_c(x)\) and \(F_b(sz)-F_c(z+h)\) are eventually negative.
Both corresponding density ratios tend to infinity, at a positive
constant times the rate \(x^{2\delta}\) or \(z^{2\delta}\).

For a line through the origin, the density ratio
\(\rho_s(x)=s f_b(sx)/f_c(x)\) satisfies
\[
(\log\rho_s)'(x)=
\frac{2x\{a+c-(a+b)s^2+\delta s^2x^2\}}
{(1+x^2)(1+s^2x^2)}.
\]
The expression in braces increases strictly with \(x^2\). Hence
\(\rho_s\) first decreases and then increases, with the first phase
possibly absent.

For a negative-intercept line, put \(z=x-h>0\), \(L=s^2\), and
\(\rho(z)=s f_b(sz)/f_c(z+h)\). Direct differentiation gives
\[
(\log\rho)'(z)=
\frac r z-\frac{2(a+b)Lz}{1+Lz^2}
-\frac r{z+h}+\frac{2(a+c)(z+h)}{1+(z+h)^2}.
\]
Multiplication by the positive denominator
\[
z(z+h)(1+Lz^2)\{1+(z+h)^2\}
\]
gives a polynomial \(P(z)=\sum_{i=0}^5 C_i z^i\), where
\[
\begin{aligned}
C_0&=rh(1+h^2),\\
C_1&=h^2(q+2\delta+3r),\\
C_2&=h\{2q+4\delta+3r-Lq(1+h^2)\},\\
C_3&=q+2\delta+r-L\{2(q-\delta)h^2+q+r\},\\
C_4&=-Lh(q-4\delta),\\
C_5&=2L\delta.
\end{aligned}
\]
In the chosen increment range,
\[
C_0\ge0,\qquad C_1>0,\qquad C_4\le0,\qquad C_5>0.
\]
Moreover, \(C_2\le0\) implies \(C_3<0\). Indeed, \(C_2\le0\) yields
\[
L\ge\frac{2q+4\delta+3r}{q(1+h^2)}.
\]
The denominator \(2(q-\delta)h^2+q+r\) is positive, and the needed
strict comparison of thresholds follows from
\[
\begin{aligned}
&(2q+4\delta+3r)\{2(q-\delta)h^2+q+r\}\\
&\hspace{1cm}-(q+2\delta+r)q(1+h^2)\\
&=\{(3q-4\delta)(q+2\delta)+r(5q-6\delta)\}h^2\\
&\quad+q^2+4qr+3r^2+2q\delta+4r\delta>0.
\end{aligned}
\]
The final inequality holds because \(0<\delta\le q/4\).

Thus the nonzero coefficients of \(P\), in ascending powers, consist
of positive coefficients, at most one block of negative coefficients,
and a positive leading coefficient. We give the elementary root-count
argument explicitly. If there is no negative coefficient, then \(P>0\).
Otherwise let \(j\in\{2,3,4\}\) be the first index with \(C_j<0\).
Every coefficient below degree five in
\[
z^{j+1}\left(\frac{P(z)}{z^j}\right)'
=\sum_{i=0}^5(i-j)C_i z^i
\]
is nonpositive, and the degree-five coefficient is positive. At least
one lower coefficient is strictly negative, because \(C_1>0\) and
\(j\ge2\). After division by \(z^5\), this expression is a positive
constant minus a nonzero sum of terms \(a_i z^{i-5}\), with \(a_i\ge0\)
and \(i<5\). It increases strictly from minus infinity to a positive
limit. Hence \((P/z^j)'\) changes sign exactly once, from negative to
positive, and \(P/z^j\) has at most two positive zeros.

The polynomial \(P\) is positive near zero and infinity. At \(d=1\),
\(C_0=0\), but \(C_1>0\), so the assertion near zero still holds.
Consequently its only possible strict sign pattern is \(+,-,+\),
with the negative phase possibly absent. The logarithmic derivative
of \(\rho\) has the same signs. Thus \(\rho\) successively increases,
decreases, and increases.

By \cref{lem:convexity-crossings}, \(F_b^{-1}\circ F_c\) is convex
for the chosen increment.
\cref{thm:connectivity} gives the associated curvature inequality.
Chaining over the partition proves \cref{eq:radial-curvature-order}
for every \(0<b<c\).

To complete the HCCT argument, use positive-rescaling invariance to
translate this order to \(H_{d,k}\). For \(k\ge d+1\),
\cref{lem:hc-boundary-curvature} gives
\[
\mathcal T_{G_C}(u)>
\mathcal T_{H_{d,d+1}}(u)
\ge\mathcal T_{H_{d,k}}(u),\qquad 0<u<1.
\]
For scalar studies, the sharper anchor \(H_{1,1}=G_C\) and the same
all-pairs order give
\(\mathcal T_{G_C}\ge\mathcal T_{H_{1,k}}\) for every \(k\ge1\),
including equality at \(k=1\). Thus all finite-degree HCCT scores
specified in \cref{thm:connectivity2} are increasing and convex.
The chi scores follow from \cref{lem:convexity-pointwise-limit}.

Every such score is nonnegative, continuous at zero with value zero,
and divergent at infinity. Apply \cref{lem:convexity-active-rank} with
\[
A_j=\widehat\Sigma_j^{-1/2}P_j,
\qquad
b_j=-\widehat\Sigma_j^{-1/2}\widehat\xi_j.
\]
The HCCT region in \cref{eq:invert3} is therefore closed and convex.
Conditional on nonemptiness, it is compact exactly when the stacked
active matrix has full column rank; otherwise it contains an affine
line. Positive definiteness of the active covariance matrices makes
this rank condition equivalent to spanning by the active row spaces
of the \(P_j\)'s. This proves \cref{thm:connectivity2} for HCCT.
\end{proof}

\subsubsection{Proof 6: Multivariate convexity via a normalized logarithmic derivative}
\label{sec:hcct-proof6}

We give another proof of \cref{eq:radial-curvature-order}, without using
its preceding proof, a small-increment restriction, or a degree-five
coefficient analysis.

\begin{proof}
Fix \(a=d/2\ge1/2\), \(r=d-1\ge0\), and \(0<b<c\). Put
\(\delta=c-b>0\), and abbreviate
\[
f_b(x)=f_{a,b}(x)=\frac{2}{B(a,b)}x^r(1+x^2)^{-a-b},
\qquad F_b=F_{a,b}.
\]
We verify \cref{lem:convexity-crossings} directly for \(F=F_b\) and
\(G=F_c\). Integration of the density tail gives
\[
1-F_b(x)\sim\frac{x^{-2b}}{bB(a,b)}.
\]
Thus \(F_b(sx)-F_c(x)\) and \(F_b(sz)-F_c(z+h)\) are eventually
negative for every \(s,h>0\). The two associated density ratios tend
to infinity, since their leading powers are \(x^{2\delta}\) and
\(z^{2\delta}\), respectively.

For a through-origin line, let
\(\rho_s(x)=s f_b(sx)/f_c(x)\). Then
\[
(\log\rho_s)'(x)=
\frac{2x\{a+c-(a+b)s^2+\delta s^2x^2\}}
{(1+x^2)(1+s^2x^2)}.
\]
The term in braces is strictly increasing in \(x^2\), so \(\rho_s\)
first decreases and then increases, with the first phase possibly
absent.

For a negative-intercept line \(s(x-h)\), put
\[
z=x-h>0,\qquad L=s^2,\qquad M=2(a+b),\qquad
\Theta(z)=\log\frac{s f_b(sz)}{f_c(z+h)}.
\]
Direct differentiation gives
\[
\Theta'(z)=\frac{rh}{z(z+h)}
-\frac{MLz}{1+Lz^2}
+\frac{(M+2\delta)(z+h)}{1+(z+h)^2}.
\]
Instead of clearing every denominator, use the positive normalization
\[
\begin{aligned}
K(z)
&=\frac{(1+Lz^2)\{1+(z+h)^2\}}z\,\Theta'(z)\\
&=\frac{(M+2\delta)h}{z}
  +\{M+2\delta-ML(1+h^2)\}\\
&\quad+Lh(2\delta-M)z+2L\delta z^2+rhE(z),
\end{aligned}
\]
where
\[
\begin{aligned}
E(z)
&=\left(\frac1{z^2}+L\right)
  \left(z+h+\frac1{z+h}\right)\\
&=\frac1z+\frac h{z^2}+\frac1{z^2(z+h)}
  +Lz+Lh+\frac L{z+h}.
\end{aligned}
\]
Every nonaffine summand in \(E\) is convex. For the product term,
\[
\left\{\frac1{z^2(z+h)}\right\}''
=\frac6{z^4(z+h)}
 +\frac4{z^3(z+h)^2}
 +\frac2{z^2(z+h)^3}>0.
\]
Therefore
\[
K''(z)=\frac{2(M+2\delta)h}{z^3}
       +4L\delta+rhE''(z)>0.
\]
Moreover, \(K(z)\to+\infty\) at zero and infinity: the positive
reciprocal term controls the first limit, and the positive quadratic
term controls the second. Hence the negative set of \(K\) is an
interval, possibly empty, and the only possible strict sign pattern
is \(+,-,+\). Since the normalization is positive, \(\Theta'\) has the
same signs. The shifted density ratio \(e^\Theta\) consequently
increases, decreases, and increases, with phases possibly absent.

The hypotheses of \cref{lem:convexity-crossings} are verified, so
\(F_b^{-1}\circ F_c\) is convex on \([0,\infty)\).
\cref{thm:connectivity} now proves \cref{eq:radial-curvature-order}
for every \(c>b>0\). No parameter partition is needed. At \(d=1\), the
additional term \(rhE\) vanishes; the remaining reciprocal, affine,
and quadratic terms are the scalar normalized-cubic structure.

For \(k\ge d+1\), positive-rescaling invariance and
\cref{lem:hc-boundary-curvature} yield
\[
\mathcal T_{G_C}(u)>
\mathcal T_{H_{d,d+1}}(u)
\ge\mathcal T_{H_{d,k}}(u),\qquad 0<u<1.
\]
For scalar studies, \(H_{1,1}=G_C\) and the independently established
all-pairs order instead give the sharper range \(k\ge1\), with the
identity transform at \(k=1\). By \cref{thm:connectivity}, the
corresponding finite-degree HCCT scores are increasing and convex.
\cref{lem:convexity-pointwise-limit} gives the chi scores.

All these scores are nonnegative, continuous at zero with value zero,
and divergent at infinity. Their compositions with the active affine
norms satisfy \cref{lem:convexity-active-rank}, using
\[
A_j=\widehat\Sigma_j^{-1/2}P_j,
\qquad
b_j=-\widehat\Sigma_j^{-1/2}\widehat\xi_j.
\]
It follows that the HCCT region in \cref{eq:invert3} is closed and
convex. If it is nonempty, it is compact if and only if the active row
spaces span the ambient parameter space; otherwise it contains an
affine line. This proves the HCCT part of \cref{thm:connectivity2}
independently of the degree-five argument.
\end{proof}

\subsubsection{Comparison of the six proof versions}\label{sec:convexity-proof-comparison}

The preceding arguments are six complete proof versions, not six unrelated
strategies. The original-style pair
\cref{sec:hcct-proof1,sec:hcct-proof4} uses beta quantiles and an
endpoint contradiction; \cref{sec:hcct-proof2,sec:hcct-proof5,sec:hcct-proof6}
use affine crossings; \cref{sec:hcct-proof3} compares the scalar score
directly with Half-Cauchy. In particular, the degree-five and normalized
arguments are two versions of the same multivariate crossing strategy.

\begin{table}[t]
\centering
\small
\renewcommand{\arraystretch}{1.15}
\begin{tabularx}{\linewidth}{@{}>{\raggedright\arraybackslash}p{0.21\linewidth}>{\raggedright\arraybackslash}X>{\raggedright\arraybackslash}X@{}}
\toprule
Version & Scope and advantage & Limitation or cost\\
\midrule
Scalar beta, \cref{sec:hcct-proof1}
& Strict curvature ordering for every real $0<k<\ell$; beta-quantile ratios
and explicit endpoint information.
& More notation than the target comparison requires. Some endpoint work
is unnecessary for the scalar contradiction.\\
Scalar crossings, \cref{sec:hcct-proof2}
& All-pairs scalar ordering without inverse-beta differentiation.
& Requires the affine-crossing criterion and a cubic sign calculation.\\
Direct scalar, \cref{sec:hcct-proof3}
& Short target proof for $k\ge1$, with a separate direct half-normal argument.
& Does not establish arbitrary-pair degree ordering or the beta lemmas.\\
Compensated beta, \cref{sec:hcct-proof4}
& Closely matches the original scalar route and gives quantitative
beta-quantile derivative information.
& Compares $b=1$ with $c>1$, not arbitrary $b<c$.
Signed radial curvature requires a compensating term.\\
Degree-five, \cref{sec:hcct-proof5}
& Radial ordering for every real $d-1<k_1<k_2$.
& Longer polynomial casework and a small-increment chaining argument.\\
Normalized derivative, \cref{sec:hcct-proof6}
& The same all-pairs radial ordering, without the degree-five case split
or small-increment restriction.
& Shares the affine-crossing strategy with the preceding version.\\
\bottomrule
\end{tabularx}
\caption{Scope and limitations of the six HCCT proof versions, in their
order of presentation. All use the stated endpoint and geometric
conditions when applied to confidence regions.}
\label{tab:hcct-proof-comparison}
\end{table}

The advantage of retaining the original beta route is its intermediate
information. Besides the scalar quantile-ratio result,
\cref{lem:hcct-beta-derivative-gap,cor:hcct-beta-complementary-ratio}
give, for $a>0,c>1$, increasing lower- and upper-tail quantile ratios,
strict concavity of $x\mapsto I_x(a,c)^{1/a}$, and
\[
 W=\frac{U_c-U_1}{V_c}
 \ \hbox{strictly decreasing from } \frac{c-1}{a+1}\ \hbox{to }0,
 \qquad
 0<U_c-U_1<\frac{c-1}{a+1}V_c.
\]
Equivalently, the logarithm of the lower-tail quantile ratio is strictly
concave as a function of $-\log(1-q_c)$.
The beta endpoint inequality in
\cref{eq:beta-endpoint-monotonicity}, although optional for the scalar
contradiction, is used in the sharpness argument below.

The multivariate compensation is essential to that proof architecture.
For $d>1$, the radial curvature is negative near $u=0$, so a ratio of
raw curvatures need not be positive. Nor can one simply compare their
positive contributions: for $a=3$, $b=1$, $c=2$, and $P_j=(j+1/2)V_j$,
\[
 \lim_{u\downarrow0}\frac{P_1(u)}{P_2(u)}
 =\frac35\,4^{1/3}<1.
\]
The positive ratio constructed in \cref{sec:hcct-proof4} controls the
additional derivative term instead.
For the shortest target argument in one dimension,
\cref{sec:hcct-proof3} is preferable; for a shorter general radial
degree-order argument, \cref{sec:hcct-proof6} is preferable.
The beta pair remains useful when the parallel scalar/multivariate
language or the stronger beta results are desired.

\subsubsection{Direct Convexity Proofs for EHMP}\label{sec:ehmp-direct-convexity}

The Pareto score is $1/(1-H)$ for a radial reference CDF $H$.
We prove its convexity directly; no Half-Cauchy comparison is used.

\begin{lemma}[Exact Pareto degree condition]\label{lem:pareto-radial-convexity}
For every real $d\ge1$ and valid $k>d-1$, the score
$G_P^{-1}\circ H_{d,k}$ is convex on $[0,\infty)$ if and only if $k\ge d$.
For $k\ge d$ its second derivative is strictly positive on $(0,\infty)$.
\end{lemma}

\begin{proof}
Positive rescaling does not affect convexity, so write the radial density as
\[
 f(x)=C x^r(1+x^2)^{-(r+\eta)/2},\qquad
 r=d-1\ge0,\qquad \eta=k-d+2>1,\qquad C>0.
\]
Let $H$ be its CDF, $S=1-H$, and $g=f^{-1/\eta}$.
Direct differentiation gives
\begin{equation}\label{eq:pareto-tangent-density}
 \frac{g''(x)}{g(x)}
 =\frac{(r+\eta)(r+\eta x^2)}
 {\eta^2x^2(1+x^2)^2}>0,\qquad x>0.
\end{equation}
Beyond the density mode, $x^2>r/\eta$, we have $g'(x)>0$.
Strict convexity gives $g(t)>g(x)+g'(x)(t-x)$ for $t>x$, and therefore
\[
 \begin{aligned}
 S(x)&=\int_x^\infty g(t)^{-\eta}\,dt\\
 &<\int_x^\infty
   \{g(x)+g'(x)(t-x)\}^{-\eta}\,dt
 =\frac{g(x)^{1-\eta}}{(\eta-1)g'(x)}.
 \end{aligned}
\]
Put $T(x)=-f'(x)/f(x)^2=\eta g'(x)g(x)^{\eta-1}$.
We obtain the stronger intermediate bound
\begin{equation}\label{eq:pareto-tail-curvature-bound}
 T(x)S(x)<\frac{\eta}{\eta-1}.
\end{equation}
The exact identity
\begin{equation}\label{eq:pareto-score-second-derivative}
 \left(\frac1{S(x)}\right)''
 =\frac{f'(x)}{S(x)^2}+\frac{2f(x)^2}{S(x)^3}
 =\frac{f(x)^2}{S(x)^3}\{2-T(x)S(x)\}
\end{equation}
now proves strict convexity beyond the mode if $\eta\ge2$, equivalently
$k\ge d$. At and below the mode $f'\ge0$, so the first expression in
\cref{eq:pareto-score-second-derivative} is strictly positive.
Continuity at zero, with score value one, extends convexity to $[0,\infty)$.

Conversely, if $d-1<k<d$, then $1<\eta<2$. The explicit density and its
derivative give
\[
 f(x)=Cx^{-\eta}\{1+O(x^{-2})\},\quad
 f'(x)=-\eta Cx^{-\eta-1}\{1+O(x^{-2})\},\quad
 S(x)=\frac{C}{\eta-1}x^{1-\eta}\{1+O(x^{-2})\}.
\]
It follows that
\begin{equation}\label{eq:radial-tail-curvature-limit}
 T(x)S(x)\longrightarrow\frac{\eta}{\eta-1}>2.
\end{equation}
Thus \cref{eq:pareto-score-second-derivative} is negative for all sufficiently
large $x$, and the score is not convex. This calculation uses the explicit
density derivative, not differentiation of an unproved tail asymptotic.
\end{proof}

\begin{lemma}[Pareto scores for chi references]\label{lem:pareto-chi-convexity}
For every real $d\ge1$, $G_P^{-1}\circ H_d$ is convex on $[0,\infty)$,
with strictly positive second derivative on $(0,\infty)$.
\end{lemma}

\begin{proof}
The chi density has the form $f(x)=Cx^r e^{-x^2/2}$, $r=d-1\ge0$.
Let $S=1-H_d$ and $\ell=-\log f$. Then
\[
 \ell'(x)=x-r/x,\qquad \ell''(x)=1+r/x^2>0.
\]
Above the mode $x>\sqrt r$, set $\lambda=\ell'(x)>0$.
Strict convexity of $\ell$ gives
$f(t)<f(x)e^{-\lambda(t-x)}$ for $t>x$, and hence
\[
 S(x)<\frac{f(x)}{\lambda},\qquad
 -\frac{f'(x)}{f(x)^2}S(x)=\frac{\lambda S(x)}{f(x)}<1<2.
\]
\cref{eq:pareto-score-second-derivative} proves strict convexity there.
At and below the mode $f'\ge0$ gives the same conclusion directly.
Extend the score continuously to one at zero.
For $d=1$ this is also a direct half-normal proof.
\end{proof}

\begin{proof}[Proof of \cref{thm:connectivity0,thm:connectivity2} for EHMP]
For each active study, \cref{lem:pareto-radial-convexity,lem:pareto-chi-convexity}
make the radial score increasing and convex under the stated
$k_j\ge d_j$ condition, including $k_j\ge1$ when $d_j=1$.
It is nonnegative, continuous at zero, and tends to infinity as the radius
increases. Apply \cref{lem:convexity-active-rank} with
$A_j=\widehat\Sigma_j^{-1/2}P_j$ and
$b_j=-\widehat\Sigma_j^{-1/2}\widehat\xi_j$, or with the corresponding
scalar location and positive scale.
This yields closed convex regions and the exact active-rank characterization
of compactness when nonempty. In one dimension the region is empty,
a singleton, or a closed bounded interval.
The argument is independent of all six HCCT proofs.
\end{proof}

\subsubsection{Proofs for \cref{sec:convexity-boundaries}}
\label{sec:convexity-boundary-proofs}

\begin{proof}[Proof of \cref{prop:convexity-integer-boundaries}]
Put $\eta=k-d+2$ as in \cref{lem:pareto-radial-convexity}.
For $d-1<k<d$, \cref{eq:radial-tail-curvature-limit} gives
\[
 (1-u)\mathcal T_{H_{d,k}}(u)\longrightarrow
 \frac{\eta}{\eta-1}>2\qquad(u\uparrow1).
\]
The corresponding Pareto expression equals two, and the Half-Cauchy
expression tends to two. Both curvature comparisons therefore fail for
$u$ sufficiently close to one; by \cref{eq:convex-transform-identity},
both radial scores then have negative second derivative.

To treat the Half-Cauchy boundary $k=d$, use scale invariance and set
\[
 C_d=\frac2{B(d/2,1/2)},\qquad q=\frac{d+1}{2},\qquad s=1-u.
\]
The scale-free density, its explicit derivative, and its survival function
satisfy, as $x\to\infty$,
\[
 \begin{aligned}
 f(x)&=C_dx^{-2}\{1-qx^{-2}+O(x^{-4})\},\\
 f'(x)&=-2C_dx^{-3}\{1-2qx^{-2}+O(x^{-4})\},\\
 1-F(x)&=\frac{C_d}{x}-\frac{C_dq}{3x^3}+O(x^{-5}).
 \end{aligned}
\]
These expansions follow from the explicit formula in
\cref{eq:radial-beta-family}; in particular the derivative expansion is
computed from that density formula. Substitution at $x=F^{-1}(u)$ yields
\[
 s\mathcal T_{F_{d/2,1/2}}(u)
 =2-\frac{d+1}{3C_d^2}s^2+O(s^4),
 \qquad
 s\mathcal T_{G_C}(u)
 =2-\frac{\pi^2}{6}s^2+O(s^4).
\]
By \cref{eq:beta-endpoint-monotonicity},
$K(t)=(t+1/2)B(t,1/2)^2$ is strictly decreasing. Hence
\[
 \frac{C_d^2}{d+1}=\frac2{K(d/2)}
\]
is strictly increasing in $d$ and equals $2/\pi^2$ at $d=1$.
For $d>1$,
$(d+1)/(3C_d^2)-\pi^2/6<0$.
Thus $\mathcal T_{G_C}(u)<\mathcal T_{H_{d,d}}(u)$ near one,
and the Half-Cauchy score is not convex.

The scalar sufficiency for all real $k\ge1$ was proved in
\cref{sec:hcct-proof1,sec:hcct-proof2,sec:hcct-proof3}.
At $d=1,k=1$, the Half-Cauchy score is exactly linear, not strictly convex.
The multivariate sufficiency for $k\ge d+1$ follows from
\cref{lem:hc-boundary-curvature} and any of the multivariate proofs.
The Pareto condition follows from \cref{lem:pareto-radial-convexity}.
For integer $d\ge2$, the only valid integer degree below $d+1$ is $k=d$,
which completes the claimed integer characterization.
\end{proof}

\begin{proof}[Proof of \cref{prop:convexity-real-degrees}]
For $d=2,k=5/2$, the scale-free parameters are $a=1,b=3/4$, and
\[
 F_{1,3/4}(x)=1-(1+x^2)^{-3/4}.
\]
Put $s=1-u$ and $y=s^{1/3}\in(0,1)$. Direct differentiation gives
\[
 s\mathcal T_{F_{1,3/4}}(u)
 =\frac{5-7y^4}{3(1-y^4)}.
\]
The difference from the intermediate Half-Cauchy bound is
\[
 \frac{4u(1-u)}{1-u^2}
 -(1-u)\mathcal T_{F_{1,3/4}}(u)
 =\frac{2-7y^3+2y^4+5y^7}{3(1-y^4)(2-y^3)}.
\]
The denominator is positive. Apply the arithmetic-geometric mean
inequality to the six quantities
\[
 \frac23,\quad\frac23,\quad\frac23,\quad
 2y^4,\quad\frac52y^7,\quad\frac52y^7.
\]
This gives
\[
 2+2y^4+5y^7
 \ge6(100/27)^{1/6}y^3>7y^3,
\]
where the strict constant inequality follows from
$6^6(100/27)=172800>117649=7^6$.
Thus the numerator is positive, and
\cref{eq:hc-boundary-curvature} gives
\[
 \mathcal T_{F_{1,3/4}}(u)
 <\frac{4u}{1-u^2}<\mathcal T_{G_C}(u).
\]
The all-pairs order in \cref{eq:radial-curvature-order} extends this
conclusion to every $k\ge5/2$.

For the general local improvement, fix $d>1$, put $a=d/2$, and restrict
$b$ initially to a compact interval $[b_-,b_+]$ with
$1/2<b_-<1<b_+<\infty$. Write $C_b=2/B(a,b)$.
Uniformly for $b$ in this interval, the explicit density gives
\[
 \begin{aligned}
 f_{a,b}(x)&=C_bx^{d-1}\{1+O(x^2)\},&
 F_{a,b}(x)&=\frac{C_b}{d}x^d\{1+O(x^2)\}
 &&(x\downarrow0),\\
 f_{a,b}(x)&=C_bx^{-2b-1}\{1+O(x^{-2})\},&
 1-F_{a,b}(x)&=\frac{C_b}{2b}x^{-2b}\{1+O(x^{-2})\}
 &&(x\to\infty).
 \end{aligned}
\]
Here $C_b$ stays positive and bounded, and the binomial expansion
remainders are uniform because $a+b$ and $b$ range over compact sets.
The exact logarithmic derivative
\[
 \frac{f_{a,b}'(x)}{f_{a,b}(x)}
 =\frac{d-1}{x}-\frac{2(a+b)x}{1+x^2}
\]
therefore yields the uniform limits
\begin{equation}\label{eq:uniform-radial-curvature-endpoints}
 \begin{aligned}
 u\mathcal T_{F_{a,b}}(u)&\longrightarrow-\frac{d-1}{d}
 &&(u\downarrow0),\\
 (1-u)\mathcal T_{F_{a,b}}(u)&\longrightarrow1+\frac1{2b}
 &&(u\uparrow1).
 \end{aligned}
\end{equation}
To see that the quantile substitutions preserve uniformity, the first
CDF expansion bounds the small quantiles above and below by constant
multiples of $u^{1/d}$, uniformly in $b$.
The survival expansion similarly sends the upper quantiles to infinity
uniformly; its curvature remainder is
$O((1-u)^{1/b_+})$.
The corresponding Half-Cauchy limits are zero and two.
Because $d>1$ and $b_->1/2$, the comparisons at both ends are uniformly
strict.

On the remaining compact interval of $u$, the strict inequality at $b=1$
from \cref{lem:hc-boundary-curvature} has a positive minimum gap.
The beta density is smooth and positive in the interior, so implicit
inversion gives joint continuity of the curvature in $(b,u)$ there.
The gap persists for all $b$ sufficiently close to one.
Combining this with \cref{eq:uniform-radial-curvature-endpoints}, choose
$\delta_d\in(0,1/2)$ so that the Half-Cauchy comparison holds for
$b\in[1-\delta_d,1+\delta_d]$.
The all-pairs order extends it to every $b\ge1-\delta_d$.
Using $k=d-1+2b$, set $\varepsilon_d=2\delta_d$ to conclude.
\end{proof}

\subsection{Blockwise Volume Comparison}

\begin{proof}[Proof of \cref{prop:block-divide-combine-volume}]
\cref{eq:block-hcct-region} follows immediately from
\cref{eq:hcct-cct-radial-identities}.  In block polar coordinates, its volume
is
\[
    s_{k-1}^{m}
    \int_{\substack{u_j\geq0\\\sum_j u_j\leq R}}
       \prod_{j=1}^m u_j^{k-1}\,du_1\cdots du_m,
    \qquad
    s_{k-1}=\frac{2\pi^{k/2}}{\Gamma(k/2)}.
\]
The Dirichlet integral equals
$R^{mk}\Gamma(k)^m/\Gamma(mk+1)$, proving
\cref{eq:block-hcct-volume}.  Its Euclidean diameter is attained by placing
the entire block-$\ell_{2,1}$ radius in a single block, proving
\cref{eq:block-hcct-diameter}.

For CCT, fixing any block at its estimate, so that $r_j=0$, makes the
corresponding CCT term equal to $-\infty$.  All remaining blocks can vary
arbitrarily.  Hence $\mathcal R_C$ contains an affine subspace of dimension
$d-k$ and is unbounded.

To show that its Lebesgue measure is infinite, it is not enough to use these
measure-zero affine subspaces; we instead construct tubes around them.  For a
large number $L$, restrict every block $j\geq2$ to
\[
    L\leq r_j\leq2L
\]
and restrict the first block to
\[
    0<r_1\leq\{4(m-1)L\}^{-1}.
\]
For all sufficiently large $L$, the negative contribution $-1/r_1$ dominates
the positive contributions from the other blocks, so this entire product set
is contained in $\mathcal R_C$.  Its volume is bounded below, up to a positive
constant independent of $L$, by
\[
    L^{(m-1)k}L^{-k}=L^{(m-2)k}.
\]
Taking disjoint dyadic shells $L=2^n$ and summing their volumes gives
infinity.  When $m=2$, each shell contributes a positive constant; when
$m>2$, the shell volumes themselves diverge.
\end{proof}

\section{Omitted Proofs for \cref{sec:hctest} and \cref{sec:furtherdis}}\label{sec:pfhctest}

This section contains the proofs for \cref{thm:gclthc,thm:computeconvhc,thm:extremecase,thm:fixed-hcct,thm:growing-hcct,lem:gaussian-quadrant}. For clarity, \cref{thm:revalidity} and \cref{thm:nval} are formally restated as \cref{thm:fixed-hcct} and \cref{cor:normal-growing}, respectively, with the proof of the latter presented in \cref{sec:pairwisebinormal}.

For the stable-limit and tail-comparison arguments, we recall the classical
domain-of-attraction criteria and their normalizer conclusion
\citep{gnedenko1954limit,nolan2020univariate}, using the stable-law
parameterization in \cref{sec:Landau}.
\begin{lemma}[Generalized CLT]\label{thm:genclt}
Let $X_1,X_2,\ldots$ be i.i.d. with a nondegenerate CDF $F$.
Attraction to a nondegenerate limit $Z$ means that, for suitable real
centers $A_n$ and normalizers $B_n>0$ with $B_n\to\infty$,
\begin{equation}\label{eq:genclt-iid-convergence}
 \frac{1}{B_n}\sum_{i=1}^nX_i-A_n
 \mathrel{\xrightarrow{\mathrm d}}Z.
\end{equation}
Here $A_n$ is the centering after division by $B_n$.
The distribution $F$ is attracted to a normal distribution if and only if
\[
 \frac{t^2\int_{|x|>t}\dif F(x)}
 {\int_{|x|<t}x^2\dif F(x)}\longrightarrow0
 \qquad(t\to\infty).
\]
For $0<\alpha<2$, $\beta\in[-1,1]$, $c>0$, and $\mu\in\mathbb R$,
it is attracted to $S(\alpha,\beta,c,\mu)$ if and only if
$V(t):=F(-t)+1-F(t)$ is positive for all sufficiently large $t$ and
\[
 \frac{V(rt)}{V(t)}\longrightarrow r^{-\alpha}
 \quad\text{for every }r>0,\qquad
 \frac{1-F(t)}{V(t)}\longrightarrow\frac{1+\beta}{2}
 \qquad(t\to\infty).
\]
The bounded tail share includes both endpoints $\beta=\pm1$.

Moreover, whenever the normalized i.i.d. sums in
\cref{eq:genclt-iid-convergence} converge to
$Z\sim S(\alpha,\beta,c,\mu)$ with $0<\alpha<2$, the same normalizers
necessarily satisfy, for every $t>0$,
\begin{equation}\label{eq:genclt-normalizer-tails}
 \begin{aligned}
 nF(-B_nt)&\longrightarrow
 \frac{D_{\alpha,c}(1-\beta)}{2t^\alpha},\\
 n\{1-F(B_nt)\}&\longrightarrow
 \frac{D_{\alpha,c}(1+\beta)}{2t^\alpha},
 \qquad
 D_{\alpha,c}:=
 \frac{2\Gamma(\alpha)\sin(\pi\alpha/2)c^\alpha}{\pi}.
 \end{aligned}
\end{equation}
In particular, $nV(B_nt)\to D_{\alpha,c}t^{-\alpha}$.
These are limits, including when one of the one-sided constants is zero.
\end{lemma}

The following weighted version is proved directly, including the case
$\alpha=1$. We use the stable-law parameterization in \cref{sec:Landau}.

\begin{lemma}[Generalized weighted CLT]\label{thm:gcltnew}
Let $X_1,X_2,\ldots$ be i.i.d. with density $f$. Suppose that, for some
$0<\alpha<2$ and $c_1,c_2\ge0$ with $c_1+c_2>0$,
\begin{equation}\label{eq:densityalpha2}
 \lim_{x\to\infty}x^{1+\alpha}f(-x)=c_1,\qquad
 \lim_{x\to\infty}x^{1+\alpha}f(x)=c_2.
\end{equation}
For each $n$, let $w_{nj}\ge0$, $1\le j\le n$, be deterministic weights
with $\sum_{j=1}^n w_{nj}=1$. Define
\begin{equation}\label{eq:gclt-normalized-weights}
 s_n=\left(\sum_{j=1}^n w_{nj}^{\alpha}\right)^{1/\alpha},\qquad
 v_{nj}=\frac{w_{nj}}{s_n},\qquad
 \delta_n=\max_{1\le j\le n}v_{nj}\longrightarrow0.
\end{equation}
Then
\[
 \frac{\sum_{j=1}^n w_{nj}X_j-A_n}{s_n}
 \mathrel{\xrightarrow{\mathrm d}}S(\alpha,\beta,c,0),
\]
where
\begin{equation}\label{eq:determinec}
 \beta=\frac{c_2-c_1}{c_1+c_2},\qquad
 c^{\alpha}=\frac{\pi(c_1+c_2)}
 {2\Gamma(1+\alpha)\sin(\pi\alpha/2)},
\end{equation}
and
\[
 A_n=\begin{cases}
 0,&0<\alpha<1,\\[1mm]
 \displaystyle\sum_{j=1}^n\operatorname{Im}\ell(w_{nj}),&\alpha=1,\\[2mm]
 \mathbb E X_1,&1<\alpha<2.
 \end{cases}
\]
Here $\ell$ is the continuous logarithm of
$\phi(u)=\mathbb E e^{iuX_1}$ in a neighborhood of zero, normalized by
$\ell(0)=0$. Such a neighborhood exists since $\phi(0)=1$.
When $\alpha=1$, $s_n=1$, so all weights eventually lie in that
neighborhood. The centers for the finitely many earlier rows may be
chosen arbitrarily, and zero weights contribute zero.
\end{lemma}

For $0<\alpha\le1$, the condition $\max_jw_{nj}\to0$ implies
\cref{eq:gclt-normalized-weights}, since $s_n\ge1$.
For $1<\alpha<2$, the unnormalized maximum tending to zero does not
in general suffice; the normalized condition above provides a sufficient
replacement.

\begin{proof}
First record the Fourier integrals
\begin{equation}\label{eq:gclt-fourier-integrals}
 \begin{aligned}
 \int_0^{\infty}\frac{1-\cos y}{y^{1+\alpha}}\dif y
 &=K_{\alpha},
 &K_{\alpha}&=\frac{\pi}
 {2\Gamma(1+\alpha)\sin(\pi\alpha/2)},\\
 \int_0^{\infty}
 \frac{\sin y-y\mathbf1_{\{\alpha>1\}}}{y^{1+\alpha}}\dif y
 &=K_{\alpha}\tan(\pi\alpha/2),&&\alpha\ne1.
 \end{aligned}
\end{equation}
To verify the constants, use
\[
 y^{-1-\alpha}=\frac1{\Gamma(1+\alpha)}
 \int_0^{\infty}r^{\alpha}e^{-ry}\dif r.
\]
Tonelli's theorem reduces the cosine integral to
$\Gamma(1+\alpha)^{-1}\int_0^{\infty}r^{\alpha-1}/(1+r^2)\dif r$.
For $\alpha<1$, absolute Fubini reduces the sine integral to
$\Gamma(1+\alpha)^{-1}\int_0^{\infty}r^{\alpha}/(1+r^2)\dif r$.
For $\alpha>1$, use $y-\sin y\ge0$ and Tonelli to obtain instead
$-\Gamma(1+\alpha)^{-1}\int_0^{\infty}r^{\alpha-2}/(1+r^2)\dif r$.
The beta integral and gamma reflection identity give
\[
 \int_0^{\infty}\frac{r^{q-1}}{1+r^2}\dif r
 =\tfrac12 B(q/2,1-q/2)
 =\frac{\pi}{2\sin(\pi q/2)},\qquad 0<q<2,
\]
which proves \cref{eq:gclt-fourier-integrals}, including the negative
sign of its compensated sine integral when $\alpha>1$.

Suppose first that $\alpha\ne1$. Put $m_{\alpha}=0$ for $\alpha<1$
and $m_{\alpha}=\mathbb E X_1$ for $\alpha>1$; the latter expectation
exists absolutely by \cref{eq:densityalpha2}.
In the integral for $\phi(u)-1-ium_{\alpha}$, split at a fixed $R$
beyond which $f(\pm x)\le Cx^{-1-\alpha}$.
The bounded part is $O(|u|)$ for $\alpha<1$ and $O(u^2)$ for
$\alpha>1$, hence is $o(|u|^{\alpha})$ in both cases.
On each tail, substitute $y=|u||x|$.
The uncompensated exponential is $O(y)$ at zero and bounded at infinity;
the compensated exponential is $O(y^2)$ at zero and $O(y)$ at infinity.
These bounds times $y^{-1-\alpha}$ are integrable in the respective
ranges. Dominated convergence and \cref{eq:gclt-fourier-integrals}
therefore give
\begin{equation}\label{eq:gclt-local-exponent}
 \ell(u)-im_{\alpha}u
 =-c^{\alpha}|u|^{\alpha}
 \{1-i\beta\operatorname{sgn}(u)\tan(\pi\alpha/2)\}+r(u),
 \qquad r(u)=o(|u|^{\alpha}).
\end{equation}
Passing from $\phi(u)-1$ to its local logarithm adds
$O(|u|^{2\alpha})$ for $\alpha<1$ and $O(u^2)$ for $\alpha>1$,
both negligible here. Set $r(0)=0$.
Since $\sum_jv_{nj}^{\alpha}=1$, for each fixed $t\ne0$,
\[
 \sum_j|r(tv_{nj})|
 \le |t|^{\alpha}
 \sup_{0<|u|\le |t|\delta_n}\frac{|r(u)|}{|u|^{\alpha}}
 \longrightarrow0.
\]
For sufficiently large $n$, all $tv_{nj}$ lie in the local-logarithm
neighborhood. By independence, the characteristic function of the centered row is
the exponential of $\sum_j\ell(tv_{nj})-itA_n/s_n$.
The mean terms cancel when $\alpha>1$ because $\sum_jv_{nj}=1/s_n$.
Thus \cref{eq:gclt-local-exponent} gives the required stable exponent.
This uses a sum of local logarithms, not an identity between the
principal logarithm of a product and the sum of principal logarithms.

Now let $\alpha=1$, and set
$\kappa=\pi(c_1+c_2)/2$ and $\Delta=c_2-c_1$.
The tail bound $f(\pm x)\le Cx^{-2}$ and
$|e^{iz}-1|\le\min(2,|z|)$ imply
\[
 |\phi(u)-1|=O\bigl(|u|\{1+|\log|u||\}\bigr),\qquad
 \ell(u)-\{\phi(u)-1\}
 =O\bigl(u^2\{1+\log^2|u|\}\bigr)=o(|u|).
\]
For fixed real $t$ and $v\downarrow0$, the cosine tail calculation gives
$\operatorname{Re}\ell(tv)/v\to-\kappa|t|$.
For the imaginary part, the cancellation before taking the limit yields
\[
 \frac{\operatorname{Im}\ell(tv)-t\operatorname{Im}\ell(v)}{v}
 \longrightarrow
 \Delta\int_0^{\infty}\frac{\sin(ty)-t\sin y}{y^2}\dif y.
\]
Indeed, replacing the logarithms by $\phi-1$ changes the quotient by
$o(1)$. The bounded part of the density integral is $O(v^3)$ before
division by $v$. After rescaling the tails, the displayed integrand is
$O(y)$ at zero and $O(y^{-2})$ at infinity, so dominated convergence
applies.
For $t>0$, at finite cutoffs,
\[
 \int_{\varepsilon}^{R}\frac{\sin(ty)-t\sin y}{y^2}\dif y
 =t\left\{\int_{t\varepsilon}^{\varepsilon}\frac{\sin z}{z^2}\dif z
 +\int_R^{tR}\frac{\sin z}{z^2}\dif z\right\}
 \longrightarrow-t\log t.
\]
The first limit uses $\sin z/z^2=1/z+O(z)$ at zero, and the second
integral tends to zero absolutely. Oddness handles $t<0$; $t=0$ is
immediate, with $t\log|t|$ defined continuously there.
Consequently,
\begin{equation}\label{eq:gclt-critical-exponent}
 \ell(tv)-it\operatorname{Im}\ell(v)
 =v\{-\kappa|t|-i\Delta t\log|t|\}+r_t(v),
 \qquad r_t(v)=o(v).
\end{equation}
At $\alpha=1$, $s_n=1$. Set $r_t(0)=0$ and use $\sum_jw_{nj}=1$ to get
\[
 \sum_j|r_t(w_{nj})|
 \le\sup_{0<v\le\delta_n}\frac{|r_t(v)|}{v}
 \longrightarrow0.
\]
Thus the centered row characteristic functions converge to
$\exp\{-\kappa|t|-i\Delta t\log|t|\}$.
Since $2\kappa\beta/\pi=\Delta$, this is the characteristic function
of $S(1,\beta,\kappa,0)$ in the stated parameterization.
The limiting characteristic functions in both cases are continuous at
zero. L\'evy's continuity theorem completes the proof.
\end{proof}

For the two laws in \cref{thm:gclthc}, a direct calculation also gives
the explicit centering constants without invoking \cref{thm:gcltnew}.

\begin{proof}[Proof of \cref{thm:gclthc}]
Let $P$ have the $\mathrm{Pareto}(1,1)$ distribution and let $H$ have
the standard Half-Cauchy distribution, with characteristic functions
$\phi_P$ and $\phi_H$, respectively.
For $t>0$, integration by parts gives
\[
 \phi_P(t)=\cos t-t\{\pi/2-\operatorname{Si}(t)\}
 +i\{\sin t-t\operatorname{Ci}(t)\},
\]
where the sine and cosine integrals are
\[
 \operatorname{Si}(t)=\int_0^t\frac{\sin u}{u}\dif u,\qquad
 \operatorname{Ci}(t)
 =C_\gamma+\log t-\int_0^t\frac{1-\cos u}{u}\dif u.
\]
The boundary terms at infinity vanish and the corresponding improper
integrals converge. Since $\operatorname{Si}(t)=t+O(t^3)$ and
$\operatorname{Ci}(t)=C_\gamma+\log t+O(t^2)$, conjugate symmetry
extends the resulting expansion to both signs of $t$:
\[
 \phi_P(t)=1-\frac\pi2|t|
 +it\{1-C_\gamma-\log|t|\}+O(t^2).
\]

To obtain the Half-Cauchy expansion, put $\eta=2/\pi$ and define,
for $x>0$,
\[
 r(x)=\frac{\eta}{1+x^2}
 -\frac{\eta\mathbf1_{\{x\ge1\}}}{x^2}.
\]
Direct integration gives
\[
 \int_0^{\infty}r(x)\dif x=1-\eta,\qquad
 \int_0^{\infty}xr(x)\dif x=0,\qquad
 \int_0^{\infty}x^2|r(x)|\dif x=\eta.
\]
The first moment converges absolutely; its integrals over $(0,1)$ and
$(1,\infty)$ are $\eta\log2/2$ and $-\eta\log2/2$.
Using $|e^{iz}-1-iz|\le z^2/2$, we obtain
\[
 \left|\phi_H(t)-1-\eta\{\phi_P(t)-1\}\right|
 =\left|\int_0^{\infty}(e^{itx}-1-itx)r(x)\dif x\right|
 \le\frac{\eta t^2}{2}.
\]

For $X=P,H$, expand the local logarithm $\ell_X=\log\phi_X$,
with $\ell_X(0)=0$:
\begin{equation}\label{eq:landau-local-exponents}
 \ell_X(t)=-\lambda_X|t|
 +i\zeta_Xt\{1-C_\gamma-\log|t|\}+R_X(t),
\end{equation}
where
\[
 (\lambda_P,\zeta_P)=(\pi/2,1),\qquad
 (\lambda_H,\zeta_H)=(1,2/\pi),\qquad
 R_X(t)=O\bigl(t^2\{1+\log^2|t|\}\bigr)=o(|t|).
\]
Set $R_X(0)=0$. Let $(X_j)$ be i.i.d. copies of $X$,
write $w_j=w_j^{(m)}$, and set
\[
 \mathcal H_m=-\sum_{j=1}^m w_j\log w_j,\qquad 0\log0:=0.
\]
For each fixed $t\ne0$, all $tw_j$ lie in the local-logarithm
neighborhood for sufficiently large $m$.
By independence and \cref{eq:landau-local-exponents}, the characteristic
function of the centered weighted sum is the exponential of
\[
 \begin{aligned}
 &\sum_{j=1}^m\ell_X(tw_j)
 -it\zeta_X(\mathcal H_m+1-C_\gamma)\\
 &\qquad=-\lambda_X|t|-i\zeta_Xt\log|t|
 +\sum_{j=1}^m R_X(tw_j).
 \end{aligned}
\]
Zero-weight terms contribute zero, and
\[
 \sum_{j=1}^m|R_X(tw_j)|
 \le |t|\sup_{0<|u|\le |t|\max_jw_j}
 \frac{|R_X(u)|}{|u|}\longrightarrow0.
\]
Since $\zeta_X=2\lambda_X/\pi$, the limiting characteristic function
is that of $S(1,1,\lambda_X,0)$ and is continuous at zero.
L\'evy's continuity theorem proves both limits.
\end{proof}
A different derivation for the case with equal weights can be found in \citet{zaliapin2005approximating}, which was utilized for the harmonic mean method in \citet{wilson2019harmonic}. Note that there is an extra $\log \frac{\pi}{2}$ in the location term of \citet{zaliapin2005approximating} because they expressed the limiting distribution in a different way as 
$$
\mathrm{Landau}\Bigl(0,\frac{\pi}{2}\Bigr)=\frac{\pi}{2}\mathrm{Landau}(0,1)+\log \frac{\pi}{2}. 
$$
Competing parameterizations of stable distributions have caused a lot of confusion in the literature. Please refer to \citet{nolan2020univariate} for a comprehensive review.

Next, we derive the density and CDF formulas for weighted sums of
Half-Cauchy and Pareto variables.

\begin{proof}[Proof of \cref{thm:computeconvhc}]
For the nonnegative weights under consideration, let
$I_+=\{j:w_j>0\}$, $q=|I_+|\ge1$, and
$W=\sum_{j\in I_+}w_j$; under the normalization in the main text,
$W=1$. Zero-weight summands have Laplace transform one and may be
omitted from all products. We derive the Half-Cauchy formulas for $x>0$
and the Pareto formulas for $x>W$. The corresponding CDFs are zero
for $x\le0$ and $x\le W$, respectively.

For $\Re z>0$, write the individual Laplace transforms as
\[
\begin{aligned}
 H(z)
 &=\frac2\pi\int_0^\infty\frac{e^{-zu}}{1+u^2}\dif u
 =-\frac2\pi\{\sin z\,\operatorname{ci}(z)
               +\cos z\,\operatorname{si}(z)\},\\
 P(z)
 &=\int_1^\infty e^{-zu}u^{-2}\dif u
 =e^{-z}-zE_1(z),\qquad
 E_1(z)=\int_1^\infty\frac{e^{-zu}}u\dif u.
\end{aligned}
\]
The first identity is given in \citet{diedhiou1998self}; the second
follows by integration by parts. We use the principal branch of
$\log z$ on $\mathbb C\setminus(-\infty,0]$. With the manuscript's
conventions $\operatorname{ci}=-\operatorname{Ci}$ and
$\operatorname{si}=\operatorname{Si}-\pi/2$, both transforms continue
analytically to this slit plane. Define
\[
 L_{\mathrm{HC}}(z)=\prod_{j\in I_+}H(w_jz),\qquad
 L_{\mathrm{Pareto}}(z)=\prod_{j\in I_+}P(w_jz).
\]
These are the transforms of the weighted-sum densities by
independence.

Each scaled density, extended by zero below its support, is integrable
and has bounded variation. These properties are preserved under
convolution and multiplication by $e^{-cx}$ on the nonnegative
half-line, for $c>0$. Fourier inversion of the exponentially tilted
density therefore gives the symmetric Bromwich inversion
\citep{bellman1966numerical}
\[
 f_{\nu,\boldsymbol w}(x)
 =\lim_{u\to\infty}\frac1{2\pi i}
   \int_{c-iu}^{c+iu}e^{xz}L_\nu(z)\dif z,
 \qquad \nu\in\{\mathrm{HC},\mathrm{Pareto}\},
\]
at the indicated interior continuity points. This argument includes
$q=1$ and does not require absolute convergence of the Bromwich
integral.

Use the contour in \cref{fig:contour2}, with
$R=(c^2+u^2)^{1/2}$ and $0<r<c$. Its oriented boundary consists of
the vertical segment from $c-iu$ to $c+iu$, the upper outer circular
arc to the upper bank of the negative real axis, that bank from
$-R$ to $-r$, the clockwise circle about zero, the lower bank from
$-r$ to $-R$, and the lower outer arc back to $c-iu$.
The bank integrals mean limits of contours inside the slit plane;
these limits exist for fixed $r,R$. Cauchy's integral theorem applies
before taking $r\downarrow0$ and $R\to\infty$.

\begin{figure}[tbp]
  \centering
  \includegraphics*[width=0.5\textwidth]{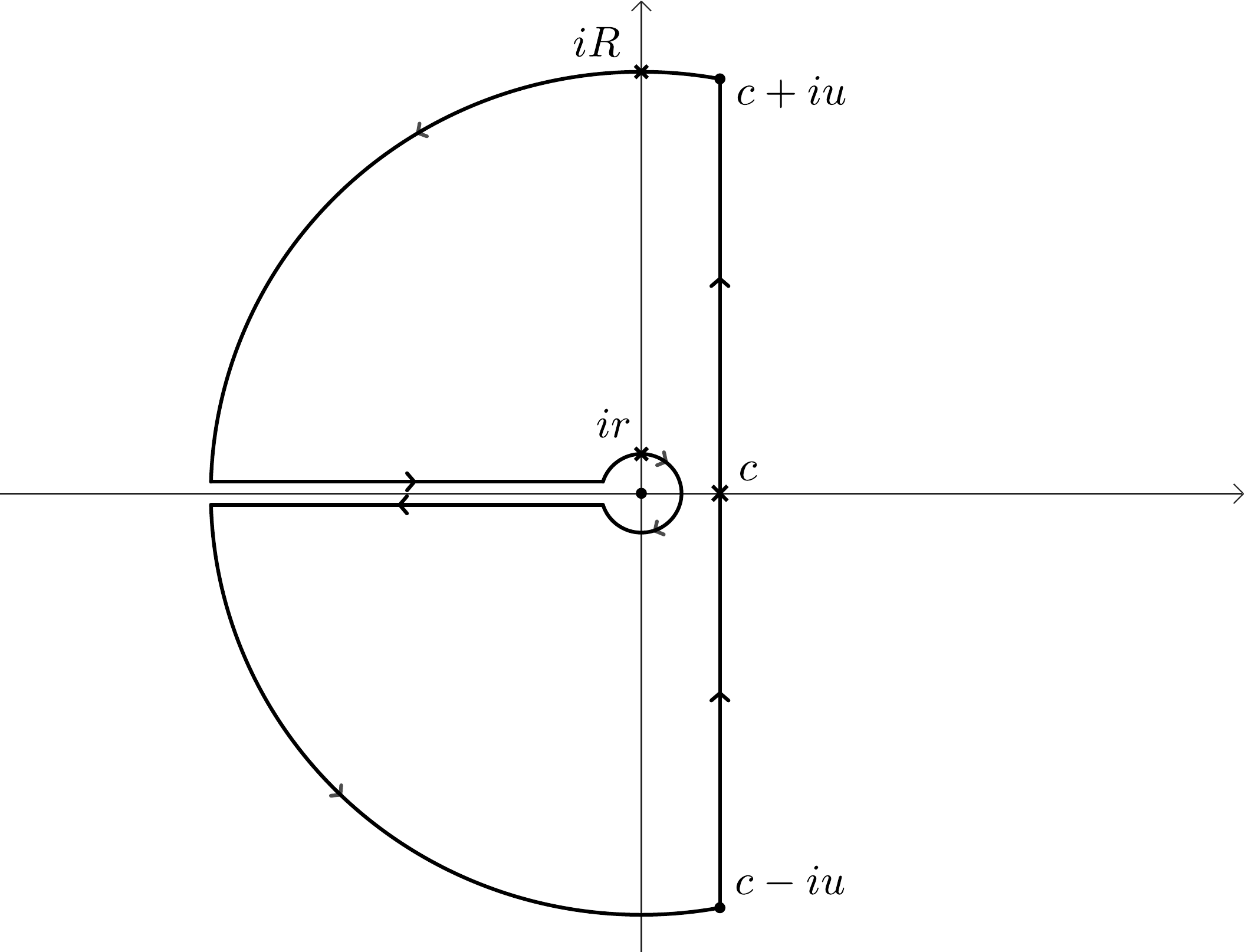}
  \caption{Contour integration.}
  \label{fig:contour2}
\end{figure}

The power series for the sine, cosine, and exponential integrals
\citep{abramowitz1968handbook} give, uniformly up to either bank as
$z\to0$,
\[
\begin{aligned}
 H(z)&=1+\frac2\pi z\{\log z+C_\gamma-1\}+O(|z|^2),\\
 P(z)&=1+z\{\log z+C_\gamma-1\}+O(|z|^2).
\end{aligned}
\]
Thus $L_\nu(z)=1+O(|z|\{1+|\log|z||\})$, and the integral of
$e^{xz}L_\nu(z)$ over the small circle is $O(r)$.

We next bound the outer arcs. Let $h(v)=2/\{\pi(1+v^2)\}$.
For $\Re z\ge0$, $z\ne0$, integration by parts with antiderivative
$-e^{-zv}/z$ gives
\[
 \begin{aligned}
 H(z)&=\int_0^\infty h(v)e^{-zv}\dif v\\
 &=\left[-\frac{h(v)e^{-zv}}z\right]_{v=0}^{v=\infty}
   +\frac1z\int_0^\infty e^{-zv}h'(v)\dif v\\
 &=\frac{h(0)}z+\frac1z\int_0^\infty e^{-zv}h'(v)\dif v.
 \end{aligned}
\]
The boundary term at infinity vanishes because $h(v)\to0$ and
$|e^{-zv}|\le1$. Since
$h'(v)=-4v/\{\pi(1+v^2)^2\}$ and
$\int_0^\infty|h'(v)|\dif v=h(0)=2/\pi$, uniformly on this half-plane,
\[
 |H(z)|\le\frac{h(0)+\int_0^\infty|h'(v)|\dif v}{|z|}
 =\frac4{\pi|z|}.
\]
For $\Re z<0$, $\Im z>0$, the principal logarithms satisfy
$\log(-z)=\log z-i\pi$. Since $\operatorname{Si}$ is odd and
$\operatorname{Ci}(z)-\log z$ is even, this gives
$\operatorname{Si}(-z)=-\operatorname{Si}(z)$ and
$\operatorname{Ci}(-z)=\operatorname{Ci}(z)-i\pi$.
Substitution in the formula for $H$ yields
$H(z)+H(-z)=2(\cos z+i\sin z)=2e^{iz}$.
Conjugation supplies the lower-half-plane identity, so
\[
 H(z)=
 \begin{cases}
  -H(-z)+2e^{iz},&\Re z<0,\ \Im z>0,\\
  -H(-z)+2e^{-iz},&\Re z<0,\ \Im z<0.
 \end{cases}
\]
Consequently, if $w_*=\min_{j\in I_+}w_j>0$, then on the left
semicircle, including its limiting bank values,
\[
 |L_{\mathrm{HC}}(z)|
 \le C_{\boldsymbol w}\{R^{-q}+e^{-w_*|\Im z|}\}.
\]
Indeed, in expanding the product of the individual bounds, the term
with no exponential factor is $O(R^{-q})$, and every other term
contains $e^{-w_j|\Im z|}$ for at least one active index.

For the Pareto transform, the expansion in
\citet[formula~5.1.51]{abramowitz1968handbook} is
\[
 E_1(z)=\frac{e^{-z}}z
        \{1-z^{-1}+O(|z|^{-2})\},
 \qquad |\arg z|\le3\pi/2-\delta,\quad \delta>0.
\]
Taking, for example, $\delta=\pi/4$ covers all arguments in the principal
slit plane and both limiting banks $|\arg z|=\pi$. The remainder is
therefore uniform on the outer arcs used here. Substitution into
$P(z)=e^{-z}-zE_1(z)$, retaining the second term because the first
cancels, yields
\[
 P(z)=\frac{e^{-z}}z\{1+O(|z|^{-1})\},\qquad
 |L_{\mathrm{Pareto}}(z)|
 \le C_{\boldsymbol w}R^{-q}e^{-W\Re z}.
\]

For every $a>0$, the left semicircle satisfies the elementary bound
\[
 \int_{\substack{|z|=R\\\Re z\le0}}e^{a\Re z}|\,\dif z|
 =2R\int_0^{\pi/2}e^{-aR\sin\theta}\dif\theta
 \le\frac{\pi}{a},
\]
using $\sin\theta\ge2\theta/\pi$ on $[0,\pi/2]$.
Also $-\Re z+|\Im z|\ge R$ there. Hence the integral in absolute
value over that semicircle is bounded by
\[
\begin{cases}
 C_{x,\boldsymbol w}R^{-q}
 +C_{\boldsymbol w}R
       e^{-R\min(x,w_*)},&\nu=\mathrm{HC},\quad x>0,\\
 C_{x,\boldsymbol w}R^{-q},
       &\nu=\mathrm{Pareto},\quad x>W,
\end{cases}
\]
where the second line uses $a=x-W$. The two remaining outer arcs
have $0\le\Re z\le c$ and total length
$2R\arcsin(c/R)=O(1)$; the right-half-plane
bound for $H$ and the uniform bound for $P$ make their contributions
$O(R^{-q})$. All outer-arc contributions therefore vanish, including
when there is only one positive weight.

For $t>0$, the same continuation identities and
$E_1(-t\pm i0)=-\operatorname{Ei}(t)\mp i\pi$ give
\[
\begin{aligned}
 H(-t-i0)&=-H(t)+2e^{it},&
 H(-t+i0)&=-H(t)+2e^{-it},\\
 P(-t-i0)&=-\operatorname{Ei}_2(t)+i\pi t,&
 P(-t+i0)&=-\operatorname{Ei}_2(t)-i\pi t.
\end{aligned}
\]
Here $\operatorname{Ei}(t)$ is the real Cauchy principal value and
$\operatorname{Ei}_2(t)=t\operatorname{Ei}(t)-e^t$, as in the main
text. Put
\[
\begin{aligned}
 A_{\mathrm{HC}}(t)
 &=\prod_{j\in I_+}\{-H(w_jt)+2e^{iw_jt}\},\\
 A_{\mathrm{Pareto}}(t)
 &=\prod_{j\in I_+}
        \{-\operatorname{Ei}_2(w_jt)+i\pi w_jt\},\\
 J_\nu(t)
 &=L_\nu(-t-i0)-L_\nu(-t+i0)
   =A_\nu(t)-\overline{A_\nu(t)}=2i\Im A_\nu(t).
\end{aligned}
\]
Expanding these finite products at zero gives
\[
\begin{aligned}
 J_{\mathrm{HC}}(t)&=4iWt
                  +O(t^2\{1+|\log t|\}),\\
 J_{\mathrm{Pareto}}(t)&=2\pi iWt
                  +O(t^2\{1+|\log t|\}).
\end{aligned}
\]
At infinity, $0<H(t)\le1$ bounds $J_{\mathrm{HC}}(t)$, and
$|J_{\mathrm{Pareto}}(t)|\le C_{\boldsymbol w}e^{Wt}t^{-q}$
follows from the Pareto bound. The cut integrals are therefore
absolutely convergent on the stated domains.

The upper bank is traversed toward zero and the lower bank away
from zero. Thus Cauchy's theorem, the vanishing circular contributions,
and Bromwich inversion yield
\begin{equation}\label{eq:laplaceres}
 \begin{aligned}
  f_{\nu,\boldsymbol w}(x)
  &=\frac1{2\pi i}\int_0^\infty e^{-xt}
    \{L_\nu(-t-i0)-L_\nu(-t+i0)\}\dif t\\
  &=\frac1\pi\int_0^\infty e^{-xt}\Im A_\nu(t)\dif t,
 \end{aligned}
\end{equation}
on the stated domains, proving \cref{eq:numint0,eq:numint0hmp}.

To derive the CDFs, the same bounds show that
\[
 \int_x^\infty\int_0^\infty e^{-yt}|J_\nu(t)|\dif t\dif y
 =\int_0^\infty\frac{e^{-xt}}t|J_\nu(t)|\dif t<\infty
\]
on the stated domains. Since
$\int_x^\infty f_{\nu,\boldsymbol w}(y)\dif y=1-F_{\nu,\boldsymbol w}(x)$,
Fubini's theorem and \cref{eq:laplaceres} give
\[
\begin{aligned}
 F_{\mathrm{HC},\boldsymbol w}(x)
 &=1-\frac1\pi\int_0^\infty
       \frac{e^{-xt}}t\Im A_{\mathrm{HC}}(t)\dif t,
       &&x>0,\\
 F_{\mathrm{Pareto},\boldsymbol w}(x)
 &=1-\frac1\pi\int_0^\infty
       \frac{e^{-xt}}t\Im A_{\mathrm{Pareto}}(t)\dif t,
       &&x>W.
\end{aligned}
\]
These are \cref{eq:numint,eq:numinthmp}.

Finally, the density integrands
$e^{-xt}J_\nu(t)/(2\pi i)$ have continuous value zero at $t=0$.
The integrands subtracted from one in the CDF formulas extend
continuously there with the precise values
\[
 \lim_{t\downarrow0}\frac{e^{-xt}J_{\mathrm{HC}}(t)}{2\pi it}
 =\frac{2W}{\pi},\qquad
 \lim_{t\downarrow0}\frac{e^{-xt}J_{\mathrm{Pareto}}(t)}{2\pi it}
 =W.
\]
If zero-weight factors are retained, their continuous values are one,
using $H(0)=1$, $\operatorname{Ei}_2(0)=-1$, and $0\log0=0$.
The contour approach follows \citet{ramsay2006distribution}.
\end{proof}

Next, we prove \cref{thm:extremecase} using \cref{thm:genclt,thm:gcltnew}.

\begin{proof}[Proof of \cref{thm:extremecase}]
Integrating the density-tail limits gives
\begin{equation}\label{eq:goodofall}
 \lim_{k\to\infty}k^\alpha F_\nu(-kt)
 =\frac{c_1}{\alpha t^\alpha},\qquad
 \lim_{k\to\infty}k^\alpha\{1-F_\nu(kt)\}
 =\frac{c_2}{\alpha t^\alpha}
 \quad\text{for every }t>0.
\end{equation}
Put $u_\alpha=(\sum_{i=1}^m w_i^\alpha)^{1/\alpha}>0$.
Let $X_{ij}$, $1\le i\le m$, $j\ge1$, be i.i.d. with CDF $F_\nu$,
and define the i.i.d. variables
\[
 Y_j=\frac{\sum_{i=1}^m w_iX_{ij}}{u_\alpha}
\]
with common CDF $G$.
For each $k$, give $X_{ij}$, $1\le j\le k$, the repeated-row weight
$w_i/k$. These weights sum to one, and their norm and normalized maximum are
\[
 s_k=\left\{k\sum_{i=1}^m(w_i/k)^\alpha\right\}^{1/\alpha}
 =k^{1/\alpha-1}u_\alpha,\qquad
 \max_{i,j}\frac{w_i/k}{s_k}
 =\frac{\max_iw_i}{k^{1/\alpha}u_\alpha}\longrightarrow0.
\]
Thus \cref{thm:gcltnew}, first regrouping the repeated-row sum and
then using equal weights $1/k$, gives
\[
 \begin{aligned}
 \frac{1}{k^{1/\alpha}}\sum_{j=1}^kY_j-a_k
 &\mathrel{\xrightarrow{\mathrm d}}S(\alpha,\beta,c,0),\\
 \frac{1}{k^{1/\alpha}}\sum_{j=1}^kX_{1j}-b_k
 &\mathrel{\xrightarrow{\mathrm d}}S(\alpha,\beta,c,0),
 \end{aligned}
\]
for suitable real normalized centers $a_k,b_k$ and the same
parameters $\beta,c$ in \cref{eq:determinec}.

Apply the normalizer conclusion in \cref{eq:genclt-normalizer-tails}
to these two i.i.d. sequences, with $B_k=k^{1/\alpha}$.
For $Q=F_\nu,G$, it gives
\[
 k\{Q(-k^{1/\alpha}t)+1-Q(k^{1/\alpha}t)\}
 \longrightarrow D_{\alpha,c}t^{-\alpha}
 =\frac{c_1+c_2}{\alpha t^\alpha}
 \quad\text{for every }t>0.
\]
The equality of constants follows from \cref{eq:determinec}.
For arbitrary $x\to\infty$, take $k=\lfloor x^\alpha\rfloor$.
Then $k^{1/\alpha}\le x<(k+1)^{1/\alpha}$ and $k/x^\alpha\to1$;
monotonicity of each tail therefore extends the preceding limits to
\[
 x^\alpha\{Q(-x)+1-Q(x)\}
 \longrightarrow\frac{c_1+c_2}{\alpha},
 \qquad Q=F_\nu,G.
\]
The tail-balance conclusion of \cref{thm:genclt} also gives
\[
 \lim_{x\to\infty}\frac{G(-x)}{1-G(x)}
 =\frac{1-\beta}{1+\beta}=\frac{c_1}{c_2},
\]
where $c_2>0$ ensures that $1+\beta>0$.
Consequently,
\[
 x^\alpha\{1-G(x)\}\longrightarrow\frac{c_2}{\alpha},\qquad
 \frac{1-F_\nu(x)}{1-G(x)}\longrightarrow1,
\]
using \cref{eq:goodofall} for the second limit.
In particular, the right tail of $G$ is regularly varying with index
$-\alpha$. Since $G(x)=F_{\nu,\boldsymbol w}(u_\alpha x)$,
\[
 \begin{aligned}
 \lim_{x\to\infty}
 \frac{1-F_\nu(x)}{1-F_{\nu,\boldsymbol w}(x)}
 &=\lim_{x\to\infty}
 \frac{1-F_\nu(x)}{1-G(x)}
 \frac{1-G(x)}{1-G(x/u_\alpha)}\\
 &=u_\alpha^{-\alpha}
 =\frac{1}{\sum_{i=1}^m w_i^\alpha}.
 \end{aligned}
\]
Under perfect dependence, all scores coincide, so their weighted sum
has CDF $F_\nu$, proving \cref{eq:extre2}.
For $m\ge2$, this limit is one for every admissible weight vector
exactly when $\alpha=1$: sufficiency uses $\sum_iw_i=1$, and necessity
uses $w_i=1/m$.
\end{proof}

Finally, we prove the relevant result from \cref{sec:furtherdis}.

\begin{proof}[Proof of \cref{thm:fixed-hcct}]
The case $m=1$ follows directly from \cref{lem:hc-tail}, so assume
$m\geq2$.  We follow the three-step notation of the current proof.

\medskip
\noindent\textit{Step I: decomposition.}
Define
\begin{align*}
  A_{i,t}
  &=\left\{
      H_i>\frac{(1+\delta_t)t}{w_i},
      \ \HCCT>t
    \right\},\\
  B_{i,t}
  &=\left\{
      H_i\leq\frac{(1+\delta_t)t}{w_i},
      \ \HCCT>t
    \right\},
\end{align*}
and
\[
  A_t=\bigcup_{i=1}^m A_{i,t},
  \qquad
  B_t=\bigcap_{i=1}^m B_{i,t}.
\]
Because every $H_i$ is nonnegative, the condition $\HCCT>t$ in
$A_{i,t}$ is redundant.  The events $A_t$ and $B_t$ are disjoint and
\begin{equation}\label{eq:AB-decomposition}
  \{\HCCT>t\}=A_t\mathbin{\dot\cup}B_t.
\end{equation}

\medskip
\noindent\textit{Step II: the no-big-jump event.}
On $\{\HCCT>t\}$, at least one index $i$ satisfies
\[
  H_i>\frac{t}{mw_i}.
\]
Therefore
\begin{equation}\label{eq:B-I1-I2}
  \mathbb P(B_t)\leq I_1+I_2,
\end{equation}
where
\begin{align}
  I_1
  &=\sum_{i=1}^m
    \mathbb P\!\left\{
      \frac{t}{mw_i}<H_i\leq\frac{(1-\delta_t)t}{w_i},
      \ \HCCT>t
    \right\},\label{eq:I1-fixed}\\
  I_2
  &=\sum_{i=1}^m
    \mathbb P\!\left\{
      \frac{(1-\delta_t)t}{w_i}<H_i
      \leq\frac{(1+\delta_t)t}{w_i}
    \right\}.\label{eq:I2-fixed}
\end{align}

For an event in the $i$th summand of $I_1$,
\[
  \sum_{j\neq i}w_jH_j>\delta_t t.
\]
Hence some $j\neq i$ satisfies
\[
  H_j>\frac{\delta_t t}{(m-1)w_j}.
\]
Using $\arctan x\leq x$ and $m-1\leq m$, we obtain
\begin{align}
  I_1
  &\leq
  \sum_{1\leq i\neq j\leq m}
  \mathbb P\!\left\{
    \frac{t}{mw_i}<H_i\leq\frac{(1-\delta_t)t}{w_i},
    \ H_j>\frac{\delta_t t}{(m-1)w_j}
  \right\}\notag\\
  &\leq
  m(m-1)
  \max_{1\leq i\neq j\leq m}
  \mathbb P\!\left(
    0<p_i<\frac{2w_i m}{\pi t},
    0<p_j<\frac{2w_j m}{\pi\delta_t t}
  \right)\notag\\
  &=o(t^{-1}).\label{eq:I1-fixed-bound}
\end{align}
The maximum must be over ordered pairs because the two thresholds have
different roles.

For $I_2$, exact marginal probabilities give
\begin{align}
  I_2
  &=\frac{2}{\pi}\sum_{i=1}^m
  \left[
    \arctan\!\left(\frac{w_i}{(1-\delta_t)t}\right)
    -
    \arctan\!\left(\frac{w_i}{(1+\delta_t)t}\right)
  \right]\notag\\
  &\leq
  \frac{4\delta_t}{\pi(1-\delta_t^2)t}
  \sum_{i=1}^m w_i
  =\frac{4\delta_t}{\pi(1-\delta_t^2)t}
  =o(t^{-1}).\label{eq:I2-fixed-bound}
\end{align}
Thus $\mathbb P(B_t)=o(t^{-1})$.

\medskip
\noindent\textit{Step III: the big-jump event.}
Bonferroni's inequalities give
\begin{equation}\label{eq:bonf-fixed}
  \sum_{i=1}^m\mathbb P(A_{i,t})
  -\sum_{1\leq i<j\leq m}
    \mathbb P(A_{i,t}\cap A_{j,t})
  \leq \mathbb P(A_t)
  \leq\sum_{i=1}^m\mathbb P(A_{i,t}).
\end{equation}
The marginal sum is
\begin{align}
  \sum_{i=1}^m\mathbb P(A_{i,t})
  &=\frac{2}{\pi}\sum_{i=1}^m
    \arctan\!\left(\frac{w_i}{(1+\delta_t)t}\right)\notag\\
  &=\frac{2}{\pi(1+\delta_t)t}+O(t^{-3})
  =\frac{2}{\pi t}+o(t^{-1}),\label{eq:marginal-sum-fixed}
\end{align}
where the $O(t^{-3})$ term is uniform because
$\sum_iw_i^3\leq\sum_iw_i=1$.

Moreover, for $i<j$,
\begin{align*}
  A_{i,t}\cap A_{j,t}
  \subseteq
  \left\{
    0<p_i<\frac{2w_i m}{\pi t},
    0<p_j<\frac{2w_j m}{\pi\delta_t t}
  \right\},
\end{align*}
because $\delta_t<1$ eventually.  Hence
\begin{equation}\label{eq:intersection-fixed}
  \sum_{1\leq i<j\leq m}
  \mathbb P(A_{i,t}\cap A_{j,t})
  =o(t^{-1}).
\end{equation}
\cref{eq:bonf-fixed,eq:marginal-sum-fixed,eq:intersection-fixed} imply
\[
  \mathbb P(A_t)=\frac{2}{\pi t}+o(t^{-1}).
\]
Together with \cref{eq:AB-decomposition} and Step II, this proves
\cref{eq:fixed-main-tail}.

Finally, let $p_1',\ldots,p_m'$ be independent exact-uniform $p$-values.
For fixed $m$,
\begin{align*}
  &\mathbb P\!\left(
    0<p_i'<\frac{2w_i m}{\pi t},
    0<p_j'<\frac{2w_j m}{\pi\delta_t t}
  \right)\\
  &\hspace{2cm}=
  \frac{4w_iw_jm^2}{\pi^2\delta_t t^2}
  =o(t^{-1}),
\end{align*}
because $t\delta_t\to\infty$.  Applying the preceding proof to the
independent vector gives \cref{eq:fixed-ind-tail}.  \cref{lem:hc-tail}
then yields \cref{eq:fixed-ratios}.
\end{proof}

\begin{proof}[Proof of \cref{thm:growing-hcct}.]
Use the definitions $A_{i,t}$, $B_{i,t}$, $A_t$, $B_t$, $I_1$, and $I_2$
from the proof of \cref{thm:fixed-hcct}.  Every bound used there is
uniform in $m$ and the normalized weights, except for the pair sums, whose
factors are now retained explicitly.  Specifically,
\begin{align}
  I_1
  &\leq m(m-1)R_t,\label{eq:growing-I1}\\
  I_2
  &\leq\frac{4\delta_t}{\pi(1-\delta_t^2)t},\label{eq:growing-I2}\\
  \sum_{1\leq i<j\leq m}
  \mathbb P(A_{i,t}\cap A_{j,t})
  &\leq\frac{m(m-1)}{2}R_t,\label{eq:growing-intersections}
\end{align}
where
\[
  R_t=
  \max_{1\leq i\neq j\leq m}
  \mathbb P\!\left(
    0<p_i<\frac{2w_i m}{\pi t},
    0<p_j<\frac{2w_j m}{\pi\delta_t t}
  \right).
\]
The marginal sum remains uniformly controlled:
\begin{equation}\label{eq:growing-marginal-sum}
  \sum_{i=1}^m\mathbb P(A_{i,t})
  =\frac{2}{\pi(1+\delta_t)t}+O(t^{-3}),
\end{equation}
because $\sum_iw_i^3\leq1$.  Thus the explicit bound
\cref{eq:master-error} remains valid with $m=m(t)$.  Conditions
\cref{eq:growing-delta,eq:growing-master-condition} prove
\cref{eq:growing-main-tail}; \cref{lem:hc-tail} gives
\cref{eq:growing-half-cauchy-ratio}.

It remains to establish the exact independence-reference tail under
\cref{eq:growing-weight-and-m}.  Let $M<\infty$ satisfy
$mw_i\leq M$ eventually and set
\begin{equation}\label{eq:eta-choice}
  \eta_t=\frac{m}{\sqrt t}.
\end{equation}
Then $\eta_t\to0$,
$t\eta_t=m\sqrt t\to\infty$, and, for independent exact-uniform
$p_i'$,
\begin{align}
  &\max_{1\leq i\neq j\leq m}
  \mathbb P\!\left(
    0<p_i'<\frac{2w_i m}{\pi t},
    0<p_j'<\frac{2w_j m}{\pi\eta_t t}
  \right)\notag\\
  &\hspace{2cm}\leq
  \frac{4M^2}{\pi^2\eta_t t^2}.
  \label{eq:ind-pair-growing}
\end{align}
Consequently,
\begin{equation}\label{eq:ind-pair-remainder}
  m^2\frac{4M^2}{\pi^2\eta_t t^2}
  =O\!\left(\frac{m}{t^{3/2}}\right)
  =o(t^{-1}).
\end{equation}
Applying the already proved growing-$m$ argument to the independent vector,
using $\eta_t$ in place of $\delta_t$, proves
\cref{eq:growing-ind-tail} and then \cref{eq:growing-both-ratios}.
\end{proof}

\begin{proof}[Proof of \cref{lem:gaussian-quadrant}.]
It suffices first to consider the positive-positive quadrant with
$0\leq\rho\leq r$.  In the bivariate normal density, put
$x=d+s$ and $y=d+u$.  The exponent satisfies
\begin{align*}
  \frac{x^2-2\rho xy+y^2}{2(1-\rho^2)}
  ={}&\frac{d^2}{1+\rho}
    +\frac{d(s+u)}{1+\rho}
    +\frac{s^2-2\rho su+u^2}{2(1-\rho^2)}.
\end{align*}
The final quadratic form is nonnegative.  Dropping it and integrating the
remaining exponentials gives
\begin{align*}
  \mathbb P(Z_1>d,Z_2>d)
  &\leq
  \frac{(1+\rho)^2}{2\pi\sqrt{1-\rho^2}}
  d^{-2}\exp\!\left(-\frac{d^2}{1+\rho}\right)\\
  &\leq C_r d^{-2}\exp\!\left(-\frac{d^2}{1+r}\right).
\end{align*}
Changing one or both signs replaces $\rho$ by $\rho$ or $-\rho$.  A
bivariate Gaussian upper-quadrant probability is nondecreasing in its
correlation, so the same bound applies uniformly whenever $|\rho|\leq r$.

Set
\[
  q_t=\frac{K}{\delta_t t},
  \qquad
  d_t=\Phi^{-1}(1-q_t).
\]
Both unequal normal thresholds are at least $d_t$.  Therefore each relevant
quadrant is bounded by \cref{eq:quadrant-bound}.  Mills' inequalities give
\[
  d_t^2\asymp\log(1/q_t)
\]
and
\[
  \exp(-d_t^2/2)\leq Cq_td_t.
\]
Writing $a=2/(1+r)$,
\begin{align*}
  d_t^{-2}\exp\!\left(-\frac{d_t^2}{1+r}\right)
  &\leq Cq_t^a d_t^{a-2}\\
  &\leq
  C(\delta_t t)^{-2/(1+r)}
  \{\log(\delta_t t)\}^{-r/(1+r)}.
\end{align*}
For two-sided $p$-values, sum this bound over the four sign quadrants.
This changes only the constant.
\end{proof}

\begin{proof}[Proof of \cref{thm:fixed-hm}.]
For $m=1$, \cref{eq:pareto-tail} gives the claim exactly.  Suppose
$m\geq2$.  Use the proof of \cref{thm:fixed-hcct} with
$Y_i=1/p_i$ and $\HM=\sum_iw_iY_i$.  On the no-big-jump event, a selected
index satisfies
\[
  \frac{t}{mw_i}<Y_i\leq\frac{(1-\delta_t)t}{w_i},
\]
and some ordered partner satisfies
\[
  Y_j>\frac{\delta_t t}{(m-1)w_j}.
\]
Therefore its ordered-pair part is at most $m(m-1)R_t^{\mathrm{HM}}$.
The shell part is exactly
\begin{align}
  &\sum_{i=1}^m
  \mathbb P\!\left\{
    \frac{(1-\delta_t)t}{w_i}<Y_i
    \leq\frac{(1+\delta_t)t}{w_i}
  \right\}\notag\\
  &\hspace{3cm}=
  \frac{2\delta_t}{(1-\delta_t^2)t}.
  \label{eq:hm-shell}
\end{align}

For the big-jump events
\[
  A_{i,t}^{\mathrm{HM}}
  =\left\{Y_i>\frac{(1+\delta_t)t}{w_i}\right\},
\]
the marginal sum is exactly
\begin{equation}\label{eq:hm-marginal-sum}
  \sum_{i=1}^m\mathbb P(A_{i,t}^{\mathrm{HM}})
  =\frac{1}{(1+\delta_t)t},
\end{equation}
for all sufficiently large $t$, and the unordered intersection sum is at
most $m(m-1)R_t^{\mathrm{HM}}/2$.  The same decomposition and Bonferroni
bounds therefore yield
\begin{equation}\label{eq:hm-master-error}
  \left|\mathbb P(\HM>t)-\frac1t\right|
  \leq
  \frac{3m(m-1)}2R_t^{\mathrm{HM}}
  +\frac{2\delta_t}{(1-\delta_t^2)t}
  +\frac{\delta_t}{(1+\delta_t)t}.
\end{equation}
This proves \cref{eq:fixed-hm-tail}.

For independent exact-uniform $p_i'$, the natural pair probability equals
\[
  \frac{w_iw_jm^2}{\delta_t t^2}=o(t^{-1}),
\]
because $t\delta_t\to\infty$.  Applying the preceding argument to the
independent vector proves \cref{eq:fixed-hm-ind-tail}, and the ratios
follow.
\end{proof}

\begin{proof}[Proof of \cref{thm:growing-hm}.]
The bound \cref{eq:hm-master-error} is uniform in $m$ and the normalized
weights once the explicit pair factor is retained.  Hence
\cref{eq:growing-hm-master} and $\delta_t\to0$ prove
\cref{eq:growing-hm-tail}.

For the independent reference, choose $M<\infty$ with $mw_i\leq M$ and
put $\eta_t=m/\sqrt t$.  The independent natural pair probability is at
most $M^2/(\eta_t t^2)$, so its accumulated contribution is
\[
  m^2\frac{M^2}{\eta_t t^2}
  =O\!\left(\frac{m}{t^{3/2}}\right)
  =o(t^{-1}).
\]
Applying the first part with $\eta_t$ proves \cref{eq:growing-hm-ind}.
\end{proof}

\section{Related Literature on Global Testing}\label{sec:global}

\vspace*{5pt}
\setlength{\parskip}{5pt}

Global testing is a statistical strategy that evaluates the overall effect across multiple studies or experiments, rather than focusing on individual outcomes. This problem is widely encountered in fields such as genetics \citep{zeggini2009meta,wu2010powerful,wang2015integrative,yoon2021powerful}, environmental science \citep{halpern2008global,smith2009public,ouyang2016improvements}, and social sciences \citep{ferreira2008global,hastings2018snap}, where researchers seek consistent patterns or associations across diverse conditions or populations. Traditionally, statisticians have combined $p$-values from individual tests to decide whether to reject a global null hypothesis. However, while $p$-value aggregation is well-studied, previous work has not addressed constructing confidence intervals or regions for combined estimates. In this paper, we propose a method for obtaining confidence sets by inverting combination tests, introducing new global testing methods that yield guaranteed convex confidence regions in common scenarios.

The essence of global testing is to synthesize information from multiple sources to make a unified inference about a global hypothesis, which posits a general effect or relationship across all studies or variables. Dependence between individual tests is often significant. For example, in genome-wide association studies (GWAS), single nucleotide polymorphisms (SNPs) are often highly correlated due to linkage disequilibrium \citep{zeggini2009meta}. Such correlations can inflate Type I error for widely used methods like Fisher’s combination test \citep{fisher1925statistical} and Stouffer Z-score test \citep{stouffer1949american}, making it crucial to use combination tests that remain valid under general dependence.

In contrast, the Bonferroni correction \citep{dunn1961multiple} is provably valid regardless of dependency structure. Designed to control the family-wise error rate (FWER), it rejects the global null only if at least one test’s $p$-value falls below \(1/m\) of the significance level. This conservative approach inspired Simes' test \citep{simes1986improved}, which forms the basis of the Benjamini–Hochberg method \citep{benjamini1995controlling} for false discovery rate (FDR) control. However, these methods are often criticized for low power \citep{o1984procedures,moran2003arguments,dmitrienko2009multiple}, especially in settings with strong positive correlation among tests.

Additionally there have been methods that address dependence by assuming specific covariance models. Brown’s method \citep{brown1975400} combines dependent $p$-values under the assumption that test statistics follow a multivariate normal distribution with a known covariance matrix. Kost’s method \citep{kost2002combining} extends this by allowing covariance matrices known up to a scalar factor. Similarly, the higher criticism test, originally developed for detecting sparse alternatives \citep{donoho2004higher}, was later generalized by \citet{barnett2017generalized} to account for known covariance structures. These methods rely on explicitly modeling dependencies across studies, whereas CCT, HMP, and our proposed methods remain robust even when dependencies are unknown.

We also emphasize that global testing methods differ from multiple testing procedures, which assess each effect independently and focus on controlling FWER or FDR (false discovery rate) due to the large number of tests. Notably, any well-calibrated combination test can be adapted into a multilevel test to control the strong-sense FWER \citep{marcus1976closed,wilson2019harmonic,wilson2020generalized,wilson2021evy}. Additionally, extensive research exists on FDR control for dependent studies, such as the Benjamini--Hochberg procedure \citep{benjamini1995controlling}, which was extended by \citet{benjamini2001control} to accommodate dependent $p$-values. %

\clearpage
\noindent\textbf{Data availability statement.}
The clinical-trial data supporting the network meta-analysis in this study are
openly available in the R package \textbf{netmeta} on the Comprehensive R
Archive Network (CRAN) at
\url{https://CRAN.R-project.org/package=netmeta}. The data are also reproduced
in the supplemental material.

\medskip
\noindent\textbf{Declaration of generative AI use.}
OpenAI GPT-5.6 Sol was used during manuscript preparation to audit the
manuscript and to propose alternative proofs. The authors take full
responsibility for the originality, accuracy, and integrity of the manuscript.\par
{\small
\setlength{\bibsep}{2pt}
\bibliographystyle{apalike}
\bibliography{biblio}
}
\end{document}